\documentclass[final,5p,times,twocolumn,authoryear]{elsarticle}

\usepackage{amssymb}
\usepackage{amsmath}
\usepackage{geometry}
\usepackage{pdflscape}

\graphicspath{{figures/}}
\usepackage{xcolor}
\usepackage{hyperref}
\hypersetup{
    colorlinks=true,
    linkcolor=black,
    filecolor=black,
    urlcolor=navy,
    citecolor=black,
}
\newcommand{\acronym}[1]{{#1}}

\newcommand{\gaia}{\textsl{Gaia}}
\newcommand{\hipparcos}{\textsl{HIPPARCOS}}
\newcommand{\dr}[1]{\acronym{DR}#1}

\newcommand{\given}{\,|\,}

\usepackage{savesym}
\savesymbol{tablenum}
\usepackage{siunitx}
\ifdefined\unit\else
  \ifdefined\NewCommandCopy
    \NewCommandCopy\unit\si
  \else
    \NewDocumentCommand\unit{O{}m}{\si[#1]{#2}}
  \fi
\fi
\restoresymbol{SIX}{tablenum}
\DeclareSIUnit\year{yr}
\DeclareSIUnit\parsec{pc}
\DeclareSIUnit\mag{mag}
\DeclareSIUnit\Msun{M_\odot}
\DeclareSIUnit\msun{M_\odot}
\DeclareSIUnit\Rsun{R_\odot}
\DeclareSIUnit\mas{mas}
\newcommand{\mas}{\unit{\milli\arcsecond}}

\newcommand{\kms}{\unit{\km\per\s}}
\newcommand{\masyr}{\unit{\mas\per\year}}
\newcommand{\kpc}{\unit{\kilo\parsec}}

\newcommand{\abun}[2]{\ensuremath{{[\mathrm{#1}/\mathrm{#2}]}}}
\newcommand{\feh}{\abun{Fe}{H}}

\newcommand{\ion}[2]{#1\,\textsc{\romannumeral #2}}
\newcommand{\lcdm}{\ensuremath{\Lambda}CDM}

\newcommand{\rper}{\ensuremath{r_{\textrm{per}}}}

\journal{New Astronomy Reviews}

\begin{document}

\begin{frontmatter}

\title{Stellar Streams in the Gaia Era}

\author[ociw]{Ana~Bonaca}
\author[cca]{Adrian~M.~Price-Whelan}

\affiliation[ociw]{organization={The Observatories of the Carnegie Institution for Science},
            addressline={813 Santa Barbara Street},
            city={Pasadena},
            postcode={91101},
            state={CA},
            country={USA}}

\affiliation[cca]{
    organization={Center for Computational Astrophysics, Flatiron Institute},
    addressline={162 Fifth Ave.},
    city={New York},
    postcode={10010},
    state={NY},
    country={USA}
}

\begin{abstract}
The hierarchical model of galaxy formation predicts that the Milky Way halo is populated by tidal debris of dwarf galaxies and globular clusters.
Due to long dynamical times, debris from the lowest mass objects remains coherent as thin and dynamically cold stellar streams for billions of years.
The \gaia\ mission, providing astrometry and spectrophotometry for billions of stars, has brought three fundamental changes to our view of stellar streams in the Milky Way.
First, more than a hundred stellar streams have been discovered and characterized using \gaia\ data.
This is an order of magnitude increase in the number of known streams, thanks to \gaia's
capacity for identifying comoving groups of stars among the field Milky Way population.
Second, \gaia\ data have revealed that density variations both along and across stellar streams are common.
Dark-matter subhalos, as well as baryonic structures were theoretically predicted to form such features, but observational evidence for density variations was uncertain before \gaia.
Third, stream kinematics are now widely available and have constrained the streams' orbits and origins.
\gaia\ has not only provided proper motions directly, but also enabled efficient spectroscopic follow-up of the proper-motion selected targets.
These discoveries have established stellar streams as a dense web of sensitive gravitational tracers in the Milky Way halo.
We expect the coming decade to bring a full mapping of the Galactic population of stellar streams, as well as develop numerical models that accurately capture their evolution within the Milky Way for a variety of cosmological models.
Perhaps most excitingly, the comparison between the two will be able to reveal the presence of dark-matter subhalos below the threshold for galaxy formation ($\lesssim10^6\,\unit{\msun}$), and provide the most stringent test of the cold dark matter paradigm on small scales.
\end{abstract}

% \begin{keyword}



% \end{keyword}

\end{frontmatter}

%%%%%%%%%%%%%%%%%%%%%%
\section{Introduction}
\label{sec:intro}

The observed distribution of galaxies on large scales gave rise to the current cosmological paradigm: a Universe dominated by dark matter in which galaxies form through hierarchical accretion of lower-mass galaxies \citep{press:1974,white:1978}.
However, it was much closer to home, in our own Milky Way galaxy that tracers of these ``proto-galactic fragments'' were first identified.
In a landmark study of halo globular clusters, \citet{searle:1978} boldly proposed they originate from lower-mass galaxies that merged with the Milky Way.
At the time, star clusters dissolving in the Milky Way disk have already been identified as comoving groups in catalogs of stars with measured 3D velocities \citep[e.g.,][and references therein]{eggen:1965}.
Similarly, tidal debris of dwarf galaxies and globular clusters in the Galactic halo form stellar streams \citep[e.g.,][]{lynden-bell:1995}.
As such, streams serve as direct evidence of hierarchical structure formation, fossil records of star and galaxy formation in the early universe, and luminous tracers in the most dark-matter-dominated region of the Galaxy.
Obtaining kinematic data needed to fully deliver on these unique insights from halo streams has been challenging historically \citep[e.g.,][]{morrison:2000}.
The \gaia\ mission has recently measured positions and motions of almost two billion stars in the Milky Way \citep{gaiamission:2016, gaiadr1, gaiadr2, gaiaedr3, gaiadr3} to fundamentally change our view of stellar streams.

We center this review on three major breakthroughs that \gaia\ brought to stellar streams.
The first is the sheer number of stream discoveries, which \gaia\ increased by an order of magnitude from $\approx10$ to $\approx100$ (\S\ref{sec:discovery}).
The second are the stream kinematics, which constrain their precise orbits, as well as
origins (\S\ref{sec:orbits}).
And finally, the detailed structure of stellar streams, resolved by \gaia\ for the first
time to suggest dynamical histories rich in perturbations (\S\ref{sec:structure}).
In these sections we also discuss the key opportunities and new research directions that these breakthroughs have made possible.
In the concluding section, we reflect more broadly on what these developments entail for the future of Galactic archaeology (\S\ref{sec:outlook}).
As for the remainder of this section, the last review dedicated to stellar streams was almost a decade ago and pre-\gaia\ \citep{newberg:2016}, while a more recent, \gaia-based review focused on the Milky Way's halo more broadly \citep{helmi:2020}, so we now proceed to outline the theoretical and historical context for \gaia's stellar stream revolution.

\begin{figure*}[t!]
\begin{center}
\includegraphics[width=1\textwidth]{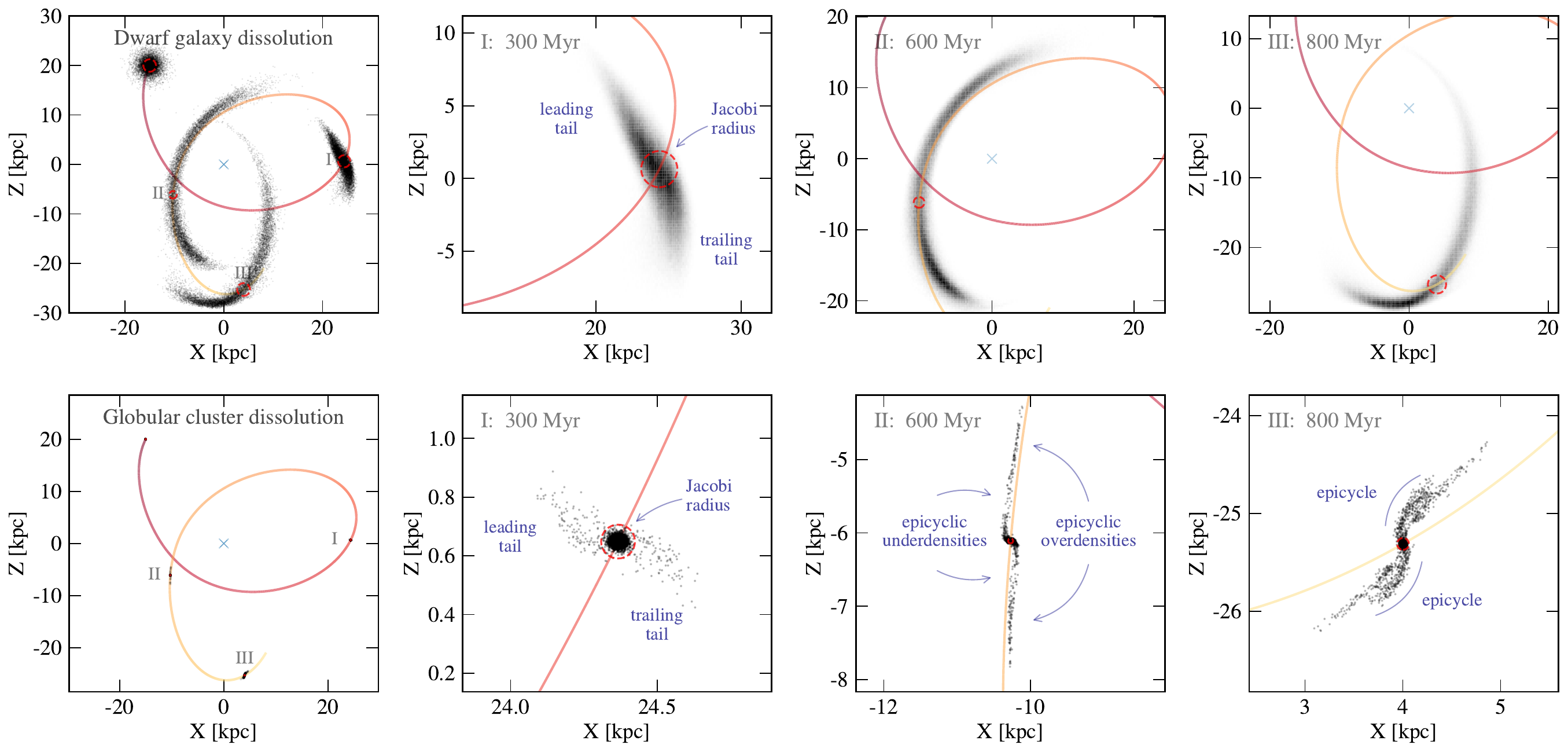}
\end{center}
\caption{%
Numerical models of a $10^8\,\unit{\msun}$ dwarf galaxy (top) and a $10^4\,\unit{\msun}$ globular cluster (bottom) tidally dissolving on the same orbit in a Milky Way-like gravitational potential to form stellar streams.
\textbf{Left column:} Overview of the first billion years of dissolution, with the progenitors initialized at $(X,Z)=(-15,20)\,\unit{kpc}$, their orbital path shown with a colored line, and their tidal debris at three subsequent snapshots shown in black here, and zoomed-in on in the remaining columns.
The Galactic center is marked with a light-blue cross.
\textbf{Column I} (300\,\unit{Myr}, past first pericenter): During the initial stage of tidal disruption, satellites develop short leading and trailing tails, while most of the stars are still bound inside the Jacobi radius (red dashed line).
\textbf{Column II} (600\,\unit{Myr}, approaching second pericenter): Tidal tails are significantly more elongated at pericenter than at apocenter, and also better aligned with the progenitor's orbit.
Dwarf galaxy tails are significantly longer than globular cluster's at any given time.
\textbf{Column III} (900\,\unit{Myr}, close to apocenter): Longer tails show significant misalignment from the progenitor's orbit at their second apocenter, and in the case of the globular cluster, have developed a full pair of epicycles.
}
\label{fig:stream_formation}
\end{figure*}

Physics driving the formation of stellar streams rests with the tidal forces of a host galaxy acting on its satellites.
When sufficiently strong, these tides pull stars out of the satellite through two stationary points \citep{bt:2008}, thus forming the leading tidal tail (stars with shorter orbital periods speeding ahead of the progenitor) and the trailing tidal tail (stars lagging behind the progenitor due to their longer orbital periods).
We illustrate the process of tidal dissolution in Figure~\ref{fig:stream_formation}, with a dwarf galaxy satellite at the top, and a globular cluster on a matching orbit at the bottom.
The dwarf galaxy of an initial mass of $10^8\,\unit{\msun}$ was represented with 50,000 particles and evolved with the Gyrfalcon tree code \citep{dehnen:2014}.
The globular cluster of an initial mass of $10^4\,\unit{\msun}$, was represented with 500,000 particles and modeled with direct N-body code PeTar \citep{wang:2020}.
The left-most panels show the satellites' orbital paths, initial positions, and tidal debris in three subsequent snapshots (labeled with roman numerals).
The remaining columns zoom into each of these snapshots.
The first snapshot, 300\,\unit{Myr} since the beginning of dissolution shows short tails of gravitationally unbound stars emanating from each progenitor.
The boundary between the bound and unbound stars is (conceptually) set by the tidal radius.
Typically, the satellite mass $m$ is much smaller than the enclosed mass of the host galaxy at its location $M(<R_0)$, so the tidal radius is approximately the Jacobi radius from the restricted three-body problem \citep[][]{szebehely:1967, valtonen:2006}:
\begin{equation}
r_J = \left(\frac{m}{3M(<R_0)}\right)^{1/3} R_0
\end{equation}
The Jacobi radius, shown with a red dashed line in Figure~\ref{fig:stream_formation}, matches well the tidal boundary of both the globular cluster and the dwarf galaxy derived through self-consistent numerical simulations.
This agreement has been used to develop fast methods for generating models of stellar streams by placing tracer particles at the progenitor's Jacobi radius, and then solving the restricted three-body problem \citep[the so-called \emph{particle-spray methods};][]{varghese:2011, lane:2012, kupper:2012, bonaca:2014, gibbons:2014, fardal:2015}.

Governed by the same underlying gravitational potential, tidal tails of a globular cluster and a dwarf galaxy on the same orbit have similar shape and orientation.
Notably, both streams delineate the progenitor's orbit at pericenter, but are misaligned at apocenter (panels II and III in Figure~\ref{fig:stream_formation}, respectively), which means that representing a stream as an orbit is not always appropriate, no matter how tempting the computational speed-ups are \citep{eyre:2009, eyre:2011, sanders:2013a, sanders:2013b}.
On the other hand, it is clear from Figure~\ref{fig:stream_formation} that a globular cluster and a dwarf galaxy produce markedly different streams: dwarf galaxy streams are longer, wider, and smoother than globular cluster streams.
The cause of this discrepancy is not just the difference in their initial masses, but also in their number density of stars.
In more diffuse dwarf galaxies, a higher fraction of stars find themselves outside of the tidal radius when it shrinks during a pericentric passage (Figure~\ref{fig:stream_formation}, panel II, top).
Combined with a higher velocity dispersion due to a higher initial mass \citep[cf.][]{simon:2007, baumgardt:2019}, the resulting tails are more massive, wider and grow faster.
In contrast, more compact globular clusters maintain a tidal radius larger than its half-mass radius even at pericenter (panel II, bottom).
As a result, their mass loss is dominated by stars escaping through close two-body encounters, i.e., the process of two-body relaxation or evaporation \citep{gnedin:1997,vesperini:1997,baumgardt:2003}.
Mass loss being driven by an internal process means that stars escape a globular cluster with a small relative velocity and at an approximately constant rate.
Furthermore, stars escape the cluster predominantly from its Lagrange points and continue orbiting the Galaxy on orbits slightly displaced from the progenitor's orbit (panel II, bottom).
In the progenitor's reference frame, the stream stars appear to be moving on epicycles (panel III, bottom).
Seen in projection, these epicycles form a regular pattern of over- and under-densities along the globular cluster tails \citep[Figure~\ref{fig:stream_formation}, panel II, bottom;][]{kupper:2008, kupper:2010, just:2009}.
This pattern is absent from dwarf galaxy tails because the episodic mass loss of kinematically hotter stars results in a fairly uniform density (Figure~\ref{fig:stream_formation}, panel II, top).

Early theoretical models of interacting galaxies showed that reproducing the various configurations of tail and bridge features seen in catalogs of interacting galaxies \citep{vorontsov-velyaminov:1959, arp:1966} is possible through gravity alone \citep{pfleiderer:1963, toomre:1972}.
Their tidal origin makes streams extremely powerful tracers of the underlying gravitation field and dark matter, which dominates the mass budget in most galaxies \citep[e.g.,][]{moster:2010, behroozi:2019}.
Theoretical studies showed that the phase space of stellar streams can be used to measure the mass and shape of a dark matter halo on large scales \citep{springel:1999, dubinski:1999}, and to measure the abundance of dark-matter subhalos on small scales \citep{johnston:2002, ibata:2002, siegal-gaskins:2008}.
Thanks to facilities optimized for detecting low surface-brightness features, dozens of tidal streams and shells have been detected throughout the Local Universe \citep[e.g.,][]{martinez-delgado:2010, martinez-delgado:2023, bilek:2020} and beyond \citep[e.g.,][]{kado-fong:2018, ren:2020}, and a fraction of them have been modeled to infer the properties of the hosts' dark matter halos \citep{amorisco:2015, vandokkum:2019, pearson:2022}.
Unfortunately, extragalactic streams provide somewhat limited constraints on the gravitational potential because they have so far been mostly detected in 2D projected sky coordinates \citep{nibauer:2023}, but in the Milky Way we have been able to measure both the 3D positions and 3D velocities \citep{koposov:2010}.
The Milky Way stellar streams are poised to answer one of the major open questions in both physics and astrophysics---the nature of dark matter.

The first stream discovered in the Milky Way---tidal tails of the Sagittarius dwarf galaxy \citep{ibata:1995, ibata:2001a, Newberg:2002, majewski:2003, belokurov:2006}---was also the first one to constrain the shape of the Milky Way's dark matter halo.
\citet{ibata:2001b} immediately observed that Sagittarius stream traces close to a great circle, and concluded that since its orbital plane hasn't been significantly torqued, the Milky Way halo must be nearly spherical.
However, radial velocities measured along the stream were consistent with models of the Sagittarius stream in a prolate halo, and ruled out oblate and spherical halo shapes \citep{helmi:2004}.
Yet still, when \citet{johnston:2005} used more numerous M giant stars to measure a small amount of the Sagittarius' orbital precession, they found it strongly disfavors prolate halo shapes and prefers a mildly oblate halo.
This discrepancy suggested that a more flexible model of the halo is needed to fit all of the available data, and indeed, a few years later a triaxial halo model resolved the discrepancy \citep{law:2009, law:2010, deg:2013}.
Radial velocities were key in producing an accurate model of the Sagittarius stream.
Such a massive stream provides hundreds of bright targets suitable for spectroscopy \citep[e.g.,][]{majewski:2004}, a subset of which can even be observed at high resolution \citep[e.g.,][]{monaco:2007}.
However, identifying spectroscopic members of the more common lower-mass streams is more challenging as it requires targeting fainter stars, and therefore using large-aperture facilities, to still yield an efficiency of only a few percent \citep{odenkirchen:2009, kuzma:2015}.
Proper motion measurements were even more limited: they were either available over wide areas and fairly uncertain, due to a short time baseline between large ground-based photometric surveys \citep[e.g.,][]{munn:2004}, or more precise, but only available over sparse fields, either ground-based with longer baselines \citep[e.g.,][]{casetti:2006}, or space-based with fortuitously placed first-epoch imaging \citep[e.g.,][]{sohn:2016}.
\emph{In summary, an accurate potential reconstruction was limited by the poor availability of stream kinematics before \gaia.}

The need for widespread stream observations was highlighted by expanding potential reconstruction to streams beyond Sagittarius.
Rather than confirm the triaxial halo shape, dynamical modeling of the Palomar~5 tidal tails \citep[Pal~5;][]{kupper:2015} and GD-1 \citep{koposov:2010}, two thin streams expected to provide more sensitive constraints on the potential \citep{lux:2013}, found that the inner halo of the Milky Way is close to spherical, or possibly mildly oblate.
When \citet{pearson:2015} simulated tidal tails of Pal~5 in the triaxial halo potential preferred by the Sagittarius stream, they found that the resulting Pal~5 model is significantly more dispersed than the observed stream \citep{odenkirchen:2001, rockosi:2002}.
Combined with the finding that a stellar disk in this potential model is unstable \citep{debattista:2013}, a triaxial halo was disfavored, at least in the inner Galaxy.
Furthermore, a joint analysis of the GD-1 and Pal~5 streams found their joint constraints are consistent with a spherical halo, but individually they indicate the halo becomes more oblate further out \citep{bovy:2016}.
This again suggests that the adopted ellipsoidal and static halo models do not accurately represent the Milky Way's halo.
Indeed, cosmological N-body simulations predict that dark matter halos change shape with radius \citep[e.g.,][]{allgood:2006, bett:2007, maccio:2007, peter:2013, chua:2019}.
In addition, recent observations suggest that the Large Magellanic Cloud is massive and on first infall \citep[e.g.,][]{kallivayalil:2006, kallivayalil:2013, besla:2007, penarrubia:2016}, in which case the Milky Way halo is in significant disequilibrium \citep[e.g.,][]{bekki:2012, gomez:2015, garavito-camargo:2019}.
Non-parametric, time-dependent expansions of the gravitational potential can capture these effects, but they require quite a number of terms \citep[e.g.,][]{lowing:2011, lilley:2018, garavito-camargo:2021}.
Multiple streams are needed to constrain such flexible models, ideally covering a wide range of locations throughout the Galaxy \citep[e.g.,][]{bh:2018}.
Deeper photometric surveys have always uncovered new stellar streams \citep[e.g.,][]{koposov:2014, martin:2014, shipp:2018, jethwa:2018}, however none are sufficiently deep to detect streams in the outer galaxy, nor cover the full sky.
\emph{In a modest number of stellar streams known before \gaia, we hit another limit to tracing dark matter in the Milky Way halo on large scales.}

At the same time, progress was being made in constraining dark matter substructure.
Numerical experiments of cold dark-matter subhalos impacting stellar streams, both in idealized and cosmological settings, demonstrated that subhalos produce a range of observable features in streams, including variations in density, enhanced dispersion, stream folds, and stream track wiggles and asymmetries \citep{carlberg:2009, yoon:2011, ngan:2015, ngan:2016, sandford:2017}.
Across these studies, a stream underdensity, or a gap, was the most ubiquitous signature of a subhalo impact, so follow-up studies estimated the expected rates of subhalo-induced stream gaps \citep{carlberg:2012b, ngan:2014, erkal:2016, banik:2018}, as well as developed theoretical frameworks for inferring parameters of the impact from properties of individual gaps \citep{carlberg:2013b, erkal:2015a, erkal:2015b, helmi:2016, sanders:2016, koppelman:2021}.
A handful of seemingly high-confidence gaps were identified using SDSS and PAndAS photometry for streams discovered in the Milky Way and M31, respectively \citep{carlberg:2011, carlberg:2012, carlberg:2013, carlberg:2016b, carlberg:2016}, in broad agreement with the expected LCDM rates.
These findings motivated deeper imaging of these streams, which confirmed several prominent, degrees-wide gaps, but also revealed a density profile that is overall more uniform than that inferred from the shallower imaging \citep{ibata:2016, erkal:2017, bovy:2017, deboer:2018, bonaca:2020}.
This underlines the importance of securely identifying stream members from the foreground Milky Way stars, whose density variations could be mistaken for signatures of dark-matter subhalo impacts \citep{thomas:2016}.
\emph{Unfortunately, deep photometry needed to map the structure of all known streams is prohibitively expensive even with today's facilities.}

By mid-2010s, we had a dozen streams discovered in the Milky Way, statistical inference methods developed and tested, but were no closer to knowing either the mass and shape of our dark matter halo, or its structure on small scales.
A prominent astronomer at an IAU Symposium 311 held in July 2014 summed up this lack of progress by asking ``Where is the beef with streams?''.
As we now know, at that moment in time the revolution was already on its way to Earth's Lagrangian L2 point: the European Space Agency's \gaia\ spacecraft launched in December 2013.
Conceived even before any stream had been discovered in the Milky Way \citep{lindegren:1993, battrick:1994, lindegren:1996}, \gaia\ was designed to deliver \emph{``a census of the accurate positions, distances, space motions (proper motions and radial velocities), and photometry of all approximately one billion objects complete to $V=20$\,mag''} \citep{perryman:2001} --- precisely the measurements needed to break through the limits of stream studies.
The first three Data Releases from the \gaia\ mission unveiled the structure of our Galaxy in unprecedented detail \citep{babusiaux:2018, helmi:2018, katz:2018, antoja:2021, smart:2021, drimmel:2023, schultheis:2023}, including dwarf galaxies accreted and phase-mixed onto the Milky Way \citep{belokurov:2018, helmi:2018b, myeong:2019, naidu:2020}, as well as moving groups and dissolved star clusters in the Solar neighborhood \citep{antoja:2018, kawata:2018, ramos:2018, meingast:2019, roser:2019}.
While these substructures all represent different stages in the tidal dissolution of Milky Way satellites and can be considered streams, this review is too short to do them all justice.
We will focus on a subset of long, coherent stellar streams accreted in the Milky Way halo, which in the \gaia\ era became our most sensitive antennae for dark matter.
Let's see how!

\begin{figure*}[t!]
\begin{center}
\includegraphics[width=1\textwidth]{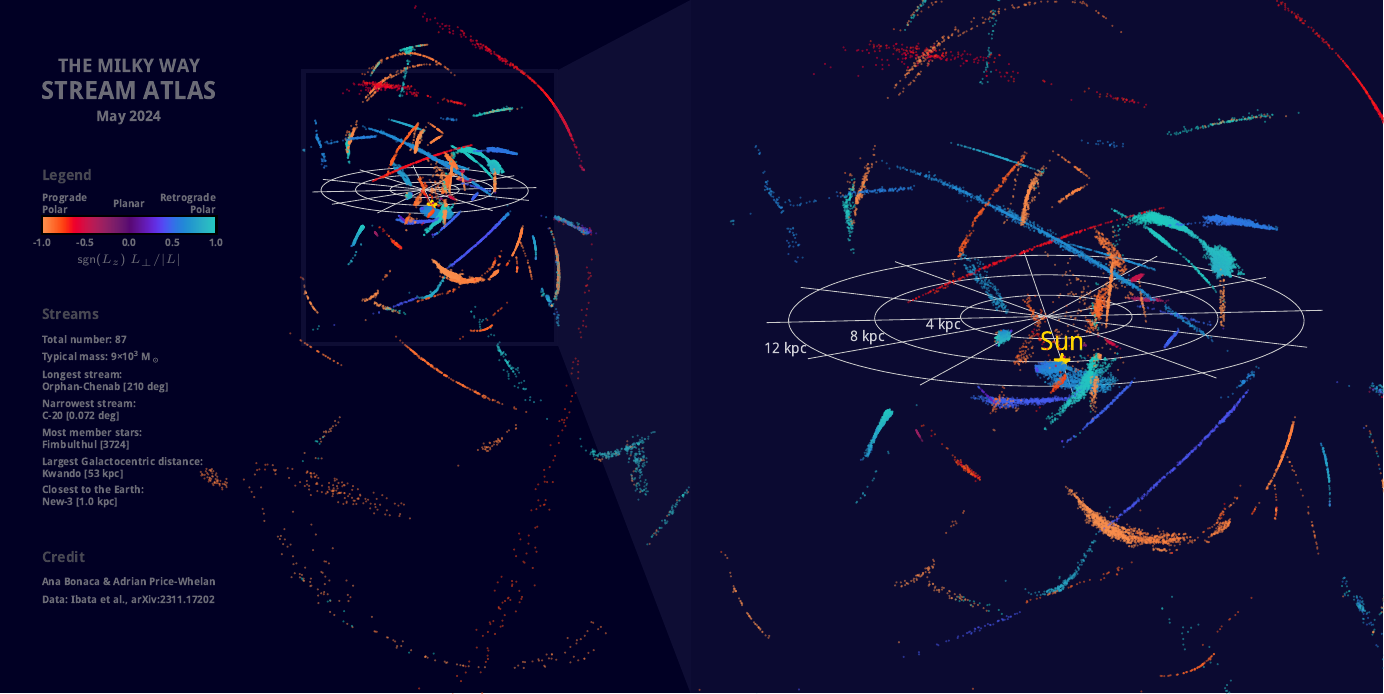}
\end{center}
\caption{%
An atlas of stellar streams discovered in the Milky Way.
Every point represents a high-probability member star of a Milky Way stream identified by \citet{ibata:2023} using the \texttt{STREAMFINDER} algorithm, whose distance from the Sun has been adopted from the best-fitting orbit (\ref{apx:stream-fit}) to improve on the uncertain distance measurements.
Streams are shown in Cartesian Galactocentric coordinates, with the left panel providing an overall view of the stream population and the right panel zooming in on the numerous streams discovered in the inner Milky Way.
A gray grid outlines the radial extent of the Milky Way's disk, while the Sun's position is marked with a yellow star.
Streams are color-coded by the fraction of the angular momentum vector residing in the disk plane ($sign(L_z)L_\perp/|L|$), with red for prograde and blue for retrograde streams.
}
\label{fig:fos_3d}
\end{figure*}

\section{Stream Discovery and Characterization in the \gaia\ Era}
\label{sec:discovery}

Stellar streams inherently have low stellar densities and their constituent stars are
typically far from the Sun, making them difficult to observe and study.
Finding and characterizing streams is therefore all about enhancing contrast over the
foreground Milky Way stars, either by filtering datasets to increase the relative
occurrence of stream stars in the filtered set, or by restricting the search to rare
tracers that are more likely to be found in streams than in the field (e.g., blue
horizontal branch stars that are common to old, metal-poor stellar populations).
The first streams and substructures around the Milky Way were discovered this way by
filtering photometric data for stars in the Galactic halo, where the background stellar
density is also inherently lower than in the disk or central galaxy.
Subsequent discoveries were made with deeper or wider-area photometry surveys, or by
combining photometric and spectroscopic information (e.g., radial velocities).
The largest explosion in the number of known streams has come only over the last few
years thanks to data from the \gaia\ Mission.
The astrometric data from \gaia\ has revolutionized our ability to find kinematic
sub-structures because it is now possible to combine photometric and kinematic data for
enormous samples of stars.
In this section, we summarize the methods used to find streams in the pre-\gaia\ era and
highlight new methods that are now possible thanks to the all-sky kinematic data
available from \gaia.
We provide a brief census and summary of the streams that have been published up to now,
but note that this is still an active area for discovery and characterization.
Figure~\ref{fig:fos_3d} provides a 3D map of the known stellar streams.
We also discuss the limitations of the \gaia\ data and the need for future astrometric
surveys to push our understanding of the Milky Way's stellar halo and its dark matter
content to fainter magnitudes and larger distances.

\subsection{Before \gaia}

The first stellar streams discovered around the Milky Way --- the Sagittarius stream
(Sgr) and the Palomar 5 stream (Pal 5) --- were found in the late 1990s and early 2000s
as over-densities of stars that connect back to their progenitor systems (the
Sagittarius dwarf galaxy and the Palomar 5 globular cluster, respectively).
Sgr and Pal 5 have since become archetypes of two major observational and interpreted
types of streams: dwarf galaxy or ``wide'' streams as compared to globular cluster or
``thin'' streams.
These discoveries, and most of the halo stellar streams found through the early 2000's,
were found using photometry.
The advent of deep, wide-area, multi-band sky surveys --- especially the Sloan Digital
Sky Survey (SDSS; \citealt{york:2000, gunn:1998, fukugita:1996}) and the Two Micron All
Sky Survey (2MASS; \citealt{skrutskie:2006}) --- were crucial for this initial phase of
stream discovery.
The combination of wide-area coverage, well-calibrated photometry
\citep{padmanabhan:2008}, and near-uniform depth across the survey footprints enabled
constructing detailed maps of the stellar density of the Milky Way halo.
Still, streams are relatively low-density features and are surpassed in number counts by
the foreground and background stellar populations in the Milky Way in sky density alone.
Stream search methods must therefore leverage careful selection of stars based on the
expected stellar populations in accreted systems using multi-band photometry.

The most effective method for using photometric data to identify streams is to apply a
``matched filter'' \citep{rockosi:2002} to the color--magnitude distribution of stars
to preferentially select or up-weight stars that are likely to be distant and low
metallicity.\footnote{The focus here is on stellar streams from accreted dwarf galaxy or
globular cluster progenitor systems, which are expected to be old and metal poor.}
Matched filters can be used as a boolean selection (e.g., select all stars within some
region of color and magnitude defined by isochrones) or as a weighting function (e.g.,
weight all stars by the probability that they are near an isochrone for an old and
metal-poor stellar population).
To search for streams, the filtered data was typically then displayed in sky
coordinates, sometimes by making movies that step the filter in distance modulus, and
maps of stellar density were smoothed and visually inspected for curving or linear
over-densities of stars.
Automated detection methods were explored, but this was the early days for computer
vision and machine learning, and most stream discoveries (even through the 2010's)
happened through visual inspection.

\begin{figure*}[t!]
    \begin{center}
    \includegraphics[width=1\textwidth]{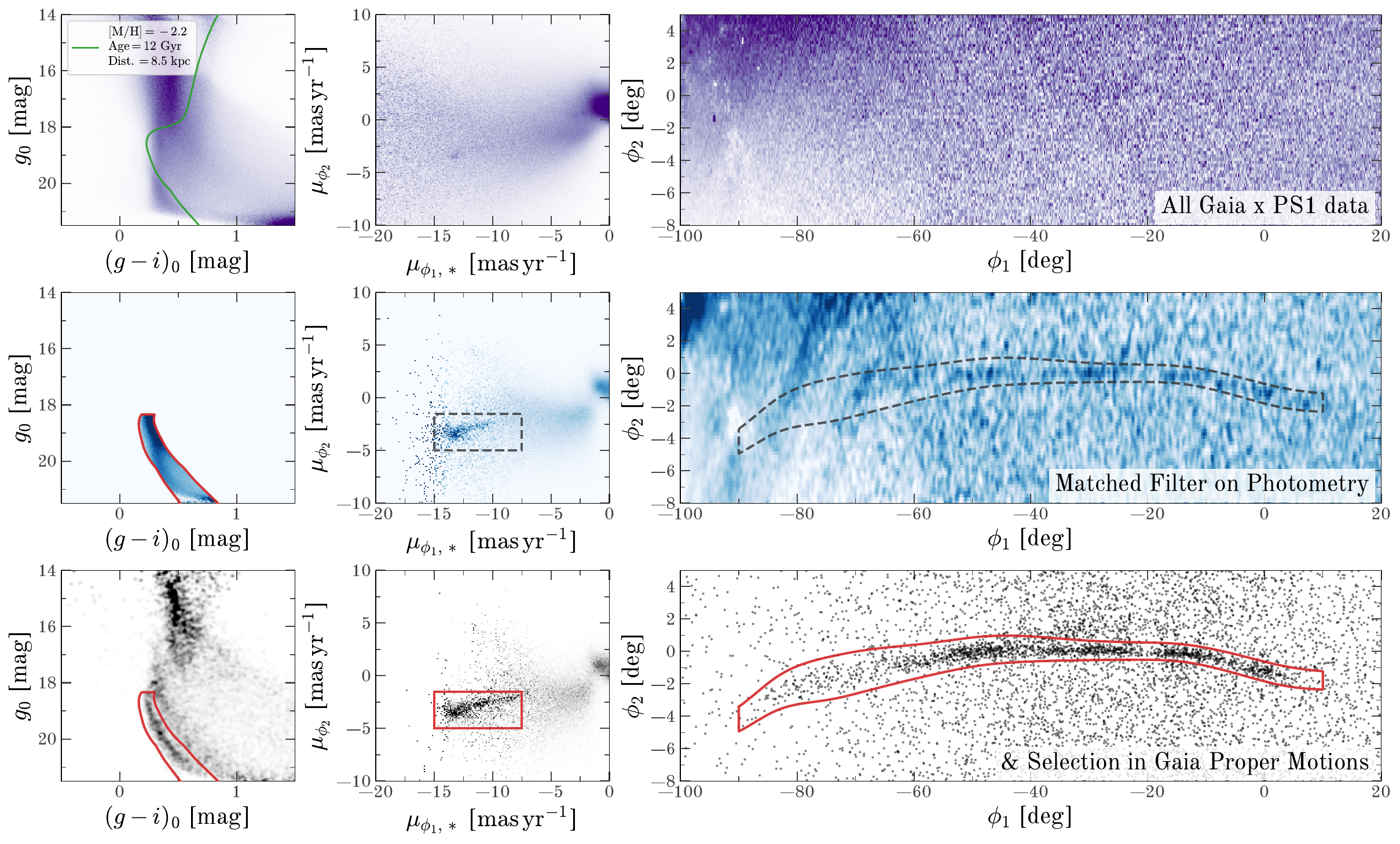}
    \end{center}
    \caption{%
    A demonstration of how matched filtering with photometry and \gaia\ astrometry can
    be used to reveal a stellar stream, here the GD-1 stream.
    In all rows, the left panel shows an extinction-corrected color--magnitude diagram
    using $g, r, i$ photometry from the Pan-STARRS (PS1) survey \citep{chambers:2016},
    visualized as a row-normalized 2D histogram.
    The center panel shows the proper motions from \gaia\ \dr{3} \citep{gaiadr3}
    transformed into a coordinate system aligned with the stream, $(\mu_{\phi_1, *},
    \mu_{\phi_2})$, where $\phi_1$ is the stream longitude and $\phi_2$ is the stream
    latitude, and the color scale is a column-normalized 2D histogram.
    The right panel shows the sky positions of stars as a column-normalized 2D
    histogram.
    \textbf{Top row:} No selections have been applied: These panels show all stars in a
    cross-match between \gaia\ \dr{3} and PS1 for the region of the sky around the GD-1
    stream.
    \textbf{Middle row:} A binary matched filter has been applied to the photometry, as
    shown by the red polygon in the left panel of this row.
    Note that already the stream is visible as an over-density of stars in
    proper-motions (center panel) and in the known stream track (dashed line in the
    right panel).
    \textbf{Bottom row:} In a given panel, selections are applied based on the red
    outlines in the other two panels.
    For example, in the left panel, stars are shown as selected within the proper motion
    selection box (red outline in the center panel) and the sky track polygon (red
    outline in the right panel).
    }
    \label{fig:gd1-demo}
\end{figure*}

Figure~\ref{fig:gd1-demo} shows an example of how binary matched filtering based on
photometry can be used to reveal a stream.
The top row of panels in Figure~\ref{fig:gd1-demo} shows all data in a cross-match
between \gaia\ and Pan-STARRS (PS1; \citealt{chambers:2016}) photometry for stars in the
sky region near the GD-1 stream \citep{grillmair:2006-gd1}: The top left panel shows the
extinction-corrected color--magnitude diagram based on PS1 photometry, the top center
panel shows the \gaia\ proper motions transformed into a coordinate system aligned with
the stream \citep{koposov:2010}, and the top right panel shows a 2D histogram of the sky
positions of all stars in this cross-match.
The over-plotted (green) line in the top left panel is an isochrone from the PARSEC
\citep{bressan:2012, chen:2015} isochrone library for an old and metal-poor stellar
population at the approximate distance of GD-1.
The middle row of panels show the same data, but now with a binary matched filter
applied based on the photometric selection shown in the middle left panel (red outlined
region).
This filter was designed by shifting the isochrone shown in the top panel by a small
amount in color.
The dashed line in the middle right panel shows the approximate location of the GD-1
stream in sky positions (middle right panel) as is now understood from selections using
\gaia\ data.
Even with only the photometric selection, the matched filter reveals a clear
over-density of stars near $\phi_2 \sim 0^\circ$ --- this is the GD-1 stream.
However, notice that photometric selection alone is not sufficient to reveal the full
extent of the stream --- we will return to this example and discuss the importance of
kinematic information for stream discovery later in this section.

Even now, many ($\sim$50) of the currently known streams were found using a matched
filter based on photometry, plus visual inspection.
For example, the GD-1 stream --- the first long, thin stream discovered without a clear
progenitor --- was initially found by \citet{grillmair:2006-gd1} using a matched filter
on SDSS photometry with visual inspection of the resulting stellar density maps.
Matched filters were also critical for many early discoveries of streams with the SDSS
data \citep[e.g.,][]{Newberg:2002, Yanny:2003, belokurov:2006, grillmair:2006-orphan}.
This method has since been applied to many large photometric surveys, such as 2MASS,
Pan-STARRS (PS1; \citealt{chambers:2016}), the Dark Energy Survey (DES;
\citealt{des:2005, des:2016}), the DECam Legacy Survey  (DECaLS; \citealt{dey:2019}),
the Pan-Andromeda Archaeological Survey (PAndAS; \citealt{mcconnachie:2009}),
and many smaller surveys.
In all cases, this has led to new stream and halo substructure discoveries.
For example, from 2MASS \citep{rocha-pinto:2003, rocha-pinto:2004}, PS1
\citep{bernard:2014, bernard:2016, navarrete:2017}, DES \citep{shipp:2018}, SLAMS
\citep{jethwa:2018}, PAndAS \citep{martin:2014}, and DECaLS \citep{shipp:2020}.
Table~\ref{tbl:stream-summary} has a list of all stellar streams that
have been reported to date.

Many of the systematic searches for streams described above were done without
considering specific progenitor systems (i.e. globular clusters or dwarf galaxies).
The discovered streams were then subsequently either found to be without a clear
progenitor system, or were associated with known systems based on their location in the
sky and/or their kinematics.
However, some searches were specifically designed to find streams associated with known
globular clusters or dwarf galaxies \citep[e.g.,][]{grillmair:1995, kuhn:1996,
leon:2000} using photometry around these systems.
With the later exceptions of the striking tidal tails around the Pal 5 globular cluster
\citep{odenkirchen:2001} and the faint stream associated with NGC 5466
\citep{grillmair:2006-ngc5466}, most of these searches were unsuccessful or only found
evidence for very low-density tidal tails based on the presence of small numbers of
extra-tidal stars around the progenitor systems \citep{grillmair:1995, leon:2000}.

The first stream discoveries and early theoretical work motivated the development of
new methods to identify lower surface brightness and more distant streams using a myriad
of techniques.
In most cases, however, searches for streams have focused on finding linear or curving
but ``thin'' stream features in projected sky coordinates.
For example, even before the first halo streams were found, it was recognized that tidal
debris from distant merger events should remain spatially coherent for many dynamical
times (gigayears, in the Milky Way halo; \citealt{johnston:1998, helmi:1999}).
From early numerical simulations, it was also noted that --- in a spherical or
nearly-spherical dark matter distribution --- the streams would phase mix almost along
great circles on the sky \citep{johnston:1996,ibata:2001b}.
This motivated the development of methods to search for over-densities of stars along
great circles in sky positions or photometrically-filtered sky positions of stars (the
``great circle cell count'', GC3, method; \citealt{johnston:1996}).
The GC3 method was used to identify and trace the extent of the Sagittarius stream using
2MASS data for M giant stars \citep{majewski:2003} and was later adapted to handle other
kinematic dimensions (the modified GC3 or ``mGC3'' method, which also uses proper motion
and radial velocity data; \citealt{mateu:2011}).
The mGC3 method has since been used to identify many candidate stellar streams using RR
Lyrae-type stars \citep{mateu:2018}.

Even in the pre-\gaia\ era, it was recognized that the kinematics of stars could be used
to search for streams and kinematic substructures in terms of dynamical invariants,
which would enable finding substructures that are not spatially coherent or so nearby
that they appear broad in sky projection.
For example, to identify potential streams or other dynamical structures in phase-space,
\citet{helmi:1999} used a small sample ($<100$) of metal-poor giant stars in the solar
neighborhood to search for kinematic over-densities of stars.
This early search for substructures in the solar neighborhood used astrometry from the
\hipparcos\ Mission \citep{perryman:1997} combined with literature measurements of
radial velocities of these stars to obtain full (6D) phase-space measurements of the
stars.
With 6D phase-space measurements, the stellar kinematics can be transformed into
integrals of motion (e.g., angular momentum) and Galactocentric velocity components,
where streams and other accreted tidal debris features are predicted to remain coherent
for longer than in position-space alone \citep{helmi:1999, sanderson:2015}.
Until recently, this was only possible with relatively small samples of stars with
precisely-measured astrometry, because the ability to identify clumps and over-densities
in dynamical or orbital quantities depends strongly on the precision of observed
kinematic coordinates and the number of stars.

\begin{figure}[t!]
    \begin{center}
    \includegraphics[width=0.5\textwidth]{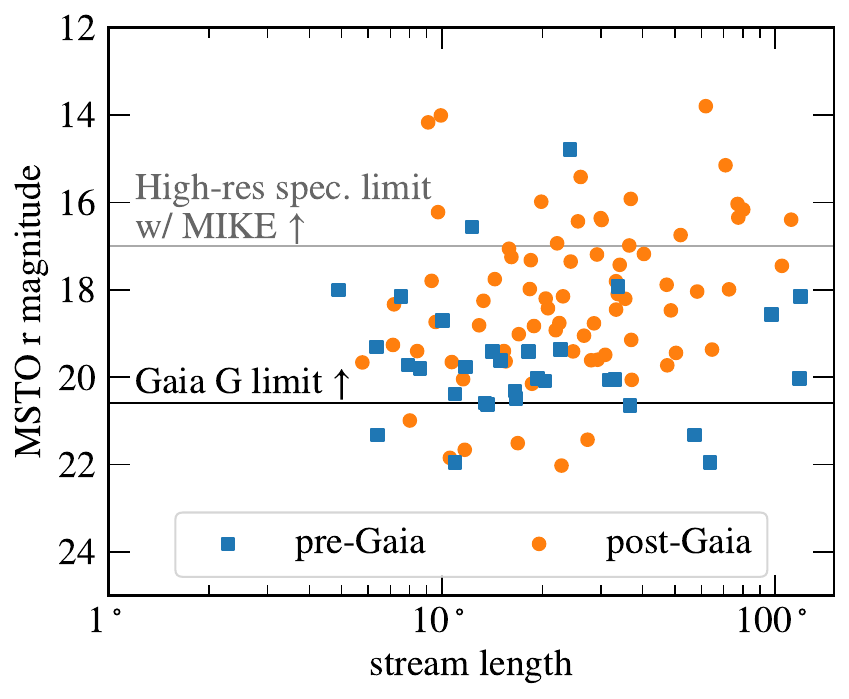}
    \end{center}
    \caption{%
    The mean apparent $r$-band magnitude of the main sequence turn-off (MSTO)
    for streams discovered before \gaia\ \dr{2} (blue squares) and since (orange circles).
    The distance moduli were determined by fitting orbits to stream stars associated
    with each of the streams in \citet{ibata:2023} (this procedure is described in
    detail in \ref{apx:stream-fit}).
    The MSTO magnitude is computed assuming all streams have a single stellar population
    with an age of $12~\unit{\giga\year}$ and a metallicity of $\feh = -2.2$, but this
    is only meant to qualitatively show the relative brightness of the MSTO for the
    known streams.
    High-resolution spectroscopic follow-up of MSTO stars in these streams is difficult
    or currently infeasible for many of these streams: The upper gray line shows the
    approximate magnitude limit for observing an FGK star with the MIKE spectrograph
    \citep{bernstein:2003} to obtain a signal-to-noise ratio of 50 at $5500~\unit{\angstrom}$ with
    a spectral resolution $R\sim \num{30000}$ with an exposure time $\lesssim
    1.5~\textrm{hours}$.
    RGB stars in the streams are typically bright enough for this type of follow-up, but
    are far less numerous than MSTO stars (globular cluster streams may only have a few
    to tens of RGB stars).
    }
    \label{fig:msto-rc-mag}
\end{figure}

Prior to the \gaia\ Mission, only a handful of the discovered streams were followed up
and characterized with subsequent observations or surveys (for example, to obtain radial
velocities or detailed element abundance measurements for the constituent stars).
This was largely due to the fact that many of the streams discovered in deep photometric
surveys were faint and distant, and therefore difficult to observe spectroscopically.
In all but a few cases, the main sequence turn-offs (MSTO) of the streams are
sufficiently faint so that only the Red Giant Branch (RGB) or Horizontal Branch (HB)
stars were practically observable.
But the RGB and HB stars in many thin streams are sparse and often distributed over
large areas of the sky, making it difficult to obtain a large sample of stream stars.
As a demonstration of this point, Figure~\ref{fig:msto-rc-mag} shows the PS1 $r$-band
magnitude of the MSTO for the mean stream distance for streams discovered before \gaia\
\dr{2} (blue square markers) compared to those found after \gaia\ \dr{2} (orange circle
markers).
Streams discovered before \gaia\ were typically fainter and more difficult to follow up
spectroscopically than those discovered after \gaia.
To obtain precise and high signal-to-noise element abundance measurements using existing
facilities, stars typically have to be brighter than $r \lesssim 18~\textrm{mag}$ --- the
MSTO for most streams known pre-\gaia\ are fainter than this limit.

The \gaia\ Mission --- especially data from the second data release in 2018 --- has
since transformed the discovery, characterization, and follow-up efficiency of stellar
streams around the Milky Way.

\subsection{After \gaia\ Data Release 2}

The \gaia\ Mission released Data Release 2 (DR2) in April 2018 \citep{gaiadr2}, which
included astrometry and mean photometry for nearly 1.7 billion stars based on the first
22 months of observations.
This data was transformative for the study of Milky Way stellar streams and stellar halo
stars.
While parallax measurements for stars in most stellar streams and throughout most of the
stellar halo were too noisy to be useful, the proper motions provided key kinematic
information for many of these stars for the first time.
For example, consider a MSTO star for an old stellar population with absolute magnitude
$M_G \sim 4~\unit{\mag}$, at a distance of $d = 10~\kpc$ with a transverse velocity
typical of the Milky Way halo of $v_{\textrm{tan}} = 100~\kms$.
This star would have a true parallax of $\varpi = 0.1~\unit{\mas}$, a total proper
motion $\mu \approx 2.1~\masyr$, and an apparent magnitude of $G = 19~\unit{\mag}$.
For a star of this apparent magnitude in \gaia\ \dr{2}, a typical parallax uncertainty
$\sigma_\varpi \sim 0.4~\unit{\mas}$ and a typical proper motion uncertainty $\sigma_\mu
\sim 0.4~\masyr$.
This means that the parallax measurement is entirely dominated by the uncertainty, but
the proper motion still has an appreciable signal-to-noise ratio of $\mu / \sigma_\mu
\sim 5$.
The ability to use proper motions to identify members of stellar streams immediately
enabled detailed looks at previously-known streams and new methods for finding streams
in the Milky Way.

Proper motion data provided by \gaia\ has proven to be invaluable for studying the
density structure of stellar streams.
While matched filtering on photometry alone can reveal the presence of a stream, the
stream stars are often still relatively low signal-to-noise features over the background
stellar density, meaning that the stream density and any variations in the density is
typically not well constrained.
Soon after \gaia\ \dr{2}, \citet{price-whelan:2018} demonstrated the power of combining
photometric and kinematic data to identify stream members and measure the density of the
GD-1 stream.
Figure~\ref{fig:gd1-demo} is a recreation of the selections performed in
\citet{price-whelan:2018} but now using data from \gaia\ \dr{3} \citep{gaiadr3}.
As mentioned above, the top row of panels in Figure~\ref{fig:gd1-demo} shows photometry
from PS1 (top left panel), proper motions from \gaia\ \dr{3} (top center panel), and sky
positions (top right panel) for stars in the sky region near the GD-1 stream.
The middle row of panels shows the same data, but now with a matched filter applied to
the photometry.
The bottom row of panels shows the same data, but now with a matched filter applied to
the photometry and a selection on the proper motions (shown as a red box in the bottom
center panel).
The bottom right panel now shows the sky positions of individual stars that pass these
selections, which reveals more of the stream (especially at $\phi_1 \lesssim -55^\circ$)
and enables quantifying the magnitude of density variations and gaps present along the
stream track.
This new view of GD-1 also revealed features of the stream that were not visible in
previous filtered characterizations of the stream \citep{grillmair:2006-gd1,
koposov:2010, deboer:2018}, which we discuss more below in Section~\ref{sec:structure}.
Many other studies have also used \gaia\ proper motions to study the structure of GD-1
\citep[e.g.,][]{malhan:2018, li-yanny:2018, huang:2019, deboer:2020, ibata:2020}, the
Orphan--Chenab stream \citep{koposov:2019}, the Helmi streams \citep{koppelman:2019},
the Jhelum stream \citep{bonaca:2019}, the Ophiuchus stream \citep{caldwell:2020}, along
with work that has modeled the density structure of many streams simultaneously
\citep[e.g.,][]{patrick:2022}.

\begin{figure}[t!]
\begin{center}
\includegraphics[width=\columnwidth]{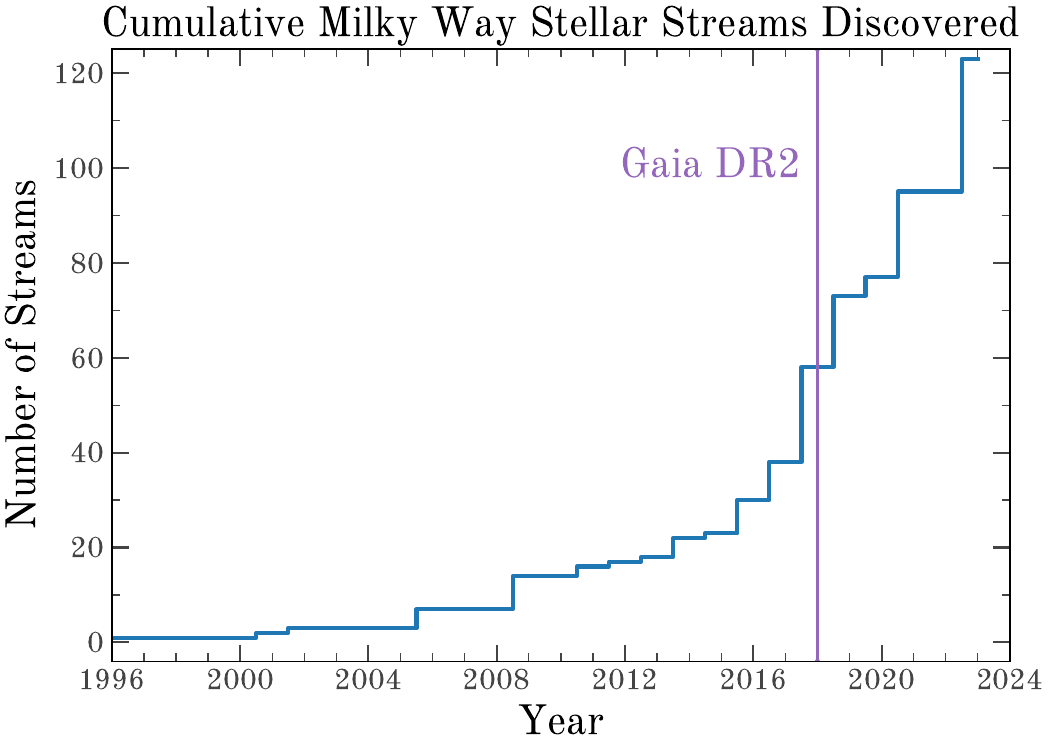}
\end{center}
\caption{%
The number of known streams has been gradually increasing since the discovery of the Sagittarius tidal tails in 1995 and entered exponential growth following \gaia\ \dr{2} in 2018.
The growth has been fairly smooth because it has been driven both by releases of more comprehensive datasets and by improved methods for finding streams.
}
\label{fig:num-streams}
\end{figure}

Over the years leading up to \gaia\ \dr{2}, searches for new streams pushed deeper using
well-calibrated photometry from surveys like PS1 and DES, resulting in a steady and
continuous increase in the number of known streams.
Figure~\ref{fig:num-streams} shows the cumulative number of Milky Way stellar streams
reported in the literature as a function of time, as collated in the \texttt{galstreams}
\citep{mateu:2023} library and combined with the recent atlas of stream data from
\citet{ibata:2023}.
However, the kinematic data from \gaia\ has enabled a new and important validation step
for previously discovered streams.
While many streams have been re-discovered and studied in detail using \gaia\ data (see
Table~\ref{tbl:stream-summary}), a number of streams that were previously reported in
the literature have not been re-identified with \gaia\ data.
This is not totally unexpected, as many of the streams discovered in the pre-\gaia\ era
pushed to fainter magnitudes and larger distances, where the \gaia\ data is noisier
and/or only RGB stars in the streams may pass the faint limit of \gaia.
These objects will require additional follow-up observations to confirm they are indeed stellar streams, and to
determine their intrinsic and orbital parameters (see the last section of
Table~\ref{tbl:stream-summary}).

The exquisite astrometry provided by \gaia\ has also enabled a new class of methods for
searching for new streams around the Milky Way \citep[e.g.,][]{malhan:2018a,
borsato:2020, necib:2020, gatto:2020, shih:2022, shih:2023, pettee:2024}, which has led
to an enormous increase in the number of known low-density stellar streams.
These methods all leverage the kinematic data now available from \gaia\ to identify
over-densities of stars in different projections or transformations of the phase-space
data.
For example, \texttt{STREAMFINDER} \citep{malhan:2018a} searches for over-densities of
stars near orbits computed assuming a model for the Milky Way's mass distribution.
This works by projecting orbits into the data space, where uncertainties and missing
data can be naturally handled by the search method.
Other methods either transform the phase-space data into a Galactocentric coordinate
system where clustering methods can search directly in the phase-space distribution
function \citep{necib:2020,gatto:2020}, or apply machine learning and anomaly detection
methods to the observed (heliocentric) data directly \citep{borsato:2020, shih:2022,
shih:2023, pettee:2024}.

The ability to use newly-accessible phase-space dimensions to search for streams has led
to the discovery of many low-density streams that have eluded photometric searches.
For example, early work using \texttt{STREAMFINDER} identified a total of 13 new streams
in \gaia\ \dr{2} data \citep{malhan:2018, ibata:2018, ibata:2019}, which are typically
closer (and therefore appear wider) and have lower surface densities than previous
photometrically-discovered streams.
This data has also enabled the discovery of streams that are not spatially coherent in
\emph{sky position} but do appear to contain comoving, metal-poor stellar populations
\citep[e.g.,]{necib:2020b, balbinot:2023}.
Efforts have also searched for kinematic substructures that may be phase-mixed or less
spatially coherent using integrals of motion or other dynamical invariants
\citep[e.g.,][]{koppelman:2019b, yuan:2020, naidu:2020, ou:2023}.

The number of known stellar streams around the Milky Way has exploded in the \gaia\ era:
over 120 streams have now been reported in the literature (see
Figure~\ref{fig:num-streams}), with more discoveries coming every year, especially
enabled by new data releases from \gaia\ and new methodological developments.
Figure~\ref{fig:sky-map} shows the on-sky footprints of all currently known stellar
streams in the Milky Way, as listed in Table~\ref{tbl:stream-summary}.
With so many new streams, significant effort has been put into obtaining follow-up
spectroscopic observations of stream stars to measure their radial velocities and
detailed element abundances.
For example, surveys like the Southern Stellar Stream Spectroscopic Survey (S5;
\citep{li:2019,li:2022}) and more targeted efforts \citep[e.g.,][]{ibata:2021} are
steadily improving the validation and characterization of the current population of
known streams.
Future data releases from \gaia\ and new spectroscopic surveys and facilities such as
the Dark Energy Spectroscopic Instrument (DESI; \citealt{desi:2016}), H3
\citep{conroy:2019}, 4MOST \citep{4most:2012}, and WEAVE \citep{weave:2012} will
continue to enable new discoveries and detailed follow-up of the known streams.

\subsection{Key opportunity: Completing the census of stellar streams}
\begin{figure*}[t!]
\begin{center}
\includegraphics[width=1\textwidth]{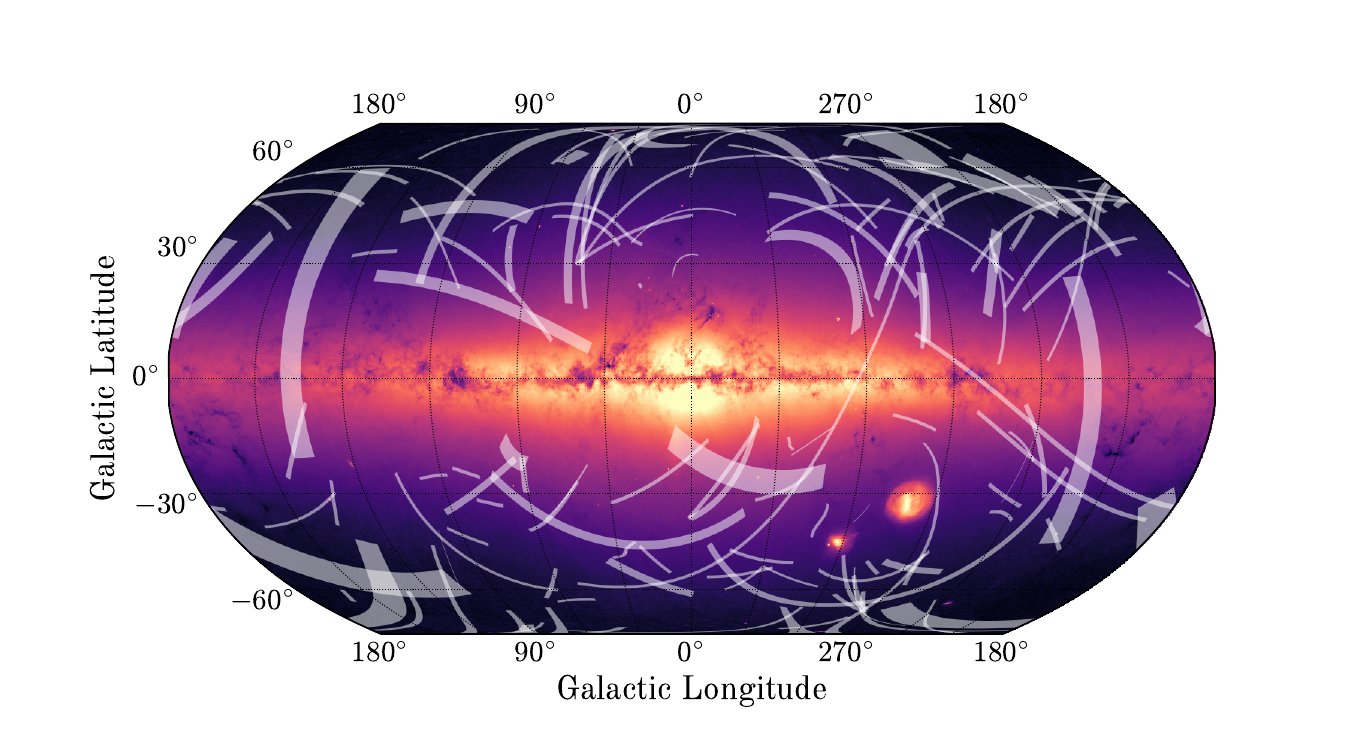}
\end{center}
\caption{%
A representation of the on-sky footprints of all known stellar streams in the Milky Way,
shown in Galactic coordinates.
The footprints are shown as white polygons that approximate the currently-known lengths
and mean widths of the streams (the streams are also listed in
Table~\ref{tbl:stream-summary}).
The background image is a map of the log-number counts of stars in \gaia\ \dr{3}.
}
\label{fig:sky-map}
\end{figure*}

\begin{description}
    \item[Stream discovery in the disk plane and the Galactic center] As is apparent in
    Figure~\ref{fig:sky-map}, most of the halo stellar streams discovered to date have
    been found at high Galactic latitudes. This is a reflection of the fact that it is
    easier identify stream stars in regions of lower background stellar density. We
    expect the distribution of halo streams to be closer to uniform suggesting that many
    stellar streams await discovery at low Galactic latitude, and toward the Galactic
    center region.
    \item[Automated discovery with quantified selection function] Recent efforts to
    discover new streams with automated methods have been successful, but the selection
    function of these methods is not well understood. Quantifying this is a critical
    step in understanding the completeness of the current stream census, and in
    interpreting the population of observed streams in relation to the initial
    population of stream progenitors.
\end{description}

%%%%%%%%%%%%%%%%%%%%%%%%%%%
\section{Stream kinematics}
\label{sec:orbits}
% AB
In the \gaia\ era, stream kinematics are no longer derived from painstaking follow-up of photometrically identified targets, but are a part of stream discovery \citep[e.g.,][]{malhan:2018b, malhan:2019, ibata:2018, ibata:2019, ibata:2021, grillmair:2019, grillmair:2022}.
Stream member stars identified using \gaia\ proper motions provide not only two components of the velocity vector, but also candidates for further selecting radial velocity members by cross-matching large spectroscopic surveys \citep[e.g.,][]{huang:2019, yang:2022, ibata:2023} or by conducting highly efficient spectroscopic follow-up \citep[e.g.,][]{li:2019, bonaca:2020b}.
Overall, 105 streams reported in Table~\ref{tbl:stream-summary} have proper motion measurements from \gaia, while 72 of those streams also have a radial velocity detection from spectroscopic follow-up surveys.

\begin{figure*}
\begin{center}
\includegraphics[width=0.95\textwidth]{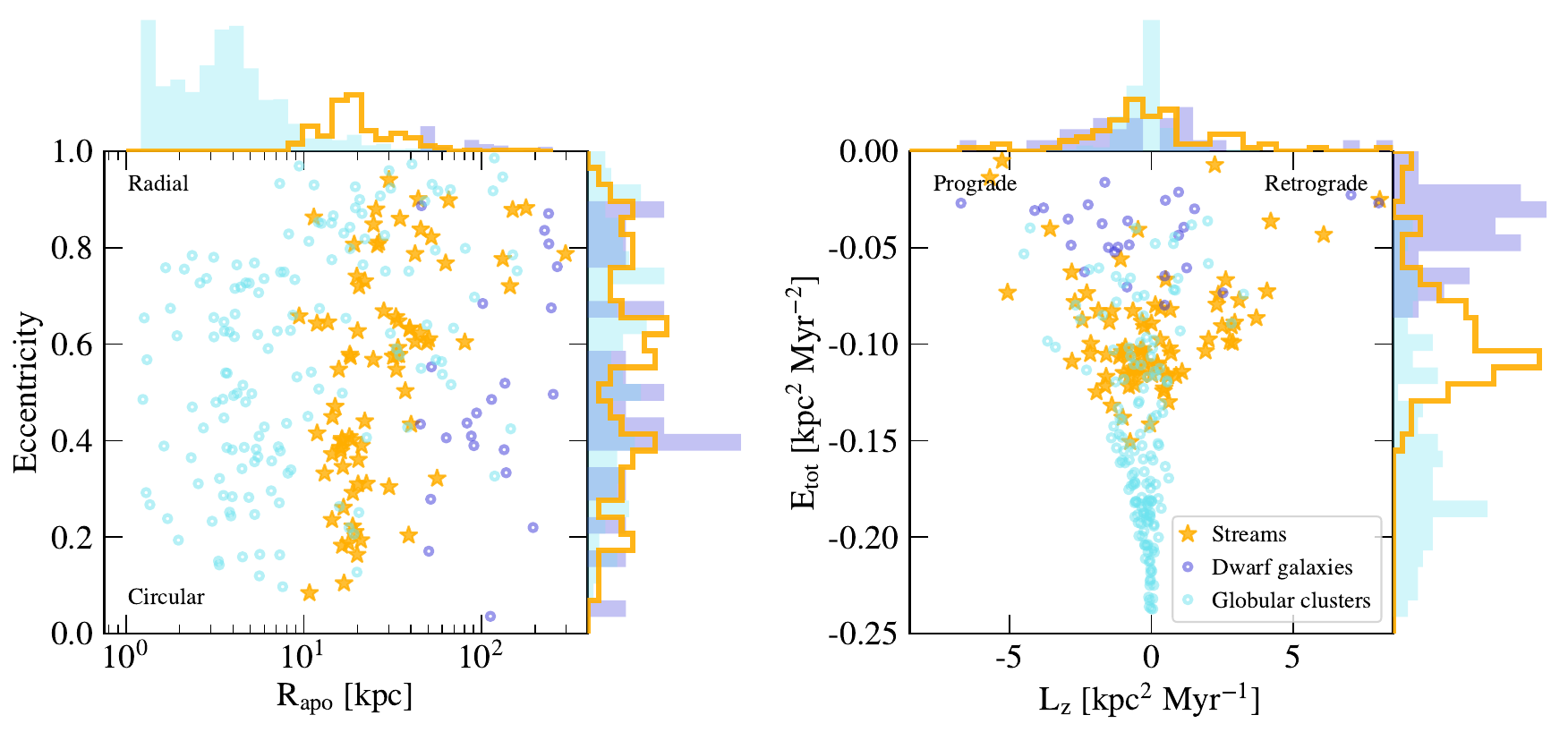}
\end{center}
\caption{%
Orbital phase space of stellar streams reported in \citet{ibata:2023} and fit in \ref{apx:stream-fit} (yellow stars), as well as globular clusters \citep[light blue circles,][]{baumgardt:2019}, and dwarf galaxies \citep[dark blue circles,][]{helmi:2018,simon:2018}.
\textbf{Left:} Orbital eccentricity as a function of apocenter, with circular orbits on the bottom and radial on top.
The distribution of stream apocenters peaks at $\approx20\,\unit{kpc}$ (top histogram), while those of globular clusters and dwarf galaxies peak at lower and higher radii, respectively, indicating additional streams may be discovered in those regions of this parameter space.
The distribution of stream eccentricities shows multiple peaks (right histogram), which likely correspond to streams of a common (accretion) origin.
\textbf{Right:} Total orbital energy, $E_{tot}$, as a function of the $z$-component of the angular momentum vector, $L_z$ (aligned with the disk angular momentum), with orbits prograde with the disk rotation on the left and retrograde orbits on right.
Both globular clusters and streams are clustered in this 2D projection of the orbital phase space, with each cluster likely corresponding to objects delivered in distinct accretion events.
}
\label{fig:phase_space}
\end{figure*}

\subsection{Orbits and the Milky Way potential}
\label{sec:mwpotential}
The wide availability of kinematics has influenced how streams are described: instead of reporting the empirical velocity gradients along a stream, they are now used to determine the stream's orbit.
Even in streams with just a handful of radial velocity measurements, the more numerous \gaia\ proper motions are sufficient to constrain an orbit in an assumed Milky Way potential (as described in \ref{apx:stream-fit}).
Streams' orbits have been useful in placing them into a broader context within the Milky Way, revealing their dynamical histories and processes likely to have produced features discussed in \S\ref{sec:structure}.
Figure~\ref{fig:phase_space} summarizes orbital properties of stellar streams from the \citet{ibata:2023} catalog (yellow stars).
The assumed underlying gravitational potential is a static and analytic \texttt{MilkyWayPotential2022} model \citep{price-whelan:2017}, and the orbit fitting procedure is described in \ref{apx:stream-fit}.
The majority of known Milky Way streams have apocenters between 10\,\unit{\kpc} and 50\,\unit{\kpc}, and thus spend most of the time beyond the Milky Way disk.
However, their wide range of eccentricities (Figure~\ref{fig:phase_space}, left) brings pericenters for half of this sample within the Solar circle.
$\approx60\%$ of the sample are on orbits prograde with the Milky Way's disk, which makes them more susceptible to perturbation from the Milky Way bar, spiral arms and molecular clouds \citep[e.g.,][]{pearson:2017, banik:2019}.
There is a long tail towards retrograde orbits, but this is likely a detection bias due to retrograde streams being more easily detected in proper motions \citep[e.g.,][]{ibata:2021}.

Quite surprisingly, orbits also revealed that some streams extend farther than previously thought and connect to another stream.
Analyzing halo RR~Lyrae in \gaia~\dr{2}, \citet{koposov:2019} traced the Orphan stream through the Galactic plane to the southern hemisphere and found it connects to the Chenab stream in sky positions, distance, and proper motions.
Similarly, \citet{li:2021} found that a single orbit connects the adjacent ATLAS and Aliqa Uma streams in both proper motions and radial velocities.
In both cases, stream kinematics were key in making these connections.
In contrast, relying on sky positions alone can be unreliable, e.g., Hermus and Phoenix streams appear to be in the same orbital plane \citep{grillmair:2016b}, but radial velocities show they are not associated \citep{martin:2018}.
As we are completing the stream census, accurately accounting for the total number of accretion events onto the Milky Way is important for testing galaxy formation models \citep{pillepich:2015, shipp:2023, wright:2023, khoperskov:2023}.

Both the Orphan--Chenab and ATLAS--Aliqa Uma complexes were misidentified as separate structures due to the effects of dynamical perturbations.
A recent passage of the massive LMC shifted the orbital planes of the Orphan and Chenab streams, and misaligned their proper motion vectors with respect to the stream path in the sky coordinates, i.e. the stream track \citep{erkal:2019}.
Similar misalignments were detected in multiple streams of the southern hemisphere \citep{shipp:2019}, and used to measure the mass of the LMC \citep[$\gtrsim10^{11}\,\unit{\msun}$;][]{erkal:2019, shipp:2021, koposov:2023}.
Lower mass satellites can also impact stellar streams if their orbits intersect \citep[e.g.,][]{dillamore:2022}.
For example, both ATLAS--Aliqa Uma and the Jhelum stream may have been perturbed by the Sagittarius dwarf galaxy \citep[respectively; see also Section~\ref{sec:structure}]{li:2019,woudenberg:2023}.
Furthermore, orbital calculations back in time show that some globular clusters may have had a flyby encounter with a dwarf galaxy \citep{garrow:2020, el-falou:2022}.
On the other hand, no known satellite passed close enough to perturb the GD-1 stream, leaving a dark-matter subhalo impact as a viable option \citep{bonaca:2019, doke:2022}.

Of course, these orbital calculations are only as good as the adopted gravitational potential model.
While multi-component analytic models of the Milky Way have been tuned to reproduce a wide range of observables \citep[e.g.,][]{bovy:2015, mcmillan:2017,price-whelan:2017}, not accounting for the realistic mass accretion history in halo models can introduce biases on the order of tens of percent \citep[e.g.,][]{bonaca:2014, dsouza:2022, arora:2022, santistevan:2024}.
Simultaneous modeling of multiple globular cluster streams in the inner $\lesssim20$\,\unit{\kpc} found an ellipsoidal halo model a reasonably good fit, and thanks to \gaia\ kinematics put strong constraints on the prolate halo shape \citep[$q_z=1.06\pm0.06$;][]{palau:2023}, and the halo virial mass \citep[$M_{200}=1.09^{+0.19}_{-0.14}\times10^{12}\,\unit{\msun}$;][see also \citealt{reino:2021}]{ibata:2023}.
However, reproducing the positions and kinematics of the more distant Sagittarius stream requires a more complex, time-dependent model with an oblate inner halo, triaxial outer halo, as well as a massive LMC on its first infall \citep{vasiliev:2021,kang:2023}.
This change in the halo shape is likely due to the LMC deforming the Milky Way's outer halo \citep[e.g.,][]{garavito-camargo:2019,shao:2021,vasiliev:2023}.
Modeling techniques that self-consistently track the evolution of the Milky Way and the LMC are currently being developed using basis function expansions \citep[e.g.,][]{sanders:2020,garavito-camargo:2021,lilleengen:2023}.

\subsection{Orbital phase space and hierarchical assembly}
\label{sec:phasespace}
Despite the ongoing perturbations to the outer Galaxy, it is often helpful to represent the Milky Way with an integrable, static and axisymmetric potential, so that positions in the orbital phase-space spanned by the total energy, $E$, and the $L_z$ component of the angular momentum are conserved \citep{noether:1918}.
In such a potential, it is possible to simplify the equations of motions using the Hamiltonian canonical variables \citep{goldstein:2002, helmi:1999, tremaine:1999, eyre:2011}:
\begin{align}
\pmb{\theta}(t) &= \pmb{\theta}_0 + \pmb{\Omega} t
\label{eq:angle}
\\
\Omega_i(\pmb{J}) &\equiv \dot{\theta}_i = \frac{\partial H}{\partial J_i}
\label{eq:frequency}
\end{align}
where $H$ is the Hamiltonian, $\pmb{\theta}$ are the angle variables (canonical coordinates), $\pmb{J}$ are actions (conjugate momenta and constants of motion), and $\Omega_i$ are the orbital frequencies (also constant).
The Hamiltonian action-angle formalism has been a popular framework for studying stellar streams because it allows for a quick calculation of orbits and generation of stream models \citep[e.g.,][]{eyre:2011,sanders:2013a,bovy:2014, vasiliev:2019}.
Orbital frequencies have also provided deeper insights into the evolution of tidal debris: an individual stream has distinct frequencies, whose separation decreases with time, making it possible to estimate the time of accretion \citep{gomez:2010}.
The orbital structure of a stellar halo can be summarized with a 2D ``frequency map'' by plotting ratios of frequencies in each of the three spatial dimensions (e.g., $\Omega_x/\Omega_z$ vs $\Omega_y/\Omega_z$), for every available orbit \citep[][]{valluri:2010,valluri:2012}.
This space easily visualizes resonances; as orbits with commensurable frequencies, resonances appear as straight lines in a frequency map.
Stream progenitors on regular, non-resonant orbits form thin and long-lived streams, while those on nearly-resonant and chaotic orbits quickly disperse \citep{price-whelan:2016a, price-whelan:2016b, yavetz:2023}, so the overall structure of the halo in the frequency space can be used to constrain the gravitational potential \citep{valluri:2012, yavetz:2021}.
\gaia\ data has enabled mapping the frequency space of the Milky Way halo for the first time, and curiously revealed that both the Sagittarius and Helmi streams populate the $\Omega_\phi:\Omega_z=1:1$ resonance \citep{koppelman:2021a, dodd:2022}.
However, despite the elegance of the action-angle approach, its application for modeling other halo streams with \gaia\ data has been limited by their complex morphologies that indicate significant perturbation by non-equilibrium processes that are difficult to capture within the framework (e.g., the recent infall of the LMC).

On the other hand, when viewed for a population of streams, the orbital phase space has produced valuable insights into the hierarchical assembly of our Galaxy.
The sky distribution of orbital poles shows that stellar streams, and globular clusters, are distributed isotropically around the Galaxy \citep{riley:2020, vasiliev:2019b}, in contrast to dwarf galaxy satellites which are preferentially distributed in an apparently rotating plane \citep[e.g.,][]{pawlowski:2012, pawlowski:2020, fritz:2018}.
The planar configuration being limited to dwarf galaxies, rather than universal for all halo objects, suggests that its origin is not due to modified gravity \citep[e.g.,][]{pawlowski:2018}, but likely a transient occurrence \citep{sawala:2023, xu:2023}, possibly due to group infall \citep{patel:2020} or induced by the LMC \citep{garavito-camargo:2021,garavito-camargo:2024}.
More direct evidence for group infall comes from the stream distribution in the $E-L_z$ space.
Easily calculated for individual stars and conserved in standard Milky Way potentials, energies and angular momenta were mapped for the Solar neighborhood with \gaia~\dr{2} and revealed clumps that correspond to dwarf galaxies accreted onto the Milky Way billions of years ago \citep[e.g.,][]{helmi:2018,belokurov:2018,myeong:2019}.
\gaia~\dr{3} improved proper motions and orbital constraints of the halo streams to allow placing them on the $E-L_z$ plane (Figure~\ref{fig:phase_space}, right).
Streams are highly clustered in the $E-L_z$ space (also seen as distinct peaks in orbital eccentricities, Figure~\ref{fig:phase_space} left), and aligned with debris from identified Milky Way progenitors \citep{bonaca:2021,malhan:2022}.
This alignment indicates that most stellar streams originate from clusters and galaxies that were brought into the Milky Way by a handful of larger dwarf galaxies, thus providing additional evidence for the hierarchical nature of galaxy formation.

Understanding the full dynamical history of a stellar stream back to its progenitor galaxy is crucial for interpreting the observed stream features (discussed in more detail in \S\,\ref{sec:structure}).
N-body simulations of ex-situ globular clusters---clusters formed in dwarf galaxies that were subsequently accreted onto the Milky Way---produce stellar streams surrounded by a range of low density features \citep[e.g., wide envelopes, or cocoons, and/or sub-streams;][]{carlberg:2018,carlberg:2020,malhan:2019a,qian:2022}.
These structures originate from stars lost before the accretion onto the Milky Way, rather than a perturbation within the Milky Way.
An example of such a group accretion event is the wide Cetus-Palca stream and the associated globular cluster NGC~5824 \citep{yuan:2019,thomas:2022}.
Three thin stellar streams share the same orbit with NGC~5824 and Cetus: Triangulum/Pisces, Willka Yaku, and C-20 \citep{bonaca:2021, yuan:2022}.
N-body reconstruction of this system suggests that Cetus-Palca is tidal debris of a $M_\star\approx10^6\,\unit{\msun}$ dwarf galaxy that accreted onto the Milky Way $\approx5\,\unit{Gyr}$ ago, at which time it also deposited its satellite globular clusters into the Galactic halo \citep{chang:2020}.
The associated thin streams are likely segments of the extended tidal tails of NGC~5824 or another fully disrupted cluster.
A similar scenario may even explain the long-standing puzzle of the bifurcation in the Sagittarius stream \citep{belokurov:2006, koposov:2012}.
Constrained N-body simulations show that the fainter, more metal-poor wrap of the Sagittarius stream \citep[e.g.,][]{ramos:2022} could have formed during a binary dwarf merger as tidal debris of the lower-mass companion \citep{davies:2024a,davies:2024b}.

Knowledge of these hierarchical relations between different streams can directly test dark matter on small scales.
Numerical simulations predict that tidal debris from the dwarf galaxy itself has a lower velocity dispersion if the galaxy is cuspy \citep{errani:2015}, due to its more compact size.
However, a globular cluster accreted in a cuspy galaxy results in a dynamically hotter stream than one from a cored galaxy of the same mass \citep{malhan:2021}, due to enhanced mass loss on a more compact orbit.
Observational data for individual streams have so far put weak constraints on the inner density slope of dark matter due to its degeneracy with the total subhalo mass \citep{malhan:2022b}.
Systems where streams of both the progenitor and the satellite have been detected provide a new opportunity to build self-consistent models that accurately constrain the inner slope of the accreted dark matter halos.

\subsection{Progenitors and early star formation}
\label{sec:progenitors}
The orbital phase space also revealed the direct progenitors of several stellar streams previously thought to be completely dissolved \citep{palau:2019,riley:2020,ibata:2019b,ibata:2021,bonaca:2021}.
These streams have originally not been connected back to the progenitor in sky coordinates, but instead coincide with an orbit of a Milky Way globular cluster or a dwarf galaxy (light and dark blue circles in Figure~\ref{fig:phase_space}, respectively).
In some cases, like between the Gj\" ol stream and NGC~3201 globular cluster, the connection is obscured because it passes through the Milky Way disk \citep{riley:2020}.
In other cases there are no obvious observational challenges to explain gaps spanning up to tens of degrees between the stream and the progenitor, nor the detection of a single tidal tail rather than a symmetric pair \citep[e.g,][]{bonaca:2021, yang:2022b}.
These peculiarities may simply be due to insufficient depth \citep[e.g., non-detections of the low-mass stars that populate cluster tails in the early phases of dissolution,][]{balbinot:2018}, but if they are confirmed as physical, they may indicate an unusual mass-loss history \citep[e.g., due to a population of black holes in the progenitor,][]{roberts:2024}, a massive external perturbation \citep[e.g., from the Galactic bar,][]{pearson:2017}, or even modified gravity \citep[causing asymmetric escape into the leading and trailing tails,][]{thomas:2018}.

Regardless, the now increased subset of stellar streams with known progenitors (see Table~\ref{tbl:stream-summary} for a complete list) are extremely valuable for gravitational potential inference.
First, a known progenitor places strong constraints on the stream's orbit \citep{bh:2018}, which reduces the number of free parameters and improves potential reconstruction \citep{palau:2023}.
Second, globular cluster progenitors produce epicyclic overdensities, which place additional constraints on the global potential when matched to the observed stream density structure \citep{kupper:2015}.
And finally, a known globular cluster progenitor predicts the exact locations of epicyclic underdensities to conclusively determine whether the observed stream gaps are due to internal evolution or external perturbation \citep[cf.][]{ibata:2020}.

Understanding whether a stream's progenitor was a dwarf galaxy or a globular cluster is important both for dynamical modeling and tracing the Milky Way's formation history.
Before \gaia, the stream width was a reliable indicator of its origin, with more massive dwarf galaxies producing wider streams \citep[e.g.,][]{belokurov:2006, bonaca:2012}.
However, \gaia\ revealed a number of wide, but apparently low-mass stellar streams (see Table~\ref{tbl:stream-summary}).
Alternatively, high-resolution spectroscopy can be used to measure whether stream members have a spread in metallicity \citep[e.g.,][]{ji:2020,chandra:2022}, which has been established as a criterion for identifying dwarf galaxies from globular clusters \citep{willman:2012}.
Multi-element abundances have also been used to validate the stream's progenitor \citep[e.g., NGC~3201 as the progenitor of the Gj\" ol stream,][]{hansen:2020}.
Detailed characterization of stellar populations and chemical abundances of streams also revealed that some reported stream candidates are actually not tidal debris of accreted satellites: the Eastern Banded Structure, the Anticenter Stream, and the Monoceros Ring appear to be disk perturbations excited during different Sagittarius passages \citep[cf.][]{deason:2018, laporte:2020}, and part of a more extended population of disk stars kicked out to high latitudes \citep{price-whelan:2015,bergemann:2018,laporte:2018}, while the Nyx stream is consistent with a dynamical substructure within the thick disk \citep[cf.][]{zucker:2021,wang:2023}.

Widely available stream spectroscopy and chemistry have also constrained the mass and structure of their dwarf galaxy progenitors.
Median metallicities of dwarf galaxy streams are typically lower than those of the bound dwarf galaxies in the Milky Way \citep{li:2022}, and their multi-element abundance patterns are consistent with low-mass satellites \citep[$\approx10^6\,\unit{\msun}$,][]{ji:2020,hawkins:2023}.
In the most massive stellar stream, Sagittarius, extensive spectroscopic coverage even mapped the internal structure of the progenitor.
First, metallicity gradients were detected along both of its tidal tails in APOGEE \citep{hayes:2020} and LAMOST \citep{zhao:2020}.
More recently, \citet{cunningham:2024} used an even larger sample of \gaia\ metallicities \citep{andrae:2023} to detect a metallicity gradient both along and across the Sagittarius tails, and map it to the radial metallicity gradient in the progenitor.
\citet{johnson:2020} additionally found a very metal poor population ($\feh<-2$) that is kinematically hotter than the more metal-rich stars, and consistent with having been stripped early from the progenitor's outskirts, possibly from a separate halo component \citep[see also][]{limberg:2023}.
A couple of streams even show complex chemistry suggesting mixed tidal debris of a dwarf galaxy and a globular cluster \citep{hansen:2021,limberg:2024}.

Metallicities of globular cluster streams themselves have resolved one of the long-standing puzzles in cluster formation: the apparent metallicity floor of $\feh\approx-2.5$ \citep[e.g.,][]{harris:1996,harris:2010,beasley:2019}.
Median metallicities of cluster streams are low \citep[$\feh\lesssim-1$,][]{martin:2022b}, with several streams below $\feh\lesssim-3$ \citep{roederer:2019,wan:2020,martin:2022a}, demonstrating that clusters can form below the metallicity floor observed in surviving clusters.
Dynamical masses of these streams are low (C-11, C-19, C-20, Phoenix, and Sylgr all have masses $\lesssim10^4\,\unit{\msun}$, see Table~\ref{tbl:stream-summary}), suggesting that only more massive clusters, which form in more massive, and therefore more metal-rich galaxies \citep[e.g.,][]{maiolino:2019}, survive until today, as theoretically predicted by \citet{kruijssen:2019}.
These low-metallicity, low-mass stellar streams provide unique laboratories for several other aspects of cluster formation.
For example, Phoenix has a large spread in the \ion{Sr}{2} abundance, likely due to pollution from a massive, fast-rotating star before the cluster fully formed \citep{casey:2021}.
C-19 on the other hand has unexpectedly large width and velocity dispersion for a cluster of its mass, possibly due to having formed in a dark matter halo instead \citep{errani:2022}.
Finally, a spread in light-element abundances has been detected in several streams \citep{martin:2022a,balbinot:2022,usman:2024}, providing evidence of multiple stellar populations even in low-mass clusters, in addition to their ubiquity among the more massive globular clusters \citep[e.g.,][]{gratton:2012}.
It is uncertain to what extent these multiple stellar populations are due to various polluting processes \citep[e.g., massive AGB stars, fast-rotating massive stars, supermassive stars,][]{bastian:2018} and diluting processes \citep[e.g., stellar mass-loss, interactive binaries, re-accretion,][]{gratton:2019} operating during cluster formation.
The interplay of these processes may depend on both the age and the mass of the cluster, so globular cluster streams---old and low-mass---provide a new test bed for developing the physical picture of cluster formation.

\subsection{Key opportunity: Mapping the entire 6D phase-space of stellar streams}
\begin{description}
\item[Validation and spectroscopic characterization]{
Kinematic information is vital for validating stellar stream discoveries.
Of the 133 known streams, 61 lack spectroscopy, and 28 have no kinematic data at all (Table~\ref{tbl:stream-summary}).
Future \gaia\ data releases will further improve proper motion precision to levels sufficient for validating most reported streams.
Spectroscopic characterization will remain challenging for distant streams with current instrumentation (see Figure~\ref{fig:msto-rc-mag}), however, targeting based on \gaia\ proper motions vastly increases efficiency in recovering stream members, so a coordinated effort should yield full kinematic data for all streams.
}
\item[Astrometry below the \gaia\ faint limit]{
The most distant known streams, and those expected to be discovered in Rubin/LSST photometric data, may only have a few members observed by \gaia.
Fortunately, LSST will itself measure proper motions over a 10-year baseline throughout the Milky Way halo at the current \gaia\ precision but 3\,mag deeper, while the space-based Roman Space Telescope has capability of further improving these proper motions by an order of magnitude \citep[][]{sanderson:2019}.
These data sets will be essential in completing the kinematic map of the Milky Way streams.
}
\end{description}

%%%%%%%%%%%%%%%%%%%%%%%%%%%%%%%%%%%%%%%%%%%%%%%
\section{Density Structures of Stellar Streams}
\label{sec:structure}

Soon after the discovery of the Sagittarius stream, it was realized that streams around
the Milky Way would be powerful tools for studying the Galactic mass distribution.
Streams are some of the few structures that provide nearly direct information about
the shapes of orbits around the Galaxy.
In a conceptual sense, streams form linear structures as the constituent stars ``sort''
themselves by orbital energy relative to their progenitor system.
The curvature of a stream through phase-space therefore enables a more direct inference
of the mass or acceleration field relative to traditional dynamical inference methods
that require phase-mixed populations to trace the underlying distribution function
\citep{bh:2018}.
It was then also realized that these long, dynamically cold structures should be very
sensitive to passing mass, providing one of the only ways to infer the presence of
otherwise star-free dark matter subhalos \citep{johnston:2002, ibata:2002, yoon:2011}.
Clearly streams should therefore encode detailed information about both the global and
small-scale properties of dark matter throughout the Galaxy.

The streams around the Milky Way, however, tell a more intricate story.
Every stream that has been studied in detail shows evidence for other dynamical
processes that have impacted the density structure of the stream.
For example, the Sagittarius stream is known to be ``bifurcated,'' with two intertwined
components of the stream separated by a small angle on the sky
\citep[e.g.,][]{belokurov:2006}.
But even low-mass streams like GD-1, Pal 5, Ophiuchus, Jhelum, Atlas--Aliqa Uma (AAU),
and more show evidence for density variations along the streams, extended envelopes of
stars around the streams, breaks, gaps, and other features that protrude from the main
stream track.

Simulations of stream formation --- motivated by the observed complexity of the streams
around the Milky Way --- have shown that many other dynamical processes (beyond tidal
stripping and phase mixing) can plausibly impact the density structure and track of a
stream.
For one, it has been shown that the slow tidal disruption of a globular cluster through
internal relaxation will eject stars into the tails with a very small velocity
dispersion.
If the cluster is on an eccentric orbit, the ejected stars will slowly phase mix away
from the orbit of the cluster, but can create over- and under-densities along the tidal
tails known as ``epicyclic over-densities'' \citep{kupper:2010}.
Other phenomena, such as interactions with time-dependent features in the Milky Way like
the galactic bar or spiral arms, the resonant structure of orbits in the galaxy, and,
excitingly, interactions with dark matter subhalos can all create density structures in
the tails of streams.

Below, we review dynamical processes that can impact streams and simulations that have
been developed to understand the diversity of density structures observed in streams.
We then review the observed kinematic structure of a few of the known streams around the
Milky Way as case studies.
The sections below are presented sequentially, but really the dynamical models and
simulations were developed contemporaneously with the discovery and study of the
structure of streams around the Milky Way.

\subsection{Expectations from Theory and Simulations}
\label{sec:structure-theory}

As discussed above (Section~\ref{sec:intro}), stellar streams form from the tidal
disruption of a progenitor system, which can be a globular cluster, dwarf galaxy, or
other bound stellar system.
As with above, we focus here on streams that form from globular cluster or low-mass
dwarf galaxy progenitors, as these are the most numerous and well-studied streams around
the Milky Way.

The phase-space structure of a stream is set by a number of internal and external
processes that are, in detail, coupled and interrelated.
However, it is conceptually useful to distinguish the dominant factors that can impact
stream formation and structure:
(1) the internal dynamics of the progenitor system, which determines the rate of stellar
ejection from the progenitor and the phase-space distribution function of the stripped
stars,
(2) the external gravitational field of the Milky Way, which determines the rate of
phase mixing, the orbit of the progenitor system, and the overall shape of the stream,
and
(3) the impact of time-dependent dynamical processes in the Milky Way, such as the bar,
spiral arms, and dark matter subhalos, which can create density variations and other
features in the stream.

\begin{figure}[t!]
    \centering
    \includegraphics[width=\columnwidth]{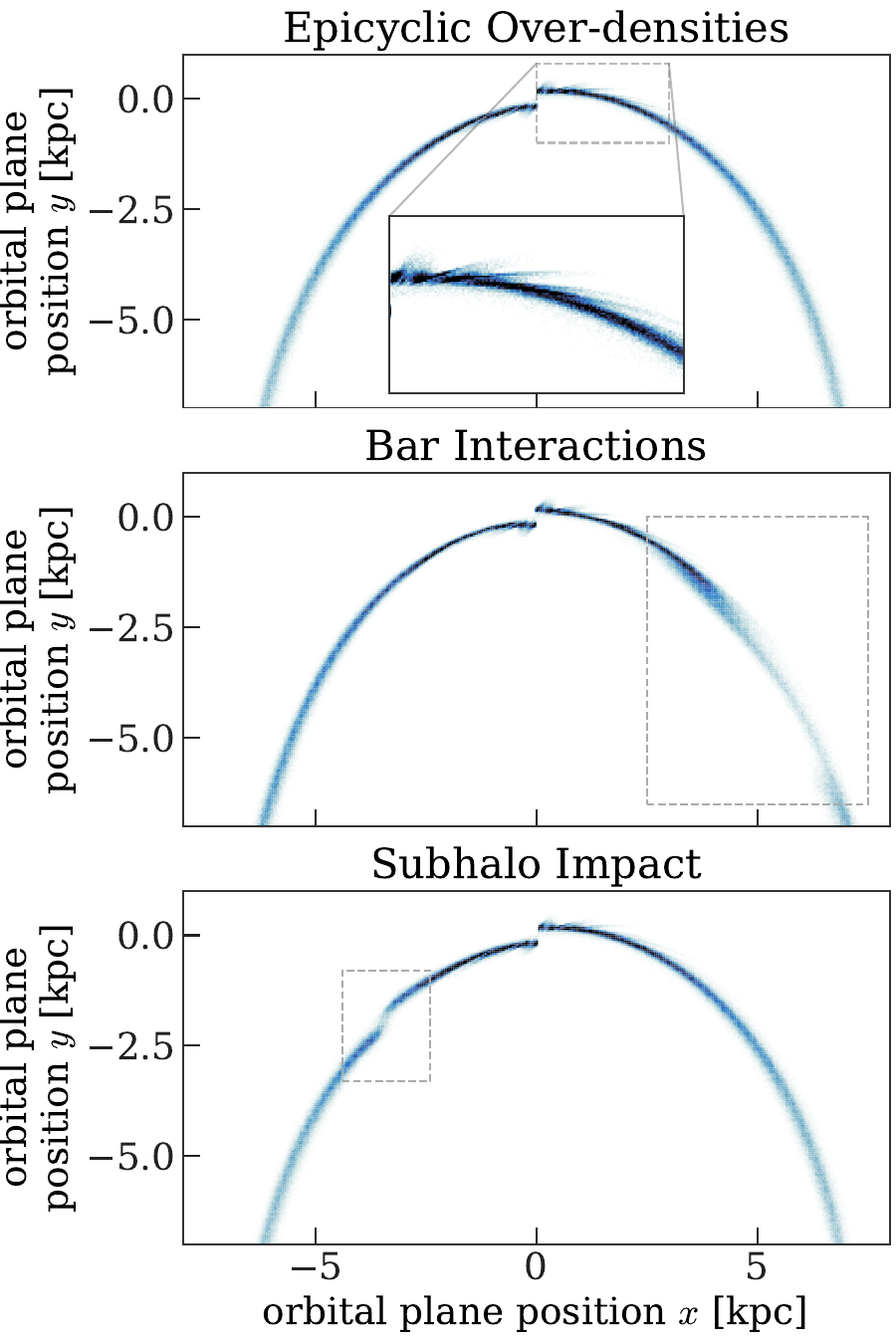}
    \caption{
        Three simulations of globular cluster-mass stellar streams using a particle
        spray scheme \citep{fardal:2015} to simulate the resulting stream morphology.
        The stream particle density is shown face-on in the instantaneous orbital plane
        of the progenitor system.
        The progenitor orbit is chosen to have a similar pericenter, apocenter, and
        eccentricity to the Palomar 5 stream.
        \textbf{Top panel:} A stream simulation in an axisymmetric mass model of the
        Milky Way using the \texttt{MilkyWayPotential2022} from \texttt{gala}
        \citep{price-whelan:2017}.
        The density structure of the stream shows ``epicyclic over-densities'' along the
        stream track \citep{kupper:2012}, which arise from the relative orbits of the
        stripped stream stars and the progenitor system.
        \textbf{Middle panel:} A stream simulation in a time-dependent mass model
        (closely matched to the mass enclosed profile of the
        \texttt{MilkyWayPotential2022} from \texttt{gala}), but with a massive ($M =
        10^{10}~\unit{\msun}$), trixial, rotating ($\Omega =
        40~\unit{\kilo\meter\per\second\per\kilo\parsec}$) bar component.
        \textbf{Bottom panel:} The same as the top panel, but with a single dark matter
        $10^7~\unit{\msun}$ subhalo passing through the stream $300~\unit{\mega\year}$
        in the past (the impact site is highlighted by the dashed box).
    }
    \label{fig:sim-streams}
\end{figure}

\subsubsection{The Orbit and Tidal Disruption of the Progenitor}

The orbit of the progenitor system around the Milky Way sets the overall morphology of a
stream and the mechanism by which stars are stripped from the progenitor.
A progenitor on a near-circular orbit will feel only weak variations of the tidal field
of the Galaxy, and will therefore be stripped of stars slowly and steadily through
internal two-body relaxation \citep[e.g.,][]{kupper:2010}.
The morphology of a stream on such an orbit will generally be more elongated along the
orbit of the progenitor, and the phase mixing of the stream will be dominated by the
internal dispersion of relative energy of the stars in the stream
\citep[e.g.,][]{hendel:2015}.
Both in the case of circular and eccentric orbits, the stripped stars will not
phase mix uniformly into the stream, but will instead form over- and under-densities
along the stream track known as epicyclic density variations \citep{kupper:2008,kupper:2010,kupper:2012,just:2009,mastrobuono-battisti:2012}.

A progenitor stellar system on a very eccentric orbit ($e \gtrsim 0.5$) will instead
experience strong variations of the Galactic tidal field, and the stripping of stars can
therefore be more episodic as the Jacobi radius shrinks significantly during orbital
pericenter.
This repeated tidal shocking can add heat to the internal dynamics of the progenitor
system and lead to more rapid mass loss \citep[e.g.,][]{gnedin:1999}.
The morphology of a stream on such an orbit will generally be more broad and diffuse, as
the stream stars primarily phase mix due to the internal spread of angular momentum of
the stripped stars.
This leads to the formation of more ``shell-like'' structures in the stream
\citep[e.g.,][]{hendel:2015}.
When the eccentricity approaches unity (i.e. a radial orbit), the tidal debris will form
a shell structure instead of a stream \citep[e.g.,][]{hernquist:1987, sanderson:2013}.

\subsubsection{The Orbital Structure of the Milky Way}

A disrupting progenitor system on a near-circular orbit in a spherical gravitational
potential will form a fairly uniform, thin stream that is primarily elongated close to
the orbit of the progenitor.
In more complex potentials --- even static axisymmetric, triaxial, or otherwise
asymmetric --- the orbital structure of the system can lead to more complex stream
morphologies and evolution \citep[e.g.,][]{amorisco:2015b}.
The first effect of a non-spherical potential is the broadening of the stream due to the
differential precession of the orbital planes of the stream stars \citep{erkal:2016b}.
More generically, this effect can also be understood in terms of the Hessian of the
potential in action--angle coordinates, and how this projects into position and velocity
coordinates \citep{sanders:2013a}: If the Hessian is dominated by a single eigenvalue,
the stream will phase mix more rapidly in the direction of the corresponding eigenvector
and thus form a nearly one-dimensional structure.
However, if the Hessian has two or three eigenvalues with comparable magnitudes, the
stream will form a more two- or three-dimensional structure as it forms.

More complex gravitational potentials (axisymmetric and beyond, but still
time-independent) also contain a richer structure of resonances and, generically,
regions of chaotic orbits that can impact the formation and evolution of streams.
This was noticed in the context of simulations of the Pal 5 stream: When a Pal 5 stream
simulation was run in a triaxial potential fitted to the kinematics of the Sagittarius
stream \citep{law:2010}, the stream did not maintain a thin structure, but instead
formed broad ``fans'' of tidal debris at the edges \citep{pearson:2015}.
Since streams are dynamically cold structures, the density evolution of a stream is very
sensitive to the local orbital structure.
In regions of strong dynamical chaos, streams will generally diffuse more rapidly
\citep{price-whelan:2016a, mestre:2020}.
When streams form on orbits near strong resonance overlap, the separations between
orbital families can lead to even more complex evolution of the debris
\citep{yavetz:2021,yavetz:2023}.

\subsubsection{Interactions with Time-Dependent Structures}

Time-dependent structures in the Milky Way, such as the bar and spiral arms, dwarf
galaxies, and low-mass dark matter subhalos, can also impact the density structure of
streams on different time and spatial scales.

The central Milky Way contains a bar structure that is thought to be a major driver of
the dynamics of the inner Galaxy \citep[e.g.,][]{blitz:1991}.
While constraining the pattern speed, size, and mass of the bar are all still active
areas of work, the bar is thought to have a total mass around $10^{10}~\unit{\msun}$, a
major-axis length of $\sim 3~\kpc$ \citep[e.g.,][]{wegg:2015}, and a pattern speed
around $40~\unit{\kms\per\kpc}$ \citep[e.g.,][]{portail:2017, sanders:2019}.
It is thought that even at the solar radius, the impact of resonances with the Galactic
bar has a significant and observable impact on the kinematics of stars
\citep[e.g.,][]{dehnen:1999, kawata:2021}.
Stellar streams that pass through the disk plane with pericentric radii $R \lesssim
10~\kpc$ therefore can be significantly impacted by the bar.
The time-varying dipole asymmetry of the bar can enhance the chaotic diffusion of orbits
in the central Galaxy and therefore cause rapid diffusion of stream debris
\citep{price-whelan:2016b}, can lead to resonant confinement of a stream and cause
truncation of a stream's length \citep{hattori:2016}, and can lead to the formation of
broad fans of tidal debris at the edges of a stream \citep{pearson:2017}.
This effect is even more pronounced for streams within the Galactic disk
\citep{thomas:2023}.
Spiral arms in the Galactic disk can similarly impact streams that pass through the disk
midplane, as they can impose time-varying dipole or quadrupole components to the gravitational
potential.
Both of these effects generally lead to perturbations to stream structure on spatial
scales comparable to the length of a stream (i.e. these effects can cause global changes
to the overall structure of a stream, or create large-scale density variations along the
stream track).

Interactions between streams and massive dwarf galaxies around the Milky Way can also
have significant effects on the global structure of streams.
For example, strong gravitational encounters between a stream and a dwarf galaxy can
cause a stream to ``fold'' or develop strong asymmetries between the two tidal tails, as
has been shown in the context of a GD-1-like stream interacting with a massive
Sagittarius-like dwarf galaxy halo \citep{dillamore:2022}.
Large-scale interactions between the LMC and streams can cause subtler morphological
changes to streams that are only observable in the context of the kinematics of stars
along a stream.
For example, prior to \gaia\ \dr{2}, it was hypothesized that the misalignment between
the elongation of the Tuc 3 stream and its orbital path was due to an interaction with
the LMC \citep{erkal:2018}, which made a clear prediction that the proper motions of the
stream stars should also be misaligned with the stream track.
This effect was later observed in the proper motion track of the Orphan--Chenab stream
\citep{koposov:2019, erkal:2019}.

Just before the discovery of the first long globular cluster stream from the Pal 5
cluster, it was realized that thin tidal streams around the Milky Way would be powerful
tools for studying the abundance and structure of low-mass dark matter subhalos
\citep{ibata:2002,johnston:2002}.
For a close, impulsive encounter between a stream and a dark matter subhalo (i.e. when
the impact parameter is comparable to the scale radius of the subhalo), the velocity
kick imparted to the stream stars can be much larger than the local velocity dispersion
of the stream stars.
As a rough demonstration of this, the local velocity dispersion of a section of a
globular cluster stream can be $\sigma_v \ll 1~\kms$.
The velocity kick $\Delta v$ imparted by a dark matter subhalo impact with a small
impact parameter --- assuming an NFW density profile subhalo and a concentration set by
its mass according to a subhalo mass--concentration relation \citep{moline:2017} ---
scales with
\begin{equation}
    \Delta v \approx 1~\kms \,
        \left(\frac{M_{\rm subhalo}}{10^7~{\rm M}_\odot}\right)^{2/3}\,
        \left(\frac{v}{100~{\rm km}~{\rm s}^{-1}}\right)^{-1}
\end{equation}
where $v$ is the relative velocity between the stream stars at the impact site and the
subhalo \citep[see also][]{erkal:2015a, sanders:2016}.
This impulse can therefore create a ``gap'' or under-density in the stream that is
potentially observable.
The resulting stream density variations can be even more striking if the subhalos are
denser or more massive \citep[see, e.g.,][]{yoon:2011, bonaca:2014}.
Encounters with dark matter subhalos in the mass range $10^6$--$10^8~\unit{\msun}$ are
therefore detectable with stellar streams.

\subsection{Observed Structures: A Few Case Studies}
\label{sec:structure-obs}

All of the streams that have been studied in detail show evidence for a complex
dynamical history in the form of density features along and around these streams.
As with above, we mainly focus here on the low-mass (likely globular-cluster origin)
streams, but even the Sagittarius stream shows a clear bifurcation in the density of
stars along the extent of the stream (see, e.g., Figure~1 in \citealt{belokurov:2006})
with several possible physical explanations \citep[e.g.,][]{penarrubia:2010,
vasiliev:2020, vasiliev:2021}.
The streams below are some of the best studied streams around the Milky Way, but
theoretical models built on the mechanisms discussed in
Section~\ref{sec:structure-theory} still struggle to quantitatively explain the observed
complexity of these streams.

\begin{figure*}[t!]
    \centering
    \includegraphics[width=\textwidth]{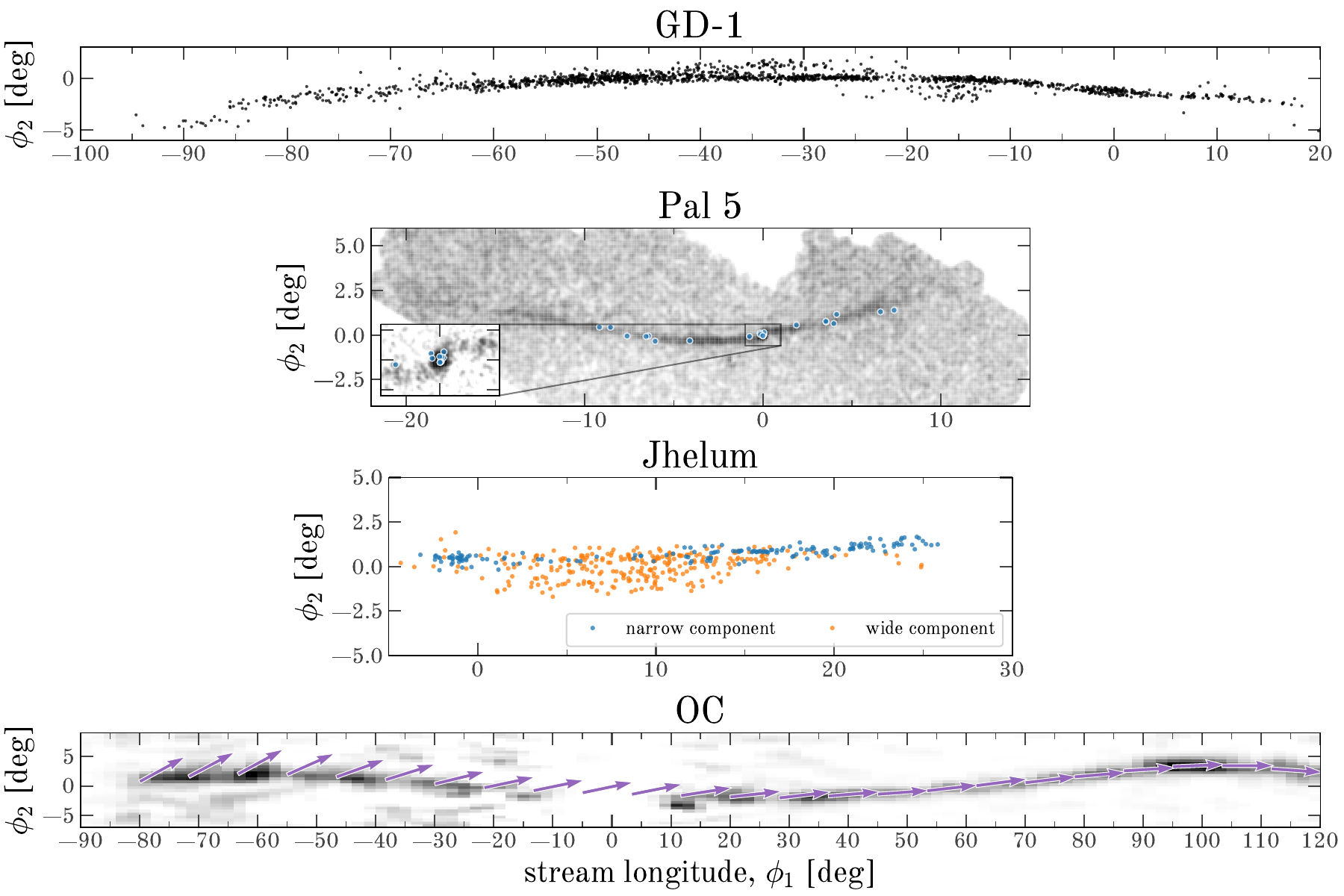}
    \caption{
        Four observed streams around the Milky Way with complex density structures,
        shown in sky coordinates aligned with each stream, where $\phi_1$ is a
        stream-aligned longitude and $\phi_2$ a stream-aligned latitude coordinate.
        \textbf{Top panel:} The GD-1 stream, which shows a clear spur of stars that
        extends from the main stream body between $\phi_1 \sim -40^\circ$ to $-25^\circ$.
        The stream stars shown are taken from high-probability members determined using
        a flexible membership model that uses astrometry from \gaia\ \dr{3} and
        photometry from PS1 (reproduced from \citealt{tavangar:2024}).
        \textbf{Middle top panel:} The Palomar 5 (Pal 5) stream, shown in stellar
        density of stars selected from DECam photometry of the stream (reproduced from
        \citealt{bonaca:2019}).
        Blue circle markers show the locations of RR Lyrae-type stars from \gaia\ \dr{2}
        that have been associated with the stream \citep{price-whelan:2019}.
        \textbf{Middle bottom panel:} High-probability members of the Jhelum stream,
        separated into members of the ``narrow'' and ``wide'' components of the stream
        (reproduced from \citealt{awad:2024}).
        \textbf{Bottom panel:} The Orphan--Chenab (OC) stream, shown in stellar density
        (under-plotted grayscale image) of member stars selected from \gaia\ \dr{3}
        (reproduced from \citealt{koposov:2023}).
        The over-plotted purple arrows show the direction of the solar reflex-corrected
        proper motion vectors in bins of stars along the stream.
        The proper motion vectors are expected to point along the stream track, but
        clearly deviate from the track for $\phi_1 \lesssim 10^\circ$.
    }
    \label{fig:four-obs-streams}
\end{figure*}

\subsubsection{GD-1}
\label{sec:gd1}

As discussed above (Section~\ref{sec:discovery}), the GD-1 stream was discovered in 2006
using photometry from the SDSS \citep{grillmair:2006-gd1}.
GD-1 is a long, thin stream (see Figure~\ref{fig:gd1-demo}) that is thought to be the
result of the complete tidal disruption of a globular cluster \citep{koposov:2010}.
Already in the discovery paper it was noted that the stream displays significant density
variations along the stream track \citep{grillmair:2006-gd1}.
The stream was later studied in detail using deeper photometry from a survey using the
Canada--France--Hawaii Telescope/Megacam, which revealed a number of over- and
under-densities along the stream, as well as deviations to the stream track
on the sky \citep{deboer:2018}.

Data from \gaia\ \dr{2} has enabled a much more detailed look at the density structure
of the GD-1 stream.
For example, Figure~\ref{fig:gd1-demo} (bottom right panel) shows the sky positions of
stars selected from \gaia\ \dr{3} using a recreation of the proper motion and
photometry-based selections used in \citep{price-whelan:2018} to select likely members
of the GD-1 stream.
The stream is clearly visible in the sky positions of individual stars, which revealed
the depth of the density variations and, for the first time, the presence of a ``spur''
of stars that extends from the main stream track at $\phi_1 \sim -50^\circ$
(\citealt{price-whelan:2018}; see also Figure~1 of \citealt{bonaca:2019b}).
Spectroscopic follow-up of stars in the spur revealed that it is co-moving with the main
stream body \citep{bonaca:2020b}.

Models of the GD-1 stream have shown that the spur is consistent with a perturbation
from a low-mass dark matter subhalo \citep{bonaca:2019, bonaca:2020b}.
However, further observations and modeling of the stream are needed to fully confirm
this interpretation.
For one, GD-1 is also thought to be surrounded by a ``cocoon'' of stars
\citep{malhan:2019a}, which could point to a more complex formation history of the
stream.
If the spur formed from a perturbation, it could have been from an interaction with a
more massive object (e.g., Sagittarius; \citealt{dillamore:2022}) or from a known object
(e.g., another globular cluster).
At the present quality of the data, these scenarios are already unlikely, but more
precise proper motions of individual stars in the spur and stream combined with more
extensive spectroscopic follow-up would help to confirm the origin of the spur and the
cocoon.
Figure~\ref{fig:four-obs-streams} (top panel) shows a more robust look at the GD-1
stream, highlighting the existence of the ``spur'' between $-40^\circ \lesssim \phi_1
\lesssim -25^\circ$: The stars shown are high-probability members of the GD-1 stream
selected using a flexible membership model that uses astrometry from \gaia\ \dr{3} and
photometry from PS1 \citep{tavangar:2024}.

\subsubsection{Ophiuchus}
\label{sec:ophiuchus}

The Ophiuchus stream is a low-mass stream that orbits in the inner galaxy ($4~\kpc
\lesssim r \lesssim 14~\kpc$), and is notable because it is unexpectedly short (length
$\sim 3~\kpc$) given the expected age of the stream and its orbit \citep{bernard:2014,
sesar:2015}.
The stream is thought to be extended almost in the line-of-sight direction
\citep{sesar:2016}, and may be associated with lower surface-density ``fans'' of tidal
debris.
Given its proximity to the inner galaxy and the time-dependent impact of the Galactic
bar, it is thought that the peculiar structure of the Ophiuchus stream may be due to
interactions with the bar or strong chaos induced by the bar \citep{sesar:2016,
price-whelan:2016b}.

Using data from \gaia\ \dr{2} combined with spectroscopic follow-up,
\citet{caldwell:2020} showed that the Ophiuchus stream extends several degrees further
than previously reported, and that there is a potential off-track, ``spur'' feature
associated with this stream as well.
New constraints on the orbit of Ophiuchus from \citet{caldwell:2020} still supports the
idea that the Galactic bar has significantly impacted the structure of the stream.
An alternate interpretation of the history of the Ophiuchus stream is that it formed
recently, and was only recently kicked onto its present-day orbit after a strong
encounter with the Sagittarius dwarf galaxy \citep{lane:2020}.

\subsubsection{Palomar 5 (Pal 5)}
\label{sec:pal5}

The Palomar 5 stream (Pal 5) was the first globular-cluster mass stream discovered
around the Milky Way using early imaging data from the SDSS \citep{odenkirchen:2001,
rockosi:2002, grillmair:2006-pal5}.
Pal 5 serves as an archetype for globular cluster streams, as it is long ($\sim
7~\kpc$), thin (width standard deviation $\sigma \sim 4~\unit{\parsec}$), has a
relatively high surface brightness, still connects to its progenitor system, and has
been studied in detail since its discovery with both targeted photometric observations
\citep{ibata:2016, bonaca:2020} and spectroscopic follow-up \citep{kuzma:2015,
ishigaki:2016, ibata:2017, kuzma:2022}.

The trailing tail of the Pal 5 stream shows an interesting pattern of small-scale
($\sim$degree-scale) density variations that have been modeled as ``gaps'' resulting
from interactions with dark matter subhalos \citep{carlberg:2012, erkal:2017}.
However, variable photometric completeness and dust extinction over the length of the
stream have made it difficult to confirm the presence of these gaps with photometric
data alone.
It has also been suggested that gaps in the Pal 5 stream could instead be formed from
interactions between the stream and the Galactic bar \citep{pearson:2017}.
Confirming and characterizing the nature of these small-scale density variations will be
of great importance for understanding the population of dark matter subhalos around the
Milky Way.

Even if small-scale features of Pal 5 require further study to confirm, the more global
properties of the stream have been well characterized and are peculiar.
For example, the two tails of the Pal 5 stream are not symmetric in the sky (the leading
tail is significantly shorter than the trailing tail).
Recent deep photometry of the stream from the Dark Energy Camera (DECam;
\citealt{flaugher:2015}) has provided a more detailed look at the global density
structure of the Pal 5 stream \citep{bonaca:2020}.
This has shown that the stream is asymmetric for the fainter parts of the stream (even
taking into account the possible extension of the stream reported in
\citealt{starkman:2020}), but is also asymmetric in the densest part of the stream near
the progenitor system \citep[see Figure 2 of][]{bonaca:2020}.
The precisely-calibrated photometry from DECam has also enabled looking at lower
surface-brightness parts of the stream, such as the ends of the leading and trailing
tails.
While the stream is no longer detected as a coherent structure in the sky positions of
stars, stream stars are detected in a wide region around the end of the leading tail
(refereed to as the fanned part of the stream, or the ``fan''; \citealt{bonaca:2020}).
This abrupt change in the width and surface density of the stream is another potential
indicator of a complex orbital structure \citep{pearson:2015, price-whelan:2016a} or
strong interactions with the Galactic bar \citep{pearson:2017, bonaca:2020}.

The Pal 5 stream is just far enough away ($d = 20.6~\kpc$; \citealt{price-whelan:2019})
such that its main sequence turn-off stars are near the faint limit of \gaia, so the
bulk of the stream's stellar population is not well resolved in \gaia\ data.
However, the stream has been detected in brighter and rarer tracers such as horizontal
branch stars, RR Lyrae-type stars, and red giant branch stars.
Using a sample of RR Lyrae stars from \gaia\ and the Pan-STARRS1 survey,
\citet{price-whelan:2019} measured the distance to the Pal 5 cluster and inferred sky,
distance, and proper-motion tracks for the stream.
This data showed that the cluster is actually about $\approx 10\%$ closer than has been
traditionally assumed based on isochrone fitting to the cluster without modeling the
dust extinction to the cluster.
The RR Lyrae population also revealed an unexpected partitioning of RR Lyrae oscillator
type between the cluster and the tails:
Most (about two-thirds) of the RR$c$-type oscillators are found in the cluster remnant
itself, whereas all but two of the RR$ab$-type oscillators are found in the stream.
Figure~\ref{fig:four-obs-streams}, middle top panel, shows the density of stars selected
to be associated with the Pal 5 stream (based on photometry) as the grayscale image, and
the locations of RR Lyrae-type stars from \gaia\ \dr{2} as blue circle markers
\citep{price-whelan:2019}.

\subsubsection{Jhelum}
\label{sec:jhelum}

The Jhelum stream \citep{shipp:2018} was discovered using photometry from the DES and
later confirmed using proper motion measurements from \gaia\ \dr{2} \citep{shipp:2019}.
It was recognized early on that the morphology of the stream is unusual, in that there
seems to be a denser, thinner portion of the stream overlapping with a wider, more
spatially uniform, and lower-density stream \citep{bonaca:2019}.
Jhelum has a measurable spread in element abundances \citep{ji:2020, li:2022} and is
therefore thought to be the remnant of a dwarf galaxy \citep{sheffield:2021}.
The unusual morphology of the stream is still not fully understood, but several possible
scenarios have been proposed to explain the coincident thin and thick components of the
stream \citep[see, e.g., the discussion Section~5 of ][]{bonaca:2019}.
For example, the stream is likely on a very radial orbit (eccentricity $e \sim 0.5$) in
with a relatively small pericenter ($\rper \sim 8~\kpc$): If this extreme orbit is the
main driver of the structure, the density structure could be a result of differential
orbital precession of the stream star orbits \citep{erkal:2016b}, or the result of a
fold caustic in the phase-space density \citep{tremaine:1999}.
It is also possible that the Galactic bar has impacted the stream through a similar
mechanism to that proposed for the Ophiuchus stream \citep{price-whelan:2016b} or Pal 5
\citep{pearson:2017}.
The orbit of the stream also brings it close to the orbit of the Sagittarius dwarf
galaxy, suggesting that the stream could have been perturbed by the Sagittarius dwarf
galaxy \citep{woudenberg:2023}.

Another possible explanation for the two components of the Jhelum stream is that the
stream formed from the joint disruption of a globular cluster and a dwarf galaxy
\citep{bonaca:2019}.
A recent characterization of the two components of the stream shows that the thin and
wide components of the stream have different velocity dispersions and element abundance
spreads \citep{awad:2024}, which is consistent with this scenario.
Figure~\ref{fig:four-obs-streams}, middle bottom panel, shows the high-probability
members of the narrow and wide components of the Jhelum stream.

\subsubsection{Orphan--Chenab (OC)}
\label{sec:oc}

The Orphan stream was discovered using photometry from the SDSS in the northern sky as a
long ($50^\circ$), bright stream with no obvious remnant progenitor system
\citep{belokurov:2006, grillmair:2006-orphan, belokurov:2007}.
To this date, no definitive association has been made between the stream and any bound
dwarf galaxies or stellar systems around the Galaxy \citep{fellhauer:2007, newberg:2010,
casey:2013, casey:2014, grillmair:2015}, suggesting that it is a fully-disrupted
remnant.
The stream remains bright as it approaches the southernmost extent of the SDSS survey
footprint, and early $N$-body simulations of the stream suggested that the stream should
extend into the southern sky \citep[e.g.,][]{sales:2008}.
First evidence for this component of the stream --- which lies on the other side of the
Galactic plane from the originally detected footprint of the Orphan stream --- was
reported using a dedicated survey of the extrapolated stream footprint with the Dark
Energy Camera \citep{grillmair:2015}.
The distance track along the northern extent of the stream was measured using RR
Lyrae-type stars, which found a steep distance gradient: the nearest part of the
detected northern Orphan stream is $\sim$22~\kpc\ and the farthest is $\sim$50~\kpc\
\citep{sesar:2013, hendel:2018}.

The Chenab stream was discovered in the DES footprint as a short (in angular size),
distant stream with a large physical width ($\sigma \sim 500~\unit{\parsec}$)
\citep{shipp:2018}. Based on its width, it was expected that this stream was formed from
a dwarf galaxy. The mean distance of the stream was estimated to be $\sim$40~\kpc.

Using RR Lyrae-type stars released as a part of \gaia\ \dr{2}, \citet{koposov:2019}
showed that the Orphan and Chenab streams were actually physically and kinematically
connected; the stream has thus become known as the Orphan--Chenab or OC stream
\citep[see also][]{fardal:2019}.
For example, the 3D positions of RR Lyrae-type stars kinematically selected to be near
the Orphan stream show a continuous over-density from the northern Galactic sky to the
southern (see Figure~1 in \citealt{koposov:2019}).

Unified, the OC stream is the second longest and second most massive stream around the
Milky Way (second to the Sagittarius stream), spanning over 100~\kpc\ in length.
Recent work using spectroscopic data from the S5 and APOGEE surveys have shown that the
progenitor of the OC stream was likely a low-mass dwarf galaxy, with a stellar mass
between $10^5$--$10^6~\unit{\msun}$ \citep{koposov:2023, hawkins:2023}, in line with
estimates from reconstructing the stellar population structure of the progenitor from
the stream \citep{mendelsohn:2022} and from its horizontal branch population \citep{prudil:2021}.

Combining precise distance estimates to RR Lyrae stars in the OC stream (exploiting the
fact that RR Lyrae are standard candles) with precise proper motions from \gaia,
\citet{koposov:2019} also showed that the orbital poles of stars in the stream drift
continuously from the northern portion of the stream (Orphan) to the southern portion
(Chenab).
See, for example, the bottom panel of Figure~\ref{fig:four-obs-streams} for another
demonstration of this: The solar reflex-corrected proper motion vectors of stars in the
stream deviate from the stream track for $\phi_1 \lesssim 10^\circ$.
In follow-up work, this kinematic misalignment was later shown to plausibly come from a
direct encounter between the stream and the Large Magellanic Cloud \citep{erkal:2019}.
The OC stream is thought to also be sensitive to details of the ongoing merger
between the LMC and the Milky Way, and therefore is a powerful object for constraining
the large-scale behavior of dark matter in the Milky Way \citep{lilleengen:2023}.

\subsection{Key opportunity: Precision kinematics to determine the origin of stream perturbations}
\label{sec:xkinematics}
Different sources of stream perturbations often produce similar morphological features.
As a result, none of the numerous features detected in the Milky Way streams have been unambiguously associated with a perturbation mechanism.
Here we discuss precision kinematics as a way forward, and illustrate how they could distinguish between two scenarios for an origin of a stream gap: a dissolved progenitor and a dark-matter subhalo impact.

\begin{figure}[t!]
\begin{center}
\includegraphics[width=\columnwidth]{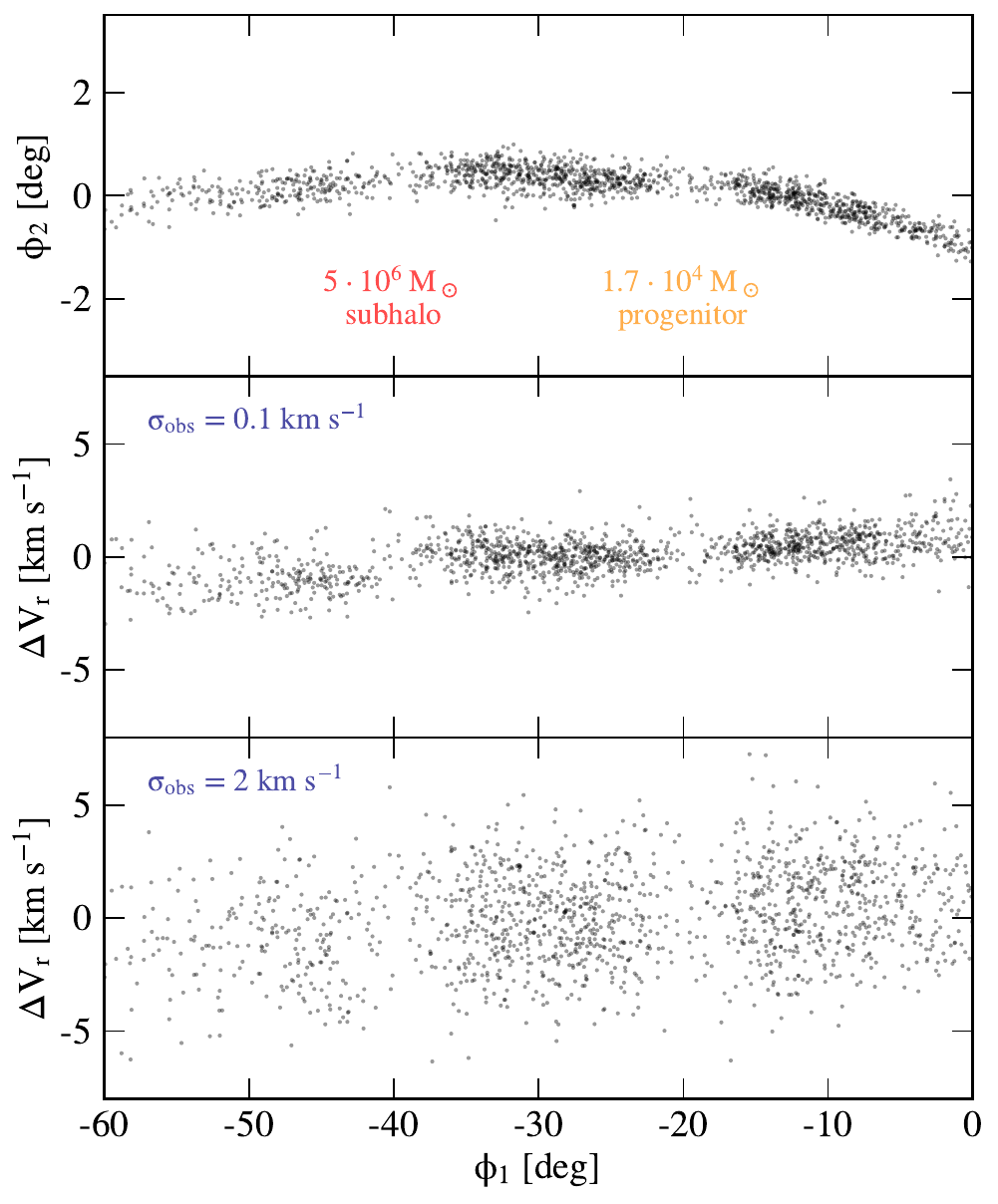}
\end{center}
\caption{%
Precise radial velocities can distinguish the origins of stream gaps.
\textbf{Top:} Sky positions of a GD-1 like stream observed with \gaia-like data.
The stream's progenitor dissolved 350\,Myr ago and produced a gap at $\phi_1\approx-20^\circ$, while the gap at $\phi_1\approx-40^\circ$ was produced by an impact of a $5\cdot10^6\,\unit{\msun}$ dark-matter subhalo 500\,Myr ago.
\textbf{Middle:} Radial velocities relative to the orbital radial velocity trend, assuming an observational uncertainty of $0.1\,\unit{\kms}$.
At this precision, the velocity profiles along the two gaps are markedly different.
\textbf{Bottom:} Radial velocities relative to the orbital radial velocity trend, assuming an observational uncertainty of $2\,\unit{\kms}$.
At this precision, the progenitor and subhalo signatures are nearly indistinguishable.
}
\label{fig:rv_signal}
\end{figure}

Figure~\ref{fig:rv_signal} shows particle-spray models of a GD-1-like stream, with the gap at $\phi_1\approx-20^\circ$ being the site where its progenitor dissolved $350\,\unit{Myr}$ ago, and the gap at $\phi_1\approx-40^\circ$ due to a $5\times10^6\,\unit{\msun}$ dark-matter subhalo fly-by $495\,\unit{Myr}$ ago.
This configuration was chosen because it produces two gaps similar to each other in sky positions (top), and also similar to the observed GD-1 \citep[e.g.,][]{price-whelan:2018}.
The middle row shows radial velocities relative to the orbital trend, convolved with a measurement uncertainty of $0.1\,\unit{\kms}$, which is comparable to the intrinsic dispersion in the outer regions of the low-mass globular clusters \citep{baumgardt:2019}, and therefore a fundamental limit for resolving velocity structures in the resulting streams.
Radial velocities along the progenitor-induced gap are similar to the orbital trend, with only slight and monotonic offsets, whereas the subhalo impact introduces non-monotonic features.
Relative radial velocities in the bottom row are convolved with a measurement uncertainty of $2\,\unit{\kms}$, typical for current wide-field spectroscopy of distant halo stars \citep[e.g.,][]{li:2019, cooper:2023}.
In this case, the non-monotonic aspect of the subhalo-induced gap is only marginally resolved due to high scatter.

This investigation suggests that the origins of stream gaps can be discovered using radial velocities measured to a precision of $\approx0.1\,\unit{\kms}$ \citep[similar to forecasts in][]{li:2019b}.
More extensive theoretical work is needed to determine subhalo detectability in the kinematic space as a function of the subhalo mass, size, impact geometry and time, as well as to identify kinematic signatures that are unique to subhalo impacts and differentiate them from other perturbation mechanisms.
On the observational side, reaching a precision of $0.1\,\unit{\kms}$ will require developing novel instrumentation that improves upon the current capabilities of wide-field, high-throughput spectrographs by an order of magnitude.
In the next section we discuss how to combine these efforts and place definitive constraints on the \lcdm\ cosmology.

%%%%%%%%%%%%%%%%%
\section{Outlook}
\label{sec:outlook}

Simple by nature, stellar streams have been excellent test beds for theories of galaxy formation and cosmology.
Their mass function indicates globular clusters are a regular product of clustered star formation (\S\ref{sec:orbits}).
Their kinematics reveal massive dark matter halos around both the Milky Way and the Large Magellanic Cloud (\S\ref{sec:orbits}).
Their mere presence confirms a basic tenet of cold dark matter cosmologies that galaxies grow at least in part hierarchically (\S\ref{sec:discovery}).
We expect that all of these aspects will be further developed in the future, but perhaps the most exciting prospect for streams is constraining the nature of dark matter (\S\ref{sec:structure}).

To that end, we propose that the goal for the next decade should be an unambiguous association of a stream perturbation with a dark subhalo ($\lesssim10^6\,\unit{\msun}$).
Such a detection would confirm the prediction of the \lcdm\ model that halos exist below the mass threshold for galaxy formation \citep{bullock:2000, benson:2002, somerville:2002, benitez-llambay:2020}, and resolve the long-standing missing satellites problem \citep{klypin:1999}.
In this section we describe priorities for transforming streams into an ultimate test of dark matter.

\subsection{Data}
\textbf{Discovering streams in the outer halo.} Despite the already large number of known streams, there is good reason to believe more remain to be discovered.
Observations of young massive clusters \citep[e.g.,][]{portegies-zwart:2010}, as well as recent models reproducing population properties of globular clusters indicate that clusters form with initial masses drawn from a simple power-law distribution \citep[e.g.,][]{choksi:2018,chen:2023}.
In this scenario, thousands of lower-mass clusters should be dissolved in a Milky Way-like galaxy, a fraction of which may be still coherent stellar streams.
This scenario is further supported by the typical mass of stellar streams ($\approx10^4\,\unit{\msun}$, see Table~\ref{tbl:stream-summary}) being lower than that of globular clusters \citep[$\approx10^5\,\unit{\msun}$,][]{baumgardt:2018}.
The outer halo is an especially important region to search for new streams because it is dark matter dominated.
Currently 70 of the 133 known thin streams reside within the extent of the stellar disk ($\lesssim15\,\unit{kpc}$, Table~\ref{tbl:stream-summary}), where baryons dominate the gravitational potential \citep[e.g.,][]{mcmillan:2017}.
Streams at distances beyond $\gtrsim20\,\unit{kpc}$ have largely eluded our detection techniques so far.
To illustrate, a GD-1-like, $10^4\,\unit{\msun}$ stream at a distance of 50\,\unit{kpc} would have a parallax of 0.02\,\unit{mas}, proper motions of $\lesssim1\,\unit{\masyr}$, and its main-sequence turn-off would be 5\,\unit{mag} fainter---pushing the stream outside of current detection capabilities.
However, the upcoming fourth and fifth data releases from the \gaia\ mission (expected in 2025 and 2030) will deliver proper motions with uncertainties less than $<0.45\,\unit{\masyr}$ and $<0.16\,\unit{\masyr}$, respectively.\footnote{\url{https://www.cosmos.esa.int/web/gaia/science-performance}}
Furthermore, the ground-based Vera Rubin Observatory's Large Survey of Space and Time (LSST, starting in 2026) will provide deep multi-band imaging capable of reaching $r\approx26.9\,\unit{mag}$ \citep[corresponding to the main-sequence turn-off at $\approx300\,\unit{kpc}$]{ivezic:2008}.
Space-based wide-field telescopes Euclid \citep[launched in 2023]{laureijs:2011} and Roman \citep[to launch in 2027]{spergel:2013} will both survey thousands of square degrees at high Galactic latitudes and reach $J\approx26.9\,\unit{mag}$ (turn-off at $\approx500\,\unit{kpc}$).

\textbf{Spectroscopic follow-up on the industrial scale.}
The wide and uniform availability of \gaia\ proper motions has transformed the field of streams: discovered new streams, their orbits, and complex structure.
However, additional spectroscopy is needed to measure radial velocities and determine which dynamical processes were at play in shaping individual streams, from revealing the wide envelopes of earlier dissolution episodes (\S\ref{sec:phasespace}) to uncovering the origin of stream gaps (\S\ref{sec:xkinematics}).
Many spectroscopic surveys of the Milky Way have been conducted in the past couple of decades, but they obtained a fairly small number of stream members because they were either limited to bright targets (e.g., APOGEE, \citealt{majewski:2017}; RAVE, \citealt{steinmetz:2020}; GALAH, \citealt{buder:2021}; \gaia, \citealt{katz:2023}), and/or employed a sparse targeting strategy that often missed the narrow streams' profiles (e.g., SEGUE, \citealt{yanny:2009}; H3, \citealt{conroy:2019}).
On the other hand, dedicated programs using wide-field, heavily multiplexed spectrographs can efficiently target stream members \citep[e.g.,][]{li:2021}.
Excitingly, several such current and soon-operational facilities have a dedicated streams program, including DESI \citep[Dark Energy Spectroscopic Instrument,][]{cooper:2023}, WEAVE \citep[William Herschel Telescope Enhanced Area Velocity Explorer,][]{jin:2023}, 4MOST \citep[4-metre Multi-Object Spectroscopic Telescope,][]{dejong:2019}, and PFS \citep[Prime Focus Spectrograph,][]{takada:2014}.
In addition, the Via Project\footnote{\url{https://via-project.org/}} is currently building twin high-resolution spectrographs for the 6.5\,\unit{m} MMT and Magellan telescopes that can efficiently survey the halo at $\lesssim0.1\,\unit{\kms}$ precision (first light expected in early 2027).
On a decade time-scale, the Mauna Kea Spectroscopic Explorer (MSE) has been proposed as a new 11.25\,\unit{m} facility, designed in part as the ultimate high-resolution follow-up of \gaia\ targets \citep{mse:2019}.
The ideal outcome of these projects would be a complete (magnitude limited) census of member stars with precise 6D kinematics and multi-element chemical abundances for all of the Milky Way streams.

\textbf{Community repository for stream data.}
The wide variety of dynamical processes affect streams on different orbits to a different extent (\S\ref{sec:structure-theory}), so the key to disentangling signatures of dark-matter subhalos will be simultaneous modeling of the entire population of streams.
So far, such studies were limited by the non-uniformity of stream observations, largely stemming from a range of selection criteria employed to identify stream members in a range of data sets.
The uniform, all-sky coverage of the \gaia\ data has for the first time allowed uniform studies across dozens of streams \citep[e.g.,][]{bonaca:2021,malhan:2022,ibata:2023}, as well as launched a community-based effort to develop procedures for uniformly identifying member stars of all known streams and produce the Community Atlas of Tidal Streams (CATS\footnote{\url{https://github.com/stellarstreams/cats}}).
Astronomy as a field has a long tradition of open catalogs (e.g., published through VizieR, \citealt{ochsenbein:2000}), and large national facilities increasingly provide support for curating community data products (e.g., through NOIRLab Data Lab, \citealt{nikutta:2020}; STScI MAST\footnote{\url{https://archive.stsci.edu/new-mission-partnerships-with-mast}}).
Specialized central repositories for both the raw observational data and derived parameters have already been developed in several research areas (e.g., the NASA Exoplanet Archive, \citealt{akeson:2013}; the Weizmann Interactive Supernova Data Repository, \citealt{yaron:2012}).
An open repository with homogeneous stream membership, phase-space information, as well as the measured width, density, and velocity profiles of streams will be a quintessential resource for constraining the small-scale structure of the Milky Way dark matter halo.
Such a repository would also facilitate coordination in follow-up of individual streams across different projects, allow the broader scientific community to engage with the stream data without the need to repeat lower-level reduction and analysis of observational data sets, and finally, bring high-fidelity stream data to the public for use in both education and outreach.

\subsection{Models}
\textbf{High-resolution simulations in different dark matter cosmologies.}
A series of N-body zoom-in simulations have shown that cold dark matter forms halos across 20 orders of magnitude, down to $10^{-6}\,\unit{\msun}$ \citep[approximately Earth mass,][]{diemand:2005, springel:2008, wangj:2020}.
Streams are sensitive to subhalos as low in mass as $10^5\,\unit{\msun}$ \citep[e.g.,][]{drlica-wagner:2019}, however simulations of Milky Way-mass halos confidently resolve subhalos only down to $\gtrsim10^7\,\unit{\msun}$ \citep[cf.][]{wetzel:2016, richings:2020, nadler:2023, mansfield:2023}.
Semi-analytic approaches can easily reach $10^5\,\unit{\msun}$ limit \citep[][]{press:1974,sheth:2002,benson:2012}, but they lack phase information for the subhalo orbits, and are therefore best suited for forecasting average properties or subhalos in relaxed halos.
The Milky Way is currently in significant disequilibrium due to an undergoing merger with the LMC \citep[e.g.,][]{besla:2007,kallivayalil:2013,vasiliev:2024}, which is expected to result in large variations of subhalo impacts on streams with different orbits \citep{barry:2023,arora:2023}.
Therefore, producing high-resolution cosmological simulations of Milky Way-like halos that resolve $10^5\,\unit{\msun}$ dark-matter subhalos is the top priority for an accurate interpretation of stream gaps.
Ideally, this effort would produce a suite of models with multiple realizations of a Milky Way-like galaxy to control for uncertainties in the Milky Way's accretion history, and for a given set of initial conditions, span a range of dark matter models beyond CDM (e.g., warm dark matter, WDM, \citealt{bode:2001}; self-interacting dark matter, SIDM, \citealt{spergel:2000}, or fuzzy dark matter, FDM, \citealt{hu:2000}).

\textbf{Stream evolution in tailored simulations of the Milky Way.}
Cosmological simulations of Milky Way-like galaxies show that the detailed structure of its stellar and dark matter halos depends sensitively on the mass accretion history \citep[e.g.,][]{bullock:2005,springel:2008,cooper:2010,hopkins:2018,bose:2020,pillepich:2023}.
There is growing observational evidence for perturbations of the Milky Way streams by the two most massive recent accretion events: the Sagittarius dwarf galaxy \citep[e.g.,][]{bonaca:2020,li:2021,woudenberg:2023} and the Large Magellanic Cloud \citep[e.g.,][]{erkal:2019, shipp:2019, lilleengen:2023, koposov:2023}.
Simulations predict that such massive mergers also increase the abundance of low-mass dark-matter subhalos, which affects their impact rates on streams \citep[e.g.,][]{barry:2023, arora:2023}.
In order to accurately infer the nature of dark matter from subhalo impacts on streams, we need to account for the Milky Way's accretion history.
Thanks to \gaia, the major accretion events onto the Milky Way have been identified \citep[e.g.,][and references therein]{kruijssen:2020}.
Most of these events have been reconstructed through idealized N-body models (e.g., GSE, \citealt{naidu:2021}; Sagittarius, \citealt{vasiliev:2021}; Cetus \citealt{chang:2020}; Helmi Streams, \citealt{koppelman:2019}; Wukong, \citealt{malhan:2021b}; Kraken, Sequoia, Thamnos, and I'itoi, \citealt{sharpe:2024}), while cosmological N-body simulations of Milky Way-like halos are now increasingly constrained to match the mass and time of the most significant mergers \citep[e.g.,][]{nadler:2020,buch:2024}.
In the coming years, we will need to combine these efforts to forecast subhalo encounter rates tailored to individual streams in the Milky Way.

Developing tailored simulations is important not only for accurately estimating the amount of subhalo perturbation each stream experiences, but also to identify effects that may confound signatures of dark matter.
These include matching the stream progenitor mass and orbital phase (to capture epicyclic overdensities), its host galaxy (to capture the possible presence of low-density cocoons), and its orbit in a time-evolving potential (to estimate the likelihood of baryonic perturbations from molecular clouds, spiral arms, Galactic bar and luminous satellites).
A self-consistent model of a stellar stream that takes into account all of these processes will require novel ways of solving multi-scale gravitational dynamics.
Promising progress has recently been made using basis function expansions to capture the structure and evolution of the gravitational potential on large scales \citep{garavito-camargo:2021,petersen:2022}, to speed up N-body computations using machine learning techniques \citep{breen:2020,villaescusa-navarro:2021}, and to approximate direct N-body dynamical models of globular clusters to high accuracy with Monte Carlo codes \citep{rodriguez:2015,kremer:2020}.

\subsection{Inference}
\textbf{Modeling individual subhalo fly-bys.}
Stellar streams in the Milky Way halo provide a unique opportunity to study individual dark-matter subhalos in exquisite detail.
Reconstruction of a subhalo impact site can constrain the subhalo mass and density profile \citep{erkal:2015b, bonaca:2019, hilmi:2024}.
With precise kinematics, it may even be possible to constrain the subhalo's orbit and current 3D position, which would allow targeted multi-messenger studies of the dark sector \citep{bonaca:2020, mirabal:2021}.
However, theoretical models currently struggle to quantitatively explain the observed complexity of stellar streams.
On the positive side, the wealth and precision of observational data promises to disentangle numerous processes that affected stellar streams.
But to deliver on this promise, theoretical models need to include all of the relevant effects; otherwise the inference becomes ineffectual \citep[e.g., the power-spectrum of the GD-1 density can be separately reproduced both by internal epicyclic overdensities and external impacts of dark-matter subhalos, molecular clouds and spiral arms, cf.][]{ibata:2020, banik:2021}.
As discussed throughout this review, stellar streams are affected by gravitational dynamics over a wide range of temporal and spatial scales, so the challenge for the next decade will be devising fast theoretical models and inference techniques that can efficiently explore this vast parameter space.

\textbf{Modeling multiple streams.}
Given the large number of known stellar streams, an alternative approach to constraining the amount of dark matter substructure in the Milky Way is from its cumulative effect on the population of stellar streams.
This statistical approach is particularly well-suited to infer the overall properties of dark-matter subhalos, such as their mass function and radial distribution \citep[e.g.,][]{carlberg:2013,banik:2021b}.
Because the expected mass function and radial profile of dark-matter subhalos are different from those of baryonic sources of perturbation, this approach complements modeling of individual stream impact sites and can break its degeneracies.
The predominant summary statistic of the subhalo impacts on a stream so far has been the power spectrum of the 1D density variations along the stream, which allows for easy extensions to multiple streams \citep[e.g.,][]{banik:2021, banik:2021b}.
However this metric cannot easily distinguish between a subhalo and other origins of density variations \citep[e.g., epicycles,][]{ibata:2020}.
On the other hand, higher-order statistics like the bispectrum, and cross-correlations with spectroscopic and astrometric data can uniquely identify subhalo perturbations \citep[e.g.,][]{bovy:2017}.
Further improvements may be possible by modeling the 2D structure of tidal streams, possibly using machine learning techniques like the wavelet scattering transforms \citep{mallat:2012}, which have recently been demonstrated as an efficient technique to infer cosmological parameters from the weak-lensing maps \citep{cheng:2020} and magnetohydrodynamic properties of turbulence in molecular clouds \citep{saydjari:2021}.

\vspace{0.5cm}

This imminent collection of data and methods will provide a conclusive answer to whether $10^6\,\unit{\msun}$ dark-matter subhalos exist in the Milky Way.
If all of the stream perturbations can be explained by more massive subhalos and baryons, this will put an upper limit on the mass of a thermal-relic dark-matter particle $m_{dm}\lesssim 25\,\unit{keV}$ \citep[Equation~2.3 in][]{drlica-wagner:2019}, and thus falsify the \lcdm\ cosmology.
If the detected stream perturbations do require impacts of $10^6\,\unit{\msun}$ subhalos, this would be first observational evidence for pure dark-matter subhalos, put a lower limit on the mass of a dark-matter particle $m_{dm}\gtrsim 25\,\unit{keV}$, and open a new research program for studying the nature of dark matter on otherwise unobservable scales.

\section*{Acknowledgements}

We thank Rodrigo Ibata for sharing the stream member data for stars in
\citet{ibata:2023} ahead of its publication.
We further thank Jake Nibauer for generating numerical models of streams presented in Figure~\ref{fig:stream_formation}.
We also thank Petra Awad, Eduardo Balbinot, Sergey Koposov, and Kiyan Tavangar for
sharing data used to recreate figures from their work.
Finally, we thank Charlie Conroy, Cecilia Mateu, and Francois Schweizer
for useful discussions and comments on this article.

\appendix

\section{Stream Summary Data}
\label{apx:stream-summary}

We compile stream data from the literature and summarize the key properties of each
stream in Table~\ref{tbl:stream-summary}.
The table includes the stream name, the sky coordinates of the origin of a stream
coordinate frame aligned with the stream, the mean heliocentric distance to the stream
$D_\textrm{hel}$, the mean Galactocentric radius $r_{\textrm{gal}}$, the angular length
and width of the stream in the sky, the width of the stream in length units, a rough
estimate of the stellar mass of the stream, and a list of references for the compiled
stream properties.
The table is split into three sections.
The first section contains streams with reported detections in \gaia\ and with
spectroscopic follow-up data.
The next section contains streams that have been detected in \gaia\ but without
spectroscopic follow-up data.
The final section contains streams without reported detections in \gaia\ that require
follow-up investigation.

The stream coordinate origins are defined as the sky coordinates of the $\phi_1, \phi_2
= (0, 0)^\circ$ point in the stream coordinate system.
For each stream, the stream coordinate system is either taken from the
\texttt{galstreams} package \citep{mateu:2023}, or is defined in this work by fitting a
great circle to stream member star sky positions (this was only used for the few newer
streams in \citet{ibata:2023} that are missing from \texttt{galstreams}).
The stream distance was again either taken from the \texttt{galstreams} package, the
discovery paper, or was estimated by fitting an orbit to the stream member stars and
taking the mean distance of the fitted orbit over the observed segment of the stream
(see Appendix~\ref{apx:stream-fit}).

The stellar masses are either taken from source papers (see the reference list), or were
estimated here (for streams with member data from \citealt{ibata:2023}).
We estimate the stellar mass of a stream from \citet{ibata:2023} by comparing the
observed number of stars down to the brightness limit of \gaia\ to the predicted number
from a single stellar population with a metallicity $\feh = -2$, $\textrm{age} =
12~\unit{\giga\year}$, and a Kroupa initial mass function \citep{Kroupa:2001} (we use
the CMD interface to the PARSEC isochrones; \citealt{bressan:2012, chen:2015}).

\clearpage
\begin{landscape}
\begin{table}
\begin{tabular}{cSSSScSccccc}
\hline \hline
{Name} & {Origin RA} & {Origin Dec.} & {$D_{\rm hel}$} & {$r_{\rm gal}$} & {Length (sky)} & {Width (sky)} & {Width} & {Stellar Mass} & {References}\\
 & \unit{\degree} & \unit{\degree} & \unit{\kilo\parsec} & \unit{\kilo\parsec} & \unit{\degree} & \unit{\degree} & \unit{\parsec} & \unit{\Msun} & \\
\hline\\
\multicolumn{10}{l}{\bf Has spectroscopic follow up and Gaia detection:}\\[1pt]
300S & 157.510 & 15.344 & 17.3 & 21.8 & 11 & 0.9 & 298 & $5 \times 10^{4}$ & (17),(41),(44) \\
AAU & 19.837 & -27.704 & 21.6 & 23.7 & 33 & 0.3 & 96 & $3 \times 10^{5}$ & (*),(16),(26),(29),(34),(41) \\
C-7 & 289.551 & -49.389 & 8.1 & 3.8 & 34 & 0.6 & 90 & $7 \times 10^{3}$ & (*),(43) \\
C-9 & 153.739 & 19.278 & 5.8 & 12.0 & 24 & 2.6 & 260 & $4 \times 10^{3}$ & (*),(43) \\
C-10 & 146.045 & 2.491 & 5.6 & 11.8 & 16 & 0.6 & 54 & $4 \times 10^{3}$ & (*),(43) \\
C-11 & 108.738 & 46.429 & 7.1 & 14.9 & 33 & 2.0 & 248 & $5 \times 10^{3}$ & (*),(43) \\
C-12 & 96.250 & -29.906 & 16.7 & 21.9 & 28 & 0.8 & 218 & $3 \times 10^{4}$ & (*),(43) \\
C-19 & 352.435 & 12.314 & 17.7 & 19.8 & 47 & 0.5 & 161 & $3 \times 10^{4}$ & (*),(43) \\
C-20 & 358.350 & 10.998 & 21.1 & 23.5 & 19 & 0.1 & 27 & $6 \times 10^{4}$ & (*),(43) \\
C-25 & 112.358 & 40.462 & 12.7 & 20.4 & 27 & 2.6 & 585 & $2 \times 10^{4}$ & (*),(43) \\
Elqui & 15.457 & -39.758 & 49.7 & 49.6 & 11 & 0.6 & 472 & $3 \times 10^{6}$ & (16),(23),(26),(29),(41) \\
Fimbulthul & 199.368 & -31.277 & 10.0 & 8.8 & 35 & 2.1 & 381 & $1 \times 10^{4}$ & (*),(22),(27),(43) \\
Fimbulthul-S & 287.874 & -55.293 & 4.2 & 5.0 & 52 & 3.4 & 264 & $4 \times 10^{3}$ & (*),(43) \\
Fjorm & 207.509 & 35.374 & 5.4 & 9.3 & 105 & 1.4 & 151 & $4 \times 10^{3}$ & (*),(22),(43) \\
GD-1 & 147.185 & 33.402 & 7.5 & 14.0 & 119 & 0.4 & 57 & $7 \times 10^{3}$ & (*),(3),(31),(35),(43) \\
Gaia-1 & 193.835 & -5.577 & 5.3 & 8.2 & 40 & 0.5 & 49 & $4 \times 10^{3}$ & (*),(43) \\
Gaia-7 & 181.920 & -17.244 & 5.1 & 8.6 & 16 & 0.5 & 45 & $4 \times 10^{3}$ & (*),(43) \\
Gaia-8 & 188.219 & -22.467 & 7.3 & 8.8 & 47 & 1.0 & 131 & $6 \times 10^{3}$ & (*),(43) \\
Gaia-9 & 237.467 & 57.270 & 4.8 & 9.4 & 36 & 2.0 & 177 & $4 \times 10^{3}$ & (*),(43) \\
Gaia-11 & 271.670 & 42.025 & 14.9 & 14.5 & 50 & 0.6 & 164 & $3 \times 10^{4}$ & (*),(43) \\
Gaia-12 & 33.367 & 23.118 & 16.2 & 22.5 & 29 & 0.4 & 104 & $3 \times 10^{4}$ & (*),(43) \\
Gjoll & 96.510 & -26.264 & 3.2 & 10.3 & 112 & 0.6 & 41 & $3 \times 10^{3}$ & (*),(22),(37),(43) \\
Gunnthra & 292.960 & -69.810 & 3.1 & 6.3 & 20 & & & & (37) \\
Hrid & 279.564 & 44.002 & 2.9 & 7.8 & 77 & 1.6 & 89 & $2 \times 10^{3}$ & (*),(37),(43) \\
Indus & 335.131 & -58.568 & 15.1 & 12.4 & 18 & 0.9 & 240 & $6 \times 10^{6}$ & (*),(16),(23),(26),(29),(38),(41) \\
Jet & 138.629 & -22.094 & 30.9 & 34.3 & 30 & 0.2 & 90 & $2 \times 10^{4}$ & (20),(40),(41) \\
Jhelum & 35.652 & -48.647 & 9.8 & 12.7 & 97 & 0.9 & 166 & $1 \times 10^{4}$ & (*),(26),(29),(36),(41),(43) \\
Kshir & 250.822 & 65.765 & 13.2 & 16.1 & 37 & 0.3 & 80 & $2 \times 10^{4}$ & (*),(24),(43) \\
Kwando & 24.155 & -12.259 & 50.0 & 53.0 & 64 & 2.8 & 2412 & $2 \times 10^{5}$ & (*),(43) \\
LMS-1 & 236.014 & 27.216 & 14.8 & 13.3 & 65 & 1.3 & 338 & $2 \times 10^{4}$ & (*),(32),(43) \\
Leiptr & 84.623 & -24.791 & 7.4 & 13.8 & 73 & 0.5 & 74 & $6 \times 10^{3}$ & (*),(22),(43) \\
M5 & 205.559 & 14.091 & 7.5 & 9.3 & 34 & 0.3 & 42 & $6 \times 10^{3}$ & (*),(43) \\
M92 & 263.639 & 42.126 & 11.8 & 12.0 & 19 & 1.1 & 223 & $1 \times 10^{4}$ & (*),(43) \\
NGC288 & 12.944 & -31.516 & 9.9 & 12.6 & 33 & 0.8 & 143 & $9 \times 10^{3}$ & (*),(43) \\
\hline \hline
\end{tabular}
\end{table}

\end{landscape}
\restoregeometry

\clearpage
\begin{landscape}
\begin{table}
\begin{tabular}{cSSSScSccccc}
\hline \hline
{Name} & {Origin RA} & {Origin Dec.} & {$D_{\rm hel}$} & {$r_{\rm gal}$} & {Length (sky)} & {Width (sky)} & {Width} & {Stellar Mass} & {References}\\
 & \unit{\degree} & \unit{\degree} & \unit{\kilo\parsec} & \unit{\kilo\parsec} & \unit{\degree} & \unit{\degree} & \unit{\parsec} & \unit{\Msun} & \\
\hline
NGC2808 & 140.461 & -65.275 & 16.6 & 16.8 & 16 & 0.4 & 123 & $3 \times 10^{4}$ & (*),(43) \\
NGC6101 & 247.347 & -72.063 & 15.0 & 10.8 & 15 & 0.2 & 44 & $2 \times 10^{4}$ & (*),(43) \\
NGC6397 & 289.562 & -50.797 & 5.3 & 4.2 & 34 & 0.9 & 87 & $4 \times 10^{3}$ & (*),(37),(43) \\
NGC7089 & 314.960 & -3.367 & 12.0 & 9.6 & 22 & 0.2 & 50 & $1 \times 10^{4}$ & (*),(43) \\
NGC7099 & 324.636 & -23.482 & 10.1 & 8.2 & 14 & 0.3 & 47 & $1 \times 10^{4}$ & (*),(43) \\
New-2 & 26.781 & -48.483 & 15.5 & 17.0 & 31 & 0.1 & 39 & $3 \times 10^{4}$ & (*),(43) \\
New-3 & 34.014 & 4.682 & 1.0 & 8.8 & 62 & 4.7 & 100 & $6 \times 10^{2}$ & (*),(43) \\
New-4 & 99.035 & -12.210 & 2.0 & 9.7 & 71 & 3.1 & 115 & $1 \times 10^{3}$ & (*),(43) \\
New-5 & 91.547 & -36.563 & 12.6 & 17.6 & 17 & 0.1 & 20 & $2 \times 10^{4}$ & (*),(43) \\
New-6 & 101.578 & -1.880 & 1.3 & 9.2 & 98 & 3.6 & 80 & $7 \times 10^{2}$ & (*),(43) \\
New-7 & 79.016 & 73.804 & 3.1 & 10.5 & 80 & 3.6 & 213 & $3 \times 10^{3}$ & (*),(43) \\
New-8 & 142.610 & 23.676 & 3.8 & 10.9 & 26 & 1.2 & 82 & $3 \times 10^{3}$ & (*),(43) \\
New-11 & 146.646 & -15.585 & 2.4 & 9.1 & 26 & 1.5 & 62 & $2 \times 10^{3}$ & (*),(43) \\
New-13 & 162.680 & -38.538 & 7.8 & 10.4 & 18 & 1.7 & 235 & $6 \times 10^{3}$ & (*),(43) \\
New-14 & 163.917 & -35.176 & 5.3 & 9.1 & 29 & 0.7 & 70 & $4 \times 10^{3}$ & (*),(43) \\
New-15 & 169.664 & -20.005 & 1.3 & 8.1 & 10 & 0.7 & 15 & $7 \times 10^{2}$ & (*),(43) \\
New-16 & 171.995 & 9.162 & 5.5 & 10.4 & 18 & 3.8 & 383 & $4 \times 10^{3}$ & (*),(43) \\
New-17 & 188.278 & -35.905 & 4.8 & 7.5 & 22 & 0.5 & 45 & $4 \times 10^{3}$ & (*),(43) \\
New-21 & 315.097 & 15.760 & 1.2 & 7.7 & 9 & 2.4 & 58 & $7 \times 10^{2}$ & (*),(43) \\
New-22 & 317.945 & 10.423 & 11.1 & 10.4 & 10 & 1.0 & 193 & $1 \times 10^{4}$ & (*),(43) \\
New-23 & 326.285 & -31.216 & 20.3 & 16.4 & 12 & 0.5 & 172 & $5 \times 10^{4}$ & (*),(43) \\
New-24 & 330.711 & 18.988 & 3.7 & 8.2 & 30 & 0.9 & 58 & $3 \times 10^{3}$ & (*),(43) \\
New-25 & 333.393 & -10.299 & 3.4 & 7.4 & 4 & 0.8 & 48 & $3 \times 10^{3}$ & (*),(43) \\
New-26 & 334.482 & -29.638 & 50.0 & 46.3 & 23 & 1.2 & 1078 & $2 \times 10^{5}$ & (*),(43) \\
New-27 & 341.363 & 30.223 & 7.2 & 11.1 & 9 & 0.4 & 52 & $5 \times 10^{3}$ & (*),(43) \\
New-28 & 342.573 & -52.387 & 42.4 & 38.9 & 12 & 0.4 & 296 & $2 \times 10^{5}$ & (*),(43) \\
OC & 192.672 & -63.952 & 18.3 & 15.5 & 210 & 0.8 & 463 & $6 \times 10^{5}$ & (*),(26),(29),(41),(42) \\
Ophiuchus & 241.852 & -7.551 & 10.8 & 5.8 & 10 & 0.2 & 48 & $1 \times 10^{4}$ & (*),(28),(41),(43),(9) \\
Pal5 & 231.977 & 1.243 & 21.0 & 16.2 & 32 & 0.3 & 94 & $5 \times 10^{4}$ & (*),(1),(10),(13),(43) \\
Palca & 21.438 & -25.809 & 36.3 & 38.2 & 57 & 0.5 & 300 & $1 \times 10^{6}$ & (16),(39),(41) \\
Phlegethon & 321.887 & -20.125 & 3.4 & 6.5 & 77 & 1.3 & 86 & $3 \times 10^{3}$ & (*),(43) \\
Phoenix & 24.686 & -48.706 & 17.9 & 18.9 & 12 & 0.2 & 53 & $3 \times 10^{4}$ & (*),(16),(23),(26),(29),(41) \\
Slidr & 161.592 & 9.752 & 2.9 & 9.4 & 37 & 2.3 & 123 & $2 \times 10^{3}$ & (*),(22),(43) \\
Sylgr & 175.497 & -4.405 & 3.6 & 8.8 & 30 & 1.4 & 91 & $3 \times 10^{3}$ & (*),(43) \\
\hline \hline
\end{tabular}
\end{table}

\end{landscape}
\restoregeometry

\clearpage
\begin{landscape}
\begin{table}
\begin{tabular}{cSSSScSccccc}
\hline \hline
{Name} & {Origin RA} & {Origin Dec.} & {$D_{\rm hel}$} & {$r_{\rm gal}$} & {Length (sky)} & {Width (sky)} & {Width} & {Stellar Mass} & {References}\\
 & \unit{\degree} & \unit{\degree} & \unit{\kilo\parsec} & \unit{\kilo\parsec} & \unit{\degree} & \unit{\degree} & \unit{\parsec} & \unit{\Msun} & \\
\hline
Tuc3 & 355.957 & -59.571 & 24.4 & 22.3 & 17 & 0.3 & 153 & $8 \times 10^{4}$ & (*),(43) \\
Turranburra & 67.141 & -22.464 & 26.3 & 31.7 & 14 & 0.6 & 288 & $2 \times 10^{6}$ & (16),(23),(41) \\
Willka Yaku & 37.366 & -61.455 & 36.3 & 36.0 & 6 & 0.2 & 127 & $1 \times 10^{5}$ & (16),(41) \\
Ylgr & 172.694 & -19.061 & 9.6 & 11.9 & 49 & 1.0 & 169 & $9 \times 10^{3}$ & (*),(22),(43) \\
\hline \\
\multicolumn{10}{l}{\bf Gaia detection:}\\[1pt]
C-4 & 226.560 & 82.656 & 3.6 & 10.1 & 14 & & & & (37) \\
C-5 & 116.658 & 35.213 & 4.3 & 12.1 & 6 & & & & (37) \\
C-8 & 341.099 & -77.442 & 3.4 & 7.0 & 10 & & & & (37) \\
C-13 & 34.112 & 44.148 & 8.4 & 15.3 & 23 & 0.7 & 98 & $7 \times 10^{3}$ & (*),(43) \\
C-22 & 200.891 & 17.327 & 8.8 & 10.7 & 13 & 0.3 & 42 & $7 \times 10^{3}$ & (*),(43) \\
C-23 & 195.236 & 44.290 & 8.6 & 12.6 & 20 & 0.5 & 73 & $7 \times 10^{3}$ & (*),(43) \\
C-24 & 153.259 & 52.464 & 20.2 & 25.8 & 37 & 0.6 & 229 & $5 \times 10^{4}$ & (*),(43) \\
Gaia-2 & 5.354 & -14.791 & 7.0 & 10.8 & 14 & & & & (37) \\
Gaia-3 & 175.619 & -25.414 & 12.0 & 13.1 & 18 & & & & (21) \\
Gaia-4 & 165.089 & -6.860 & 11.1 & 14.5 & 9 & & & & (21) \\
Gaia-5 & 145.024 & 33.280 & 19.6 & 25.6 & 24 & & & & (21) \\
Gaia-6 & 213.832 & 32.689 & 9.5 & 11.3 & 21 & 0.4 & 75 & $9 \times 10^{3}$ & (*),(43) \\
Gaia-10 & 156.263 & 15.506 & 14.9 & 19.7 & 25 & 0.3 & 89 & $2 \times 10^{4}$ & (*),(43) \\
Hydrus & 44.929 & -79.803 & 14.4 & 13.8 & 6 & 0.1 & 23 & $2 \times 10^{4}$ & (*),(43) \\
M2 & 321.251 & -2.031 & 11.1 & 9.8 & 22 & & & & (37) \\
NGC1261 & 35.187 & -57.796 & 37.1 & 37.0 & 27 & 0.6 & 366 & $2 \times 10^{5}$ & (*),(43) \\
NGC1261a & 49.394 & -53.523 & 26.9 & 28.3 & 17 & 1.0 & 462 & $1 \times 10^{5}$ & (*),(43) \\
NGC1261b & 49.512 & -52.529 & 46.2 & 47.3 & 11 & 0.5 & 442 & $2 \times 10^{5}$ & (*),(43) \\
NGC1851 & 77.652 & -39.005 & 17.6 & 21.9 & 8 & 0.5 & 158 & $3 \times 10^{4}$ & (*),(43) \\
NGC2298 & 100.671 & -35.185 & 11.3 & 16.4 & 13 & 0.4 & 73 & $1 \times 10^{4}$ & (*),(43) \\
NGC5466 & 214.015 & 27.035 & 14.7 & 14.9 & 23 & 0.3 & 77 & $2 \times 10^{4}$ & (*),(43) \\
NGC7492 & 347.922 & -15.285 & 31.1 & 30.1 & 8 & 0.5 & 263 & $1 \times 10^{5}$ & (*),(43) \\
New-1 & 7.353 & -35.979 & 11.1 & 12.8 & 29 & 0.4 & 75 & $1 \times 10^{4}$ & (*),(43) \\
New-9 & 142.965 & 35.226 & 1.1 & 8.9 & 5 & 0.8 & 15 & $6 \times 10^{2}$ & (*),(43) \\
New-10 & 144.782 & 30.388 & 9.2 & 15.7 & 7 & 0.8 & 133 & $8 \times 10^{3}$ & (*),(43) \\
New-12 & 153.629 & 36.690 & 14.1 & 19.8 & 7 & 0.3 & 66 & $2 \times 10^{4}$ & (*),(43) \\
New-18 & 209.199 & 20.865 & 3.7 & 8.0 & 9 & 0.4 & 27 & $3 \times 10^{3}$ & (*),(43) \\
New-19 & 211.686 & 23.998 & 12.7 & 13.1 & 11 & 1.6 & 385 & $2 \times 10^{4}$ & (*),(43) \\
New-20 & 223.163 & 35.173 & 16.7 & 16.7 & 6 & 1.7 & 498 & $3 \times 10^{4}$ & (*),(43) \\
Ravi & 338.595 & -51.987 & 22.9 & 19.7 & 17 & 0.7 & 288 & $1 \times 10^{4}$ & (16) \\
\hline \hline
\end{tabular}
\end{table}

\end{landscape}
\restoregeometry

\clearpage
\begin{landscape}
\begin{table}
\begin{tabular}{cSSSScSccccc}
\hline \hline
{Name} & {Origin RA} & {Origin Dec.} & {$D_{\rm hel}$} & {$r_{\rm gal}$} & {Length (sky)} & {Width (sky)} & {Width} & {Stellar Mass} & {References}\\
 & \unit{\degree} & \unit{\degree} & \unit{\kilo\parsec} & \unit{\kilo\parsec} & \unit{\degree} & \unit{\degree} & \unit{\parsec} & \unit{\Msun} & \\
\hline
Svol & 240.935 & 25.173 & 7.7 & 7.9 & 58 & 0.7 & 97 & $7 \times 10^{3}$ & (*),(43) \\
Turbio & 27.941 & -53.500 & 16.6 & 17.6 & 15 & 0.2 & 72 & $4 \times 10^{3}$ & (16) \\
Wambelong & 84.434 & -40.084 & 15.1 & 19.5 & 14 & 0.4 & 106 & $2 \times 10^{3}$ & (16) \\
\hline \\
\multicolumn{10}{l}{\bf Needs follow-up observations:}\\[1pt]
20.0-1 & 298.182 & -29.913 & 26.8 & 20.0 & 37 & 1.8 & 641 & $1 \times 10^{5}$ & (19) \\
Acheron & 244.605 & 10.300 & 3.7 & 6.1 & 37 & 0.9 & 60 & & (2) \\
Alpheus & 24.955 & -56.989 & 1.8 & 8.0 & 24 & 3.2 & 100 & & (7) \\
Aquarius & 344.906 & -8.951 & 4.1 & 8.1 & 12 & & & & (4) \\
Cocytos & 221.211 & 10.529 & 11.0 & 9.7 & 75 & 0.7 & 140 & & (2) \\
Corvus & 178.207 & -14.396 & 8.7 & 11.0 & 72 & 1.6 & 313 & $1 \times 10^{5}$ & (19) \\
Eridanus & 66.185 & -21.188 & 95.0 & 100.0 & 1 & & & & (15) \\
Hermus & 241.511 & 27.792 & 19.6 & 17.3 & 46 & 0.7 & 220 & & (8) \\
Hyllus & 245.959 & 22.374 & 20.8 & 17.4 & 23 & 1.2 & 180 & & (8) \\
Lethe & 214.173 & 25.393 & 13.0 & 13.3 & 81 & 0.4 & 95 & & (2) \\
Molonglo & 2.075 & -18.516 & 20.0 & 21.0 & 19 & 0.5 & 170 & & (14) \\
Murrumbidgee & 20.107 & -40.089 & 20.0 & 21.3 & 118 & 0.4 & 125 & & (14) \\
NGC6362 & 262.989 & -66.741 & 7.6 & 5.1 & 4 & & & & (33) \\
Orinoco & 11.454 & -24.616 & 20.6 & 22.2 & 20 & 0.7 & 240 & & (12),(14) \\
PS1-A & 29.260 & -4.208 & 7.9 & 13.6 & 5 & 0.5 & 63 & $4 \times 10^{2}$ & (11) \\
PS1-B & 148.373 & -11.147 & 14.5 & 18.7 & 10 & 0.5 & 112 & $2 \times 10^{3}$ & (11) \\
PS1-C & 332.539 & 14.951 & 17.4 & 17.5 & 8 & 0.3 & 99 & $2 \times 10^{3}$ & (11) \\
PS1-D & 139.676 & 0.829 & 22.9 & 28.1 & 45 & 0.9 & 350 & $2 \times 10^{4}$ & (11) \\
PS1-E & 173.064 & 55.605 & 12.6 & 17.7 & 25 & 0.6 & 140 & $1 \times 10^{3}$ & (11) \\
Pal13 & 347.192 & 13.567 & 23.6 & 24.8 & 11 & 0.2 & 103 & $3 \times 10^{3}$ & (30) \\
Pal15 & 255.116 & -0.845 & 38.4 & 31.6 & 1 & & & & (15) \\
Parallel & 174.539 & 4.145 & 14.3 & 16.9 & 37 & & & & (18) \\
Pegasus & 330.710 & 24.566 & 18.0 & 18.7 & 9 & 0.4 & 112 & $4 \times 10^{3}$ & (25) \\
Perpendicular & 183.405 & 16.674 & 15.3 & 17.6 & 19 & & & & (18) \\
Sangarius & 132.868 & 7.422 & 21.0 & 27.1 & 50 & 1.0 & 350 & & (12) \\
Scamander & 148.139 & 11.963 & 21.0 & 26.1 & 65 & 1.0 & 350 & & (12) \\
Styx & 226.381 & 23.908 & 45.0 & 42.3 & 60 & 3.3 & 2600 & & (2) \\
Tri-Pis & 22.502 & 29.494 & 26.0 & 31.4 & 13 & 0.2 & 75 & & (5),(6) \\
\hline \hline
\end{tabular}
\label{tbl:stream-summary}
\caption{(1): \citet{odenkirchen:2009}, (2): \citet{grillmair:2009}, (3): \citet{koposov:2010}, (4): \citet{williams:2011}, (5): \citet{bonaca:2012}, (6): \citet{martin:2013}, (7): \citet{grillmair:2013}, (8): \citet{grillmair:2014}, (9): \citet{sesar:2015}, (10): \citet{kuzma:2015}, (11): \citet{bernard:2016}, (12): \citet{grillmair:2017}, (13): \citet{ibata:2017}, (14): \citet{grillmair:2017b}, (15): \citet{myeong:2017}, (16): \citet{shipp:2018}, (17): \citet{fu:2018}, (18): \citet{weiss:2018}, (19): \citet{mateu:2018}, (20): \citet{jethwa:2018}, (21): \citet{malhan:2018}, (22): \citet{ibata:2019}, (23): \citet{shipp:2019}, (24): \citet{malhan:2019}, (25): \citet{perottoni:2019}, (26): \citet{li:2019}, (27): \citet{ibata:2019b}, (28): \citet{caldwell:2020}, (29): \citet{ji:2020}, (30): \citet{shipp:2020}, (31): \citet{bonaca:2020}, (32): \citet{yuan:2020}, (33): \citet{sollima:2020}, (34): \citet{li:2021}, (35): \citet{gialluca:2021}, (36): \citet{sheffield:2021}, (37): \citet{ibata:2021}, (38): \citet{hansen:2021}, (39): \citet{thomas:2022}, (40): \citet{ferguson:2022}, (41): \citet{li:2022}, (42): \citet{koposov:2023}, (43): \citet{ibata:2023}, (44): \citet{usman:2024}, (*): this work}
\end{table}

\end{landscape}
\restoregeometry

\clearpage

\section{Fitting Orbits to Streams}
\label{apx:stream-fit}

We fit orbits to the 87 streams in the catalog of stream stars from \citet{ibata:2023}
for demonstrative visualizations shown in this article.

For a given stream, we first determine a heliocentric, spherical coordinate system that
is approximately aligned with the stream track.
In this coordinate system, the stream longitude is $\phi_1$ and the stream latitude is
$\phi_2$.
We initialize this process by computing the eigenvectors of the covariance matrix of the
positions of the stream stars projected onto the unit sphere.
We project the unit vectors of the stream star positions onto the largest eigenvector
and use the endpoints along this axis to define an initial great circle coordinate
system.
We then adjust this great circle coordinate system by allowing small rotations around
the great circle $x$ and $y$ axes to minimize an objective function $F$ that computes
the sum of the squared stream latitudes, $\phi_2$: $F(\phi_2) = \sum_i \phi_{2,i}^2$ for
all stars in a stream, indexed by $i$.

To fit an orbit to each stream, we fix the model for the Milky Way's gravitational
potential to the \texttt{MilkyWayPotential2022} in \texttt{gala}
\citep{price-whelan:2017} and optimize over the orbital initial conditions.
We specify the orbital initial conditions in heliocentric coordinates in the rotated
stream coordinate system described above.
In this coordinate system, we define the orbital initial conditions at $\phi_1=0^\circ$,
so that the free parameters are the stream longitude, $\phi_2$, the distance, $D$, the
proper motion components, $(\mu_{\phi_1}, \mu_{\phi_2})$, and the line-of-sight
velocity, $v_r$.
For a given setting of the parameters $(\phi_2, D, \mu_{\phi_1}, \mu_{\phi_2}, v_r)$, we
integrate the orbit forward and backward in time, $\tau$, for an amount set by the
observed angular extent of the stream, $\Delta \phi_1$ divided by the transverse angular
velocity, $\sqrt{\mu_{\phi_1}^2 + \mu_{\phi_2}^2}$, and multiplied by two.
If the orbit wraps (over $360^\circ$), we truncate the orbit segment at $\phi_1 \pm
180^\circ$.

We then compute the log-likelihood of the orbit given the stream stars by assuming each
coordinate dimension is independent and normally distributed with a variance set by the
observational uncertainties.
In detail, we interpolate the orbit to the $\phi_1$ values of the stream stars using
cubic interpolation and compute the log-likelihood as:
\begin{equation}
\begin{split}
    \ln \mathcal{L} = \sum_i
    &\mathcal{N}(\tilde{\phi}_2(\phi_{1, i}) \given \phi_{2, i}, \sigma_{\phi_{2}}^2) \\
    \times \, &\mathcal{N}(\tilde{D}(\phi_{1, i})^{-1} \given \varpi_{i}, \sigma_{\varpi}^2) \\
    \times \, &\mathcal{N}(\tilde{\mu}_{\phi_1}(\phi_{1, i}) \given \mu_{\phi_1, i}, \sigma_{\mu_{\phi_1}}^2) \\
    \times \, &\mathcal{N}(\tilde{\mu}_{\phi_2}(\phi_{1, i}) \given \mu_{\phi_2, i}, \sigma_{\mu_{\phi_2}}^2) \\
    \times \, &\mathcal{N}(\tilde{v}_r(\phi_{1, i}) \given v_{r, i}, \sigma_{v_r}^2)
\end{split}
\end{equation}
where the tilde indicates the model predicted coordinates from the orbit and
interpolation step, evaluated at each stream star $\phi_{1, i}$, the distance component
is compared to the observed parallaxes $\varpi$, $\mathcal{N}(x \given \mu, \sigma^2)$
is the normal distribution with mean $\mu$ and variance $\sigma^2$, and the sum is over
all observed stars in the stream indexed by $i$.

We optimize the log-likelihood $\ln \mathcal{L}$ using the L-BFGS-B implementation in
\texttt{scipy.optimize} \citep{scipy}, with bounds on the parameters such that $\phi_2
\in (-5, 5)^\circ$, $D \in (1, 50)~\kpc$, $\mu_{\phi_1} \in (-100, 100)~\masyr$,
$\mu_{\phi_2} \in (-100, 100)~\masyr$, and $v_r \in (-500, 500)~\kms$.
We start the optimization from six different initial distance values between $1~\kpc$
and $40~\kpc$ (spaced logarithmically) and take the optimized parameters from the run
with the highest log-likelihood value.
This process failed for five streams (New-6, New-18, NGC~6397, NGC~7099, and Orphan), possibly due to them not being well-described by a single orbit in a static gravitational potential because they are close to the orbital apocenter, or in the case of Orphan, because of the active perturbation from the LMC.

\bibliographystyle{model2-names-astronomy}
\bibliography{refs}

\begin{thebibliography}{467}
\expandafter\ifx\csname natexlab\endcsname\relax\def\natexlab#1{#1}\fi
\providecommand{\url}[1]{\texttt{#1}}
\providecommand{\href}[2]{#2}
\providecommand{\path}[1]{#1}
\providecommand{\DOIprefix}{doi:}
\providecommand{\ArXivprefix}{arXiv:}
\providecommand{\URLprefix}{URL: }
\providecommand{\Pubmedprefix}{pmid:}
\providecommand{\doi}[1]{\href{http://dx.doi.org/#1}{\path{#1}}}
\providecommand{\Pubmed}[1]{\href{pmid:#1}{\path{#1}}}
\providecommand{\bibinfo}[2]{#2}
\ifx\xfnm\relax \def\xfnm[#1]{\unskip,\space#1}\fi
%Type = Article
\bibitem[{{Akeson} et~al.(2013){Akeson}, {Chen}, {Ciardi}, {Crane}, {Good},
  {Harbut}, {Jackson}, {Kane}, {Laity}, {Leifer}, {Lynn}, {McElroy}, {Papin},
  {Plavchan}, {Ram{\'\i}rez}, {Rey}, {von Braun}, {Wittman}, {Abajian}, {Ali},
  {Beichman}, {Beekley}, {Berriman}, {Berukoff}, {Bryden}, {Chan}, {Groom},
  {Lau}, {Payne}, {Regelson}, {Saucedo}, {Schmitz}, {Stauffer}, {Wyatt} and
  {Zhang}}]{akeson:2013}
{Akeson}, R.~L. et~al., \bibinfo{year}{2013}.
\newblock \bibinfo{title}{{The NASA Exoplanet Archive: Data and Tools for
  Exoplanet Research}}.
\newblock \bibinfo{journal}{\pasp} \bibinfo{volume}{125}, \bibinfo{pages}{989}.
\newblock \DOIprefix\doi{10.1086/672273},
  \href{http://arxiv.org/abs/1307.2944}{{\tt arXiv:1307.2944}}.
%Type = Article
\bibitem[{{Allgood} et~al.(2006){Allgood}, {Flores}, {Primack}, {Kravtsov},
  {Wechsler}, {Faltenbacher} and {Bullock}}]{allgood:2006}
\bibinfo{author}{{Allgood}, B.}, \bibinfo{author}{{Flores}, R.A.},
  \bibinfo{author}{{Primack}, J.R.}, \bibinfo{author}{{Kravtsov}, A.V.},
  \bibinfo{author}{{Wechsler}, R.H.}, \bibinfo{author}{{Faltenbacher}, A.},
  \bibinfo{author}{{Bullock}, J.S.}, \bibinfo{year}{2006}.
\newblock \bibinfo{title}{{The shape of dark matter haloes: dependence on mass,
  redshift, radius and formation}}.
\newblock \bibinfo{journal}{\mnras} \bibinfo{volume}{367},
  \bibinfo{pages}{1781--1796}.
\newblock \DOIprefix\doi{10.1111/j.1365-2966.2006.10094.x},
  \href{http://arxiv.org/abs/astro-ph/0508497}{{\tt arXiv:astro-ph/0508497}}.
%Type = Article
\bibitem[{{Amorisco}(2015)}]{amorisco:2015b}
\bibinfo{author}{{Amorisco}, N.C.}, \bibinfo{year}{2015}.
\newblock \bibinfo{title}{{On feathers, bifurcations and shells: the dynamics
  of tidal streams across the mass scale}}.
\newblock \bibinfo{journal}{\mnras} \bibinfo{volume}{450},
  \bibinfo{pages}{575--591}.
\newblock \DOIprefix\doi{10.1093/mnras/stv648},
  \href{http://arxiv.org/abs/1410.0360}{{\tt arXiv:1410.0360}}.
%Type = Article
\bibitem[{{Amorisco} et~al.(2015){Amorisco}, {Martinez-Delgado} and
  {Schedler}}]{amorisco:2015}
\bibinfo{author}{{Amorisco}, N.C.}, \bibinfo{author}{{Martinez-Delgado}, D.},
  \bibinfo{author}{{Schedler}, J.}, \bibinfo{year}{2015}.
\newblock \bibinfo{title}{{A dwarf galaxy's transformation and a massive
  galaxy's edge: autopsy of kill and killer in NGC 1097}}.
\newblock \bibinfo{journal}{arXiv e-prints} ,
  \bibinfo{pages}{arXiv:1504.03697}\DOIprefix\doi{10.48550/arXiv.1504.03697},
  \href{http://arxiv.org/abs/1504.03697}{{\tt arXiv:1504.03697}}.
%Type = Article
\bibitem[{{Andrae} et~al.(2023){Andrae}, {Rix} and {Chandra}}]{andrae:2023}
\bibinfo{author}{{Andrae}, R.}, \bibinfo{author}{{Rix}, H.W.},
  \bibinfo{author}{{Chandra}, V.}, \bibinfo{year}{2023}.
\newblock \bibinfo{title}{{Robust Data-driven Metallicities for 175 Million
  Stars from Gaia XP Spectra}}.
\newblock \bibinfo{journal}{\apjs} \bibinfo{volume}{267}, \bibinfo{pages}{8}.
\newblock \DOIprefix\doi{10.3847/1538-4365/acd53e},
  \href{http://arxiv.org/abs/2302.02611}{{\tt arXiv:2302.02611}}.
%Type = Article
\bibitem[{{Antoja} et~al.(2018){Antoja}, {Helmi}, {Romero-G{\'o}mez}, {Katz},
  {Babusiaux}, {Drimmel}, {Evans}, {Figueras}, {Poggio}, {Reyl{\'e}}, {Robin},
  {Seabroke} and {Soubiran}}]{antoja:2018}
{Antoja}, T. et~al., \bibinfo{year}{2018}.
\newblock \bibinfo{title}{{A dynamically young and perturbed Milky Way disk}}.
\newblock \bibinfo{journal}{\nat} \bibinfo{volume}{561},
  \bibinfo{pages}{360--362}.
\newblock \DOIprefix\doi{10.1038/s41586-018-0510-7},
  \href{http://arxiv.org/abs/1804.10196}{{\tt arXiv:1804.10196}}.
%Type = Article
\bibitem[{{Arora} et~al.(2023){Arora}, {Garavito-Camargo}, {Sanderson},
  {Cunningham}, {Wetzel}, {Panithanpaisal} and {Barry}}]{arora:2023}
\bibinfo{author}{{Arora}, A.}, \bibinfo{author}{{Garavito-Camargo}, N.},
  \bibinfo{author}{{Sanderson}, R.E.}, \bibinfo{author}{{Cunningham}, E.C.},
  \bibinfo{author}{{Wetzel}, A.}, \bibinfo{author}{{Panithanpaisal}, N.},
  \bibinfo{author}{{Barry}, M.}, \bibinfo{year}{2023}.
\newblock \bibinfo{title}{{LMC-driven anisotropic boosts in stream--subhalo
  interactions}}.
\newblock \bibinfo{journal}{arXiv e-prints} ,
  \bibinfo{pages}{arXiv:2309.15998}\DOIprefix\doi{10.48550/arXiv.2309.15998},
  \href{http://arxiv.org/abs/2309.15998}{{\tt arXiv:2309.15998}}.
%Type = Article
\bibitem[{{Arora} et~al.(2022){Arora}, {Sanderson}, {Panithanpaisal},
  {Cunningham}, {Wetzel} and {Garavito-Camargo}}]{arora:2022}
\bibinfo{author}{{Arora}, A.}, \bibinfo{author}{{Sanderson}, R.E.},
  \bibinfo{author}{{Panithanpaisal}, N.}, \bibinfo{author}{{Cunningham}, E.C.},
  \bibinfo{author}{{Wetzel}, A.}, \bibinfo{author}{{Garavito-Camargo}, N.},
  \bibinfo{year}{2022}.
\newblock \bibinfo{title}{{On the Stability of Tidal Streams in Action Space}}.
\newblock \bibinfo{journal}{\apj} \bibinfo{volume}{939}, \bibinfo{pages}{2}.
\newblock \DOIprefix\doi{10.3847/1538-4357/ac93fb},
  \href{http://arxiv.org/abs/2207.13481}{{\tt arXiv:2207.13481}}.
%Type = Article
\bibitem[{{Arp}(1966)}]{arp:1966}
\bibinfo{author}{{Arp}, H.}, \bibinfo{year}{1966}.
\newblock \bibinfo{title}{{Atlas of Peculiar Galaxies}}.
\newblock \bibinfo{journal}{\apjs} \bibinfo{volume}{14}, \bibinfo{pages}{1}.
\newblock \DOIprefix\doi{10.1086/190147}.
%Type = Article
\bibitem[{{Awad} et~al.(2024){Awad}, {Canducci}, {Balbinot}, {Viswanathan},
  {Woudenberg}, {Koop}, {Peletier}, {Ti{\v{n}}o}, {Starkenburg}, {Smith} and
  {Bunte}}]{awad:2024}
{Awad}, P. et~al., \bibinfo{year}{2024}.
\newblock \bibinfo{title}{{Swarming in stellar streams: Unveiling the structure
  of the Jhelum stream with ant colony-inspired computation}}.
\newblock \bibinfo{journal}{\aap} \bibinfo{volume}{683}, \bibinfo{pages}{A14}.
\newblock \DOIprefix\doi{10.1051/0004-6361/202347848}.
%Type = Article
\bibitem[{{Balbinot} et~al.(2022){Balbinot}, {Cabrera-Ziri} and
  {Lardo}}]{balbinot:2022}
\bibinfo{author}{{Balbinot}, E.}, \bibinfo{author}{{Cabrera-Ziri}, I.},
  \bibinfo{author}{{Lardo}, C.}, \bibinfo{year}{2022}.
\newblock \bibinfo{title}{{Evidence for C and Mg variations in the GD-1 stellar
  stream}}.
\newblock \bibinfo{journal}{\mnras} \bibinfo{volume}{515},
  \bibinfo{pages}{5802--5812}.
\newblock \DOIprefix\doi{10.1093/mnras/stac1953},
  \href{http://arxiv.org/abs/2111.12626}{{\tt arXiv:2111.12626}}.
%Type = Article
\bibitem[{{Balbinot} and {Gieles}(2018)}]{balbinot:2018}
\bibinfo{author}{{Balbinot}, E.}, \bibinfo{author}{{Gieles}, M.},
  \bibinfo{year}{2018}.
\newblock \bibinfo{title}{{The devil is in the tails: the role of globular
  cluster mass evolution on stream properties}}.
\newblock \bibinfo{journal}{\mnras} \bibinfo{volume}{474},
  \bibinfo{pages}{2479--2492}.
\newblock \DOIprefix\doi{10.1093/mnras/stx2708},
  \href{http://arxiv.org/abs/1702.02543}{{\tt arXiv:1702.02543}}.
%Type = Article
\bibitem[{{Balbinot} et~al.(2023){Balbinot}, {Helmi}, {Callingham}, {Matsuno},
  {Dodd} and {Ruiz-Lara}}]{balbinot:2023}
\bibinfo{author}{{Balbinot}, E.}, \bibinfo{author}{{Helmi}, A.},
  \bibinfo{author}{{Callingham}, T.}, \bibinfo{author}{{Matsuno}, T.},
  \bibinfo{author}{{Dodd}, E.}, \bibinfo{author}{{Ruiz-Lara}, T.},
  \bibinfo{year}{2023}.
\newblock \bibinfo{title}{{ED-2: A cold but not so narrow stellar stream
  crossing the solar neighbourhood}}.
\newblock \bibinfo{journal}{\aap} \bibinfo{volume}{678}, \bibinfo{pages}{A115}.
\newblock \DOIprefix\doi{10.1051/0004-6361/202347076},
  \href{http://arxiv.org/abs/2306.02756}{{\tt arXiv:2306.02756}}.
%Type = Article
\bibitem[{{Banik} et~al.(2018){Banik}, {Bertone}, {Bovy} and
  {Bozorgnia}}]{banik:2018}
\bibinfo{author}{{Banik}, N.}, \bibinfo{author}{{Bertone}, G.},
  \bibinfo{author}{{Bovy}, J.}, \bibinfo{author}{{Bozorgnia}, N.},
  \bibinfo{year}{2018}.
\newblock \bibinfo{title}{{Probing the nature of dark matter particles with
  stellar streams}}.
\newblock \bibinfo{journal}{\jcap} \bibinfo{volume}{2018},
  \bibinfo{pages}{061}.
\newblock \DOIprefix\doi{10.1088/1475-7516/2018/07/061},
  \href{http://arxiv.org/abs/1804.04384}{{\tt arXiv:1804.04384}}.
%Type = Article
\bibitem[{{Banik} and {Bovy}(2019)}]{banik:2019}
\bibinfo{author}{{Banik}, N.}, \bibinfo{author}{{Bovy}, J.},
  \bibinfo{year}{2019}.
\newblock \bibinfo{title}{{Effects of baryonic and dark matter substructure on
  the Pal 5 stream}}.
\newblock \bibinfo{journal}{\mnras} \bibinfo{volume}{484},
  \bibinfo{pages}{2009--2020}.
\newblock \DOIprefix\doi{10.1093/mnras/stz142},
  \href{http://arxiv.org/abs/1809.09640}{{\tt arXiv:1809.09640}}.
%Type = Article
\bibitem[{{Banik} et~al.(2021a){Banik}, {Bovy}, {Bertone}, {Erkal} and {de
  Boer}}]{banik:2021}
\bibinfo{author}{{Banik}, N.}, \bibinfo{author}{{Bovy}, J.},
  \bibinfo{author}{{Bertone}, G.}, \bibinfo{author}{{Erkal}, D.},
  \bibinfo{author}{{de Boer}, T.J.L.}, \bibinfo{year}{2021}a.
\newblock \bibinfo{title}{{Evidence of a population of dark subhaloes from Gaia
  and Pan-STARRS observations of the GD-1 stream}}.
\newblock \bibinfo{journal}{\mnras} \bibinfo{volume}{502},
  \bibinfo{pages}{2364--2380}.
\newblock \DOIprefix\doi{10.1093/mnras/stab210},
  \href{http://arxiv.org/abs/1911.02662}{{\tt arXiv:1911.02662}}.
%Type = Article
\bibitem[{{Banik} et~al.(2021b){Banik}, {Bovy}, {Bertone}, {Erkal} and {de
  Boer}}]{banik:2021b}
\bibinfo{author}{{Banik}, N.}, \bibinfo{author}{{Bovy}, J.},
  \bibinfo{author}{{Bertone}, G.}, \bibinfo{author}{{Erkal}, D.},
  \bibinfo{author}{{de Boer}, T.J.L.}, \bibinfo{year}{2021}b.
\newblock \bibinfo{title}{{Novel constraints on the particle nature of dark
  matter from stellar streams}}.
\newblock \bibinfo{journal}{\jcap} \bibinfo{volume}{2021},
  \bibinfo{pages}{043}.
\newblock \DOIprefix\doi{10.1088/1475-7516/2021/10/043},
  \href{http://arxiv.org/abs/1911.02663}{{\tt arXiv:1911.02663}}.
%Type = Article
\bibitem[{{Barry} et~al.(2023){Barry}, {Wetzel}, {Chapman}, {Samuel},
  {Sanderson} and {Arora}}]{barry:2023}
\bibinfo{author}{{Barry}, M.}, \bibinfo{author}{{Wetzel}, A.},
  \bibinfo{author}{{Chapman}, S.}, \bibinfo{author}{{Samuel}, J.},
  \bibinfo{author}{{Sanderson}, R.}, \bibinfo{author}{{Arora}, A.},
  \bibinfo{year}{2023}.
\newblock \bibinfo{title}{{The dark side of FIRE: predicting the population of
  dark matter subhaloes around Milky Way-mass galaxies}}.
\newblock \bibinfo{journal}{\mnras} \bibinfo{volume}{523},
  \bibinfo{pages}{428--440}.
\newblock \DOIprefix\doi{10.1093/mnras/stad1395},
  \href{http://arxiv.org/abs/2303.05527}{{\tt arXiv:2303.05527}}.
%Type = Article
\bibitem[{{Bastian} and {Lardo}(2018)}]{bastian:2018}
\bibinfo{author}{{Bastian}, N.}, \bibinfo{author}{{Lardo}, C.},
  \bibinfo{year}{2018}.
\newblock \bibinfo{title}{{Multiple Stellar Populations in Globular Clusters}}.
\newblock \bibinfo{journal}{\araa} \bibinfo{volume}{56},
  \bibinfo{pages}{83--136}.
\newblock \DOIprefix\doi{10.1146/annurev-astro-081817-051839},
  \href{http://arxiv.org/abs/1712.01286}{{\tt arXiv:1712.01286}}.
%Type = Book
\bibitem[{{Battrick}(1994)}]{battrick:1994}
\bibinfo{author}{{Battrick}, B.}, \bibinfo{year}{1994}.
\newblock \bibinfo{title}{{Horizon 2000 Plus. European Space Science in the
  21st Century}}.
%Type = Article
\bibitem[{{Baumgardt} and {Hilker}(2018)}]{baumgardt:2018}
\bibinfo{author}{{Baumgardt}, H.}, \bibinfo{author}{{Hilker}, M.},
  \bibinfo{year}{2018}.
\newblock \bibinfo{title}{{A catalogue of masses, structural parameters, and
  velocity dispersion profiles of 112 Milky Way globular clusters}}.
\newblock \bibinfo{journal}{\mnras} \bibinfo{volume}{478},
  \bibinfo{pages}{1520--1557}.
\newblock \DOIprefix\doi{10.1093/mnras/sty1057},
  \href{http://arxiv.org/abs/1804.08359}{{\tt arXiv:1804.08359}}.
%Type = Article
\bibitem[{{Baumgardt} et~al.(2019){Baumgardt}, {Hilker}, {Sollima} and
  {Bellini}}]{baumgardt:2019}
\bibinfo{author}{{Baumgardt}, H.}, \bibinfo{author}{{Hilker}, M.},
  \bibinfo{author}{{Sollima}, A.}, \bibinfo{author}{{Bellini}, A.},
  \bibinfo{year}{2019}.
\newblock \bibinfo{title}{{Mean proper motions, space orbits, and velocity
  dispersion profiles of Galactic globular clusters derived from Gaia DR2
  data}}.
\newblock \bibinfo{journal}{\mnras} \bibinfo{volume}{482},
  \bibinfo{pages}{5138--5155}.
\newblock \DOIprefix\doi{10.1093/mnras/sty2997},
  \href{http://arxiv.org/abs/1811.01507}{{\tt arXiv:1811.01507}}.
%Type = Article
\bibitem[{{Baumgardt} and {Makino}(2003)}]{baumgardt:2003}
\bibinfo{author}{{Baumgardt}, H.}, \bibinfo{author}{{Makino}, J.},
  \bibinfo{year}{2003}.
\newblock \bibinfo{title}{{Dynamical evolution of star clusters in tidal
  fields}}.
\newblock \bibinfo{journal}{\mnras} \bibinfo{volume}{340},
  \bibinfo{pages}{227--246}.
\newblock \DOIprefix\doi{10.1046/j.1365-8711.2003.06286.x},
  \href{http://arxiv.org/abs/astro-ph/0211471}{{\tt arXiv:astro-ph/0211471}}.
%Type = Article
\bibitem[{{Beasley} et~al.(2019){Beasley}, {Leaman}, {Gallart}, {Larsen},
  {Battaglia}, {Monelli} and {Pedreros}}]{beasley:2019}
\bibinfo{author}{{Beasley}, M.A.}, \bibinfo{author}{{Leaman}, R.},
  \bibinfo{author}{{Gallart}, C.}, \bibinfo{author}{{Larsen}, S.S.},
  \bibinfo{author}{{Battaglia}, G.}, \bibinfo{author}{{Monelli}, M.},
  \bibinfo{author}{{Pedreros}, M.H.}, \bibinfo{year}{2019}.
\newblock \bibinfo{title}{{An old, metal-poor globular cluster in Sextans A and
  the metallicity floor of globular cluster systems}}.
\newblock \bibinfo{journal}{\mnras} \bibinfo{volume}{487},
  \bibinfo{pages}{1986--1993}.
\newblock \DOIprefix\doi{10.1093/mnras/stz1349},
  \href{http://arxiv.org/abs/1904.01084}{{\tt arXiv:1904.01084}}.
%Type = Article
\bibitem[{{Behroozi} et~al.(2019){Behroozi}, {Wechsler}, {Hearin} and
  {Conroy}}]{behroozi:2019}
\bibinfo{author}{{Behroozi}, P.}, \bibinfo{author}{{Wechsler}, R.H.},
  \bibinfo{author}{{Hearin}, A.P.}, \bibinfo{author}{{Conroy}, C.},
  \bibinfo{year}{2019}.
\newblock \bibinfo{title}{{UNIVERSEMACHINE: The correlation between galaxy
  growth and dark matter halo assembly from z = 0-10}}.
\newblock \bibinfo{journal}{\mnras} \bibinfo{volume}{488},
  \bibinfo{pages}{3143--3194}.
\newblock \DOIprefix\doi{10.1093/mnras/stz1182},
  \href{http://arxiv.org/abs/1806.07893}{{\tt arXiv:1806.07893}}.
%Type = Article
\bibitem[{{Bekki}(2012)}]{bekki:2012}
\bibinfo{author}{{Bekki}, K.}, \bibinfo{year}{2012}.
\newblock \bibinfo{title}{{The influences of the Magellanic Clouds on the
  Galaxy: pole shift, warp and star formation history}}.
\newblock \bibinfo{journal}{\mnras} \bibinfo{volume}{422},
  \bibinfo{pages}{1957--1974}.
\newblock \DOIprefix\doi{10.1111/j.1365-2966.2012.20621.x},
  \href{http://arxiv.org/abs/1202.0358}{{\tt arXiv:1202.0358}}.
%Type = Article
\bibitem[{{Belokurov} et~al.(2018){Belokurov}, {Erkal}, {Evans}, {Koposov} and
  {Deason}}]{belokurov:2018}
\bibinfo{author}{{Belokurov}, V.}, \bibinfo{author}{{Erkal}, D.},
  \bibinfo{author}{{Evans}, N.W.}, \bibinfo{author}{{Koposov}, S.E.},
  \bibinfo{author}{{Deason}, A.J.}, \bibinfo{year}{2018}.
\newblock \bibinfo{title}{{Co-formation of the disc and the stellar halo}}.
\newblock \bibinfo{journal}{\mnras} \bibinfo{volume}{478},
  \bibinfo{pages}{611--619}.
\newblock \DOIprefix\doi{10.1093/mnras/sty982},
  \href{http://arxiv.org/abs/1802.03414}{{\tt arXiv:1802.03414}}.
%Type = Article
\bibitem[{Belokurov et~al.(2007)Belokurov, Evans, Irwin, {Lynden-Bell}, Yanny,
  Vidrih, Gilmore, Seabroke, Zucker, Wilkinson, Hewett, Bramich, Fellhauer,
  Newberg, Wyse, Beers, Bell, Barentine, Brinkmann, Cole, Pan and
  York}]{belokurov:2007}
Belokurov, V. et~al., \bibinfo{year}{2007}.
\newblock \bibinfo{title}{An {{Orphan}} in the ``{{Field}} of {{Streams}}''}.
\newblock \bibinfo{journal}{The Astrophysical Journal} \bibinfo{volume}{658},
  \bibinfo{pages}{337--344}.
\newblock \DOIprefix\doi{10.1086/511302}.
%Type = Article
\bibitem[{{Belokurov} et~al.(2006){Belokurov}, {Zucker}, {Evans}, {Gilmore},
  {Vidrih}, {Bramich}, {Newberg}, {Wyse}, {Irwin}, {Fellhauer}, {Hewett},
  {Walton}, {Wilkinson}, {Cole}, {Yanny}, {Rockosi}, {Beers}, {Bell},
  {Brinkmann}, {Ivezi{\'c}} and {Lupton}}]{belokurov:2006}
{Belokurov}, V. et~al., \bibinfo{year}{2006}.
\newblock \bibinfo{title}{{The Field of Streams: Sagittarius and Its
  Siblings}}.
\newblock \bibinfo{journal}{\apjl} \bibinfo{volume}{642},
  \bibinfo{pages}{L137--L140}.
\newblock \DOIprefix\doi{10.1086/504797},
  \href{http://arxiv.org/abs/astro-ph/0605025}{{\tt arXiv:astro-ph/0605025}}.
%Type = Article
\bibitem[{{Benitez-Llambay} and {Frenk}(2020)}]{benitez-llambay:2020}
\bibinfo{author}{{Benitez-Llambay}, A.}, \bibinfo{author}{{Frenk}, C.},
  \bibinfo{year}{2020}.
\newblock \bibinfo{title}{{The detailed structure and the onset of galaxy
  formation in low-mass gaseous dark matter haloes}}.
\newblock \bibinfo{journal}{\mnras} \bibinfo{volume}{498},
  \bibinfo{pages}{4887--4900}.
\newblock \DOIprefix\doi{10.1093/mnras/staa2698},
  \href{http://arxiv.org/abs/2004.06124}{{\tt arXiv:2004.06124}}.
%Type = Article
\bibitem[{{Benson}(2012)}]{benson:2012}
\bibinfo{author}{{Benson}, A.J.}, \bibinfo{year}{2012}.
\newblock \bibinfo{title}{{G ALACTICUS: A semi-analytic model of galaxy
  formation}}.
\newblock \bibinfo{journal}{\na} \bibinfo{volume}{17},
  \bibinfo{pages}{175--197}.
\newblock \DOIprefix\doi{10.1016/j.newast.2011.07.004},
  \href{http://arxiv.org/abs/1008.1786}{{\tt arXiv:1008.1786}}.
%Type = Article
\bibitem[{{Benson} et~al.(2002){Benson}, {Lacey}, {Baugh}, {Cole} and
  {Frenk}}]{benson:2002}
\bibinfo{author}{{Benson}, A.J.}, \bibinfo{author}{{Lacey}, C.G.},
  \bibinfo{author}{{Baugh}, C.M.}, \bibinfo{author}{{Cole}, S.},
  \bibinfo{author}{{Frenk}, C.S.}, \bibinfo{year}{2002}.
\newblock \bibinfo{title}{{The effects of photoionization on galaxy formation -
  I. Model and results at z=0}}.
\newblock \bibinfo{journal}{\mnras} \bibinfo{volume}{333},
  \bibinfo{pages}{156--176}.
\newblock \DOIprefix\doi{10.1046/j.1365-8711.2002.05387.x},
  \href{http://arxiv.org/abs/astro-ph/0108217}{{\tt arXiv:astro-ph/0108217}}.
%Type = Article
\bibitem[{{Bergemann} et~al.(2018){Bergemann}, {Sesar}, {Cohen}, {Serenelli},
  {Sheffield}, {Li}, {Casagrande}, {Johnston}, {Laporte}, {Price-Whelan},
  {Sch{\"o}nrich} and {Gould}}]{bergemann:2018}
{Bergemann}, M. et~al., \bibinfo{year}{2018}.
\newblock \bibinfo{title}{{Two chemically similar stellar overdensities on
  opposite sides of the plane of the Galactic disk}}.
\newblock \bibinfo{journal}{\nat} \bibinfo{volume}{555},
  \bibinfo{pages}{334--337}.
\newblock \DOIprefix\doi{10.1038/nature25490},
  \href{http://arxiv.org/abs/1803.00563}{{\tt arXiv:1803.00563}}.
%Type = Article
\bibitem[{{Bernard} et~al.(2014){Bernard}, {Ferguson}, {Schlafly}, {Abbas},
  {Bell}, {Deacon}, {Martin}, {Rix}, {Sesar}, {Slater}, {Penarrubia}, {Wyse},
  {Burgett}, {Chambers}, {Draper}, {Hodapp}, {Kaiser}, {Kudritzki}, {Magnier},
  {Metcalfe}, {Morgan}, {Price}, {Tonry}, {Wainscoat} and
  {Waters}}]{bernard:2014}
{Bernard}, E.~J. et~al., \bibinfo{year}{2014}.
\newblock \bibinfo{title}{{Serendipitous discovery of a thin stellar stream
  near the Galactic bulge in the Pan-STARRS1 3{\ensuremath{\pi}} Survey.}}
\newblock \bibinfo{journal}{\mnras} \bibinfo{volume}{443},
  \bibinfo{pages}{L84--L88}.
\newblock \DOIprefix\doi{10.1093/mnrasl/slu089},
  \href{http://arxiv.org/abs/1405.6645}{{\tt arXiv:1405.6645}}.
%Type = Article
\bibitem[{{Bernard} et~al.(2016){Bernard}, {Ferguson}, {Schlafly}, {Martin},
  {Rix}, {Bell}, {Finkbeiner}, {Goldman}, {Mart{\'\i}nez-Delgado}, {Sesar},
  {Wyse}, {Burgett}, {Chambers}, {Draper}, {Hodapp}, {Kaiser}, {Kudritzki},
  {Magnier}, {Metcalfe}, {Wainscoat} and {Waters}}]{bernard:2016}
{Bernard}, E.~J. et~al., \bibinfo{year}{2016}.
\newblock \bibinfo{title}{{A Synoptic Map of Halo Substructures from the
  Pan-STARRS1 3{\ensuremath{\pi}} Survey}}.
\newblock \bibinfo{journal}{\mnras} \bibinfo{volume}{463},
  \bibinfo{pages}{1759--1768}.
\newblock \DOIprefix\doi{10.1093/mnras/stw2134},
  \href{http://arxiv.org/abs/1607.06088}{{\tt arXiv:1607.06088}}.
%Type = Inproceedings
\bibitem[{{Bernstein} et~al.(2003){Bernstein}, {Shectman}, {Gunnels},
  {Mochnacki} and {Athey}}]{bernstein:2003}
\bibinfo{author}{{Bernstein}, R.}, \bibinfo{author}{{Shectman}, S.A.},
  \bibinfo{author}{{Gunnels}, S.M.}, \bibinfo{author}{{Mochnacki}, S.},
  \bibinfo{author}{{Athey}, A.E.}, \bibinfo{year}{2003}.
\newblock \bibinfo{title}{{MIKE: A Double Echelle Spectrograph for the Magellan
  Telescopes at Las Campanas Observatory}}, in: \bibinfo{editor}{{Iye}, M.},
  \bibinfo{editor}{{Moorwood}, A.F.M.} (Eds.), \bibinfo{booktitle}{Instrument
  Design and Performance for Optical/Infrared Ground-based Telescopes}, volume
  \bibinfo{volume}{4841} of \textit{\bibinfo{series}{Society of Photo-Optical
  Instrumentation Engineers (SPIE) Conference Series}}. pp.
  \bibinfo{pages}{1694--1704}.
\newblock \DOIprefix\doi{10.1117/12.461502}.
%Type = Article
\bibitem[{{Besla} et~al.(2007){Besla}, {Kallivayalil}, {Hernquist},
  {Robertson}, {Cox}, {van der Marel} and {Alcock}}]{besla:2007}
\bibinfo{author}{{Besla}, G.}, \bibinfo{author}{{Kallivayalil}, N.},
  \bibinfo{author}{{Hernquist}, L.}, \bibinfo{author}{{Robertson}, B.},
  \bibinfo{author}{{Cox}, T.J.}, \bibinfo{author}{{van der Marel}, R.P.},
  \bibinfo{author}{{Alcock}, C.}, \bibinfo{year}{2007}.
\newblock \bibinfo{title}{{Are the Magellanic Clouds on Their First Passage
  about the Milky Way?}}
\newblock \bibinfo{journal}{\apj} \bibinfo{volume}{668},
  \bibinfo{pages}{949--967}.
\newblock \DOIprefix\doi{10.1086/521385},
  \href{http://arxiv.org/abs/astro-ph/0703196}{{\tt arXiv:astro-ph/0703196}}.
%Type = Article
\bibitem[{{Bett} et~al.(2007){Bett}, {Eke}, {Frenk}, {Jenkins}, {Helly} and
  {Navarro}}]{bett:2007}
\bibinfo{author}{{Bett}, P.}, \bibinfo{author}{{Eke}, V.},
  \bibinfo{author}{{Frenk}, C.S.}, \bibinfo{author}{{Jenkins}, A.},
  \bibinfo{author}{{Helly}, J.}, \bibinfo{author}{{Navarro}, J.},
  \bibinfo{year}{2007}.
\newblock \bibinfo{title}{{The spin and shape of dark matter haloes in the
  Millennium simulation of a {\ensuremath{\Lambda}} cold dark matter
  universe}}.
\newblock \bibinfo{journal}{\mnras} \bibinfo{volume}{376},
  \bibinfo{pages}{215--232}.
\newblock \DOIprefix\doi{10.1111/j.1365-2966.2007.11432.x},
  \href{http://arxiv.org/abs/astro-ph/0608607}{{\tt arXiv:astro-ph/0608607}}.
%Type = Article
\bibitem[{{B{\'\i}lek} et~al.(2020){B{\'\i}lek}, {Duc}, {Cuillandre}, {Gwyn},
  {Cappellari}, {Bekaert}, {Bonfini}, {Bitsakis}, {Paudel}, {Krajnovi{\'c}},
  {Durrell} and {Marleau}}]{bilek:2020}
{B{\'\i}lek}, M. et~al., \bibinfo{year}{2020}.
\newblock \bibinfo{title}{{Census and classification of low-surface-brightness
  structures in nearby early-type galaxies from the MATLAS survey}}.
\newblock \bibinfo{journal}{\mnras} \bibinfo{volume}{498},
  \bibinfo{pages}{2138--2166}.
\newblock \DOIprefix\doi{10.1093/mnras/staa2248},
  \href{http://arxiv.org/abs/2007.13772}{{\tt arXiv:2007.13772}}.
%Type = Book
\bibitem[{{Binney} and {Tremaine}(2008)}]{bt:2008}
\bibinfo{author}{{Binney}, J.}, \bibinfo{author}{{Tremaine}, S.},
  \bibinfo{year}{2008}.
\newblock \bibinfo{title}{{Galactic Dynamics: Second Edition}}.
%Type = Article
\bibitem[{{Blitz} and {Spergel}(1991)}]{blitz:1991}
\bibinfo{author}{{Blitz}, L.}, \bibinfo{author}{{Spergel}, D.N.},
  \bibinfo{year}{1991}.
\newblock \bibinfo{title}{{Direct Evidence for a Bar at the Galactic Center}}.
\newblock \bibinfo{journal}{\apj} \bibinfo{volume}{379}, \bibinfo{pages}{631}.
\newblock \DOIprefix\doi{10.1086/170535}.
%Type = Article
\bibitem[{{Bode} et~al.(2001){Bode}, {Ostriker} and {Turok}}]{bode:2001}
\bibinfo{author}{{Bode}, P.}, \bibinfo{author}{{Ostriker}, J.P.},
  \bibinfo{author}{{Turok}, N.}, \bibinfo{year}{2001}.
\newblock \bibinfo{title}{{Halo Formation in Warm Dark Matter Models}}.
\newblock \bibinfo{journal}{\apj} \bibinfo{volume}{556},
  \bibinfo{pages}{93--107}.
\newblock \DOIprefix\doi{10.1086/321541},
  \href{http://arxiv.org/abs/astro-ph/0010389}{{\tt arXiv:astro-ph/0010389}}.
%Type = Article
\bibitem[{{Bonaca} et~al.(2020a){Bonaca}, {Conroy}, {Hogg}, {Cargile},
  {Caldwell}, {Naidu}, {Price-Whelan}, {Speagle} and {Johnson}}]{bonaca:2020b}
{Bonaca}, A. et~al., \bibinfo{year}{2020}a.
\newblock \bibinfo{title}{{High-resolution Spectroscopy of the GD-1 Stellar
  Stream Localizes the Perturber near the Orbital Plane of Sagittarius}}.
\newblock \bibinfo{journal}{\apjl} \bibinfo{volume}{892}, \bibinfo{pages}{L37}.
\newblock \DOIprefix\doi{10.3847/2041-8213/ab800c},
  \href{http://arxiv.org/abs/2001.07215}{{\tt arXiv:2001.07215}}.
%Type = Article
\bibitem[{{Bonaca} et~al.(2019a){Bonaca}, {Conroy}, {Price-Whelan} and
  {Hogg}}]{bonaca:2019}
\bibinfo{author}{{Bonaca}, A.}, \bibinfo{author}{{Conroy}, C.},
  \bibinfo{author}{{Price-Whelan}, A.M.}, \bibinfo{author}{{Hogg}, D.W.},
  \bibinfo{year}{2019}a.
\newblock \bibinfo{title}{{Multiple Components of the Jhelum Stellar Stream}}.
\newblock \bibinfo{journal}{\apjl} \bibinfo{volume}{881}, \bibinfo{pages}{L37}.
\newblock \DOIprefix\doi{10.3847/2041-8213/ab36ba},
  \href{http://arxiv.org/abs/1906.02748}{{\tt arXiv:1906.02748}}.
%Type = Article
\bibitem[{{Bonaca} et~al.(2012){Bonaca}, {Geha} and
  {Kallivayalil}}]{bonaca:2012}
\bibinfo{author}{{Bonaca}, A.}, \bibinfo{author}{{Geha}, M.},
  \bibinfo{author}{{Kallivayalil}, N.}, \bibinfo{year}{2012}.
\newblock \bibinfo{title}{{A Cold Milky Way Stellar Stream in the Direction of
  Triangulum}}.
\newblock \bibinfo{journal}{\apjl} \bibinfo{volume}{760}, \bibinfo{pages}{L6}.
\newblock \DOIprefix\doi{10.1088/2041-8205/760/1/L6},
  \href{http://arxiv.org/abs/1209.5391}{{\tt arXiv:1209.5391}}.
%Type = Article
\bibitem[{{Bonaca} et~al.(2014){Bonaca}, {Geha}, {K{\"u}pper}, {Diemand},
  {Johnston} and {Hogg}}]{bonaca:2014}
\bibinfo{author}{{Bonaca}, A.}, \bibinfo{author}{{Geha}, M.},
  \bibinfo{author}{{K{\"u}pper}, A.H.W.}, \bibinfo{author}{{Diemand}, J.},
  \bibinfo{author}{{Johnston}, K.V.}, \bibinfo{author}{{Hogg}, D.W.},
  \bibinfo{year}{2014}.
\newblock \bibinfo{title}{{Milky Way Mass and Potential Recovery Using Tidal
  Streams in a Realistic Halo}}.
\newblock \bibinfo{journal}{\apj} \bibinfo{volume}{795}, \bibinfo{pages}{94}.
\newblock \DOIprefix\doi{10.1088/0004-637X/795/1/94},
  \href{http://arxiv.org/abs/1406.6063}{{\tt arXiv:1406.6063}}.
%Type = Article
\bibitem[{{Bonaca} and {Hogg}(2018)}]{bh:2018}
\bibinfo{author}{{Bonaca}, A.}, \bibinfo{author}{{Hogg}, D.W.},
  \bibinfo{year}{2018}.
\newblock \bibinfo{title}{{The Information Content in Cold Stellar Streams}}.
\newblock \bibinfo{journal}{\apj} \bibinfo{volume}{867}, \bibinfo{pages}{101}.
\newblock \DOIprefix\doi{10.3847/1538-4357/aae4da},
  \href{http://arxiv.org/abs/1804.06854}{{\tt arXiv:1804.06854}}.
%Type = Article
\bibitem[{{Bonaca} et~al.(2019b){Bonaca}, {Hogg}, {Price-Whelan} and
  {Conroy}}]{bonaca:2019b}
\bibinfo{author}{{Bonaca}, A.}, \bibinfo{author}{{Hogg}, D.W.},
  \bibinfo{author}{{Price-Whelan}, A.M.}, \bibinfo{author}{{Conroy}, C.},
  \bibinfo{year}{2019}b.
\newblock \bibinfo{title}{{The Spur and the Gap in GD-1: Dynamical Evidence for
  a Dark Substructure in the Milky Way Halo}}.
\newblock \bibinfo{journal}{\apj} \bibinfo{volume}{880}, \bibinfo{pages}{38}.
\newblock \DOIprefix\doi{10.3847/1538-4357/ab2873},
  \href{http://arxiv.org/abs/1811.03631}{{\tt arXiv:1811.03631}}.
%Type = Article
\bibitem[{{Bonaca} et~al.(2021){Bonaca}, {Naidu}, {Conroy}, {Caldwell},
  {Cargile}, {Han}, {Johnson}, {Kruijssen}, {Myeong}, {Speagle}, {Ting} and
  {Zaritsky}}]{bonaca:2021}
{Bonaca}, A. et~al., \bibinfo{year}{2021}.
\newblock \bibinfo{title}{{Orbital Clustering Identifies the Origins of
  Galactic Stellar Streams}}.
\newblock \bibinfo{journal}{\apjl} \bibinfo{volume}{909}, \bibinfo{pages}{L26}.
\newblock \DOIprefix\doi{10.3847/2041-8213/abeaa9},
  \href{http://arxiv.org/abs/2012.09171}{{\tt arXiv:2012.09171}}.
%Type = Article
\bibitem[{{Bonaca} et~al.(2020b){Bonaca}, {Pearson}, {Price-Whelan}, {Dey},
  {Geha}, {Kallivayalil}, {Moustakas}, {Mu{\~n}oz}, {Myers}, {Schlegel} and
  {Valdes}}]{bonaca:2020}
{Bonaca}, A. et~al., \bibinfo{year}{2020}b.
\newblock \bibinfo{title}{{Variations in the Width, Density, and Direction of
  the Palomar 5 Tidal Tails}}.
\newblock \bibinfo{journal}{\apj} \bibinfo{volume}{889}, \bibinfo{pages}{70}.
\newblock \DOIprefix\doi{10.3847/1538-4357/ab5afe},
  \href{http://arxiv.org/abs/1910.00592}{{\tt arXiv:1910.00592}}.
%Type = Article
\bibitem[{{Borsato} et~al.(2020){Borsato}, {Martell} and
  {Simpson}}]{borsato:2020}
\bibinfo{author}{{Borsato}, N.W.}, \bibinfo{author}{{Martell}, S.L.},
  \bibinfo{author}{{Simpson}, J.D.}, \bibinfo{year}{2020}.
\newblock \bibinfo{title}{{Identifying stellar streams in Gaia DR2 with data
  mining techniques}}.
\newblock \bibinfo{journal}{\mnras} \bibinfo{volume}{492},
  \bibinfo{pages}{1370--1384}.
\newblock \DOIprefix\doi{10.1093/mnras/stz3479},
  \href{http://arxiv.org/abs/1907.02527}{{\tt arXiv:1907.02527}}.
%Type = Article
\bibitem[{{Bose} et~al.(2020){Bose}, {Deason}, {Belokurov} and
  {Frenk}}]{bose:2020}
\bibinfo{author}{{Bose}, S.}, \bibinfo{author}{{Deason}, A.J.},
  \bibinfo{author}{{Belokurov}, V.}, \bibinfo{author}{{Frenk}, C.S.},
  \bibinfo{year}{2020}.
\newblock \bibinfo{title}{{The little things matter: relating the abundance of
  ultrafaint satellites to the hosts' assembly history}}.
\newblock \bibinfo{journal}{\mnras} \bibinfo{volume}{495},
  \bibinfo{pages}{743--757}.
\newblock \DOIprefix\doi{10.1093/mnras/staa1199},
  \href{http://arxiv.org/abs/1909.04039}{{\tt arXiv:1909.04039}}.
%Type = Article
\bibitem[{{Bovy}(2014)}]{bovy:2014}
\bibinfo{author}{{Bovy}, J.}, \bibinfo{year}{2014}.
\newblock \bibinfo{title}{{Dynamical Modeling of Tidal Streams}}.
\newblock \bibinfo{journal}{\apj} \bibinfo{volume}{795}, \bibinfo{pages}{95}.
\newblock \DOIprefix\doi{10.1088/0004-637X/795/1/95},
  \href{http://arxiv.org/abs/1401.2985}{{\tt arXiv:1401.2985}}.
%Type = Article
\bibitem[{{Bovy}(2015)}]{bovy:2015}
\bibinfo{author}{{Bovy}, J.}, \bibinfo{year}{2015}.
\newblock \bibinfo{title}{{galpy: A python Library for Galactic Dynamics}}.
\newblock \bibinfo{journal}{\apjs} \bibinfo{volume}{216}, \bibinfo{pages}{29}.
\newblock \DOIprefix\doi{10.1088/0067-0049/216/2/29},
  \href{http://arxiv.org/abs/1412.3451}{{\tt arXiv:1412.3451}}.
%Type = Article
\bibitem[{{Bovy} et~al.(2016){Bovy}, {Bahmanyar}, {Fritz} and
  {Kallivayalil}}]{bovy:2016}
\bibinfo{author}{{Bovy}, J.}, \bibinfo{author}{{Bahmanyar}, A.},
  \bibinfo{author}{{Fritz}, T.K.}, \bibinfo{author}{{Kallivayalil}, N.},
  \bibinfo{year}{2016}.
\newblock \bibinfo{title}{{The Shape of the Inner Milky Way Halo from
  Observations of the Pal 5 and GD--1 Stellar Streams}}.
\newblock \bibinfo{journal}{\apj} \bibinfo{volume}{833}, \bibinfo{pages}{31}.
\newblock \DOIprefix\doi{10.3847/1538-4357/833/1/31},
  \href{http://arxiv.org/abs/1609.01298}{{\tt arXiv:1609.01298}}.
%Type = Article
\bibitem[{{Bovy} et~al.(2017){Bovy}, {Erkal} and {Sanders}}]{bovy:2017}
\bibinfo{author}{{Bovy}, J.}, \bibinfo{author}{{Erkal}, D.},
  \bibinfo{author}{{Sanders}, J.L.}, \bibinfo{year}{2017}.
\newblock \bibinfo{title}{{Linear perturbation theory for tidal streams and the
  small-scale CDM power spectrum}}.
\newblock \bibinfo{journal}{\mnras} \bibinfo{volume}{466},
  \bibinfo{pages}{628--668}.
\newblock \DOIprefix\doi{10.1093/mnras/stw3067},
  \href{http://arxiv.org/abs/1606.03470}{{\tt arXiv:1606.03470}}.
%Type = Article
\bibitem[{{Breen} et~al.(2020){Breen}, {Foley}, {Boekholt} and {Portegies
  Zwart}}]{breen:2020}
\bibinfo{author}{{Breen}, P.G.}, \bibinfo{author}{{Foley}, C.N.},
  \bibinfo{author}{{Boekholt}, T.}, \bibinfo{author}{{Portegies Zwart}, S.},
  \bibinfo{year}{2020}.
\newblock \bibinfo{title}{{Newton versus the machine: solving the chaotic
  three-body problem using deep neural networks}}.
\newblock \bibinfo{journal}{\mnras} \bibinfo{volume}{494},
  \bibinfo{pages}{2465--2470}.
\newblock \DOIprefix\doi{10.1093/mnras/staa713},
  \href{http://arxiv.org/abs/1910.07291}{{\tt arXiv:1910.07291}}.
%Type = Article
\bibitem[{{Bressan} et~al.(2012){Bressan}, {Marigo}, {Girardi}, {Salasnich},
  {Dal Cero}, {Rubele} and {Nanni}}]{bressan:2012}
\bibinfo{author}{{Bressan}, A.}, \bibinfo{author}{{Marigo}, P.},
  \bibinfo{author}{{Girardi}, L.}, \bibinfo{author}{{Salasnich}, B.},
  \bibinfo{author}{{Dal Cero}, C.}, \bibinfo{author}{{Rubele}, S.},
  \bibinfo{author}{{Nanni}, A.}, \bibinfo{year}{2012}.
\newblock \bibinfo{title}{{PARSEC: stellar tracks and isochrones with the
  PAdova and TRieste Stellar Evolution Code}}.
\newblock \bibinfo{journal}{\mnras} \bibinfo{volume}{427},
  \bibinfo{pages}{127--145}.
\newblock \DOIprefix\doi{10.1111/j.1365-2966.2012.21948.x},
  \href{http://arxiv.org/abs/1208.4498}{{\tt arXiv:1208.4498}}.
%Type = Article
\bibitem[{{Buch} et~al.(2024){Buch}, {Nadler}, {Wechsler} and
  {Mao}}]{buch:2024}
\bibinfo{author}{{Buch}, D.}, \bibinfo{author}{{Nadler}, E.O.},
  \bibinfo{author}{{Wechsler}, R.H.}, \bibinfo{author}{{Mao}, Y.Y.},
  \bibinfo{year}{2024}.
\newblock \bibinfo{title}{{Milky Way-est: Cosmological Zoom-in Simulations with
  Large Magellanic Cloud and Gaia-Sausage-Enceladus Analogs}}.
\newblock \bibinfo{journal}{arXiv e-prints} ,
  \bibinfo{pages}{arXiv:2404.08043}\DOIprefix\doi{10.48550/arXiv.2404.08043},
  \href{http://arxiv.org/abs/2404.08043}{{\tt arXiv:2404.08043}}.
%Type = Article
\bibitem[{{Buder} et~al.(2021){Buder}, {Sharma}, {Kos}, {Amarsi}, {Nordlander},
  {Lind}, {Martell}, {Asplund}, {Bland-Hawthorn}, {Casey}, {de Silva},
  {D'Orazi}, {Freeman}, {Hayden}, {Lewis}, {Lin}, {Schlesinger}, {Simpson},
  {Stello}, {Zucker}, {Zwitter}, {Beeson}, {Buck}, {Casagrande}, {Clark},
  {{\v{C}}otar}, {da Costa}, {de Grijs}, {Feuillet}, {Horner}, {Kafle},
  {Khanna}, {Kobayashi}, {Liu}, {Montet}, {Nandakumar}, {Nataf}, {Ness},
  {Spina}, {Tepper-Garc{\'\i}a}, {Ting}, {Traven}, {Vogrin{\v{c}}i{\v{c}}},
  {Wittenmyer}, {Wyse}, {{\v{Z}}erjal} and {Galah Collaboration}}]{buder:2021}
{Buder}, S. et~al., \bibinfo{year}{2021}.
\newblock \bibinfo{title}{{The GALAH+ survey: Third data release}}.
\newblock \bibinfo{journal}{\mnras} \bibinfo{volume}{506},
  \bibinfo{pages}{150--201}.
\newblock \DOIprefix\doi{10.1093/mnras/stab1242},
  \href{http://arxiv.org/abs/2011.02505}{{\tt arXiv:2011.02505}}.
%Type = Article
\bibitem[{{Bullock} and {Johnston}(2005)}]{bullock:2005}
\bibinfo{author}{{Bullock}, J.S.}, \bibinfo{author}{{Johnston}, K.V.},
  \bibinfo{year}{2005}.
\newblock \bibinfo{title}{{Tracing Galaxy Formation with Stellar Halos. I.
  Methods}}.
\newblock \bibinfo{journal}{\apj} \bibinfo{volume}{635},
  \bibinfo{pages}{931--949}.
\newblock \DOIprefix\doi{10.1086/497422},
  \href{http://arxiv.org/abs/astro-ph/0506467}{{\tt arXiv:astro-ph/0506467}}.
%Type = Article
\bibitem[{{Bullock} et~al.(2000){Bullock}, {Kravtsov} and
  {Weinberg}}]{bullock:2000}
\bibinfo{author}{{Bullock}, J.S.}, \bibinfo{author}{{Kravtsov}, A.V.},
  \bibinfo{author}{{Weinberg}, D.H.}, \bibinfo{year}{2000}.
\newblock \bibinfo{title}{{Reionization and the Abundance of Galactic
  Satellites}}.
\newblock \bibinfo{journal}{\apj} \bibinfo{volume}{539},
  \bibinfo{pages}{517--521}.
\newblock \DOIprefix\doi{10.1086/309279},
  \href{http://arxiv.org/abs/astro-ph/0002214}{{\tt arXiv:astro-ph/0002214}}.
%Type = Article
\bibitem[{{Caldwell} et~al.(2020){Caldwell}, {Bonaca}, {Price-Whelan}, {Sesar}
  and {Walker}}]{caldwell:2020}
\bibinfo{author}{{Caldwell}, N.}, \bibinfo{author}{{Bonaca}, A.},
  \bibinfo{author}{{Price-Whelan}, A.M.}, \bibinfo{author}{{Sesar}, B.},
  \bibinfo{author}{{Walker}, M.G.}, \bibinfo{year}{2020}.
\newblock \bibinfo{title}{{A Larger Extent for the Ophiuchus Stream}}.
\newblock \bibinfo{journal}{\aj} \bibinfo{volume}{159}, \bibinfo{pages}{287}.
\newblock \DOIprefix\doi{10.3847/1538-3881/ab8cbf},
  \href{http://arxiv.org/abs/2004.14350}{{\tt arXiv:2004.14350}}.
%Type = Article
\bibitem[{{Carlberg}(2009)}]{carlberg:2009}
\bibinfo{author}{{Carlberg}, R.G.}, \bibinfo{year}{2009}.
\newblock \bibinfo{title}{{Star Stream Folding by Dark Galactic Subhalos}}.
\newblock \bibinfo{journal}{\apjl} \bibinfo{volume}{705},
  \bibinfo{pages}{L223--L226}.
\newblock \DOIprefix\doi{10.1088/0004-637X/705/2/L223},
  \href{http://arxiv.org/abs/0908.4345}{{\tt arXiv:0908.4345}}.
%Type = Article
\bibitem[{{Carlberg}(2012)}]{carlberg:2012b}
\bibinfo{author}{{Carlberg}, R.G.}, \bibinfo{year}{2012}.
\newblock \bibinfo{title}{{Dark Matter Sub-halo Counts via Star Stream
  Crossings}}.
\newblock \bibinfo{journal}{\apj} \bibinfo{volume}{748}, \bibinfo{pages}{20}.
\newblock \DOIprefix\doi{10.1088/0004-637X/748/1/20},
  \href{http://arxiv.org/abs/1109.6022}{{\tt arXiv:1109.6022}}.
%Type = Article
\bibitem[{{Carlberg}(2013)}]{carlberg:2013b}
\bibinfo{author}{{Carlberg}, R.G.}, \bibinfo{year}{2013}.
\newblock \bibinfo{title}{{The Dynamics of Star Stream Gaps}}.
\newblock \bibinfo{journal}{\apj} \bibinfo{volume}{775}, \bibinfo{pages}{90}.
\newblock \DOIprefix\doi{10.1088/0004-637X/775/2/90},
  \href{http://arxiv.org/abs/1307.1929}{{\tt arXiv:1307.1929}}.
%Type = Article
\bibitem[{{Carlberg}(2016)}]{carlberg:2016}
\bibinfo{author}{{Carlberg}, R.G.}, \bibinfo{year}{2016}.
\newblock \bibinfo{title}{{Modeling GD-1 Gaps in a Milky Way Potential}}.
\newblock \bibinfo{journal}{\apj} \bibinfo{volume}{820}, \bibinfo{pages}{45}.
\newblock \DOIprefix\doi{10.3847/0004-637X/820/1/45},
  \href{http://arxiv.org/abs/1512.01620}{{\tt arXiv:1512.01620}}.
%Type = Article
\bibitem[{{Carlberg}(2018)}]{carlberg:2018}
\bibinfo{author}{{Carlberg}, R.G.}, \bibinfo{year}{2018}.
\newblock \bibinfo{title}{{Globular Clusters in a Cosmological N-body
  Simulation}}.
\newblock \bibinfo{journal}{\apj} \bibinfo{volume}{861}, \bibinfo{pages}{69}.
\newblock \DOIprefix\doi{10.3847/1538-4357/aac88a},
  \href{http://arxiv.org/abs/1706.01938}{{\tt arXiv:1706.01938}}.
%Type = Article
\bibitem[{{Carlberg}(2020)}]{carlberg:2020}
\bibinfo{author}{{Carlberg}, R.G.}, \bibinfo{year}{2020}.
\newblock \bibinfo{title}{{The Density Structure of Simulated Stellar
  Streams}}.
\newblock \bibinfo{journal}{\apj} \bibinfo{volume}{889}, \bibinfo{pages}{107}.
\newblock \DOIprefix\doi{10.3847/1538-4357/ab61f0},
  \href{http://arxiv.org/abs/1811.10084}{{\tt arXiv:1811.10084}}.
%Type = Article
\bibitem[{{Carlberg} and {Grillmair}(2013)}]{carlberg:2013}
\bibinfo{author}{{Carlberg}, R.G.}, \bibinfo{author}{{Grillmair}, C.J.},
  \bibinfo{year}{2013}.
\newblock \bibinfo{title}{{Gaps in the GD-1 Star Stream}}.
\newblock \bibinfo{journal}{\apj} \bibinfo{volume}{768}, \bibinfo{pages}{171}.
\newblock \DOIprefix\doi{10.1088/0004-637X/768/2/171},
  \href{http://arxiv.org/abs/1303.4342}{{\tt arXiv:1303.4342}}.
%Type = Article
\bibitem[{{Carlberg} and {Grillmair}(2016)}]{carlberg:2016b}
\bibinfo{author}{{Carlberg}, R.G.}, \bibinfo{author}{{Grillmair}, C.J.},
  \bibinfo{year}{2016}.
\newblock \bibinfo{title}{{Velocity Variations in the Phoenix-Hermus Star
  Stream}}.
\newblock \bibinfo{journal}{\apj} \bibinfo{volume}{830}, \bibinfo{pages}{135}.
\newblock \DOIprefix\doi{10.3847/0004-637X/830/2/135},
  \href{http://arxiv.org/abs/1606.05769}{{\tt arXiv:1606.05769}}.
%Type = Article
\bibitem[{{Carlberg} et~al.(2012){Carlberg}, {Grillmair} and
  {Hetherington}}]{carlberg:2012}
\bibinfo{author}{{Carlberg}, R.G.}, \bibinfo{author}{{Grillmair}, C.J.},
  \bibinfo{author}{{Hetherington}, N.}, \bibinfo{year}{2012}.
\newblock \bibinfo{title}{{The Pal 5 Star Stream Gaps}}.
\newblock \bibinfo{journal}{\apj} \bibinfo{volume}{760}, \bibinfo{pages}{75}.
\newblock \DOIprefix\doi{10.1088/0004-637X/760/1/75},
  \href{http://arxiv.org/abs/1209.1741}{{\tt arXiv:1209.1741}}.
%Type = Article
\bibitem[{{Carlberg} et~al.(2011){Carlberg}, {Richer}, {McConnachie}, {Irwin},
  {Ibata}, {Dotter}, {Chapman}, {Fardal}, {Ferguson}, {Lewis}, {Navarro},
  {Puzia} and {Valls-Gabaud}}]{carlberg:2011}
{Carlberg}, R.~G. et~al., \bibinfo{year}{2011}.
\newblock \bibinfo{title}{{Density Variations in the NW Star Stream of M31}}.
\newblock \bibinfo{journal}{\apj} \bibinfo{volume}{731}, \bibinfo{pages}{124}.
\newblock \DOIprefix\doi{10.1088/0004-637X/731/2/124},
  \href{http://arxiv.org/abs/1102.3501}{{\tt arXiv:1102.3501}}.
%Type = Article
\bibitem[{{Casetti-Dinescu} et~al.(2006){Casetti-Dinescu}, {Majewski},
  {Girard}, {Carlin}, {van Altena}, {Patterson} and {Law}}]{casetti:2006}
\bibinfo{author}{{Casetti-Dinescu}, D.I.}, \bibinfo{author}{{Majewski}, S.R.},
  \bibinfo{author}{{Girard}, T.M.}, \bibinfo{author}{{Carlin}, J.L.},
  \bibinfo{author}{{van Altena}, W.F.}, \bibinfo{author}{{Patterson}, R.J.},
  \bibinfo{author}{{Law}, D.R.}, \bibinfo{year}{2006}.
\newblock \bibinfo{title}{{A Deep Proper-Motion Survey in Kapteyn Selected
  Areas. I. Survey Description and First Results for Stars in the Tidal Tail of
  Sagittarius and in the Monoceros Ring}}.
\newblock \bibinfo{journal}{\aj} \bibinfo{volume}{132},
  \bibinfo{pages}{2082--2098}.
\newblock \DOIprefix\doi{10.1086/508274},
  \href{http://arxiv.org/abs/astro-ph/0608115}{{\tt arXiv:astro-ph/0608115}}.
%Type = Article
\bibitem[{{Casey} et~al.(2013){Casey}, {Da Costa}, {Keller} and
  {Maunder}}]{casey:2013}
\bibinfo{author}{{Casey}, A.R.}, \bibinfo{author}{{Da Costa}, G.},
  \bibinfo{author}{{Keller}, S.C.}, \bibinfo{author}{{Maunder}, E.},
  \bibinfo{year}{2013}.
\newblock \bibinfo{title}{{Hunting the Parent of the Orphan Stream: Identifying
  Stream Members from Low-resolution Spectroscopy}}.
\newblock \bibinfo{journal}{\apj} \bibinfo{volume}{764}, \bibinfo{pages}{39}.
\newblock \DOIprefix\doi{10.1088/0004-637X/764/1/39},
  \href{http://arxiv.org/abs/1301.1343}{{\tt arXiv:1301.1343}}.
%Type = Article
\bibitem[{{Casey} et~al.(2021){Casey}, {Ji}, {Hansen}, {Li}, {Koposov}, {Da
  Costa}, {Bland-Hawthorn}, {Cullinane}, {Erkal}, {Lewis}, {Kuehn}, {Mackey},
  {Martell}, {Pace}, {Simpson} and {Zucker}}]{casey:2021}
{Casey}, A.~R. et~al., \bibinfo{year}{2021}.
\newblock \bibinfo{title}{{Signature of a Massive Rotating Metal-poor Star
  Imprinted in the Phoenix Stellar Stream}}.
\newblock \bibinfo{journal}{\apj} \bibinfo{volume}{921}, \bibinfo{pages}{67}.
\newblock \DOIprefix\doi{10.3847/1538-4357/ac1346},
  \href{http://arxiv.org/abs/2109.03948}{{\tt arXiv:2109.03948}}.
%Type = Article
\bibitem[{{Casey} et~al.(2014){Casey}, {Keller}, {Da Costa}, {Frebel} and
  {Maunder}}]{casey:2014}
\bibinfo{author}{{Casey}, A.R.}, \bibinfo{author}{{Keller}, S.C.},
  \bibinfo{author}{{Da Costa}, G.}, \bibinfo{author}{{Frebel}, A.},
  \bibinfo{author}{{Maunder}, E.}, \bibinfo{year}{2014}.
\newblock \bibinfo{title}{{Hunting the Parent of the Orphan Stream. II. The
  First High-resolution Spectroscopic Study}}.
\newblock \bibinfo{journal}{\apj} \bibinfo{volume}{784}, \bibinfo{pages}{19}.
\newblock \DOIprefix\doi{10.1088/0004-637X/784/1/19},
  \href{http://arxiv.org/abs/1309.3563}{{\tt arXiv:1309.3563}}.
%Type = Article
\bibitem[{{Chambers} et~al.(2016){Chambers}, {Magnier}, {Metcalfe},
  {Flewelling}, {Huber}, {Waters}, {Denneau}, {Draper}, {Farrow}, {Finkbeiner},
  {Holmberg}, {Koppenhoefer}, {Price}, {Rest}, {Saglia}, {Schlafly}, {Smartt},
  {Sweeney}, {Wainscoat}, {Burgett}, {Chastel}, {Grav}, {Heasley}, {Hodapp},
  {Jedicke}, {Kaiser}, {Kudritzki}, {Luppino}, {Lupton}, {Monet}, {Morgan},
  {Onaka}, {Shiao}, {Stubbs}, {Tonry}, {White}, {Ba{\~n}ados}, {Bell},
  {Bender}, {Bernard}, {Boegner}, {Boffi}, {Botticella}, {Calamida},
  {Casertano}, {Chen}, {Chen}, {Cole}, {Deacon}, {Frenk}, {Fitzsimmons},
  {Gezari}, {Gibbs}, {Goessl}, {Goggia}, {Gourgue}, {Goldman}, {Grant},
  {Grebel}, {Hambly}, {Hasinger}, {Heavens}, {Heckman}, {Henderson}, {Henning},
  {Holman}, {Hopp}, {Ip}, {Isani}, {Jackson}, {Keyes}, {Koekemoer}, {Kotak},
  {Le}, {Liska}, {Long}, {Lucey}, {Liu}, {Martin}, {Masci}, {McLean}, {Mindel},
  {Misra}, {Morganson}, {Murphy}, {Obaika}, {Narayan}, {Nieto-Santisteban},
  {Norberg}, {Peacock}, {Pier}, {Postman}, {Primak}, {Rae}, {Rai}, {Riess},
  {Riffeser}, {Rix}, {R{\"o}ser}, {Russel}, {Rutz}, {Schilbach}, {Schultz},
  {Scolnic}, {Strolger}, {Szalay}, {Seitz}, {Small}, {Smith}, {Soderblom},
  {Taylor}, {Thomson}, {Taylor}, {Thakar}, {Thiel}, {Thilker}, {Unger},
  {Urata}, {Valenti}, {Wagner}, {Walder}, {Walter}, {Watters}, {Werner},
  {Wood-Vasey} and {Wyse}}]{chambers:2016}
{Chambers}, K.~C. et~al., \bibinfo{year}{2016}.
\newblock \bibinfo{title}{{The Pan-STARRS1 Surveys}}.
\newblock \bibinfo{journal}{arXiv e-prints} ,
  \bibinfo{pages}{arXiv:1612.05560}\DOIprefix\doi{10.48550/arXiv.1612.05560},
  \href{http://arxiv.org/abs/1612.05560}{{\tt arXiv:1612.05560}}.
%Type = Article
\bibitem[{{Chandra} et~al.(2022){Chandra}, {Conroy}, {Caldwell}, {Bonaca},
  {Naidu}, {Zaritsky}, {Cargile}, {Han}, {Johnson}, {Speagle}, {Ting} and
  {Woody}}]{chandra:2022}
{Chandra}, V. et~al., \bibinfo{year}{2022}.
\newblock \bibinfo{title}{{A Ghost in Bo{\"o}tes: The Least-Luminous Disrupted
  Dwarf Galaxy}}.
\newblock \bibinfo{journal}{\apj} \bibinfo{volume}{940}, \bibinfo{pages}{127}.
\newblock \DOIprefix\doi{10.3847/1538-4357/ac9b4b},
  \href{http://arxiv.org/abs/2207.13717}{{\tt arXiv:2207.13717}}.
%Type = Article
\bibitem[{{Chang} et~al.(2020){Chang}, {Yuan}, {Xue}, {Simion}, {Kang}, {Li},
  {Zhao} and {Zhao}}]{chang:2020}
\bibinfo{author}{{Chang}, J.}, \bibinfo{author}{{Yuan}, Z.},
  \bibinfo{author}{{Xue}, X.X.}, \bibinfo{author}{{Simion}, I.T.},
  \bibinfo{author}{{Kang}, X.}, \bibinfo{author}{{Li}, T.S.},
  \bibinfo{author}{{Zhao}, J.K.}, \bibinfo{author}{{Zhao}, G.},
  \bibinfo{year}{2020}.
\newblock \bibinfo{title}{{Is NGC 5824 the Core of the Progenitor of the Cetus
  Stream?}}
\newblock \bibinfo{journal}{\apj} \bibinfo{volume}{905}, \bibinfo{pages}{100}.
\newblock \DOIprefix\doi{10.3847/1538-4357/abc338},
  \href{http://arxiv.org/abs/2003.02378}{{\tt arXiv:2003.02378}}.
%Type = Article
\bibitem[{{Chen} et~al.(2015){Chen}, {Bressan}, {Girardi}, {Marigo}, {Kong} and
  {Lanza}}]{chen:2015}
\bibinfo{author}{{Chen}, Y.}, \bibinfo{author}{{Bressan}, A.},
  \bibinfo{author}{{Girardi}, L.}, \bibinfo{author}{{Marigo}, P.},
  \bibinfo{author}{{Kong}, X.}, \bibinfo{author}{{Lanza}, A.},
  \bibinfo{year}{2015}.
\newblock \bibinfo{title}{{PARSEC evolutionary tracks of massive stars up to
  350 M$_{{\ensuremath{\odot}}}$ at metallicities 0.0001 {\ensuremath{\leq}} Z
  {\ensuremath{\leq}} 0.04}}.
\newblock \bibinfo{journal}{\mnras} \bibinfo{volume}{452},
  \bibinfo{pages}{1068--1080}.
\newblock \DOIprefix\doi{10.1093/mnras/stv1281},
  \href{http://arxiv.org/abs/1506.01681}{{\tt arXiv:1506.01681}}.
%Type = Article
\bibitem[{{Chen} and {Gnedin}(2023)}]{chen:2023}
\bibinfo{author}{{Chen}, Y.}, \bibinfo{author}{{Gnedin}, O.Y.},
  \bibinfo{year}{2023}.
\newblock \bibinfo{title}{{Formation of globular clusters in dwarf galaxies of
  the Local Group}}.
\newblock \bibinfo{journal}{\mnras} \bibinfo{volume}{522},
  \bibinfo{pages}{5638--5653}.
\newblock \DOIprefix\doi{10.1093/mnras/stad1328},
  \href{http://arxiv.org/abs/2301.08218}{{\tt arXiv:2301.08218}}.
%Type = Article
\bibitem[{{Cheng} et~al.(2020){Cheng}, {Ting}, {M{\'e}nard} and
  {Bruna}}]{cheng:2020}
\bibinfo{author}{{Cheng}, S.}, \bibinfo{author}{{Ting}, Y.S.},
  \bibinfo{author}{{M{\'e}nard}, B.}, \bibinfo{author}{{Bruna}, J.},
  \bibinfo{year}{2020}.
\newblock \bibinfo{title}{{A new approach to observational cosmology using the
  scattering transform}}.
\newblock \bibinfo{journal}{\mnras} \bibinfo{volume}{499},
  \bibinfo{pages}{5902--5914}.
\newblock \DOIprefix\doi{10.1093/mnras/staa3165},
  \href{http://arxiv.org/abs/2006.08561}{{\tt arXiv:2006.08561}}.
%Type = Article
\bibitem[{{Choksi} et~al.(2018){Choksi}, {Gnedin} and {Li}}]{choksi:2018}
\bibinfo{author}{{Choksi}, N.}, \bibinfo{author}{{Gnedin}, O.Y.},
  \bibinfo{author}{{Li}, H.}, \bibinfo{year}{2018}.
\newblock \bibinfo{title}{{Formation of globular cluster systems: from dwarf
  galaxies to giants}}.
\newblock \bibinfo{journal}{\mnras} \bibinfo{volume}{480},
  \bibinfo{pages}{2343--2356}.
\newblock \DOIprefix\doi{10.1093/mnras/sty1952},
  \href{http://arxiv.org/abs/1801.03515}{{\tt arXiv:1801.03515}}.
%Type = Article
\bibitem[{{Chua} et~al.(2019){Chua}, {Pillepich}, {Vogelsberger} and
  {Hernquist}}]{chua:2019}
\bibinfo{author}{{Chua}, K.T.E.}, \bibinfo{author}{{Pillepich}, A.},
  \bibinfo{author}{{Vogelsberger}, M.}, \bibinfo{author}{{Hernquist}, L.},
  \bibinfo{year}{2019}.
\newblock \bibinfo{title}{{Shape of dark matter haloes in the Illustris
  simulation: effects of baryons}}.
\newblock \bibinfo{journal}{\mnras} \bibinfo{volume}{484},
  \bibinfo{pages}{476--493}.
\newblock \DOIprefix\doi{10.1093/mnras/sty3531},
  \href{http://arxiv.org/abs/1809.07255}{{\tt arXiv:1809.07255}}.
%Type = Article
\bibitem[{{Conroy} et~al.(2019){Conroy}, {Bonaca}, {Cargile}, {Johnson},
  {Caldwell}, {Naidu}, {Zaritsky}, {Fabricant}, {Moran}, {Rhee},
  {Szentgyorgyi}, {Berlind}, {Calkins}, {Kattner} and {Ly}}]{conroy:2019}
{Conroy}, C. et~al., \bibinfo{year}{2019}.
\newblock \bibinfo{title}{{Mapping the Stellar Halo with the H3 Spectroscopic
  Survey}}.
\newblock \bibinfo{journal}{\apj} \bibinfo{volume}{883}, \bibinfo{pages}{107}.
\newblock \DOIprefix\doi{10.3847/1538-4357/ab38b8},
  \href{http://arxiv.org/abs/1907.07684}{{\tt arXiv:1907.07684}}.
%Type = Article
\bibitem[{{Cooper} et~al.(2010){Cooper}, {Cole}, {Frenk}, {White}, {Helly},
  {Benson}, {De Lucia}, {Helmi}, {Jenkins}, {Navarro}, {Springel} and
  {Wang}}]{cooper:2010}
{Cooper}, A.~P. et~al., \bibinfo{year}{2010}.
\newblock \bibinfo{title}{{Galactic stellar haloes in the CDM model}}.
\newblock \bibinfo{journal}{\mnras} \bibinfo{volume}{406},
  \bibinfo{pages}{744--766}.
\newblock \DOIprefix\doi{10.1111/j.1365-2966.2010.16740.x},
  \href{http://arxiv.org/abs/0910.3211}{{\tt arXiv:0910.3211}}.
%Type = Article
\bibitem[{{Cooper} et~al.(2023){Cooper}, {Koposov}, {Allende Prieto}, {Manser},
  {Kizhuprakkat}, {Myers}, {Dey}, {G{\"a}nsicke}, {Li}, {Rockosi}, {Valluri},
  {Najita}, {Deason}, {Raichoor}, {Wang}, {Ting}, {Kim}, {Carrillo}, {Wang},
  {Beraldo e Silva}, {Han}, {Ding}, {S{\'a}nchez-Conde}, {Aguilar}, {Ahlen},
  {Bailey}, {Belokurov}, {Brooks}, {Cunha}, {Dawson}, {de la Macorra}, {Doel},
  {Eisenstein}, {Fagrelius}, {Fanning}, {Font-Ribera}, {Forero-Romero},
  {Gazta{\~n}aga}, {Gontcho a Gontcho}, {Guy}, {Honscheid}, {Kehoe}, {Kisner},
  {Kremin}, {Landriau}, {Levi}, {Martini}, {Meisner}, {Miquel}, {Moustakas},
  {Nie}, {Palanque-Delabrouille}, {Percival}, {Poppett}, {Prada}, {Rehemtulla},
  {Schlafly}, {Schlegel}, {Schubnell}, {Sharples}, {Tarl{\'e}}, {Wechsler},
  {Weinberg}, {Zhou} and {Zou}}]{cooper:2023}
{Cooper}, A.~P. et~al., \bibinfo{year}{2023}.
\newblock \bibinfo{title}{{Overview of the DESI Milky Way Survey}}.
\newblock \bibinfo{journal}{\apj} \bibinfo{volume}{947}, \bibinfo{pages}{37}.
\newblock \DOIprefix\doi{10.3847/1538-4357/acb3c0},
  \href{http://arxiv.org/abs/2208.08514}{{\tt arXiv:2208.08514}}.
%Type = Article
\bibitem[{{Cunningham} et~al.(2023){Cunningham}, {Hunt}, {Price-Whelan},
  {Johnston}, {Ness}, {Lu}, {Escala} and {Stelea}}]{cunningham:2024}
\bibinfo{author}{{Cunningham}, E.C.}, \bibinfo{author}{{Hunt}, J.A.S.},
  \bibinfo{author}{{Price-Whelan}, A.M.}, \bibinfo{author}{{Johnston}, K.V.},
  \bibinfo{author}{{Ness}, M.K.}, \bibinfo{author}{{Lu}, Y.},
  \bibinfo{author}{{Escala}, I.}, \bibinfo{author}{{Stelea}, I.A.},
  \bibinfo{year}{2023}.
\newblock \bibinfo{title}{{Chemical Cartography of the Sagittarius Stream with
  Gaia}}.
\newblock \bibinfo{journal}{arXiv e-prints} ,
  \bibinfo{pages}{arXiv:2307.08730}\DOIprefix\doi{10.48550/arXiv.2307.08730},
  \href{http://arxiv.org/abs/2307.08730}{{\tt arXiv:2307.08730}}.
%Type = Inproceedings
\bibitem[{{Dalton} et~al.(2012){Dalton}, {Trager}, {Abrams}, {Carter},
  {Bonifacio}, {Aguerri}, {MacIntosh}, {Evans}, {Lewis}, {Navarro}, {Agocs},
  {Dee}, {Rousset}, {Tosh}, {Middleton}, {Pragt}, {Terrett}, {Brock}, {Benn},
  {Verheijen}, {Cano Infantes}, {Bevil}, {Steele}, {Mottram}, {Bates},
  {Gribbin}, {Rey}, {Rodriguez}, {Delgado}, {Guinouard}, {Walton}, {Irwin},
  {Jagourel}, {Stuik}, {Gerlofsma}, {Roelfsma}, {Skillen}, {Ridings},
  {Balcells}, {Daban}, {Gouvret}, {Venema} and {Girard}}]{weave:2012}
{Dalton}, G. et~al., \bibinfo{year}{2012}.
\newblock \bibinfo{title}{{WEAVE: the next generation wide-field spectroscopy
  facility for the William Herschel Telescope}}, in: \bibinfo{editor}{{McLean},
  I.S.}, \bibinfo{editor}{{Ramsay}, S.K.}, \bibinfo{editor}{{Takami}, H.}
  (Eds.), \bibinfo{booktitle}{Ground-based and Airborne Instrumentation for
  Astronomy IV}, volume \bibinfo{volume}{8446} of
  \textit{\bibinfo{series}{Society of Photo-Optical Instrumentation Engineers
  (SPIE) Conference Series}}. p. \bibinfo{pages}{84460P}.
\newblock \DOIprefix\doi{10.1117/12.925950}.
%Type = Article
\bibitem[{{Dark Energy Survey Collaboration} et~al.(2016){Dark Energy Survey
  Collaboration}, {Abbott}, {Abdalla}, {Aleksi{\'c}}, {Allam}, {Amara},
  {Bacon}, {Balbinot}, {Banerji}, {Bechtol}, {Benoit-L{\'e}vy}, {Bernstein},
  {Bertin}, {Blazek}, {Bonnett}, {Bridle}, {Brooks}, {Brunner}, {Buckley-Geer},
  {Burke}, {Caminha}, {Capozzi}, {Carlsen}, {Carnero-Rosell}, {Carollo},
  {Carrasco-Kind}, {Carretero}, {Castander}, {Clerkin}, {Collett}, {Conselice},
  {Crocce}, {Cunha}, {D'Andrea}, {da Costa}, {Davis}, {Desai}, {Diehl},
  {Dietrich}, {Dodelson}, {Doel}, {Drlica-Wagner}, {Estrada}, {Etherington},
  {Evrard}, {Fabbri}, {Finley}, {Flaugher}, {Foley}, {Fosalba}, {Frieman},
  {Garc{\'\i}a-Bellido}, {Gaztanaga}, {Gerdes}, {Giannantonio}, {Goldstein},
  {Gruen}, {Gruendl}, {Guarnieri}, {Gutierrez}, {Hartley}, {Honscheid}, {Jain},
  {James}, {Jeltema}, {Jouvel}, {Kessler}, {King}, {Kirk}, {Kron}, {Kuehn},
  {Kuropatkin}, {Lahav}, {Li}, {Lima}, {Lin}, {Maia}, {Makler}, {Manera},
  {Maraston}, {Marshall}, {Martini}, {McMahon}, {Melchior}, {Merson}, {Miller},
  {Miquel}, {Mohr}, {Morice-Atkinson}, {Naidoo}, {Neilsen}, {Nichol}, {Nord},
  {Ogando}, {Ostrovski}, {Palmese}, {Papadopoulos}, {Peiris}, {Peoples},
  {Percival}, {Plazas}, {Reed}, {Refregier}, {Romer}, {Roodman}, {Ross},
  {Rozo}, {Rykoff}, {Sadeh}, {Sako}, {S{\'a}nchez}, {Sanchez}, {Santiago},
  {Scarpine}, {Schubnell}, {Sevilla-Noarbe}, {Sheldon}, {Smith}, {Smith},
  {Soares-Santos}, {Sobreira}, {Soumagnac}, {Suchyta}, {Sullivan}, {Swanson},
  {Tarle}, {Thaler}, {Thomas}, {Thomas}, {Tucker}, {Vieira}, {Vikram},
  {Walker}, {Wechsler}, {Weller}, {Wester}, {Whiteway}, {Wilcox}, {Yanny},
  {Zhang} and {Zuntz}}]{des:2016}
{Dark Energy Survey Collaboration}et~al., \bibinfo{year}{2016}.
\newblock \bibinfo{title}{{The Dark Energy Survey: more than dark energy - an
  overview}}.
\newblock \bibinfo{journal}{\mnras} \bibinfo{volume}{460},
  \bibinfo{pages}{1270--1299}.
\newblock \DOIprefix\doi{10.1093/mnras/stw641},
  \href{http://arxiv.org/abs/1601.00329}{{\tt arXiv:1601.00329}}.
%Type = Article
\bibitem[{{Davies} et~al.(2024){Davies}, {Belokurov}, {Monty} and
  {Evans}}]{davies:2024a}
\bibinfo{author}{{Davies}, E.Y.}, \bibinfo{author}{{Belokurov}, V.},
  \bibinfo{author}{{Monty}, S.}, \bibinfo{author}{{Evans}, N.W.},
  \bibinfo{year}{2024}.
\newblock \bibinfo{title}{{Disrupted dwarf binary merger as the possible origin
  of NGC 2419 and Sagittarius stream substructure}}.
\newblock \bibinfo{journal}{\mnras} \bibinfo{volume}{529},
  \bibinfo{pages}{L73--L77}.
\newblock \DOIprefix\doi{10.1093/mnrasl/slae001},
  \href{http://arxiv.org/abs/2308.01958}{{\tt arXiv:2308.01958}}.
%Type = Article
\bibitem[{{Davies} et~al.(2023){Davies}, {Monty}, {Belokurov} and
  {Dillamore}}]{davies:2024b}
\bibinfo{author}{{Davies}, E.Y.}, \bibinfo{author}{{Monty}, S.},
  \bibinfo{author}{{Belokurov}, V.}, \bibinfo{author}{{Dillamore}, A.M.},
  \bibinfo{year}{2023}.
\newblock \bibinfo{title}{{Hints of a disrupted binary dwarf galaxy in the
  Sagittarius stream}}.
\newblock \bibinfo{journal}{arXiv e-prints} ,
  \bibinfo{pages}{arXiv:2312.08424}\DOIprefix\doi{10.48550/arXiv.2312.08424},
  \href{http://arxiv.org/abs/2312.08424}{{\tt arXiv:2312.08424}}.
%Type = Article
\bibitem[{{de Boer} et~al.(2018){de Boer}, {Belokurov}, {Koposov}, {Ferrarese},
  {Erkal}, {C{\^o}t{\'e}} and {Navarro}}]{deboer:2018}
\bibinfo{author}{{de Boer}, T.J.L.}, \bibinfo{author}{{Belokurov}, V.},
  \bibinfo{author}{{Koposov}, S.E.}, \bibinfo{author}{{Ferrarese}, L.},
  \bibinfo{author}{{Erkal}, D.}, \bibinfo{author}{{C{\^o}t{\'e}}, P.},
  \bibinfo{author}{{Navarro}, J.F.}, \bibinfo{year}{2018}.
\newblock \bibinfo{title}{{A deeper look at the GD1 stream: density variations
  and wiggles}}.
\newblock \bibinfo{journal}{\mnras} \bibinfo{volume}{477},
  \bibinfo{pages}{1893--1902}.
\newblock \DOIprefix\doi{10.1093/mnras/sty677},
  \href{http://arxiv.org/abs/1801.08948}{{\tt arXiv:1801.08948}}.
%Type = Article
\bibitem[{{de Boer} et~al.(2020){de Boer}, {Erkal} and {Gieles}}]{deboer:2020}
\bibinfo{author}{{de Boer}, T.J.L.}, \bibinfo{author}{{Erkal}, D.},
  \bibinfo{author}{{Gieles}, M.}, \bibinfo{year}{2020}.
\newblock \bibinfo{title}{{A closer look at the spur, blob, wiggle, and gaps in
  GD-1}}.
\newblock \bibinfo{journal}{\mnras} \bibinfo{volume}{494},
  \bibinfo{pages}{5315--5332}.
\newblock \DOIprefix\doi{10.1093/mnras/staa917},
  \href{http://arxiv.org/abs/1911.05745}{{\tt arXiv:1911.05745}}.
%Type = Article
\bibitem[{{de Jong} et~al.(2019){de Jong}, {Agertz}, {Berbel}, {Aird},
  {Alexander}, {Amarsi}, {Anders}, {Andrae}, {Ansarinejad}, {Ansorge},
  {Antilogus}, {Anwand-Heerwart}, {Arentsen}, {Arnadottir}, {Asplund}, {Auger},
  {Azais}, {Baade}, {Baker}, {Baker}, {Balbinot}, {Baldry}, {Banerji},
  {Barden}, {Barklem}, {Barth{\'e}l{\'e}my-Mazot}, {Battistini}, {Bauer},
  {Bell}, {Bellido-Tirado}, {Bellstedt}, {Belokurov}, {Bensby}, {Bergemann},
  {Bestenlehner}, {Bielby}, {Bilicki}, {Blake}, {Bland-Hawthorn}, {Boeche},
  {Boland}, {Boller}, {Bongard}, {Bongiorno}, {Bonifacio}, {Boudon}, {Brooks},
  {Brown}, {Brown}, {Br{\"u}ggen}, {Brynnel}, {Brzeski}, {Buchert},
  {Buschkamp}, {Caffau}, {Caillier}, {Carrick}, {Casagrande}, {Case}, {Casey},
  {Cesarini}, {Cescutti}, {Chapuis}, {Chiappini}, {Childress}, {Christlieb},
  {Church}, {Cioni}, {Cluver}, {Colless}, {Collett}, {Comparat}, {Cooper},
  {Couch}, {Courbin}, {Croom}, {Croton}, {Daguis{\'e}}, {Dalton}, {Davies},
  {Davis}, {de Laverny}, {Deason}, {Dionies}, {Disseau}, {Doel}, {D{\"o}scher},
  {Driver}, {Dwelly}, {Eckert}, {Edge}, {Edvardsson}, {Youssoufi}, {Elhaddad},
  {Enke}, {Erfanianfar}, {Farrell}, {Fechner}, {Feiz}, {Feltzing}, {Ferreras},
  {Feuerstein}, {Feuillet}, {Finoguenov}, {Ford}, {Fotopoulou}, {Fouesneau},
  {Frenk}, {Frey}, {Gaessler}, {Geier}, {Gentile Fusillo}, {Gerhard},
  {Giannantonio}, {Giannone}, {Gibson}, {Gillingham},
  {Gonz{\'a}lez-Fern{\'a}ndez}, {Gonzalez-Solares}, {Gottloeber}, {Gould},
  {Grebel}, {Gueguen}, {Guiglion}, {Haehnelt}, {Hahn}, {Hansen}, {Hartman},
  {Hauptner}, {Hawkins}, {Haynes}, {Haynes}, {Heiter}, {Helmi}, {Aguayo},
  {Hewett}, {Hinton}, {Hobbs}, {Hoenig}, {Hofman}, {Hook}, {Hopgood},
  {Hopkins}, {Hourihane}, {Howes}, {Howlett}, {Huet}, {Irwin}, {Iwert},
  {Jablonka}, {Jahn}, {Jahnke}, {Jarno}, {Jin}, {Jofre}, {Johl}, {Jones},
  {J{\"o}nsson}, {Jordan}, {Karovicova}, {Khalatyan}, {Kelz}, {Kennicutt},
  {King}, {Kitaura}, {Klar}, {Klauser}, {Kneib}, {Koch}, {Koposov},
  {Kordopatis}, {Korn}, {Kosmalski}, {Kotak}, {Kovalev}, {Kreckel}, {Kripak},
  {Krumpe}, {Kuijken}, {Kunder}, {Kushniruk}, {Lam}, {Lamer}, {Laurent},
  {Lawrence}, {Lehmitz}, {Lemasle}, {Lewis}, {Li}, {Lidman}, {Lind}, {Liske},
  {Lizon}, {Loveday}, {Ludwig}, {McDermid}, {Maguire}, {Mainieri}, {Mali},
  {Mandel}, {Mandel}, {Mannering}, {Martell}, {Martinez Delgado}, {Matijevic},
  {McGregor}, {McMahon}, {McMillan}, {Mena}, {Merloni}, {Meyer}, {Michel},
  {Micheva}, {Migniau}, {Minchev}, {Monari}, {Muller}, {Murphy},
  {Muthukrishna}, {Nandra}, {Navarro}, {Ness}, {Nichani}, {Nichol}, {Nicklas},
  {Niederhofer}, {Norberg}, {Obreschkow}, {Oliver}, {Owers}, {Pai},
  {Pankratow}, {Parkinson}, {Paschke}, {Paterson}, {Pecontal}, {Parry},
  {Phillips}, {Pillepich}, {Pinard}, {Pirard}, {Piskunov}, {Plank},
  {Pl{\"u}schke}, {Pons}, {Popesso}, {Power}, {Pragt}, {Pramskiy}, {Pryer},
  {Quattri}, {Queiroz}, {Quirrenbach}, {Rahurkar}, {Raichoor}, {Ramstedt},
  {Rau}, {Recio-Blanco}, {Reiss}, {Renaud}, {Revaz}, {Rhode}, {Richard},
  {Richter}, {Rix}, {Robotham}, {Roelfsema}, {Romaniello}, {Rosario},
  {Rothmaier}, {Roukema}, {Ruchti}, {Rupprecht}, {Rybizki}, {Ryde}, {Saar},
  {Sadler}, {Sahl{\'e}n}, {Salvato}, {Sassolas}, {Saunders}, {Saviauk},
  {Sbordone}, {Schmidt}, {Schnurr}, {Scholz}, {Schwope}, {Seifert}, {Shanks},
  {Sheinis}, {Sivov}, {Sk{\'u}lad{\'o}ttir}, {Smartt}, {Smedley}, {Smith},
  {Smith}, {Sorce}, {Spitler}, {Starkenburg}, {Steinmetz}, {Stilz}, {Storm},
  {Sullivan}, {Sutherland}, {Swann}, {Tamone}, {Taylor}, {Teillon}, {Tempel},
  {ter Horst}, {Thi}, {Tolstoy}, {Trager}, {Traven}, {Tremblay}, {Tresse},
  {Valentini}, {van de Weygaert}, {van den Ancker}, {Veljanoski}, {Venkatesan},
  {Wagner}, {Wagner}, {Walcher}, {Waller}, {Walton}, {Wang}, {Winkler},
  {Wisotzki}, {Worley}, {Worseck}, {Xiang}, {Xu}, {Yong}, {Zhao}, {Zheng},
  {Zscheyge} and {Zucker}}]{dejong:2019}
{de Jong}, R.~S. et~al., \bibinfo{year}{2019}.
\newblock \bibinfo{title}{{4MOST: Project overview and information for the
  First Call for Proposals}}.
\newblock \bibinfo{journal}{The Messenger} \bibinfo{volume}{175},
  \bibinfo{pages}{3--11}.
\newblock \DOIprefix\doi{10.18727/0722-6691/5117},
  \href{http://arxiv.org/abs/1903.02464}{{\tt arXiv:1903.02464}}.
%Type = Inproceedings
\bibitem[{{de Jong} et~al.(2012){de Jong}, {Bellido-Tirado}, {Chiappini},
  {Depagne}, {Haynes}, {Johl}, {Schnurr}, {Schwope}, {Walcher}, {Dionies},
  {Haynes}, {Kelz}, {Kitaura}, {Lamer}, {Minchev}, {M{\"u}ller}, {Nuza},
  {Olaya}, {Piffl}, {Popow}, {Steinmetz}, {Ural}, {Williams}, {Winkler},
  {Wisotzki}, {Ansorge}, {Banerji}, {Gonzalez Solares}, {Irwin}, {Kennicutt},
  {King}, {McMahon}, {Koposov}, {Parry}, {Sun}, {Walton}, {Finger}, {Iwert},
  {Krumpe}, {Lizon}, {Vincenzo}, {Amans}, {Bonifacio}, {Cohen}, {Francois},
  {Jagourel}, {Mignot}, {Royer}, {Sartoretti}, {Bender}, {Grupp}, {Hess},
  {Lang-Bardl}, {Muschielok}, {B{\"o}hringer}, {Boller}, {Bongiorno}, {Brusa},
  {Dwelly}, {Merloni}, {Nandra}, {Salvato}, {Pragt}, {Navarro}, {Gerlofsma},
  {Roelfsema}, {Dalton}, {Middleton}, {Tosh}, {Boeche}, {Caffau}, {Christlieb},
  {Grebel}, {Hansen}, {Koch}, {Ludwig}, {Quirrenbach}, {Sbordone}, {Seifert},
  {Thimm}, {Trifonov}, {Helmi}, {Trager}, {Feltzing}, {Korn} and
  {Boland}}]{4most:2012}
{de Jong}, R.~S. et~al., \bibinfo{year}{2012}.
\newblock \bibinfo{title}{{4MOST: 4-metre multi-object spectroscopic
  telescope}}, in: \bibinfo{editor}{{McLean}, I.S.}, \bibinfo{editor}{{Ramsay},
  S.K.}, \bibinfo{editor}{{Takami}, H.} (Eds.),
  \bibinfo{booktitle}{Ground-based and Airborne Instrumentation for Astronomy
  IV}, volume \bibinfo{volume}{8446} of \textit{\bibinfo{series}{Society of
  Photo-Optical Instrumentation Engineers (SPIE) Conference Series}}. p.
  \bibinfo{pages}{84460T}.
\newblock \DOIprefix\doi{10.1117/12.926239},
  \href{http://arxiv.org/abs/1206.6885}{{\tt arXiv:1206.6885}}.
%Type = Article
\bibitem[{{Deason} et~al.(2018){Deason}, {Belokurov} and
  {Koposov}}]{deason:2018}
\bibinfo{author}{{Deason}, A.J.}, \bibinfo{author}{{Belokurov}, V.},
  \bibinfo{author}{{Koposov}, S.E.}, \bibinfo{year}{2018}.
\newblock \bibinfo{title}{{Cresting the wave: proper motions of the Eastern
  Banded Structure}}.
\newblock \bibinfo{journal}{\mnras} \bibinfo{volume}{473},
  \bibinfo{pages}{2428--2433}.
\newblock \DOIprefix\doi{10.1093/mnras/stx2528},
  \href{http://arxiv.org/abs/1709.08633}{{\tt arXiv:1709.08633}}.
%Type = Article
\bibitem[{{Debattista} et~al.(2013){Debattista}, {Ro{\v{s}}kar}, {Valluri},
  {Quinn}, {Moore} and {Wadsley}}]{debattista:2013}
\bibinfo{author}{{Debattista}, V.P.}, \bibinfo{author}{{Ro{\v{s}}kar}, R.},
  \bibinfo{author}{{Valluri}, M.}, \bibinfo{author}{{Quinn}, T.},
  \bibinfo{author}{{Moore}, B.}, \bibinfo{author}{{Wadsley}, J.},
  \bibinfo{year}{2013}.
\newblock \bibinfo{title}{{What's up in the Milky Way? The orientation of the
  disc relative to the triaxial halo}}.
\newblock \bibinfo{journal}{\mnras} \bibinfo{volume}{434},
  \bibinfo{pages}{2971--2981}.
\newblock \DOIprefix\doi{10.1093/mnras/stt1217},
  \href{http://arxiv.org/abs/1301.2670}{{\tt arXiv:1301.2670}}.
%Type = Article
\bibitem[{{Deg} and {Widrow}(2013)}]{deg:2013}
\bibinfo{author}{{Deg}, N.}, \bibinfo{author}{{Widrow}, L.},
  \bibinfo{year}{2013}.
\newblock \bibinfo{title}{{The Sagittarius stream and halo triaxiality}}.
\newblock \bibinfo{journal}{\mnras} \bibinfo{volume}{428},
  \bibinfo{pages}{912--922}.
\newblock \DOIprefix\doi{10.1093/mnras/sts089},
  \href{http://arxiv.org/abs/1209.6614}{{\tt arXiv:1209.6614}}.
%Type = Article
\bibitem[{{Dehnen}(1999)}]{dehnen:1999}
\bibinfo{author}{{Dehnen}, W.}, \bibinfo{year}{1999}.
\newblock \bibinfo{title}{{The Pattern Speed of the Galactic Bar}}.
\newblock \bibinfo{journal}{\apjl} \bibinfo{volume}{524},
  \bibinfo{pages}{L35--L38}.
\newblock \DOIprefix\doi{10.1086/312299},
  \href{http://arxiv.org/abs/astro-ph/9908105}{{\tt arXiv:astro-ph/9908105}}.
%Type = Misc
\bibitem[{{Dehnen}(2014)}]{dehnen:2014}
\bibinfo{author}{{Dehnen}, W.}, \bibinfo{year}{2014}.
\newblock \bibinfo{title}{{gyrfalcON: N-body code}}.
\newblock \bibinfo{howpublished}{Astrophysics Source Code Library, record
  ascl:1402.031}.
\newblock \href{http://arxiv.org/abs/1402.031}{{\tt arXiv:1402.031}}.
%Type = Article
\bibitem[{{DESI Collaboration} et~al.(2016){DESI Collaboration}, {Aghamousa},
  {Aguilar}, {Ahlen}, {Alam}, {Allen}, {Allende Prieto}, {Annis}, {Bailey},
  {Balland}, {Ballester}, {Baltay}, {Beaufore}, {Bebek}, {Beers}, {Bell},
  {Bernal}, {Besuner}, {Beutler}, {Blake}, {Bleuler}, {Blomqvist}, {Blum},
  {Bolton}, {Briceno}, {Brooks}, {Brownstein}, {Buckley-Geer}, {Burden},
  {Burtin}, {Busca}, {Cahn}, {Cai}, {Cardiel-Sas}, {Carlberg}, {Carton},
  {Casas}, {Castander}, {Cervantes-Cota}, {Claybaugh}, {Close}, {Coker},
  {Cole}, {Comparat}, {Cooper}, {Cousinou}, {Crocce}, {Cuby}, {Cunningham},
  {Davis}, {Dawson}, {de la Macorra}, {De Vicente}, {Delubac}, {Derwent},
  {Dey}, {Dhungana}, {Ding}, {Doel}, {Duan}, {Ealet}, {Edelstein},
  {Eftekharzadeh}, {Eisenstein}, {Elliott}, {Escoffier}, {Evatt}, {Fagrelius},
  {Fan}, {Fanning}, {Farahi}, {Farihi}, {Favole}, {Feng}, {Fernandez},
  {Findlay}, {Finkbeiner}, {Fitzpatrick}, {Flaugher}, {Flender}, {Font-Ribera},
  {Forero-Romero}, {Fosalba}, {Frenk}, {Fumagalli}, {Gaensicke}, {Gallo},
  {Garcia-Bellido}, {Gaztanaga}, {Pietro Gentile Fusillo}, {Gerard},
  {Gershkovich}, {Giannantonio}, {Gillet}, {Gonzalez-de-Rivera},
  {Gonzalez-Perez}, {Gott}, {Graur}, {Gutierrez}, {Guy}, {Habib}, {Heetderks},
  {Heetderks}, {Heitmann}, {Hellwing}, {Herrera}, {Ho}, {Holland}, {Honscheid},
  {Huff}, {Hutchinson}, {Huterer}, {Hwang}, {Illa Laguna}, {Ishikawa},
  {Jacobs}, {Jeffrey}, {Jelinsky}, {Jennings}, {Jiang}, {Jimenez}, {Johnson},
  {Joyce}, {Jullo}, {Juneau}, {Kama}, {Karcher}, {Karkar}, {Kehoe}, {Kennamer},
  {Kent}, {Kilbinger}, {Kim}, {Kirkby}, {Kisner}, {Kitanidis}, {Kneib},
  {Koposov}, {Kovacs}, {Koyama}, {Kremin}, {Kron}, {Kronig}, {Kueter-Young},
  {Lacey}, {Lafever}, {Lahav}, {Lambert}, {Lampton}, {Landriau}, {Lang},
  {Lauer}, {Le Goff}, {Le Guillou}, {Le Van Suu}, {Lee}, {Lee}, {Leitner},
  {Lesser}, {Levi}, {L'Huillier}, {Li}, {Liang}, {Lin}, {Linder}, {Loebman},
  {Luki{\'c}}, {Ma}, {MacCrann}, {Magneville}, {Makarem}, {Manera}, {Manser},
  {Marshall}, {Martini}, {Massey}, {Matheson}, {McCauley}, {McDonald},
  {McGreer}, {Meisner}, {Metcalfe}, {Miller}, {Miquel}, {Moustakas}, {Myers},
  {Naik}, {Newman}, {Nichol}, {Nicola}, {Nicolati da Costa}, {Nie}, {Niz},
  {Norberg}, {Nord}, {Norman}, {Nugent}, {O'Brien}, {Oh}, {Olsen}, {Padilla},
  {Padmanabhan}, {Padmanabhan}, {Palanque-Delabrouille}, {Palmese},
  {Pappalardo}, {P{\^a}ris}, {Park}, {Patej}, {Peacock}, {Peiris}, {Peng},
  {Percival}, {Perruchot}, {Pieri}, {Pogge}, {Pollack}, {Poppett}, {Prada},
  {Prakash}, {Probst}, {Rabinowitz}, {Raichoor}, {Ree}, {Refregier}, {Regal},
  {Reid}, {Reil}, {Rezaie}, {Rockosi}, {Roe}, {Ronayette}, {Roodman}, {Ross},
  {Ross}, {Rossi}, {Rozo}, {Ruhlmann-Kleider}, {Rykoff}, {Sabiu}, {Samushia},
  {Sanchez}, {Sanchez}, {Schlegel}, {Schneider}, {Schubnell}, {Secroun},
  {Seljak}, {Seo}, {Serrano}, {Shafieloo}, {Shan}, {Sharples}, {Sholl},
  {Shourt}, {Silber}, {Silva}, {Sirk}, {Slosar}, {Smith}, {Smoot}, {Som},
  {Song}, {Sprayberry}, {Staten}, {Stefanik}, {Tarle}, {Sien Tie}, {Tinker},
  {Tojeiro}, {Valdes}, {Valenzuela}, {Valluri}, {Vargas-Magana}, {Verde},
  {Walker}, {Wang}, {Wang}, {Weaver}, {Weaverdyck}, {Wechsler}, {Weinberg},
  {White}, {Yang}, {Yeche}, {Zhang}, {Zhao}, {Zheng}, {Zhou}, {Zhou}, {Zhu},
  {Zou} and {Zu}}]{desi:2016}
{DESI Collaboration}et~al., \bibinfo{year}{2016}.
\newblock \bibinfo{title}{{The DESI Experiment Part I: Science,Targeting, and
  Survey Design}}.
\newblock \bibinfo{journal}{arXiv e-prints} ,
  \bibinfo{pages}{arXiv:1611.00036}\DOIprefix\doi{10.48550/arXiv.1611.00036},
  \href{http://arxiv.org/abs/1611.00036}{{\tt arXiv:1611.00036}}.
%Type = Article
\bibitem[{{Dey} et~al.(2019){Dey}, {Schlegel}, {Lang}, {Blum}, {Burleigh},
  {Fan}, {Findlay}, {Finkbeiner}, {Herrera}, {Juneau}, {Landriau}, {Levi},
  {McGreer}, {Meisner}, {Myers}, {Moustakas}, {Nugent}, {Patej}, {Schlafly},
  {Walker}, {Valdes}, {Weaver}, {Y{\`e}che}, {Zou}, {Zhou}, {Abareshi},
  {Abbott}, {Abolfathi}, {Aguilera}, {Alam}, {Allen}, {Alvarez}, {Annis},
  {Ansarinejad}, {Aubert}, {Beechert}, {Bell}, {BenZvi}, {Beutler}, {Bielby},
  {Bolton}, {Brice{\~n}o}, {Buckley-Geer}, {Butler}, {Calamida}, {Carlberg},
  {Carter}, {Casas}, {Castander}, {Choi}, {Comparat}, {Cukanovaite}, {Delubac},
  {DeVries}, {Dey}, {Dhungana}, {Dickinson}, {Ding}, {Donaldson}, {Duan},
  {Duckworth}, {Eftekharzadeh}, {Eisenstein}, {Etourneau}, {Fagrelius},
  {Farihi}, {Fitzpatrick}, {Font-Ribera}, {Fulmer}, {G{\"a}nsicke},
  {Gaztanaga}, {George}, {Gerdes}, {Gontcho}, {Gorgoni}, {Green}, {Guy},
  {Harmer}, {Hernandez}, {Honscheid}, {Huang}, {James}, {Jannuzi}, {Jiang},
  {Joyce}, {Karcher}, {Karkar}, {Kehoe}, {Kneib}, {Kueter-Young}, {Lan},
  {Lauer}, {Le Guillou}, {Le Van Suu}, {Lee}, {Lesser}, {Perreault Levasseur},
  {Li}, {Mann}, {Marshall}, {Mart{\'\i}nez-V{\'a}zquez}, {Martini}, {du Mas des
  Bourboux}, {McManus}, {Meier}, {M{\'e}nard}, {Metcalfe},
  {Mu{\~n}oz-Guti{\'e}rrez}, {Najita}, {Napier}, {Narayan}, {Newman}, {Nie},
  {Nord}, {Norman}, {Olsen}, {Paat}, {Palanque-Delabrouille}, {Peng},
  {Poppett}, {Poremba}, {Prakash}, {Rabinowitz}, {Raichoor}, {Rezaie},
  {Robertson}, {Roe}, {Ross}, {Ross}, {Rudnick}, {Safonova}, {Saha},
  {S{\'a}nchez}, {Savary}, {Schweiker}, {Scott}, {Seo}, {Shan}, {Silva},
  {Slepian}, {Soto}, {Sprayberry}, {Staten}, {Stillman}, {Stupak}, {Summers},
  {Sien Tie}, {Tirado}, {Vargas-Maga{\~n}a}, {Vivas}, {Wechsler}, {Williams},
  {Yang}, {Yang}, {Yapici}, {Zaritsky}, {Zenteno}, {Zhang}, {Zhang}, {Zhou} and
  {Zhou}}]{dey:2019}
{Dey}, A. et~al., \bibinfo{year}{2019}.
\newblock \bibinfo{title}{{Overview of the DESI Legacy Imaging Surveys}}.
\newblock \bibinfo{journal}{\aj} \bibinfo{volume}{157}, \bibinfo{pages}{168}.
\newblock \DOIprefix\doi{10.3847/1538-3881/ab089d},
  \href{http://arxiv.org/abs/1804.08657}{{\tt arXiv:1804.08657}}.
%Type = Article
\bibitem[{{Diemand} et~al.(2005){Diemand}, {Moore} and {Stadel}}]{diemand:2005}
\bibinfo{author}{{Diemand}, J.}, \bibinfo{author}{{Moore}, B.},
  \bibinfo{author}{{Stadel}, J.}, \bibinfo{year}{2005}.
\newblock \bibinfo{title}{{Earth-mass dark-matter haloes as the first
  structures in the early Universe}}.
\newblock \bibinfo{journal}{\nat} \bibinfo{volume}{433},
  \bibinfo{pages}{389--391}.
\newblock \DOIprefix\doi{10.1038/nature03270},
  \href{http://arxiv.org/abs/astro-ph/0501589}{{\tt arXiv:astro-ph/0501589}}.
%Type = Article
\bibitem[{{Dillamore} et~al.(2022){Dillamore}, {Belokurov}, {Evans} and
  {Price-Whelan}}]{dillamore:2022}
\bibinfo{author}{{Dillamore}, A.M.}, \bibinfo{author}{{Belokurov}, V.},
  \bibinfo{author}{{Evans}, N.W.}, \bibinfo{author}{{Price-Whelan}, A.M.},
  \bibinfo{year}{2022}.
\newblock \bibinfo{title}{{The impact of a massive Sagittarius dSph on
  GD-1-like streams}}.
\newblock \bibinfo{journal}{\mnras} \bibinfo{volume}{516},
  \bibinfo{pages}{1685--1703}.
\newblock \DOIprefix\doi{10.1093/mnras/stac2311},
  \href{http://arxiv.org/abs/2205.13547}{{\tt arXiv:2205.13547}}.
%Type = Article
\bibitem[{{Dodd} et~al.(2022){Dodd}, {Helmi} and {Koppelman}}]{dodd:2022}
\bibinfo{author}{{Dodd}, E.}, \bibinfo{author}{{Helmi}, A.},
  \bibinfo{author}{{Koppelman}, H.H.}, \bibinfo{year}{2022}.
\newblock \bibinfo{title}{{Substructures, resonances, and debris streams. A new
  constraint on the inner shape of the Galactic dark halo}}.
\newblock \bibinfo{journal}{\aap} \bibinfo{volume}{659}, \bibinfo{pages}{A61}.
\newblock \DOIprefix\doi{10.1051/0004-6361/202141354},
  \href{http://arxiv.org/abs/2105.09957}{{\tt arXiv:2105.09957}}.
%Type = Article
\bibitem[{{Doke} and {Hattori}(2022)}]{doke:2022}
\bibinfo{author}{{Doke}, Y.}, \bibinfo{author}{{Hattori}, K.},
  \bibinfo{year}{2022}.
\newblock \bibinfo{title}{{Probability of Forming Gaps in the GD-1 Stream by
  Close Encounters of Globular Clusters}}.
\newblock \bibinfo{journal}{\apj} \bibinfo{volume}{941}, \bibinfo{pages}{129}.
\newblock \DOIprefix\doi{10.3847/1538-4357/aca090},
  \href{http://arxiv.org/abs/2203.15481}{{\tt arXiv:2203.15481}}.
%Type = Article
\bibitem[{{Drlica-Wagner} et~al.(2019){Drlica-Wagner}, {Mao}, {Adhikari},
  {Armstrong}, {Banerjee}, {Banik}, {Bechtol}, {Bird}, {Boddy}, {Bonaca},
  {Bovy}, {Buckley}, {Bulbul}, {Chang}, {Chapline}, {Cohen-Tanugi}, {Cuoco},
  {Cyr-Racine}, {Dawson}, {D{\'\i}az Rivero}, {Dvorkin}, {Erkal}, {Fassnacht},
  {Garc{\'\i}a-Bellido}, {Giannotti}, {Gluscevic}, {Golovich}, {Hendel},
  {Hezaveh}, {Horiuchi}, {Jee}, {Kaplinghat}, {Keeton}, {Koposov}, {Lam}, {Li},
  {Lu}, {Mandelbaum}, {McDermott}, {McNanna}, {Medford}, {Meyer}, {Marc},
  {Murgia}, {Nadler}, {Necib}, {Nuss}, {Pace}, {Peter}, {Polin},
  {Prescod-Weinstein}, {Read}, {Rosenfeld}, {Shipp}, {Simon}, {Slatyer},
  {Straniero}, {Strigari}, {Tollerud}, {Tyson}, {Wang}, {Wechsler}, {Wittman},
  {Yu}, {Zaharijas}, {Ali-Ha{\"\i}moud}, {Annis}, {Birrer}, {Biswas}, {Blazek},
  {Brooks}, {Buckley-Geer}, {Caputo}, {Charles}, {Digel}, {Dodelson},
  {Flaugher}, {Frieman}, {Gawiser}, {Hearin}, {Hlo{\v{z}}ek}, {Jain},
  {Jeltema}, {Koushiappas}, {Lisanti}, {LoVerde}, {Mishra-Sharma}, {Newman},
  {Nord}, {Nourbakhsh}, {Ritz}, {Robertson}, {S{\'a}nchez-Conde}, {Slosar},
  {Tait}, {Verma}, {Vilalta}, {Walter}, {Yanny} and
  {Zentner}}]{drlica-wagner:2019}
{Drlica-Wagner}, A. et~al., \bibinfo{year}{2019}.
\newblock \bibinfo{title}{{Probing the Fundamental Nature of Dark Matter with
  the Large Synoptic Survey Telescope}}.
\newblock \bibinfo{journal}{arXiv e-prints} ,
  \bibinfo{pages}{arXiv:1902.01055}\DOIprefix\doi{10.48550/arXiv.1902.01055},
  \href{http://arxiv.org/abs/1902.01055}{{\tt arXiv:1902.01055}}.
%Type = Article
\bibitem[{{D'Souza} and {Bell}(2022)}]{dsouza:2022}
\bibinfo{author}{{D'Souza}, R.}, \bibinfo{author}{{Bell}, E.F.},
  \bibinfo{year}{2022}.
\newblock \bibinfo{title}{{Uncertainties associated with the backward
  integration of dwarf satellites using simple parametric potentials}}.
\newblock \bibinfo{journal}{\mnras} \bibinfo{volume}{512},
  \bibinfo{pages}{739--760}.
\newblock \DOIprefix\doi{10.1093/mnras/stac404},
  \href{http://arxiv.org/abs/2202.05707}{{\tt arXiv:2202.05707}}.
%Type = Article
\bibitem[{{Dubinski} et~al.(1999){Dubinski}, {Mihos} and
  {Hernquist}}]{dubinski:1999}
\bibinfo{author}{{Dubinski}, J.}, \bibinfo{author}{{Mihos}, J.C.},
  \bibinfo{author}{{Hernquist}, L.}, \bibinfo{year}{1999}.
\newblock \bibinfo{title}{{Constraining Dark Halo Potentials with Tidal
  Tails}}.
\newblock \bibinfo{journal}{\apj} \bibinfo{volume}{526},
  \bibinfo{pages}{607--622}.
\newblock \DOIprefix\doi{10.1086/308024},
  \href{http://arxiv.org/abs/astro-ph/9902217}{{\tt arXiv:astro-ph/9902217}}.
%Type = Incollection
\bibitem[{{Eggen}(1965)}]{eggen:1965}
\bibinfo{author}{{Eggen}, O.J.}, \bibinfo{year}{1965}.
\newblock \bibinfo{title}{{Moving Groups of Stars}}, in:
  \bibinfo{editor}{{Blaauw}, A.}, \bibinfo{editor}{{Schmidt}, M.} (Eds.),
  \bibinfo{booktitle}{Galactic structure. Edited by Adriaan Blaauw and Maarten
  Schmidt. Published by the University of Chicago Press}, p.
  \bibinfo{pages}{111}.
%Type = Article
\bibitem[{{El-Falou} and {Webb}(2022)}]{el-falou:2022}
\bibinfo{author}{{El-Falou}, N.}, \bibinfo{author}{{Webb}, J.J.},
  \bibinfo{year}{2022}.
\newblock \bibinfo{title}{{The effect of dwarf galaxies on the tidal tails of
  globular clusters}}.
\newblock \bibinfo{journal}{\mnras} \bibinfo{volume}{510},
  \bibinfo{pages}{2437--2447}.
\newblock \DOIprefix\doi{10.1093/mnras/stab3505},
  \href{http://arxiv.org/abs/2111.13715}{{\tt arXiv:2111.13715}}.
%Type = Article
\bibitem[{{Erkal} and {Belokurov}(2015a)}]{erkal:2015a}
\bibinfo{author}{{Erkal}, D.}, \bibinfo{author}{{Belokurov}, V.},
  \bibinfo{year}{2015}a.
\newblock \bibinfo{title}{{Forensics of subhalo-stream encounters: the three
  phases of gap growth}}.
\newblock \bibinfo{journal}{\mnras} \bibinfo{volume}{450},
  \bibinfo{pages}{1136--1149}.
\newblock \DOIprefix\doi{10.1093/mnras/stv655},
  \href{http://arxiv.org/abs/1412.6035}{{\tt arXiv:1412.6035}}.
%Type = Article
\bibitem[{{Erkal} and {Belokurov}(2015b)}]{erkal:2015b}
\bibinfo{author}{{Erkal}, D.}, \bibinfo{author}{{Belokurov}, V.},
  \bibinfo{year}{2015}b.
\newblock \bibinfo{title}{{Properties of dark subhaloes from gaps in tidal
  streams}}.
\newblock \bibinfo{journal}{\mnras} \bibinfo{volume}{454},
  \bibinfo{pages}{3542--3558}.
\newblock \DOIprefix\doi{10.1093/mnras/stv2122},
  \href{http://arxiv.org/abs/1507.05625}{{\tt arXiv:1507.05625}}.
%Type = Article
\bibitem[{{Erkal} et~al.(2016a){Erkal}, {Belokurov}, {Bovy} and
  {Sanders}}]{erkal:2016}
\bibinfo{author}{{Erkal}, D.}, \bibinfo{author}{{Belokurov}, V.},
  \bibinfo{author}{{Bovy}, J.}, \bibinfo{author}{{Sanders}, J.L.},
  \bibinfo{year}{2016}a.
\newblock \bibinfo{title}{{The number and size of subhalo-induced gaps in
  stellar streams}}.
\newblock \bibinfo{journal}{\mnras} \bibinfo{volume}{463},
  \bibinfo{pages}{102--119}.
\newblock \DOIprefix\doi{10.1093/mnras/stw1957},
  \href{http://arxiv.org/abs/1606.04946}{{\tt arXiv:1606.04946}}.
%Type = Article
\bibitem[{{Erkal} et~al.(2019){Erkal}, {Belokurov}, {Laporte}, {Koposov}, {Li},
  {Grillmair}, {Kallivayalil}, {Price-Whelan}, {Evans}, {Hawkins}, {Hendel},
  {Mateu}, {Navarro}, {del Pino}, {Slater}, {Sohn} and {Orphan Aspen Treasury
  Collaboration}}]{erkal:2019}
{Erkal}, D. et~al., \bibinfo{year}{2019}.
\newblock \bibinfo{title}{{The total mass of the Large Magellanic Cloud from
  its perturbation on the Orphan stream}}.
\newblock \bibinfo{journal}{\mnras} \bibinfo{volume}{487},
  \bibinfo{pages}{2685--2700}.
\newblock \DOIprefix\doi{10.1093/mnras/stz1371},
  \href{http://arxiv.org/abs/1812.08192}{{\tt arXiv:1812.08192}}.
%Type = Article
\bibitem[{{Erkal} et~al.(2017){Erkal}, {Koposov} and {Belokurov}}]{erkal:2017}
\bibinfo{author}{{Erkal}, D.}, \bibinfo{author}{{Koposov}, S.E.},
  \bibinfo{author}{{Belokurov}, V.}, \bibinfo{year}{2017}.
\newblock \bibinfo{title}{{A sharper view of Pal 5's tails: discovery of stream
  perturbations with a novel non-parametric technique}}.
\newblock \bibinfo{journal}{\mnras} \bibinfo{volume}{470},
  \bibinfo{pages}{60--84}.
\newblock \DOIprefix\doi{10.1093/mnras/stx1208},
  \href{http://arxiv.org/abs/1609.01282}{{\tt arXiv:1609.01282}}.
%Type = Article
\bibitem[{{Erkal} et~al.(2018){Erkal}, {Li}, {Koposov}, {Belokurov},
  {Balbinot}, {Bechtol}, {Buncher}, {Drlica-Wagner}, {Kuehn}, {Marshall},
  {Mart{\'\i}nez-V{\'a}zquez}, {Pace}, {Shipp}, {Simon}, {Stringer}, {Vivas},
  {Wechsler}, {Yanny}, {Abdalla}, {Allam}, {Annis}, {Avila}, {Bertin},
  {Brooks}, {Buckley-Geer}, {Burke}, {Carnero Rosell}, {Carrasco Kind},
  {Carretero}, {D'Andrea}, {da Costa}, {Davis}, {De Vicente}, {Doel}, {Eifler},
  {Evrard}, {Flaugher}, {Frieman}, {Garc{\'\i}a-Bellido}, {Gaztanaga},
  {Gerdes}, {Gruen}, {Gruendl}, {Gschwend}, {Gutierrez}, {Hartley},
  {Hollowood}, {Honscheid}, {James}, {Krause}, {Maia}, {March}, {Menanteau},
  {Miquel}, {Ogando}, {Plazas}, {Sanchez}, {Santiago}, {Scarpine}, {Schindler},
  {Sevilla-Noarbe}, {Smith}, {Smith}, {Soares-Santos}, {Sobreira}, {Suchyta},
  {Swanson}, {Tarle}, {Tucker} and {Walker}}]{erkal:2018}
{Erkal}, D. et~al., \bibinfo{year}{2018}.
\newblock \bibinfo{title}{{Modelling the Tucana III stream - a close passage
  with the LMC}}.
\newblock \bibinfo{journal}{\mnras} \bibinfo{volume}{481},
  \bibinfo{pages}{3148--3159}.
\newblock \DOIprefix\doi{10.1093/mnras/sty2518},
  \href{http://arxiv.org/abs/1804.07762}{{\tt arXiv:1804.07762}}.
%Type = Article
\bibitem[{{Erkal} et~al.(2016b){Erkal}, {Sanders} and
  {Belokurov}}]{erkal:2016b}
\bibinfo{author}{{Erkal}, D.}, \bibinfo{author}{{Sanders}, J.L.},
  \bibinfo{author}{{Belokurov}, V.}, \bibinfo{year}{2016}b.
\newblock \bibinfo{title}{{Stray, swing and scatter: angular momentum evolution
  of orbits and streams in aspherical potentials}}.
\newblock \bibinfo{journal}{\mnras} \bibinfo{volume}{461},
  \bibinfo{pages}{1590--1604}.
\newblock \DOIprefix\doi{10.1093/mnras/stw1400},
  \href{http://arxiv.org/abs/1603.08922}{{\tt arXiv:1603.08922}}.
%Type = Article
\bibitem[{{Errani} et~al.(2022){Errani}, {Navarro}, {Ibata}, {Martin}, {Yuan},
  {Aguado}, {Bonifacio}, {Caffau}, {Gonz{\'a}lez Hern{\'a}ndez}, {Malhan},
  {S{\'a}nchez-Janssen}, {Sestito}, {Starkenburg}, {Thomas} and
  {Venn}}]{errani:2022}
{Errani}, R. et~al., \bibinfo{year}{2022}.
\newblock \bibinfo{title}{{The Pristine survey - XVIII. C-19: tidal debris of a
  dark matter-dominated globular cluster?}}
\newblock \bibinfo{journal}{\mnras} \bibinfo{volume}{514},
  \bibinfo{pages}{3532--3540}.
\newblock \DOIprefix\doi{10.1093/mnras/stac1516},
  \href{http://arxiv.org/abs/2203.02513}{{\tt arXiv:2203.02513}}.
%Type = Article
\bibitem[{{Errani} et~al.(2015){Errani}, {Penarrubia} and
  {Tormen}}]{errani:2015}
\bibinfo{author}{{Errani}, R.}, \bibinfo{author}{{Penarrubia}, J.},
  \bibinfo{author}{{Tormen}, G.}, \bibinfo{year}{2015}.
\newblock \bibinfo{title}{{Constraining the distribution of dark matter in
  dwarf spheroidal galaxies with stellar tidal streams.}}
\newblock \bibinfo{journal}{\mnras} \bibinfo{volume}{449},
  \bibinfo{pages}{L46--L50}.
\newblock \DOIprefix\doi{10.1093/mnrasl/slv012},
  \href{http://arxiv.org/abs/1501.04968}{{\tt arXiv:1501.04968}}.
%Type = Article
\bibitem[{{Eyre} and {Binney}(2009)}]{eyre:2009}
\bibinfo{author}{{Eyre}, A.}, \bibinfo{author}{{Binney}, J.},
  \bibinfo{year}{2009}.
\newblock \bibinfo{title}{{Locating the orbits delineated by tidal streams}}.
\newblock \bibinfo{journal}{\mnras} \bibinfo{volume}{400},
  \bibinfo{pages}{548--560}.
\newblock \DOIprefix\doi{10.1111/j.1365-2966.2009.15494.x},
  \href{http://arxiv.org/abs/0907.0360}{{\tt arXiv:0907.0360}}.
%Type = Article
\bibitem[{{Eyre} and {Binney}(2011)}]{eyre:2011}
\bibinfo{author}{{Eyre}, A.}, \bibinfo{author}{{Binney}, J.},
  \bibinfo{year}{2011}.
\newblock \bibinfo{title}{{The mechanics of tidal streams}}.
\newblock \bibinfo{journal}{\mnras} \bibinfo{volume}{413},
  \bibinfo{pages}{1852--1874}.
\newblock \DOIprefix\doi{10.1111/j.1365-2966.2011.18270.x},
  \href{http://arxiv.org/abs/1011.3672}{{\tt arXiv:1011.3672}}.
%Type = Article
\bibitem[{{Fardal} et~al.(2015){Fardal}, {Huang} and {Weinberg}}]{fardal:2015}
\bibinfo{author}{{Fardal}, M.A.}, \bibinfo{author}{{Huang}, S.},
  \bibinfo{author}{{Weinberg}, M.D.}, \bibinfo{year}{2015}.
\newblock \bibinfo{title}{{Generation of mock tidal streams}}.
\newblock \bibinfo{journal}{\mnras} \bibinfo{volume}{452},
  \bibinfo{pages}{301--319}.
\newblock \DOIprefix\doi{10.1093/mnras/stv1198},
  \href{http://arxiv.org/abs/1410.1861}{{\tt arXiv:1410.1861}}.
%Type = Article
\bibitem[{{Fardal} et~al.(2019){Fardal}, {van der Marel}, {Sohn} and {del Pino
  Molina}}]{fardal:2019}
\bibinfo{author}{{Fardal}, M.A.}, \bibinfo{author}{{van der Marel}, R.P.},
  \bibinfo{author}{{Sohn}, S.T.}, \bibinfo{author}{{del Pino Molina}, A.},
  \bibinfo{year}{2019}.
\newblock \bibinfo{title}{{The course of the Orphan Stream in the Northern
  Galactic hemisphere traced with Gaia DR2}}.
\newblock \bibinfo{journal}{\mnras} \bibinfo{volume}{486},
  \bibinfo{pages}{936--949}.
\newblock \DOIprefix\doi{10.1093/mnras/stz749},
  \href{http://arxiv.org/abs/1812.06066}{{\tt arXiv:1812.06066}}.
%Type = Article
\bibitem[{{Fellhauer} et~al.(2007){Fellhauer}, {Evans}, {Belokurov}, {Zucker},
  {Yanny}, {Wilkinson}, {Gilmore}, {Irwin}, {Bramich}, {Vidrih}, {Hewett} and
  {Beers}}]{fellhauer:2007}
{Fellhauer}, M. et~al., \bibinfo{year}{2007}.
\newblock \bibinfo{title}{{Is Ursa Major II the progenitor of the Orphan
  Stream?}}
\newblock \bibinfo{journal}{\mnras} \bibinfo{volume}{375},
  \bibinfo{pages}{1171--1179}.
\newblock \DOIprefix\doi{10.1111/j.1365-2966.2006.11404.x},
  \href{http://arxiv.org/abs/astro-ph/0611157}{{\tt arXiv:astro-ph/0611157}}.
%Type = Article
\bibitem[{{Ferguson} et~al.(2022){Ferguson}, {Shipp}, {Drlica-Wagner}, {Li},
  {Cerny}, {Tavangar}, {Pace}, {Marshall}, {Riley}, {Adam{\'o}w}, {Carlin},
  {Choi}, {Erkal}, {James}, {Koposov}, {Kuropatkin},
  {Mart{\'\i}nez-V{\'a}zquez}, {Mau}, {Mutlu-Pakdil}, {Olsen}, {Sakowska},
  {Stringfellow}, {Yanny} and {Yanny}}]{ferguson:2022}
{Ferguson}, P.~S. et~al., \bibinfo{year}{2022}.
\newblock \bibinfo{title}{{DELVE-ing into the Jet: A Thin Stellar Stream on a
  Retrograde Orbit at 30 kpc}}.
\newblock \bibinfo{journal}{\aj} \bibinfo{volume}{163}, \bibinfo{pages}{18}.
\newblock \DOIprefix\doi{10.3847/1538-3881/ac3492},
  \href{http://arxiv.org/abs/2104.11755}{{\tt arXiv:2104.11755}}.
%Type = Article
\bibitem[{{Flaugher} et~al.(2015){Flaugher}, {Diehl}, {Honscheid}, {Abbott},
  {Alvarez}, {Angstadt}, {Annis}, {Antonik}, {Ballester}, {Beaufore},
  {Bernstein}, {Bernstein}, {Bigelow}, {Bonati}, {Boprie}, {Brooks},
  {Buckley-Geer}, {Campa}, {Cardiel-Sas}, {Castander}, {Castilla}, {Cease},
  {Cela-Ruiz}, {Chappa}, {Chi}, {Cooper}, {da Costa}, {Dede}, {Derylo},
  {DePoy}, {de Vicente}, {Doel}, {Drlica-Wagner}, {Eiting}, {Elliott}, {Emes},
  {Estrada}, {Fausti Neto}, {Finley}, {Flores}, {Frieman}, {Gerdes},
  {Gladders}, {Gregory}, {Gutierrez}, {Hao}, {Holland}, {Holm}, {Huffman},
  {Jackson}, {James}, {Jonas}, {Karcher}, {Karliner}, {Kent}, {Kessler},
  {Kozlovsky}, {Kron}, {Kubik}, {Kuehn}, {Kuhlmann}, {Kuk}, {Lahav}, {Lathrop},
  {Lee}, {Levi}, {Lewis}, {Li}, {Mandrichenko}, {Marshall}, {Martinez},
  {Merritt}, {Miquel}, {Mu{\~n}oz}, {Neilsen}, {Nichol}, {Nord}, {Ogando},
  {Olsen}, {Palaio}, {Patton}, {Peoples}, {Plazas}, {Rauch}, {Reil}, {Rheault},
  {Roe}, {Rogers}, {Roodman}, {Sanchez}, {Scarpine}, {Schindler}, {Schmidt},
  {Schmitt}, {Schubnell}, {Schultz}, {Schurter}, {Scott}, {Serrano}, {Shaw},
  {Smith}, {Soares-Santos}, {Stefanik}, {Stuermer}, {Suchyta}, {Sypniewski},
  {Tarle}, {Thaler}, {Tighe}, {Tran}, {Tucker}, {Walker}, {Wang}, {Watson},
  {Weaverdyck}, {Wester}, {Woods}, {Yanny} and {DES
  Collaboration}}]{flaugher:2015}
{Flaugher}, B. et~al., \bibinfo{year}{2015}.
\newblock \bibinfo{title}{{The Dark Energy Camera}}.
\newblock \bibinfo{journal}{\aj} \bibinfo{volume}{150}, \bibinfo{pages}{150}.
\newblock \DOIprefix\doi{10.1088/0004-6256/150/5/150},
  \href{http://arxiv.org/abs/1504.02900}{{\tt arXiv:1504.02900}}.
%Type = Article
\bibitem[{{Fritz} et~al.(2018){Fritz}, {Battaglia}, {Pawlowski},
  {Kallivayalil}, {van der Marel}, {Sohn}, {Brook} and {Besla}}]{fritz:2018}
\bibinfo{author}{{Fritz}, T.K.}, \bibinfo{author}{{Battaglia}, G.},
  \bibinfo{author}{{Pawlowski}, M.S.}, \bibinfo{author}{{Kallivayalil}, N.},
  \bibinfo{author}{{van der Marel}, R.}, \bibinfo{author}{{Sohn}, S.T.},
  \bibinfo{author}{{Brook}, C.}, \bibinfo{author}{{Besla}, G.},
  \bibinfo{year}{2018}.
\newblock \bibinfo{title}{{Gaia DR2 proper motions of dwarf galaxies within 420
  kpc. Orbits, Milky Way mass, tidal influences, planar alignments, and group
  infall}}.
\newblock \bibinfo{journal}{\aap} \bibinfo{volume}{619}, \bibinfo{pages}{A103}.
\newblock \DOIprefix\doi{10.1051/0004-6361/201833343},
  \href{http://arxiv.org/abs/1805.00908}{{\tt arXiv:1805.00908}}.
%Type = Article
\bibitem[{{Fu} et~al.(2018){Fu}, {Simon}, {Shetrone}, {Bovy}, {Beers},
  {Fern{\'a}ndez-Trincado}, {Placco}, {Zamora}, {Allende Prieto},
  {Garc{\'\i}a-Hern{\'a}ndez}, {Harding}, {Ivans}, {Lane}, {Nitschelm},
  {Roman-Lopes} and {Sobeck}}]{fu:2018}
{Fu}, S.~W. et~al., \bibinfo{year}{2018}.
\newblock \bibinfo{title}{{The Origin of the 300 km s$^{-1}$ Stream near Segue
  1}}.
\newblock \bibinfo{journal}{\apj} \bibinfo{volume}{866}, \bibinfo{pages}{42}.
\newblock \DOIprefix\doi{10.3847/1538-4357/aad9f9},
  \href{http://arxiv.org/abs/1804.08622}{{\tt arXiv:1804.08622}}.
%Type = Article
\bibitem[{{Fukugita} et~al.(1996){Fukugita}, {Ichikawa}, {Gunn}, {Doi},
  {Shimasaku} and {Schneider}}]{fukugita:1996}
\bibinfo{author}{{Fukugita}, M.}, \bibinfo{author}{{Ichikawa}, T.},
  \bibinfo{author}{{Gunn}, J.E.}, \bibinfo{author}{{Doi}, M.},
  \bibinfo{author}{{Shimasaku}, K.}, \bibinfo{author}{{Schneider}, D.P.},
  \bibinfo{year}{1996}.
\newblock \bibinfo{title}{{The Sloan Digital Sky Survey Photometric System}}.
\newblock \bibinfo{journal}{\aj} \bibinfo{volume}{111}, \bibinfo{pages}{1748}.
\newblock \DOIprefix\doi{10.1086/117915}.
%Type = Article
\bibitem[{{Gaia Collaboration} et~al.(2021a){Gaia Collaboration}, {Antoja},
  {McMillan}, {Kordopatis}, {Ramos}, {Helmi}, {Balbinot}, {Cantat-Gaudin},
  {Chemin}, {Figueras}, {Jordi}, {Khanna}, {Romero-G{\'o}mez}, {Seabroke},
  {Brown}, {Vallenari}, {Prusti}, {de Bruijne}, {Babusiaux}, {Biermann},
  {Creevey}, {Evans}, {Eyer}, {Hutton}, {Jansen}, {Klioner}, {Lammers},
  {Lindegren}, {Luri}, {Mignard}, {Panem}, {Pourbaix}, {Randich}, {Sartoretti},
  {Soubiran}, {Walton}, {Arenou}, {Bailer-Jones}, {Bastian}, {Cropper},
  {Drimmel}, {Katz}, {Lattanzi}, {van Leeuwen}, {Bakker}, {Casta{\~n}eda}, {De
  Angeli}, {Ducourant}, {Fabricius}, {Fouesneau}, {Fr{\'e}mat}, {Guerra},
  {Guerrier}, {Guiraud}, {Jean-Antoine Piccolo}, {Masana}, {Messineo},
  {Mowlavi}, {Nicolas}, {Nienartowicz}, {Pailler}, {Panuzzo}, {Riclet}, {Roux},
  {Sordo}, {Tanga}, {Th{\'e}venin}, {Gracia-Abril}, {Portell}, {Teyssier},
  {Altmann}, {Andrae}, {Bellas-Velidis}, {Benson}, {Berthier}, {Blomme},
  {Brugaletta}, {Burgess}, {Busso}, {Carry}, {Cellino}, {Cheek}, {Clementini},
  {Damerdji}, {Davidson}, {Delchambre}, {Dell'Oro},
  {Fern{\'a}ndez-Hern{\'a}ndez}, {Galluccio}, {Garc{\'\i}a-Lario},
  {Garcia-Reinaldos}, {Gonz{\'a}lez-N{\'u}{\~n}ez}, {Gosset}, {Haigron},
  {Halbwachs}, {Hambly}, {Harrison}, {Hatzidimitriou}, {Heiter},
  {Hern{\'a}ndez}, {Hestroffer}, {Hodgkin}, {Holl}, {Jan{\ss}en}, {Jevardat de
  Fombelle}, {Jordan}, {Krone-Martins}, {Lanzafame}, {L{\"o}ffler}, {Lorca},
  {Manteiga}, {Marchal}, {Marrese}, {Moitinho}, {Mora}, {Muinonen}, {Osborne},
  {Pancino}, {Pauwels}, {Recio-Blanco}, {Richards}, {Riello}, {Rimoldini},
  {Robin}, {Roegiers}, {Rybizki}, {Sarro}, {Siopis}, {Smith}, {Sozzetti},
  {Ulla}, {Utrilla}, {van Leeuwen}, {van Reeven}, {Abbas}, {Abreu Aramburu},
  {Accart}, {Aerts}, {Aguado}, {Ajaj}, {Altavilla}, {{\'A}lvarez}, {{\'A}lvarez
  Cid-Fuentes}, {Alves}, {Anderson}, {Varela}, {Audard}, {Baines}, {Baker},
  {Balaguer-N{\'u}{\~n}ez}, {Balog}, {Barache}, {Barbato}, {Barros}, {Barstow},
  {Bartolom{\'e}}, {Bassilana}, {Bauchet}, {Baudesson-Stella}, {Becciani},
  {Bellazzini}, {Bernet}, {Bertone}, {Bianchi}, {Blanco-Cuaresma}, {Boch},
  {Bombrun}, {Bossini}, {Bouquillon}, {Bragaglia}, {Bramante}, {Breedt},
  {Bressan}, {Brouillet}, {Bucciarelli}, {Burlacu}, {Busonero}, {Butkevich},
  {Buzzi}, {Caffau}, {Cancelliere}, {C{\'a}novas}, {Carballo}, {Carlucci},
  {Carnerero}, {Carrasco}, {Casamiquela}, {Castellani}, {Castro-Ginard},
  {Castro Sampol}, {Chaoul}, {Charlot}, {Chiavassa}, {Cioni}, {Comoretto},
  {Cooper}, {Cornez}, {Cowell}, {Crifo}, {Crosta}, {Crowley}, {Dafonte},
  {Dapergolas}, {David}, {David}, {de Laverny}, {De Luise}, {De March}, {De
  Ridder}, {de Souza}, {de Teodoro}, {de Torres}, {del Peloso}, {del Pozo},
  {Delgado}, {Delgado}, {Delisle}, {Di Matteo}, {Diakite}, {Diener},
  {Distefano}, {Dolding}, {Eappachen}, {Enke}, {Esquej}, {Fabre}, {Fabrizio},
  {Faigler}, {Fedorets}, {Fernique}, {Fienga}, {Fouron}, {Fragkoudi}, {Fraile},
  {Franke}, {Gai}, {Garabato}, {Garcia-Gutierrez}, {Garc{\'\i}a-Torres},
  {Garofalo}, {Gavras}, {Gerlach}, {Geyer}, {Giacobbe}, {Gilmore}, {Girona},
  {Giuffrida}, {Gomez}, {Gonzalez-Santamaria}, {Gonz{\'a}lez-Vidal}, {Granvik},
  {Guti{\'e}rrez-S{\'a}nchez}, {Guy}, {Hauser}, {Haywood}, {Hidalgo}, {Hilger},
  {H{\l}adczuk}, {Hobbs}, {Holland}, {Huckle}, {Jasniewicz}, {Jonker},
  {Juaristi Campillo}, {Julbe}, {Karbevska}, {Kervella}, {Kochoska},
  {Kontizas}, {Korn}, {Kostrzewa-Rutkowska}, {Kruszy{\'n}ska}, {Lambert},
  {Lanza}, {Lasne}, {Le Campion}, {Le Fustec}, {Lebreton}, {Lebzelter},
  {Leccia}, {Leclerc}, {Lecoeur-Taibi}, {Liao}, {Licata}, {Lindstr{\o}m},
  {Lister}, {Livanou}, {Lobel}, {Madrero Pardo}, {Managau}, {Mann}, {Marchant},
  {Marconi}, {Marcos Santos}, {Marinoni}, {Marocco}, {Marshall}, {Martin Polo},
  {Mart{\'\i}n-Fleitas}, {Masip}, {Massari}, {Mastrobuono-Battisti}, {Mazeh},
  {Messina}, {Michalik}, {Millar}, {Mints}, {Molina}, {Molinaro}, {Moln{\'a}r},
  {Montegriffo}, {Mor}, {Morbidelli}, {Morel}, {Morris}, {Mulone}, {Munoz},
  {Muraveva}, {Murphy}, {Musella}, {Noval}, {Ord{\'e}novic}, {Orr{\`u}},
  {Osinde}, {Pagani}, {Pagano}, {Palaversa}, {Palicio}, {Panahi}, {Pawlak},
  {Pe{\~n}alosa Esteller}, {Penttil{\"a}}, {Piersimoni}, {Pineau}, {Plachy},
  {Plum}, {Poggio}, {Poretti}, {Poujoulet}, {Pr{\v{s}}a}, {Pulone}, {Racero},
  {Ragaini}, {Rainer}, {Raiteri}, {Rambaux}, {Ramos-Lerate}, {Re Fiorentin},
  {Regibo}, {Reyl{\'e}}, {Ripepi}, {Riva}, {Rixon}, {Robichon}, {Robin},
  {Roelens}, {Rohrbasser}, {Rowell}, {Royer}, {Rybicki}, {Sadowski},
  {Sagrist{\`a} Sell{\'e}s}, {Sahlmann}, {Salgado}, {Salguero}, {Samaras},
  {Sanchez Gimenez}, {Sanna}, {Santove{\~n}a}, {Sarasso}, {Schultheis},
  {Sciacca}, {Segol}, {Segovia}, {S{\'e}gransan}, {Semeux}, {Siddiqui},
  {Siebert}, {Siltala}, {Slezak}, {Smart}, {Solano}, {Solitro}, {Souami},
  {Souchay}, {Spagna}, {Spoto}, {Steele}, {Steidelm{\"u}ller}, {Stephenson},
  {S{\"u}veges}, {Szabados}, {Szegedi-Elek}, {Taris}, {Tauran}, {Taylor},
  {Teixeira}, {Thuillot}, {Tonello}, {Torra}, {Torra}, {Turon}, {Unger},
  {Vaillant}, {van Dillen}, {Vanel}, {Vecchiato}, {Viala}, {Vicente},
  {Voutsinas}, {Weiler}, {Wevers}, {Wyrzykowski}, {Yoldas}, {Yvard}, {Zhao},
  {Zorec}, {Zucker}, {Zurbach} and {Zwitter}}]{antoja:2021}
{Gaia Collaboration}et~al., \bibinfo{year}{2021}a.
\newblock \bibinfo{title}{{Gaia Early Data Release 3. The Galactic
  anticentre}}.
\newblock \bibinfo{journal}{\aap} \bibinfo{volume}{649}, \bibinfo{pages}{A8}.
\newblock \DOIprefix\doi{10.1051/0004-6361/202039714},
  \href{http://arxiv.org/abs/2101.05811}{{\tt arXiv:2101.05811}}.
%Type = Article
\bibitem[{{Gaia Collaboration} et~al.(2018a){Gaia Collaboration}, {Babusiaux},
  {van Leeuwen}, {Barstow}, {Jordi}, {Vallenari}, {Bossini}, {Bressan},
  {Cantat-Gaudin}, {van Leeuwen}, {Brown}, {Prusti}, {de Bruijne},
  {Bailer-Jones}, {Biermann}, {Evans}, {Eyer}, {Jansen}, {Klioner}, {Lammers},
  {Lindegren}, {Luri}, {Mignard}, {Panem}, {Pourbaix}, {Randich}, {Sartoretti},
  {Siddiqui}, {Soubiran}, {Walton}, {Arenou}, {Bastian}, {Cropper}, {Drimmel},
  {Katz}, {Lattanzi}, {Bakker}, {Cacciari}, {Casta{\~n}eda}, {Chaoul}, {Cheek},
  {De Angeli}, {Fabricius}, {Guerra}, {Holl}, {Masana}, {Messineo}, {Mowlavi},
  {Nienartowicz}, {Panuzzo}, {Portell}, {Riello}, {Seabroke}, {Tanga},
  {Th{\'e}venin}, {Gracia-Abril}, {Comoretto}, {Garcia-Reinaldos}, {Teyssier},
  {Altmann}, {Andrae}, {Audard}, {Bellas-Velidis}, {Benson}, {Berthier},
  {Blomme}, {Burgess}, {Busso}, {Carry}, {Cellino}, {Clementini}, {Clotet},
  {Creevey}, {Davidson}, {De Ridder}, {Delchambre}, {Dell'Oro}, {Ducourant},
  {Fern{\'a}ndez-Hern{\'a}ndez}, {Fouesneau}, {Fr{\'e}mat}, {Galluccio},
  {Garc{\'\i}a-Torres}, {Gonz{\'a}lez-N{\'u}{\~n}ez}, {Gonz{\'a}lez-Vidal},
  {Gosset}, {Guy}, {Halbwachs}, {Hambly}, {Harrison}, {Hern{\'a}ndez},
  {Hestroffer}, {Hodgkin}, {Hutton}, {Jasniewicz}, {Jean-Antoine-Piccolo},
  {Jordan}, {Korn}, {Krone-Martins}, {Lanzafame}, {Lebzelter}, {L{\"o}ffler},
  {Manteiga}, {Marrese}, {Mart{\'\i}n-Fleitas}, {Moitinho}, {Mora}, {Muinonen},
  {Osinde}, {Pancino}, {Pauwels}, {Petit}, {Recio-Blanco}, {Richards},
  {Rimoldini}, {Robin}, {Sarro}, {Siopis}, {Smith}, {Sozzetti}, {S{\"u}veges},
  {Torra}, {van Reeven}, {Abbas}, {Abreu Aramburu}, {Accart}, {Aerts},
  {Altavilla}, {{\'A}lvarez}, {Alvarez}, {Alves}, {Anderson}, {Andrei},
  {Anglada Varela}, {Antiche}, {Antoja}, {Arcay}, {Astraatmadja}, {Bach},
  {Baker}, {Balaguer-N{\'u}{\~n}ez}, {Balm}, {Barache}, {Barata}, {Barbato},
  {Barblan}, {Barklem}, {Barrado}, {Barros}, {Bartholom{\'e} Mu{\~n}oz},
  {Bassilana}, {Becciani}, {Bellazzini}, {Berihuete}, {Bertone}, {Bianchi},
  {Bienaym{\'e}}, {Blanco-Cuaresma}, {Boch}, {Boeche}, {Bombrun}, {Borrachero},
  {Bouquillon}, {Bourda}, {Bragaglia}, {Bramante}, {Breddels}, {Brouillet},
  {Br{\"u}semeister}, {Brugaletta}, {Bucciarelli}, {Burlacu}, {Busonero},
  {Butkevich}, {Buzzi}, {Caffau}, {Cancelliere}, {Cannizzaro}, {Carballo},
  {Carlucci}, {Carrasco}, {Casamiquela}, {Castellani}, {Castro-Ginard},
  {Charlot}, {Chemin}, {Chiavassa}, {Cocozza}, {Costigan}, {Cowell}, {Crifo},
  {Crosta}, {Crowley}, {Cuypers}, {Dafonte}, {Damerdji}, {Dapergolas}, {David},
  {David}, {de Laverny}, {De Luise}, {De March}, {de Martino}, {de Souza}, {de
  Torres}, {Debosscher}, {del Pozo}, {Delbo}, {Delgado}, {Delgado}, {Diakite},
  {Diener}, {Distefano}, {Dolding}, {Drazinos}, {Dur{\'a}n}, {Edvardsson},
  {Enke}, {Eriksson}, {Esquej}, {Eynard Bontemps}, {Fabre}, {Fabrizio},
  {Faigler}, {Falc{\~a}o}, {Farr{\`a}s Casas}, {Federici}, {Fedorets},
  {Fernique}, {Figueras}, {Filippi}, {Findeisen}, {Fonti}, {Fraile}, {Fraser},
  {Fr{\'e}zouls}, {Gai}, {Galleti}, {Garabato}, {Garc{\'\i}a-Sedano},
  {Garofalo}, {Garralda}, {Gavel}, {Gavras}, {Gerssen}, {Geyer}, {Giacobbe},
  {Gilmore}, {Girona}, {Giuffrida}, {Glass}, {Gomes}, {Granvik}, {Gueguen},
  {Guerrier}, {Guiraud}, {Guti{\'e}}, {Haigron}, {Hatzidimitriou}, {Hauser},
  {Haywood}, {Heiter}, {Helmi}, {Heu}, {Hilger}, {Hobbs}, {Hofmann}, {Holland},
  {Huckle}, {Hypki}, {Icardi}, {Jan{\ss}en}, {Jevardat de Fombelle}, {Jonker},
  {Juh{\'a}sz}, {Julbe}, {Karampelas}, {Kewley}, {Klar}, {Kochoska}, {Kohley},
  {Kolenberg}, {Kontizas}, {Kontizas}, {Koposov}, {Kordopatis},
  {Kostrzewa-Rutkowska}, {Koubsky}, {Lambert}, {Lanza}, {Lasne}, {Lavigne}, {Le
  Fustec}, {Le Poncin-Lafitte}, {Lebreton}, {Leccia}, {Leclerc},
  {Lecoeur-Taibi}, {Lenhardt}, {Leroux}, {Liao}, {Licata}, {Lindstr{\o}m},
  {Lister}, {Livanou}, {Lobel}, {L{\'o}pez}, {Managau}, {Mann}, {Mantelet},
  {Marchal}, {Marchant}, {Marconi}, {Marinoni}, {Marschalk{\'o}}, {Marshall},
  {Martino}, {Marton}, {Mary}, {Massari}, {Matijevi{\v{c}}}, {Mazeh},
  {McMillan}, {Messina}, {Michalik}, {Millar}, {Molina}, {Molinaro},
  {Moln{\'a}r}, {Montegriffo}, {Mor}, {Morbidelli}, {Morel}, {Morris},
  {Mulone}, {Muraveva}, {Musella}, {Nelemans}, {Nicastro}, {Noval},
  {O'Mullane}, {Ord{\'e}novic}, {Ord{\'o}{\~n}ez-Blanco}, {Osborne}, {Pagani},
  {Pagano}, {Pailler}, {Palacin}, {Palaversa}, {Panahi}, {Pawlak},
  {Piersimoni}, {Pineau}, {Plachy}, {Plum}, {Poggio}, {Poujoulet},
  {Pr{\v{s}}a}, {Pulone}, {Racero}, {Ragaini}, {Rambaux}, {Ramos-Lerate},
  {Regibo}, {Reyl{\'e}}, {Riclet}, {Ripepi}, {Riva}, {Rivard}, {Rixon},
  {Roegiers}, {Roelens}, {Romero-G{\'o}mez}, {Rowell}, {Royer}, {Ruiz-Dern},
  {Sadowski}, {Sagrist{\`a} Sell{\'e}s}, {Sahlmann}, {Salgado}, {Salguero},
  {Sanna}, {Santana-Ros}, {Sarasso}, {Savietto}, {Schultheis}, {Sciacca},
  {Segol}, {Segovia}, {S{\'e}gransan}, {Shih}, {Siltala}, {Silva}, {Smart},
  {Smith}, {Solano}, {Solitro}, {Sordo}, {Soria Nieto}, {Souchay}, {Spagna},
  {Spoto}, {Stampa}, {Steele}, {Steidelm{\"u}ller}, {Stephenson}, {Stoev},
  {Suess}, {Surdej}, {Szabados}, {Szegedi-Elek}, {Tapiador}, {Taris}, {Tauran},
  {Taylor}, {Teixeira}, {Terrett}, {Teyssandier}, {Thuillot}, {Titarenko},
  {Torra Clotet}, {Turon}, {Ulla}, {Utrilla}, {Uzzi}, {Vaillant}, {Valentini},
  {Valette}, {van Elteren}, {Van Hemelryck}, {Vaschetto}, {Vecchiato},
  {Veljanoski}, {Viala}, {Vicente}, {Vogt}, {von Essen}, {Voss}, {Votruba},
  {Voutsinas}, {Walmsley}, {Weiler}, {Wertz}, {Wevers}, {Wyrzykowski},
  {Yoldas}, {{\v{Z}}erjal}, {Ziaeepour}, {Zorec}, {Zschocke}, {Zucker},
  {Zurbach} and {Zwitter}}]{babusiaux:2018}
{Gaia Collaboration}et~al., \bibinfo{year}{2018}a.
\newblock \bibinfo{title}{{Gaia Data Release 2. Observational
  Hertzsprung-Russell diagrams}}.
\newblock \bibinfo{journal}{\aap} \bibinfo{volume}{616}, \bibinfo{pages}{A10}.
\newblock \DOIprefix\doi{10.1051/0004-6361/201832843},
  \href{http://arxiv.org/abs/1804.09378}{{\tt arXiv:1804.09378}}.
%Type = Article
\bibitem[{{Gaia Collaboration} et~al.(2018b){Gaia Collaboration}, {Brown},
  {Vallenari}, {Prusti}, {de Bruijne}, {Babusiaux}, {Bailer-Jones}, {Biermann},
  {Evans}, {Eyer}, {Jansen}, {Jordi}, {Klioner}, {Lammers}, {Lindegren},
  {Luri}, {Mignard}, {Panem}, {Pourbaix}, {Randich}, {Sartoretti}, {Siddiqui},
  {Soubiran}, {van Leeuwen}, {Walton}, {Arenou}, {Bastian}, {Cropper},
  {Drimmel}, {Katz}, {Lattanzi}, {Bakker}, {Cacciari}, {Casta{\~n}eda},
  {Chaoul}, {Cheek}, {De Angeli}, {Fabricius}, {Guerra}, {Holl}, {Masana},
  {Messineo}, {Mowlavi}, {Nienartowicz}, {Panuzzo}, {Portell}, {Riello},
  {Seabroke}, {Tanga}, {Th{\'e}venin}, {Gracia-Abril}, {Comoretto},
  {Garcia-Reinaldos}, {Teyssier}, {Altmann}, {Andrae}, {Audard},
  {Bellas-Velidis}, {Benson}, {Berthier}, {Blomme}, {Burgess}, {Busso},
  {Carry}, {Cellino}, {Clementini}, {Clotet}, {Creevey}, {Davidson}, {De
  Ridder}, {Delchambre}, {Dell'Oro}, {Ducourant},
  {Fern{\'a}ndez-Hern{\'a}ndez}, {Fouesneau}, {Fr{\'e}mat}, {Galluccio},
  {Garc{\'\i}a-Torres}, {Gonz{\'a}lez-N{\'u}{\~n}ez}, {Gonz{\'a}lez-Vidal},
  {Gosset}, {Guy}, {Halbwachs}, {Hambly}, {Harrison}, {Hern{\'a}ndez},
  {Hestroffer}, {Hodgkin}, {Hutton}, {Jasniewicz}, {Jean-Antoine-Piccolo},
  {Jordan}, {Korn}, {Krone-Martins}, {Lanzafame}, {Lebzelter}, {L{\"o}ffler},
  {Manteiga}, {Marrese}, {Mart{\'\i}n-Fleitas}, {Moitinho}, {Mora}, {Muinonen},
  {Osinde}, {Pancino}, {Pauwels}, {Petit}, {Recio-Blanco}, {Richards},
  {Rimoldini}, {Robin}, {Sarro}, {Siopis}, {Smith}, {Sozzetti}, {S{\"u}veges},
  {Torra}, {van Reeven}, {Abbas}, {Abreu Aramburu}, {Accart}, {Aerts},
  {Altavilla}, {{\'A}lvarez}, {Alvarez}, {Alves}, {Anderson}, {Andrei},
  {Anglada Varela}, {Antiche}, {Antoja}, {Arcay}, {Astraatmadja}, {Bach},
  {Baker}, {Balaguer-N{\'u}{\~n}ez}, {Balm}, {Barache}, {Barata}, {Barbato},
  {Barblan}, {Barklem}, {Barrado}, {Barros}, {Barstow}, {Bartholom{\'e}
  Mu{\~n}oz}, {Bassilana}, {Becciani}, {Bellazzini}, {Berihuete}, {Bertone},
  {Bianchi}, {Bienaym{\'e}}, {Blanco-Cuaresma}, {Boch}, {Boeche}, {Bombrun},
  {Borrachero}, {Bossini}, {Bouquillon}, {Bourda}, {Bragaglia}, {Bramante},
  {Breddels}, {Bressan}, {Brouillet}, {Br{\"u}semeister}, {Brugaletta},
  {Bucciarelli}, {Burlacu}, {Busonero}, {Butkevich}, {Buzzi}, {Caffau},
  {Cancelliere}, {Cannizzaro}, {Cantat-Gaudin}, {Carballo}, {Carlucci},
  {Carrasco}, {Casamiquela}, {Castellani}, {Castro-Ginard}, {Charlot},
  {Chemin}, {Chiavassa}, {Cocozza}, {Costigan}, {Cowell}, {Crifo}, {Crosta},
  {Crowley}, {Cuypers}, {Dafonte}, {Damerdji}, {Dapergolas}, {David}, {David},
  {de Laverny}, {De Luise}, {De March}, {de Martino}, {de Souza}, {de Torres},
  {Debosscher}, {del Pozo}, {Delbo}, {Delgado}, {Delgado}, {Di Matteo},
  {Diakite}, {Diener}, {Distefano}, {Dolding}, {Drazinos}, {Dur{\'a}n},
  {Edvardsson}, {Enke}, {Eriksson}, {Esquej}, {Eynard Bontemps}, {Fabre},
  {Fabrizio}, {Faigler}, {Falc{\~a}o}, {Farr{\`a}s Casas}, {Federici},
  {Fedorets}, {Fernique}, {Figueras}, {Filippi}, {Findeisen}, {Fonti},
  {Fraile}, {Fraser}, {Fr{\'e}zouls}, {Gai}, {Galleti}, {Garabato},
  {Garc{\'\i}a-Sedano}, {Garofalo}, {Garralda}, {Gavel}, {Gavras}, {Gerssen},
  {Geyer}, {Giacobbe}, {Gilmore}, {Girona}, {Giuffrida}, {Glass}, {Gomes},
  {Granvik}, {Gueguen}, {Guerrier}, {Guiraud}, {Guti{\'e}rrez-S{\'a}nchez},
  {Haigron}, {Hatzidimitriou}, {Hauser}, {Haywood}, {Heiter}, {Helmi}, {Heu},
  {Hilger}, {Hobbs}, {Hofmann}, {Holland}, {Huckle}, {Hypki}, {Icardi},
  {Jan{\ss}en}, {Jevardat de Fombelle}, {Jonker}, {Juh{\'a}sz}, {Julbe},
  {Karampelas}, {Kewley}, {Klar}, {Kochoska}, {Kohley}, {Kolenberg},
  {Kontizas}, {Kontizas}, {Koposov}, {Kordopatis}, {Kostrzewa-Rutkowska},
  {Koubsky}, {Lambert}, {Lanza}, {Lasne}, {Lavigne}, {Le Fustec}, {Le
  Poncin-Lafitte}, {Lebreton}, {Leccia}, {Leclerc}, {Lecoeur-Taibi},
  {Lenhardt}, {Leroux}, {Liao}, {Licata}, {Lindstr{\o}m}, {Lister}, {Livanou},
  {Lobel}, {L{\'o}pez}, {Managau}, {Mann}, {Mantelet}, {Marchal}, {Marchant},
  {Marconi}, {Marinoni}, {Marschalk{\'o}}, {Marshall}, {Martino}, {Marton},
  {Mary}, {Massari}, {Matijevi{\v{c}}}, {Mazeh}, {McMillan}, {Messina},
  {Michalik}, {Millar}, {Molina}, {Molinaro}, {Moln{\'a}r}, {Montegriffo},
  {Mor}, {Morbidelli}, {Morel}, {Morris}, {Mulone}, {Muraveva}, {Musella},
  {Nelemans}, {Nicastro}, {Noval}, {O'Mullane}, {Ord{\'e}novic},
  {Ord{\'o}{\~n}ez-Blanco}, {Osborne}, {Pagani}, {Pagano}, {Pailler},
  {Palacin}, {Palaversa}, {Panahi}, {Pawlak}, {Piersimoni}, {Pineau}, {Plachy},
  {Plum}, {Poggio}, {Poujoulet}, {Pr{\v{s}}a}, {Pulone}, {Racero}, {Ragaini},
  {Rambaux}, {Ramos-Lerate}, {Regibo}, {Reyl{\'e}}, {Riclet}, {Ripepi}, {Riva},
  {Rivard}, {Rixon}, {Roegiers}, {Roelens}, {Romero-G{\'o}mez}, {Rowell},
  {Royer}, {Ruiz-Dern}, {Sadowski}, {Sagrist{\`a} Sell{\'e}s}, {Sahlmann},
  {Salgado}, {Salguero}, {Sanna}, {Santana-Ros}, {Sarasso}, {Savietto},
  {Schultheis}, {Sciacca}, {Segol}, {Segovia}, {S{\'e}gransan}, {Shih},
  {Siltala}, {Silva}, {Smart}, {Smith}, {Solano}, {Solitro}, {Sordo}, {Soria
  Nieto}, {Souchay}, {Spagna}, {Spoto}, {Stampa}, {Steele},
  {Steidelm{\"u}ller}, {Stephenson}, {Stoev}, {Suess}, {Surdej}, {Szabados},
  {Szegedi-Elek}, {Tapiador}, {Taris}, {Tauran}, {Taylor}, {Teixeira},
  {Terrett}, {Teyssandier}, {Thuillot}, {Titarenko}, {Torra Clotet}, {Turon},
  {Ulla}, {Utrilla}, {Uzzi}, {Vaillant}, {Valentini}, {Valette}, {van Elteren},
  {Van Hemelryck}, {van Leeuwen}, {Vaschetto}, {Vecchiato}, {Veljanoski},
  {Viala}, {Vicente}, {Vogt}, {von Essen}, {Voss}, {Votruba}, {Voutsinas},
  {Walmsley}, {Weiler}, {Wertz}, {Wevers}, {Wyrzykowski}, {Yoldas},
  {{\v{Z}}erjal}, {Ziaeepour}, {Zorec}, {Zschocke}, {Zucker}, {Zurbach} and
  {Zwitter}}]{gaiadr2}
{Gaia Collaboration}et~al., \bibinfo{year}{2018}b.
\newblock \bibinfo{title}{{Gaia Data Release 2. Summary of the contents and
  survey properties}}.
\newblock \bibinfo{journal}{\aap} \bibinfo{volume}{616}, \bibinfo{pages}{A1}.
\newblock \DOIprefix\doi{10.1051/0004-6361/201833051},
  \href{http://arxiv.org/abs/1804.09365}{{\tt arXiv:1804.09365}}.
%Type = Article
\bibitem[{{Gaia Collaboration} et~al.(2021b){Gaia Collaboration}, {Brown},
  {Vallenari}, {Prusti}, {de Bruijne}, {Babusiaux}, {Biermann}, {Creevey},
  {Evans}, {Eyer}, {Hutton}, {Jansen}, {Jordi}, {Klioner}, {Lammers},
  {Lindegren}, {Luri}, {Mignard}, {Panem}, {Pourbaix}, {Randich}, {Sartoretti},
  {Soubiran}, {Walton}, {Arenou}, {Bailer-Jones}, {Bastian}, {Cropper},
  {Drimmel}, {Katz}, {Lattanzi}, {van Leeuwen}, {Bakker}, {Cacciari},
  {Casta{\~n}eda}, {De Angeli}, {Ducourant}, {Fabricius}, {Fouesneau},
  {Fr{\'e}mat}, {Guerra}, {Guerrier}, {Guiraud}, {Jean-Antoine Piccolo},
  {Masana}, {Messineo}, {Mowlavi}, {Nicolas}, {Nienartowicz}, {Pailler},
  {Panuzzo}, {Riclet}, {Roux}, {Seabroke}, {Sordo}, {Tanga}, {Th{\'e}venin},
  {Gracia-Abril}, {Portell}, {Teyssier}, {Altmann}, {Andrae}, {Bellas-Velidis},
  {Benson}, {Berthier}, {Blomme}, {Brugaletta}, {Burgess}, {Busso}, {Carry},
  {Cellino}, {Cheek}, {Clementini}, {Damerdji}, {Davidson}, {Delchambre},
  {Dell'Oro}, {Fern{\'a}ndez-Hern{\'a}ndez}, {Galluccio}, {Garc{\'\i}a-Lario},
  {Garcia-Reinaldos}, {Gonz{\'a}lez-N{\'u}{\~n}ez}, {Gosset}, {Haigron},
  {Halbwachs}, {Hambly}, {Harrison}, {Hatzidimitriou}, {Heiter},
  {Hern{\'a}ndez}, {Hestroffer}, {Hodgkin}, {Holl}, {Jan{\ss}en}, {Jevardat de
  Fombelle}, {Jordan}, {Krone-Martins}, {Lanzafame}, {L{\"o}ffler}, {Lorca},
  {Manteiga}, {Marchal}, {Marrese}, {Moitinho}, {Mora}, {Muinonen}, {Osborne},
  {Pancino}, {Pauwels}, {Petit}, {Recio-Blanco}, {Richards}, {Riello},
  {Rimoldini}, {Robin}, {Roegiers}, {Rybizki}, {Sarro}, {Siopis}, {Smith},
  {Sozzetti}, {Ulla}, {Utrilla}, {van Leeuwen}, {van Reeven}, {Abbas}, {Abreu
  Aramburu}, {Accart}, {Aerts}, {Aguado}, {Ajaj}, {Altavilla}, {{\'A}lvarez},
  {{\'A}lvarez Cid-Fuentes}, {Alves}, {Anderson}, {Anglada Varela}, {Antoja},
  {Audard}, {Baines}, {Baker}, {Balaguer-N{\'u}{\~n}ez}, {Balbinot}, {Balog},
  {Barache}, {Barbato}, {Barros}, {Barstow}, {Bartolom{\'e}}, {Bassilana},
  {Bauchet}, {Baudesson-Stella}, {Becciani}, {Bellazzini}, {Bernet}, {Bertone},
  {Bianchi}, {Blanco-Cuaresma}, {Boch}, {Bombrun}, {Bossini}, {Bouquillon},
  {Bragaglia}, {Bramante}, {Breedt}, {Bressan}, {Brouillet}, {Bucciarelli},
  {Burlacu}, {Busonero}, {Butkevich}, {Buzzi}, {Caffau}, {Cancelliere},
  {C{\'a}novas}, {Cantat-Gaudin}, {Carballo}, {Carlucci}, {Carnerero},
  {Carrasco}, {Casamiquela}, {Castellani}, {Castro-Ginard}, {Castro Sampol},
  {Chaoul}, {Charlot}, {Chemin}, {Chiavassa}, {Cioni}, {Comoretto}, {Cooper},
  {Cornez}, {Cowell}, {Crifo}, {Crosta}, {Crowley}, {Dafonte}, {Dapergolas},
  {David}, {David}, {de Laverny}, {De Luise}, {De March}, {De Ridder}, {de
  Souza}, {de Teodoro}, {de Torres}, {del Peloso}, {del Pozo}, {Delbo},
  {Delgado}, {Delgado}, {Delisle}, {Di Matteo}, {Diakite}, {Diener},
  {Distefano}, {Dolding}, {Eappachen}, {Edvardsson}, {Enke}, {Esquej}, {Fabre},
  {Fabrizio}, {Faigler}, {Fedorets}, {Fernique}, {Fienga}, {Figueras},
  {Fouron}, {Fragkoudi}, {Fraile}, {Franke}, {Gai}, {Garabato},
  {Garcia-Gutierrez}, {Garc{\'\i}a-Torres}, {Garofalo}, {Gavras}, {Gerlach},
  {Geyer}, {Giacobbe}, {Gilmore}, {Girona}, {Giuffrida}, {Gomel}, {Gomez},
  {Gonzalez-Santamaria}, {Gonz{\'a}lez-Vidal}, {Granvik},
  {Guti{\'e}rrez-S{\'a}nchez}, {Guy}, {Hauser}, {Haywood}, {Helmi}, {Hidalgo},
  {Hilger}, {H{\l}adczuk}, {Hobbs}, {Holland}, {Huckle}, {Jasniewicz},
  {Jonker}, {Juaristi Campillo}, {Julbe}, {Karbevska}, {Kervella}, {Khanna},
  {Kochoska}, {Kontizas}, {Kordopatis}, {Korn}, {Kostrzewa-Rutkowska},
  {Kruszy{\'n}ska}, {Lambert}, {Lanza}, {Lasne}, {Le Campion}, {Le Fustec},
  {Lebreton}, {Lebzelter}, {Leccia}, {Leclerc}, {Lecoeur-Taibi}, {Liao},
  {Licata}, {Lindstr{\o}m}, {Lister}, {Livanou}, {Lobel}, {Madrero Pardo},
  {Managau}, {Mann}, {Marchant}, {Marconi}, {Marcos Santos}, {Marinoni},
  {Marocco}, {Marshall}, {Martin Polo}, {Mart{\'\i}n-Fleitas}, {Masip},
  {Massari}, {Mastrobuono-Battisti}, {Mazeh}, {McMillan}, {Messina},
  {Michalik}, {Millar}, {Mints}, {Molina}, {Molinaro}, {Moln{\'a}r},
  {Montegriffo}, {Mor}, {Morbidelli}, {Morel}, {Morris}, {Mulone}, {Munoz},
  {Muraveva}, {Murphy}, {Musella}, {Noval}, {Ord{\'e}novic}, {Orr{\`u}},
  {Osinde}, {Pagani}, {Pagano}, {Palaversa}, {Palicio}, {Panahi}, {Pawlak},
  {Pe{\~n}alosa Esteller}, {Penttil{\"a}}, {Piersimoni}, {Pineau}, {Plachy},
  {Plum}, {Poggio}, {Poretti}, {Poujoulet}, {Pr{\v{s}}a}, {Pulone}, {Racero},
  {Ragaini}, {Rainer}, {Raiteri}, {Rambaux}, {Ramos}, {Ramos-Lerate}, {Re
  Fiorentin}, {Regibo}, {Reyl{\'e}}, {Ripepi}, {Riva}, {Rixon}, {Robichon},
  {Robin}, {Roelens}, {Rohrbasser}, {Romero-G{\'o}mez}, {Rowell}, {Royer},
  {Rybicki}, {Sadowski}, {Sagrist{\`a} Sell{\'e}s}, {Sahlmann}, {Salgado},
  {Salguero}, {Samaras}, {Sanchez Gimenez}, {Sanna}, {Santove{\~n}a},
  {Sarasso}, {Schultheis}, {Sciacca}, {Segol}, {Segovia}, {S{\'e}gransan},
  {Semeux}, {Shahaf}, {Siddiqui}, {Siebert}, {Siltala}, {Slezak}, {Smart},
  {Solano}, {Solitro}, {Souami}, {Souchay}, {Spagna}, {Spoto}, {Steele},
  {Steidelm{\"u}ller}, {Stephenson}, {S{\"u}veges}, {Szabados}, {Szegedi-Elek},
  {Taris}, {Tauran}, {Taylor}, {Teixeira}, {Thuillot}, {Tonello}, {Torra},
  {Torra}, {Turon}, {Unger}, {Vaillant}, {van Dillen}, {Vanel}, {Vecchiato},
  {Viala}, {Vicente}, {Voutsinas}, {Weiler}, {Wevers}, {Wyrzykowski}, {Yoldas},
  {Yvard}, {Zhao}, {Zorec}, {Zucker}, {Zurbach} and {Zwitter}}]{gaiaedr3}
{Gaia Collaboration}et~al., \bibinfo{year}{2021}b.
\newblock \bibinfo{title}{{Gaia Early Data Release 3. Summary of the contents
  and survey properties}}.
\newblock \bibinfo{journal}{\aap} \bibinfo{volume}{649}, \bibinfo{pages}{A1}.
\newblock \DOIprefix\doi{10.1051/0004-6361/202039657},
  \href{http://arxiv.org/abs/2012.01533}{{\tt arXiv:2012.01533}}.
%Type = Article
\bibitem[{{Gaia Collaboration} et~al.(2016a){Gaia Collaboration}, {Brown},
  {Vallenari}, {Prusti}, {de Bruijne}, {Mignard}, {Drimmel}, {Babusiaux},
  {Bailer-Jones}, {Bastian}, {Biermann}, {Evans}, {Eyer}, {Jansen}, {Jordi},
  {Katz}, {Klioner}, {Lammers}, {Lindegren}, {Luri}, {O'Mullane}, {Panem},
  {Pourbaix}, {Randich}, {Sartoretti}, {Siddiqui}, {Soubiran}, {Valette}, {van
  Leeuwen}, {Walton}, {Aerts}, {Arenou}, {Cropper}, {H{\o}g}, {Lattanzi},
  {Grebel}, {Holland}, {Huc}, {Passot}, {Perryman}, {Bramante}, {Cacciari},
  {Casta{\~n}eda}, {Chaoul}, {Cheek}, {De Angeli}, {Fabricius}, {Guerra},
  {Hern{\'a}ndez}, {Jean-Antoine-Piccolo}, {Masana}, {Messineo}, {Mowlavi},
  {Nienartowicz}, {Ord{\'o}{\~n}ez-Blanco}, {Panuzzo}, {Portell}, {Richards},
  {Riello}, {Seabroke}, {Tanga}, {Th{\'e}venin}, {Torra}, {Els},
  {Gracia-Abril}, {Comoretto}, {Garcia-Reinaldos}, {Lock}, {Mercier},
  {Altmann}, {Andrae}, {Astraatmadja}, {Bellas-Velidis}, {Benson}, {Berthier},
  {Blomme}, {Busso}, {Carry}, {Cellino}, {Clementini}, {Cowell}, {Creevey},
  {Cuypers}, {Davidson}, {De Ridder}, {de Torres}, {Delchambre}, {Dell'Oro},
  {Ducourant}, {Fr{\'e}mat}, {Garc{\'\i}a-Torres}, {Gosset}, {Halbwachs},
  {Hambly}, {Harrison}, {Hauser}, {Hestroffer}, {Hodgkin}, {Huckle}, {Hutton},
  {Jasniewicz}, {Jordan}, {Kontizas}, {Korn}, {Lanzafame}, {Manteiga},
  {Moitinho}, {Muinonen}, {Osinde}, {Pancino}, {Pauwels}, {Petit},
  {Recio-Blanco}, {Robin}, {Sarro}, {Siopis}, {Smith}, {Smith}, {Sozzetti},
  {Thuillot}, {van Reeven}, {Viala}, {Abbas}, {Abreu Aramburu}, {Accart},
  {Aguado}, {Allan}, {Allasia}, {Altavilla}, {{\'A}lvarez}, {Alves},
  {Anderson}, {Andrei}, {Anglada Varela}, {Antiche}, {Antoja}, {Ant{\'o}n},
  {Arcay}, {Bach}, {Baker}, {Balaguer-N{\'u}{\~n}ez}, {Barache}, {Barata},
  {Barbier}, {Barblan}, {Barrado y Navascu{\'e}s}, {Barros}, {Barstow},
  {Becciani}, {Bellazzini}, {Bello Garc{\'\i}a}, {Belokurov}, {Bendjoya},
  {Berihuete}, {Bianchi}, {Bienaym{\'e}}, {Billebaud}, {Blagorodnova},
  {Blanco-Cuaresma}, {Boch}, {Bombrun}, {Borrachero}, {Bouquillon}, {Bourda},
  {Bouy}, {Bragaglia}, {Breddels}, {Brouillet}, {Br{\"u}semeister},
  {Bucciarelli}, {Burgess}, {Burgon}, {Burlacu}, {Busonero}, {Buzzi}, {Caffau},
  {Cambras}, {Campbell}, {Cancelliere}, {Cantat-Gaudin}, {Carlucci},
  {Carrasco}, {Castellani}, {Charlot}, {Charnas}, {Chiavassa}, {Clotet},
  {Cocozza}, {Collins}, {Costigan}, {Crifo}, {Cross}, {Crosta}, {Crowley},
  {Dafonte}, {Damerdji}, {Dapergolas}, {David}, {David}, {De Cat}, {de Felice},
  {de Laverny}, {De Luise}, {De March}, {de Martino}, {de Souza}, {Debosscher},
  {del Pozo}, {Delbo}, {Delgado}, {Delgado}, {Di Matteo}, {Diakite},
  {Distefano}, {Dolding}, {Dos Anjos}, {Drazinos}, {Duran}, {Dzigan},
  {Edvardsson}, {Enke}, {Evans}, {Eynard Bontemps}, {Fabre}, {Fabrizio},
  {Faigler}, {Falc{\~a}o}, {Farr{\`a}s Casas}, {Federici}, {Fedorets},
  {Fern{\'a}ndez-Hern{\'a}ndez}, {Fernique}, {Fienga}, {Figueras}, {Filippi},
  {Findeisen}, {Fonti}, {Fouesneau}, {Fraile}, {Fraser}, {Fuchs}, {Gai},
  {Galleti}, {Galluccio}, {Garabato}, {Garc{\'\i}a-Sedano}, {Garofalo},
  {Garralda}, {Gavras}, {Gerssen}, {Geyer}, {Gilmore}, {Girona}, {Giuffrida},
  {Gomes}, {Gonz{\'a}lez-Marcos}, {Gonz{\'a}lez-N{\'u}{\~n}ez},
  {Gonz{\'a}lez-Vidal}, {Granvik}, {Guerrier}, {Guillout}, {Guiraud},
  {G{\'u}rpide}, {Guti{\'e}rrez-S{\'a}nchez}, {Guy}, {Haigron},
  {Hatzidimitriou}, {Haywood}, {Heiter}, {Helmi}, {Hobbs}, {Hofmann}, {Holl},
  {Holland}, {Hunt}, {Hypki}, {Icardi}, {Irwin}, {Jevardat de Fombelle},
  {Jofr{\'e}}, {Jonker}, {Jorissen}, {Julbe}, {Karampelas}, {Kochoska},
  {Kohley}, {Kolenberg}, {Kontizas}, {Koposov}, {Kordopatis}, {Koubsky},
  {Krone-Martins}, {Kudryashova}, {Kull}, {Bachchan}, {Lacoste-Seris}, {Lanza},
  {Lavigne}, {Le Poncin-Lafitte}, {Lebreton}, {Lebzelter}, {Leccia}, {Leclerc},
  {Lecoeur-Taibi}, {Lemaitre}, {Lenhardt}, {Leroux}, {Liao}, {Licata},
  {Lindstr{\o}m}, {Lister}, {Livanou}, {Lobel}, {L{\"o}ffler}, {L{\'o}pez},
  {Lorenz}, {MacDonald}, {Magalh{\~a}es Fernandes}, {Managau}, {Mann},
  {Mantelet}, {Marchal}, {Marchant}, {Marconi}, {Marinoni}, {Marrese},
  {Marschalk{\'o}}, {Marshall}, {Mart{\'\i}n-Fleitas}, {Martino}, {Mary},
  {Matijevi{\v{c}}}, {Mazeh}, {McMillan}, {Messina}, {Michalik}, {Millar},
  {Miranda}, {Molina}, {Molinaro}, {Molinaro}, {Moln{\'a}r}, {Moniez},
  {Montegriffo}, {Mor}, {Mora}, {Morbidelli}, {Morel}, {Morgenthaler},
  {Morris}, {Mulone}, {Muraveva}, {Musella}, {Narbonne}, {Nelemans},
  {Nicastro}, {Noval}, {Ord{\'e}novic}, {Ordieres-Mer{\'e}}, {Osborne},
  {Pagani}, {Pagano}, {Pailler}, {Palacin}, {Palaversa}, {Parsons}, {Pecoraro},
  {Pedrosa}, {Pentik{\"a}inen}, {Pichon}, {Piersimoni}, {Pineau}, {Plachy},
  {Plum}, {Poujoulet}, {Pr{\v{s}}a}, {Pulone}, {Ragaini}, {Rago}, {Rambaux},
  {Ramos-Lerate}, {Ranalli}, {Rauw}, {Read}, {Regibo}, {Reyl{\'e}}, {Ribeiro},
  {Rimoldini}, {Ripepi}, {Riva}, {Rixon}, {Roelens}, {Romero-G{\'o}mez},
  {Rowell}, {Royer}, {Ruiz-Dern}, {Sadowski}, {Sagrist{\`a} Sell{\'e}s},
  {Sahlmann}, {Salgado}, {Salguero}, {Sarasso}, {Savietto}, {Schultheis},
  {Sciacca}, {Segol}, {Segovia}, {Segransan}, {Shih}, {Smareglia}, {Smart},
  {Solano}, {Solitro}, {Sordo}, {Soria Nieto}, {Souchay}, {Spagna}, {Spoto},
  {Stampa}, {Steele}, {Steidelm{\"u}ller}, {Stephenson}, {Stoev}, {Suess},
  {S{\"u}veges}, {Surdej}, {Szabados}, {Szegedi-Elek}, {Tapiador}, {Taris},
  {Tauran}, {Taylor}, {Teixeira}, {Terrett}, {Tingley}, {Trager}, {Turon},
  {Ulla}, {Utrilla}, {Valentini}, {van Elteren}, {Van Hemelryck}, {van
  Leeuwen}, {Varadi}, {Vecchiato}, {Veljanoski}, {Via}, {Vicente}, {Vogt},
  {Voss}, {Votruba}, {Voutsinas}, {Walmsley}, {Weiler}, {Weingrill}, {Wevers},
  {Wyrzykowski}, {Yoldas}, {{\v{Z}}erjal}, {Zucker}, {Zurbach}, {Zwitter},
  {Alecu}, {Allen}, {Allende Prieto}, {Amorim}, {Anglada-Escud{\'e}},
  {Arsenijevic}, {Azaz}, {Balm}, {Beck}, {Bernstein}, {Bigot}, {Bijaoui},
  {Blasco}, {Bonfigli}, {Bono}, {Boudreault}, {Bressan}, {Brown}, {Brunet},
  {Bunclark}, {Buonanno}, {Butkevich}, {Carret}, {Carrion}, {Chemin},
  {Ch{\'e}reau}, {Corcione}, {Darmigny}, {de Boer}, {de Teodoro}, {de Zeeuw},
  {Delle Luche}, {Domingues}, {Dubath}, {Fodor}, {Fr{\'e}zouls}, {Fries},
  {Fustes}, {Fyfe}, {Gallardo}, {Gallegos}, {Gardiol}, {Gebran}, {Gomboc},
  {G{\'o}mez}, {Grux}, {Gueguen}, {Heyrovsky}, {Hoar}, {Iannicola}, {Isasi
  Parache}, {Janotto}, {Joliet}, {Jonckheere}, {Keil}, {Kim}, {Klagyivik},
  {Klar}, {Knude}, {Kochukhov}, {Kolka}, {Kos}, {Kutka}, {Lainey}, {LeBouquin},
  {Liu}, {Loreggia}, {Makarov}, {Marseille}, {Martayan}, {Martinez-Rubi},
  {Massart}, {Meynadier}, {Mignot}, {Munari}, {Nguyen}, {Nordlander}, {Ocvirk},
  {O'Flaherty}, {Olias Sanz}, {Ortiz}, {Osorio}, {Oszkiewicz}, {Ouzounis},
  {Palmer}, {Park}, {Pasquato}, {Peltzer}, {Peralta}, {P{\'e}turaud},
  {Pieniluoma}, {Pigozzi}, {Poels}, {Prat}, {Prod'homme}, {Raison}, {Rebordao},
  {Risquez}, {Rocca-Volmerange}, {Rosen}, {Ruiz-Fuertes}, {Russo}, {Sembay},
  {Serraller Vizcaino}, {Short}, {Siebert}, {Silva}, {Sinachopoulos}, {Slezak},
  {Soffel}, {Sosnowska}, {Strai{\v{z}}ys}, {ter Linden}, {Terrell}, {Theil},
  {Tiede}, {Troisi}, {Tsalmantza}, {Tur}, {Vaccari}, {Vachier}, {Valles}, {Van
  Hamme}, {Veltz}, {Virtanen}, {Wallut}, {Wichmann}, {Wilkinson}, {Ziaeepour}
  and {Zschocke}}]{gaiadr1}
{Gaia Collaboration}et~al., \bibinfo{year}{2016}a.
\newblock \bibinfo{title}{{Gaia Data Release 1. Summary of the astrometric,
  photometric, and survey properties}}.
\newblock \bibinfo{journal}{\aap} \bibinfo{volume}{595}, \bibinfo{pages}{A2}.
\newblock \DOIprefix\doi{10.1051/0004-6361/201629512},
  \href{http://arxiv.org/abs/1609.04172}{{\tt arXiv:1609.04172}}.
%Type = Article
\bibitem[{{Gaia Collaboration} et~al.(2023a){Gaia Collaboration}, {Drimmel},
  {Romero-G{\'o}mez}, {Chemin}, {Ramos}, {Poggio}, {Ripepi}, {Andrae},
  {Blomme}, {Cantat-Gaudin}, {Castro-Ginard}, {Clementini}, {Figueras},
  {Fouesneau}, {Fr{\'e}mat}, {Jardine}, {Khanna}, {Lobel}, {Marshall},
  {Muraveva}, {Brown}, {Vallenari}, {Prusti}, {de Bruijne}, {Arenou},
  {Babusiaux}, {Biermann}, {Creevey}, {Ducourant}, {Evans}, {Eyer}, {Guerra},
  {Hutton}, {Jordi}, {Klioner}, {Lammers}, {Lindegren}, {Luri}, {Mignard},
  {Panem}, {Pourbaix}, {Randich}, {Sartoretti}, {Soubiran}, {Tanga}, {Walton},
  {Bailer-Jones}, {Bastian}, {Jansen}, {Katz}, {Lattanzi}, {van Leeuwen},
  {Bakker}, {Cacciari}, {Casta{\~n}eda}, {De Angeli}, {Fabricius}, {Galluccio},
  {Guerrier}, {Heiter}, {Masana}, {Messineo}, {Mowlavi}, {Nicolas},
  {Nienartowicz}, {Pailler}, {Panuzzo}, {Riclet}, {Roux}, {Seabroke}, {Sordo},
  {Th{\'e}venin}, {Gracia-Abril}, {Portell}, {Teyssier}, {Altmann}, {Audard},
  {Bellas-Velidis}, {Benson}, {Berthier}, {Burgess}, {Busonero}, {Busso},
  {C{\'a}novas}, {Carry}, {Cellino}, {Cheek}, {Damerdji}, {Davidson}, {de
  Teodoro}, {Nu{\~n}ez Campos}, {Delchambre}, {Dell'Oro}, {Esquej},
  {Fern{\'a}ndez-Hern{\'a}ndez}, {Fraile}, {Garabato}, {Garc{\'\i}a-Lario},
  {Gosset}, {Haigron}, {Halbwachs}, {Hambly}, {Harrison}, {Hern{\'a}ndez},
  {Hestroffer}, {Hodgkin}, {Holl}, {Jan{\ss}en}, {Jevardat de Fombelle},
  {Jordan}, {Krone-Martins}, {Lanzafame}, {L{\"o}ffler}, {Marchal}, {Marrese},
  {Moitinho}, {Muinonen}, {Osborne}, {Pancino}, {Pauwels}, {Recio-Blanco},
  {Reyl{\'e}}, {Riello}, {Rimoldini}, {Roegiers}, {Rybizki}, {Sarro}, {Siopis},
  {Smith}, {Sozzetti}, {Utrilla}, {van Leeuwen}, {Abbas}, {{\'A}brah{\'a}m},
  {Abreu Aramburu}, {Aerts}, {Aguado}, {Ajaj}, {Aldea-Montero}, {Altavilla},
  {{\'A}lvarez}, {Alves}, {Anders}, {Anderson}, {Anglada Varela}, {Antoja},
  {Baines}, {Baker}, {Balaguer-N{\'u}{\~n}ez}, {Balbinot}, {Balog}, {Barache},
  {Barbato}, {Barros}, {Barstow}, {Bartolom{\'e}}, {Bassilana}, {Bauchet},
  {Becciani}, {Bellazzini}, {Berihuete}, {Bernet}, {Bertone}, {Bianchi},
  {Binnenfeld}, {Blanco-Cuaresma}, {Boch}, {Bombrun}, {Bossini}, {Bouquillon},
  {Bragaglia}, {Bramante}, {Breedt}, {Bressan}, {Brouillet}, {Brugaletta},
  {Bucciarelli}, {Burlacu}, {Butkevich}, {Buzzi}, {Caffau}, {Cancelliere},
  {Carballo}, {Carlucci}, {Carnerero}, {Carrasco}, {Casamiquela}, {Castellani},
  {Chaoul}, {Charlot}, {Chiaramida}, {Chiavassa}, {Chornay}, {Comoretto},
  {Contursi}, {Cooper}, {Cornez}, {Cowell}, {Crifo}, {Cropper}, {Crosta},
  {Crowley}, {Dafonte}, {Dapergolas}, {David}, {de Laverny}, {De Luise}, {De
  March}, {De Ridder}, {de Souza}, {de Torres}, {del Peloso}, {del Pozo},
  {Delbo}, {Delgado}, {Delisle}, {Demouchy}, {Dharmawardena}, {Di Matteo},
  {Diakite}, {Diener}, {Distefano}, {Dolding}, {Enke}, {Fabre}, {Fabrizio},
  {Faigler}, {Fedorets}, {Fernique}, {Fournier}, {Fouron}, {Fragkoudi}, {Gai},
  {Garcia-Gutierrez}, {Garcia-Reinaldos}, {Garc{\'\i}a-Torres}, {Garofalo},
  {Gavel}, {Gavras}, {Gerlach}, {Geyer}, {Giacobbe}, {Gilmore}, {Girona},
  {Giuffrida}, {Gomel}, {Gomez}, {Gonz{\'a}lez-N{\'u}{\~n}ez},
  {Gonz{\'a}lez-Santamar{\'\i}a}, {Gonz{\'a}lez-Vidal}, {Granvik}, {Guillout},
  {Guiraud}, {Guti{\'e}rrez-S{\'a}nchez}, {Guy}, {Hatzidimitriou}, {Hauser},
  {Haywood}, {Helmer}, {Helmi}, {Sarmiento}, {Hidalgo}, {H{\l}adczuk}, {Hobbs},
  {Holland}, {Huckle}, {Jasniewicz}, {Jean-Antoine Piccolo},
  {Jim{\'e}nez-Arranz}, {Juaristi Campillo}, {Julbe}, {Karbevska}, {Kervella},
  {Kordopatis}, {Korn}, {K{\'o}sp{\'a}l}, {Kostrzewa-Rutkowska},
  {Kruszy{\'n}ska}, {Kun}, {Laizeau}, {Lambert}, {Lanza}, {Lasne}, {Le
  Campion}, {Lebreton}, {Lebzelter}, {Leccia}, {Leclerc}, {Lecoeur-Taibi},
  {Liao}, {Licata}, {Lindstr{\o}m}, {Lister}, {Livanou}, {Lorca}, {Loup},
  {Madrero Pardo}, {Magdaleno Romeo}, {Managau}, {Mann}, {Manteiga},
  {Marchant}, {Marconi}, {Marcos}, {Marcos Santos}, {Mar{\'\i}n Pina},
  {Marinoni}, {Marocco}, {Martin Polo}, {Mart{\'\i}n-Fleitas}, {Marton},
  {Mary}, {Masip}, {Massari}, {Mastrobuono-Battisti}, {Mazeh}, {McMillan},
  {Messina}, {Michalik}, {Millar}, {Mints}, {Molina}, {Molinaro}, {Moln{\'a}r},
  {Monari}, {Mongui{\'o}}, {Montegriffo}, {Montero}, {Mor}, {Mora},
  {Morbidelli}, {Morel}, {Morris}, {Murphy}, {Musella}, {Nagy}, {Noval},
  {Oca{\~n}a}, {Ogden}, {Ordenovic}, {Osinde}, {Pagani}, {Pagano}, {Palaversa},
  {Palicio}, {Pallas-Quintela}, {Panahi}, {Payne-Wardenaar}, {Pe{\~n}alosa
  Esteller}, {Penttil{\"a}}, {Pichon}, {Piersimoni}, {Pineau}, {Plachy},
  {Plum}, {Pr{\v{s}}a}, {Pulone}, {Racero}, {Ragaini}, {Rainer}, {Raiteri},
  {Ramos-Lerate}, {Re Fiorentin}, {Regibo}, {Richards}, {Rios Diaz}, {Riva},
  {Rix}, {Rixon}, {Robichon}, {Robin}, {Robin}, {Roelens}, {Rogues},
  {Rohrbasser}, {Rowell}, {Royer}, {Ruz Mieres}, {Rybicki}, {Sadowski},
  {S{\'a}ez N{\'u}{\~n}ez}, {Sagrist{\`a} Sell{\'e}s}, {Sahlmann}, {Salguero},
  {Samaras}, {Sanchez Gimenez}, {Sanna}, {Santove{\~n}a}, {Sarasso},
  {Schultheis}, {Sciacca}, {Segol}, {Segovia}, {S{\'e}gransan}, {Semeux},
  {Shahaf}, {Siddiqui}, {Siebert}, {Siltala}, {Silvelo}, {Slezak}, {Slezak},
  {Smart}, {Snaith}, {Solano}, {Solitro}, {Souami}, {Souchay}, {Spagna},
  {Spina}, {Spoto}, {Steele}, {Steidelm{\"u}ller}, {Stephenson}, {S{\"u}veges},
  {Surdej}, {Szabados}, {Szegedi-Elek}, {Taris}, {Taylor}, {Teixeira},
  {Tolomei}, {Tonello}, {Torra}, {Torra}, {Torralba Elipe}, {Trabucchi},
  {Tsounis}, {Turon}, {Ulla}, {Unger}, {Vaillant}, {van Dillen}, {van Reeven},
  {Vanel}, {Vecchiato}, {Viala}, {Vicente}, {Voutsinas}, {Weiler}, {Wevers},
  {Wyrzykowski}, {Yoldas}, {Yvard}, {Zhao}, {Zorec}, {Zucker} and
  {Zwitter}}]{drimmel:2023}
{Gaia Collaboration}et~al., \bibinfo{year}{2023}a.
\newblock \bibinfo{title}{{Gaia Data Release 3. Mapping the asymmetric disc of
  the Milky Way}}.
\newblock \bibinfo{journal}{\aap} \bibinfo{volume}{674}, \bibinfo{pages}{A37}.
\newblock \DOIprefix\doi{10.1051/0004-6361/202243797},
  \href{http://arxiv.org/abs/2206.06207}{{\tt arXiv:2206.06207}}.
%Type = Article
\bibitem[{{Gaia Collaboration} et~al.(2018c){Gaia Collaboration}, {Helmi}, {van
  Leeuwen}, {McMillan}, {Massari}, {Antoja}, {Robin}, {Lindegren}, {Bastian},
  {Arenou}, {Babusiaux}, {Biermann}, {Breddels}, {Hobbs}, {Jordi}, {Pancino},
  {Reyl{\'e}}, {Veljanoski}, {Brown}, {Vallenari}, {Prusti}, {de Bruijne},
  {Bailer-Jones}, {Evans}, {Eyer}, {Jansen}, {Klioner}, {Lammers}, {Luri},
  {Mignard}, {Panem}, {Pourbaix}, {Randich}, {Sartoretti}, {Siddiqui},
  {Soubiran}, {Walton}, {Cropper}, {Drimmel}, {Katz}, {Lattanzi}, {Bakker},
  {Cacciari}, {Casta{\~n}eda}, {Chaoul}, {Cheek}, {De Angeli}, {Fabricius},
  {Guerra}, {Holl}, {Masana}, {Messineo}, {Mowlavi}, {Nienartowicz}, {Panuzzo},
  {Portell}, {Riello}, {Seabroke}, {Tanga}, {Th{\'e}venin}, {Gracia-Abril},
  {Comoretto}, {Garcia-Reinaldos}, {Teyssier}, {Altmann}, {Andrae}, {Audard},
  {Bellas-Velidis}, {Benson}, {Berthier}, {Blomme}, {Burgess}, {Busso},
  {Carry}, {Cellino}, {Clementini}, {Clotet}, {Creevey}, {Davidson}, {De
  Ridder}, {Delchambre}, {Dell'Oro}, {Ducourant},
  {Fern{\'a}ndez-Hern{\'a}ndez}, {Fouesneau}, {Fr{\'e}mat}, {Galluccio},
  {Garc{\'\i}a-Torres}, {Gonz{\'a}lez-N{\'u}{\~n}ez}, {Gonz{\'a}lez-Vidal},
  {Gosset}, {Guy}, {Halbwachs}, {Hambly}, {Harrison}, {Hern{\'a}ndez},
  {Hestroffer}, {Hodgkin}, {Hutton}, {Jasniewicz}, {Jean-Antoine-Piccolo},
  {Jordan}, {Korn}, {Krone-Martins}, {Lanzafame}, {Lebzelter}, {L{\"o}ffler},
  {Manteiga}, {Marrese}, {Mart{\'\i}n-Fleitas}, {Moitinho}, {Mora}, {Muinonen},
  {Osinde}, {Pauwels}, {Petit}, {Recio-Blanco}, {Richards}, {Rimoldini},
  {Sarro}, {Siopis}, {Smith}, {Sozzetti}, {S{\"u}veges}, {Torra}, {van Reeven},
  {Abbas}, {Abreu Aramburu}, {Accart}, {Aerts}, {Altavilla}, {{\'A}lvarez},
  {Alvarez}, {Alves}, {Anderson}, {Andrei}, {Anglada Varela}, {Antiche},
  {Arcay}, {Astraatmadja}, {Bach}, {Baker}, {Balaguer-N{\'u}{\~n}ez}, {Balm},
  {Barache}, {Barata}, {Barbato}, {Barblan}, {Barklem}, {Barrado}, {Barros},
  {Barstow}, {Bartholom{\'e} Mu{\~n}oz}, {Bassilana}, {Becciani}, {Bellazzini},
  {Berihuete}, {Bertone}, {Bianchi}, {Bienaym{\'e}}, {Blanco-Cuaresma}, {Boch},
  {Boeche}, {Bombrun}, {Borrachero}, {Bossini}, {Bouquillon}, {Bourda},
  {Bragaglia}, {Bramante}, {Bressan}, {Brouillet}, {Br{\"u}semeister},
  {Brugaletta}, {Bucciarelli}, {Burlacu}, {Busonero}, {Butkevich}, {Buzzi},
  {Caffau}, {Cancelliere}, {Cannizzaro}, {Cantat-Gaudin}, {Carballo},
  {Carlucci}, {Carrasco}, {Casamiquela}, {Castellani}, {Castro-Ginard},
  {Charlot}, {Chemin}, {Chiavassa}, {Cocozza}, {Costigan}, {Cowell}, {Crifo},
  {Crosta}, {Crowley}, {Cuypers}, {Dafonte}, {Damerdji}, {Dapergolas}, {David},
  {David}, {de Laverny}, {De Luise}, {De March}, {de Martino}, {de Souza}, {de
  Torres}, {Debosscher}, {del Pozo}, {Delbo}, {Delgado}, {Delgado}, {Di
  Matteo}, {Diakite}, {Diener}, {Distefano}, {Dolding}, {Drazinos},
  {Dur{\'a}n}, {Edvardsson}, {Enke}, {Eriksson}, {Esquej}, {Eynard Bontemps},
  {Fabre}, {Fabrizio}, {Faigler}, {Falc{\~a}o}, {Farr{\`a}s Casas}, {Federici},
  {Fedorets}, {Fernique}, {Figueras}, {Filippi}, {Findeisen}, {Fonti},
  {Fraile}, {Fraser}, {Fr{\'e}zouls}, {Gai}, {Galleti}, {Garabato},
  {Garc{\'\i}a-Sedano}, {Garofalo}, {Garralda}, {Gavel}, {Gavras}, {Gerssen},
  {Geyer}, {Giacobbe}, {Gilmore}, {Girona}, {Giuffrida}, {Glass}, {Gomes},
  {Granvik}, {Gueguen}, {Guerrier}, {Guiraud}, {Guti{\'e}rrez-S{\'a}nchez},
  {Hofmann}, {Holland}, {Huckle}, {Hypki}, {Icardi}, {Jan{\ss}en}, {Jevardat de
  Fombelle}, {Jonker}, {Juh{\'a}sz}, {Julbe}, {Karampelas}, {Kewley}, {Klar},
  {Kochoska}, {Kohley}, {Kolenberg}, {Kontizas}, {Kontizas}, {Koposov},
  {Kordopatis}, {Kostrzewa-Rutkowska}, {Koubsky}, {Lambert}, {Lanza}, {Lasne},
  {Lavigne}, {Le Fustec}, {Le Poncin-Lafitte}, {Lebreton}, {Leccia}, {Leclerc},
  {Lecoeur-Taibi}, {Lenhardt}, {Leroux}, {Liao}, {Licata}, {Lindstr{\o}m},
  {Lister}, {Livanou}, {Lobel}, {L{\'o}pez}, {Managau}, {Mann}, {Mantelet},
  {Marchal}, {Marchant}, {Marconi}, {Marinoni}, {Marschalk{\'o}}, {Marshall},
  {Martino}, {Marton}, {Mary}, {Matijevi{\v{c}}}, {Mazeh}, {Messina},
  {Michalik}, {Millar}, {Molina}, {Molinaro}, {Moln{\'a}r}, {Montegriffo},
  {Mor}, {Morbidelli}, {Morel}, {Morris}, {Mulone}, {Muraveva}, {Musella},
  {Nelemans}, {Nicastro}, {Noval}, {O'Mullane}, {Ord{\'e}novic},
  {Ord{\'o}{\~n}ez-Blanco}, {Osborne}, {Pagani}, {Pagano}, {Pailler},
  {Palacin}, {Palaversa}, {Panahi}, {Pawlak}, {Piersimoni}, {Pineau}, {Plachy},
  {Plum}, {Poggio}, {Poujoulet}, {Pr{\v{s}}a}, {Pulone}, {Racero}, {Ragaini},
  {Rambaux}, {Ramos-Lerate}, {Regibo}, {Riclet}, {Ripepi}, {Riva}, {Rivard},
  {Rixon}, {Roegiers}, {Roelens}, {Romero-G{\'o}mez}, {Rowell}, {Royer},
  {Ruiz-Dern}, {Sadowski}, {Sagrist{\`a} Sell{\'e}s}, {Sahlmann}, {Salgado},
  {Salguero}, {Sanna}, {Santana-Ros}, {Sarasso}, {Savietto}, {Schultheis},
  {Sciacca}, {Segol}, {Segovia}, {S{\'e}gransan}, {Shih}, {Siltala}, {Silva},
  {Smart}, {Smith}, {Solano}, {Solitro}, {Sordo}, {Soria Nieto}, {Souchay},
  {Spagna}, {Spoto}, {Stampa}, {Steele}, {Steidelm{\"u}ller}, {Stephenson},
  {Stoev}, {Suess}, {Surdej}, {Szabados}, {Szegedi-Elek}, {Tapiador}, {Taris},
  {Tauran}, {Taylor}, {Teixeira}, {Terrett}, {Teyssandier}, {Thuillot},
  {Titarenko}, {Torra Clotet}, {Turon}, {Ulla}, {Utrilla}, {Uzzi}, {Vaillant},
  {Valentini}, {Valette}, {van Elteren}, {Van Hemelryck}, {van Leeuwen},
  {Vaschetto}, {Vecchiato}, {Viala}, {Vicente}, {Vogt}, {von Essen}, {Voss},
  {Votruba}, {Voutsinas}, {Walmsley}, {Weiler}, {Wertz}, {Wevems},
  {Wyrzykowski}, {Yoldas}, {{\v{Z}}erjal}, {Ziaeepour}, {Zorec}, {Zschocke},
  {Zucker}, {Zurbach} and {Zwitter}}]{helmi:2018}
{Gaia Collaboration}et~al., \bibinfo{year}{2018}c.
\newblock \bibinfo{title}{{Gaia Data Release 2. Kinematics of globular clusters
  and dwarf galaxies around the Milky Way}}.
\newblock \bibinfo{journal}{\aap} \bibinfo{volume}{616}, \bibinfo{pages}{A12}.
\newblock \DOIprefix\doi{10.1051/0004-6361/201832698},
  \href{http://arxiv.org/abs/1804.09381}{{\tt arXiv:1804.09381}}.
%Type = Article
\bibitem[{{Gaia Collaboration} et~al.(2018d){Gaia Collaboration}, {Katz},
  {Antoja}, {Romero-G{\'o}mez}, {Drimmel}, {Reyl{\'e}}, {Seabroke}, {Soubiran},
  {Babusiaux}, {Di Matteo}, {Figueras}, {Poggio}, {Robin}, {Evans}, {Brown},
  {Vallenari}, {Prusti}, {de Bruijne}, {Bailer-Jones}, {Biermann}, {Eyer},
  {Jansen}, {Jordi}, {Klioner}, {Lammers}, {Lindegren}, {Luri}, {Mignard},
  {Panem}, {Pourbaix}, {Randich}, {Sartoretti}, {Siddiqui}, {van Leeuwen},
  {Walton}, {Arenou}, {Bastian}, {Cropper}, {Lattanzi}, {Bakker}, {Cacciari},
  {Casta n}, {Chaoul}, {Cheek}, {De Angeli}, {Fabricius}, {Guerra}, {Holl},
  {Masana}, {Messineo}, {Mowlavi}, {Nienartowicz}, {Panuzzo}, {Portell},
  {Riello}, {Tanga}, {Th{\'e}venin}, {Gracia-Abril}, {Comoretto},
  {Garcia-Reinaldos}, {Teyssier}, {Altmann}, {Andrae}, {Audard},
  {Bellas-Velidis}, {Benson}, {Berthier}, {Blomme}, {Burgess}, {Busso},
  {Carry}, {Cellino}, {Clementini}, {Clotet}, {Creevey}, {Davidson}, {De
  Ridder}, {Delchambre}, {Dell'Oro}, {Ducourant},
  {Fern{\'a}ndez-Hern{\'a}ndez}, {Fouesneau}, {Fr{\'e}mat}, {Galluccio},
  {Garc{\'\i}a-Torres}, {Gonz{\'a}lez-N{\'u}{\~n}ez}, {Gonz{\'a}lez-Vidal},
  {Gosset}, {Guy}, {Halbwachs}, {Hambly}, {Harrison}, {Hern{\'a}ndez},
  {Hestroffer}, {Hodgkin}, {Hutton}, {Jasniewicz}, {Jean-Antoine-Piccolo},
  {Jordan}, {Korn}, {Krone-Martins}, {Lanzafame}, {Lebzelter}, {L{\"o}ffler},
  {Manteiga}, {Marrese}, {Mart{\'\i}n-Fleitas}, {Moitinho}, {Mora}, {Muinonen},
  {Osinde}, {Pancino}, {Pauwels}, {Petit}, {Recio-Blanco}, {Richards},
  {Rimoldini}, {Sarro}, {Siopis}, {Smith}, {Sozzetti}, {S{\"u}veges}, {Torra},
  {van Reeven}, {Abbas}, {Abreu Aramburu}, {Accart}, {Aerts}, {Altavilla},
  {{\'A}lvarez}, {Alvarez}, {Alves}, {Anderson}, {Andrei}, {Anglada Varela},
  {Antiche}, {Arcay}, {Astraatmadja}, {Bach}, {Baker},
  {Balaguer-N{\'u}{\~n}ez}, {Balm}, {Barache}, {Barata}, {Barbato}, {Barblan},
  {Barklem}, {Barrado}, {Barros}, {Barstow}, {Bartholom{\'e} Mu{\~n}oz},
  {Bassilana}, {Becciani}, {Bellazzini}, {Berihuete}, {Bertone}, {Bianchi},
  {Bienaym{\'e}}, {Blanco-Cuaresma}, {Boch}, {Boeche}, {Bombrun}, {Borrachero},
  {Bossini}, {Bouquillon}, {Bourda}, {Bragaglia}, {Bramante}, {Breddels},
  {Bressan}, {Brouillet}, {Br{\"u}semeister}, {Brugaletta}, {Bucciarelli},
  {Burlacu}, {Busonero}, {Butkevich}, {Buzzi}, {Caffau}, {Cancelliere},
  {Cannizzaro}, {Cantat-Gaudin}, {Carballo}, {Carlucci}, {Carrasco},
  {Casamiquela}, {Castellani}, {Castro-Ginard}, {Charlot}, {Chemin},
  {Chiavassa}, {Cocozza}, {Costigan}, {Cowell}, {Crifo}, {Crosta}, {Crowley},
  {Cuypers}, {Dafonte}, {Damerdji}, {Dapergolas}, {David}, {David}, {de
  Laverny}, {De Luise}, {De March}, {de Souza}, {de Torres}, {Debosscher}, {del
  Pozo}, {Delbo}, {Delgado}, {Delgado}, {Diakite}, {Diener}, {Distefano},
  {Dolding}, {Drazinos}, {Dur{\'a}n}, {Edvardsson}, {Enke}, {Eriksson},
  {Esquej}, {Eynard Bontemps}, {Fabre}, {Fabrizio}, {Faigler}, {Falc a},
  {Farr{\`a}s Casas}, {Federici}, {Fedorets}, {Fernique}, {Filippi},
  {Findeisen}, {Fonti}, {Fraile}, {Fraser}, {Fr{\'e}zouls}, {Gai}, {Galleti},
  {Garabato}, {Garc{\'\i}a-Sedano}, {Garofalo}, {Garralda}, {Gavel}, {Gavras},
  {Gerssen}, {Geyer}, {Giacobbe}, {Gilmore}, {Girona}, {Giuffrida}, {Glass},
  {Gomes}, {Granvik}, {Gueguen}, {Guerrier}, {Guiraud}, {Guti{\'e}}, {Haigron},
  {Hatzidimitriou}, {Hauser}, {Haywood}, {Heiter}, {Helmi}, {Heu}, {Hilger},
  {Hobbs}, {Hofmann}, {Holland}, {Huckle}, {Hypki}, {Icardi}, {Jan{\ss}en},
  {Jevardat de Fombelle}, {Jonker}, {Juh{\'a}sz}, {Julbe}, {Karampelas},
  {Kewley}, {Klar}, {Kochoska}, {Kohley}, {Kolenberg}, {Kontizas}, {Kontizas},
  {Koposov}, {Kordopatis}, {Kostrzewa-Rutkowska}, {Koubsky}, {Lambert},
  {Lanza}, {Lasne}, {Lavigne}, {Le Fustec}, {Le Poncin-Lafitte}, {Lebreton},
  {Leccia}, {Leclerc}, {Lecoeur-Taibi}, {Lenhardt}, {Leroux}, {Liao}, {Licata},
  {Lindstr{\o}m}, {Lister}, {Livanou}, {Lobel}, {L{\'o}pez}, {Managau}, {Mann},
  {Mantelet}, {Marchal}, {Marchant}, {Marconi}, {Marinoni}, {Marschalk{\'o}},
  {Marshall}, {Martino}, {Marton}, {Mary}, {Massari}, {Matijevi{\v{c}}},
  {Mazeh}, {McMillan}, {Messina}, {Michalik}, {Millar}, {Molina}, {Molinaro},
  {Moln{\'a}r}, {Montegriffo}, {Mor}, {Morbidelli}, {Morel}, {Morris},
  {Mulone}, {Muraveva}, {Musella}, {Nelemans}, {Nicastro}, {Noval},
  {O'Mullane}, {Ord{\'e}novic}, {Ord{\'o}{\~n}ez-Blanco}, {Osborne}, {Pagani},
  {Pagano}, {Pailler}, {Palacin}, {Palaversa}, {Panahi}, {Pawlak},
  {Piersimoni}, {Pineau}, {Plachy}, {Plum}, {Poujoulet}, {Pr{\v{s}}a},
  {Pulone}, {Racero}, {Ragaini}, {Rambaux}, {Ramos-Lerate}, {Regibo}, {Riclet},
  {Ripepi}, {Riva}, {Rivard}, {Rixon}, {Roegiers}, {Roelens}, {Rowell},
  {Royer}, {Ruiz-Dern}, {Sadowski}, {Sagrist{\`a} Sell{\'e}s}, {Sahlmann},
  {Salgado}, {Salguero}, {Sanna}, {Santana-Ros}, {Sarasso}, {Savietto},
  {Schultheis}, {Sciacca}, {Segol}, {Segovia}, {S{\'e}gransan}, {Shih},
  {Siltala}, {Silva}, {Smart}, {Smith}, {Solano}, {Solitro}, {Sordo}, {Soria
  Nieto}, {Souchay}, {Spagna}, {Spoto}, {Stampa}, {Steele},
  {Steidelm{\"u}ller}, {Stephenson}, {Stoev}, {Suess}, {Surdej}, {Szabados},
  {Szegedi-Elek}, {Tapiador}, {Taris}, {Tauran}, {Taylor}, {Teixeira},
  {Terrett}, {Teyssandier}, {Thuillot}, {Titarenko}, {Torra Clotet}, {Turon},
  {Ulla}, {Utrilla}, {Uzzi}, {Vaillant}, {Valentini}, {Valette}, {van Elteren},
  {Van Hemelryck}, {van Leeuwen}, {Vaschetto}, {Vecchiato}, {Veljanoski},
  {Viala}, {Vicente}, {Vogt}, {von Essen}, {Voss}, {Votruba}, {Voutsinas},
  {Walmsley}, {Weiler}, {Wertz}, {Wevers}, {Wyrzykowski}, {Yoldas},
  {{\v{Z}}erjal}, {Ziaeepour}, {Zorec}, {Zschocke}, {Zucker}, {Zurbach} and
  {Zwitter}}]{katz:2018}
{Gaia Collaboration}et~al., \bibinfo{year}{2018}d.
\newblock \bibinfo{title}{{Gaia Data Release 2. Mapping the Milky Way disc
  kinematics}}.
\newblock \bibinfo{journal}{\aap} \bibinfo{volume}{616}, \bibinfo{pages}{A11}.
\newblock \DOIprefix\doi{10.1051/0004-6361/201832865},
  \href{http://arxiv.org/abs/1804.09380}{{\tt arXiv:1804.09380}}.
%Type = Article
\bibitem[{{Gaia Collaboration} et~al.(2016b){Gaia Collaboration}, {Prusti}, {de
  Bruijne}, {Brown}, {Vallenari}, {Babusiaux}, {Bailer-Jones}, {Bastian},
  {Biermann}, {Evans}, {Eyer}, {Jansen}, {Jordi}, {Klioner}, {Lammers},
  {Lindegren}, {Luri}, {Mignard}, {Milligan}, {Panem}, {Poinsignon},
  {Pourbaix}, {Randich}, {Sarri}, {Sartoretti}, {Siddiqui}, {Soubiran},
  {Valette}, {van Leeuwen}, {Walton}, {Aerts}, {Arenou}, {Cropper}, {Drimmel},
  {H{\o}g}, {Katz}, {Lattanzi}, {O'Mullane}, {Grebel}, {Holland}, {Huc},
  {Passot}, {Bramante}, {Cacciari}, {Casta{\~n}eda}, {Chaoul}, {Cheek}, {De
  Angeli}, {Fabricius}, {Guerra}, {Hern{\'a}ndez}, {Jean-Antoine-Piccolo},
  {Masana}, {Messineo}, {Mowlavi}, {Nienartowicz}, {Ord{\'o}{\~n}ez-Blanco},
  {Panuzzo}, {Portell}, {Richards}, {Riello}, {Seabroke}, {Tanga},
  {Th{\'e}venin}, {Torra}, {Els}, {Gracia-Abril}, {Comoretto},
  {Garcia-Reinaldos}, {Lock}, {Mercier}, {Altmann}, {Andrae}, {Astraatmadja},
  {Bellas-Velidis}, {Benson}, {Berthier}, {Blomme}, {Busso}, {Carry},
  {Cellino}, {Clementini}, {Cowell}, {Creevey}, {Cuypers}, {Davidson}, {De
  Ridder}, {de Torres}, {Delchambre}, {Dell'Oro}, {Ducourant}, {Fr{\'e}mat},
  {Garc{\'\i}a-Torres}, {Gosset}, {Halbwachs}, {Hambly}, {Harrison}, {Hauser},
  {Hestroffer}, {Hodgkin}, {Huckle}, {Hutton}, {Jasniewicz}, {Jordan},
  {Kontizas}, {Korn}, {Lanzafame}, {Manteiga}, {Moitinho}, {Muinonen},
  {Osinde}, {Pancino}, {Pauwels}, {Petit}, {Recio-Blanco}, {Robin}, {Sarro},
  {Siopis}, {Smith}, {Smith}, {Sozzetti}, {Thuillot}, {van Reeven}, {Viala},
  {Abbas}, {Abreu Aramburu}, {Accart}, {Aguado}, {Allan}, {Allasia},
  {Altavilla}, {{\'A}lvarez}, {Alves}, {Anderson}, {Andrei}, {Anglada Varela},
  {Antiche}, {Antoja}, {Ant{\'o}n}, {Arcay}, {Atzei}, {Ayache}, {Bach},
  {Baker}, {Balaguer-N{\'u}{\~n}ez}, {Barache}, {Barata}, {Barbier}, {Barblan},
  {Baroni}, {Barrado y Navascu{\'e}s}, {Barros}, {Barstow}, {Becciani},
  {Bellazzini}, {Bellei}, {Bello Garc{\'\i}a}, {Belokurov}, {Bendjoya},
  {Berihuete}, {Bianchi}, {Bienaym{\'e}}, {Billebaud}, {Blagorodnova},
  {Blanco-Cuaresma}, {Boch}, {Bombrun}, {Borrachero}, {Bouquillon}, {Bourda},
  {Bouy}, {Bragaglia}, {Breddels}, {Brouillet}, {Br{\"u}semeister},
  {Bucciarelli}, {Budnik}, {Burgess}, {Burgon}, {Burlacu}, {Busonero}, {Buzzi},
  {Caffau}, {Cambras}, {Campbell}, {Cancelliere}, {Cantat-Gaudin}, {Carlucci},
  {Carrasco}, {Castellani}, {Charlot}, {Charnas}, {Charvet}, {Chassat},
  {Chiavassa}, {Clotet}, {Cocozza}, {Collins}, {Collins}, {Costigan}, {Crifo},
  {Cross}, {Crosta}, {Crowley}, {Dafonte}, {Damerdji}, {Dapergolas}, {David},
  {David}, {De Cat}, {de Felice}, {de Laverny}, {De Luise}, {De March}, {de
  Martino}, {de Souza}, {Debosscher}, {del Pozo}, {Delbo}, {Delgado},
  {Delgado}, {di Marco}, {Di Matteo}, {Diakite}, {Distefano}, {Dolding}, {Dos
  Anjos}, {Drazinos}, {Dur{\'a}n}, {Dzigan}, {Ecale}, {Edvardsson}, {Enke},
  {Erdmann}, {Escolar}, {Espina}, {Evans}, {Eynard Bontemps}, {Fabre},
  {Fabrizio}, {Faigler}, {Falc{\~a}o}, {Farr{\`a}s Casas}, {Faye}, {Federici},
  {Fedorets}, {Fern{\'a}ndez-Hern{\'a}ndez}, {Fernique}, {Fienga}, {Figueras},
  {Filippi}, {Findeisen}, {Fonti}, {Fouesneau}, {Fraile}, {Fraser}, {Fuchs},
  {Furnell}, {Gai}, {Galleti}, {Galluccio}, {Garabato}, {Garc{\'\i}a-Sedano},
  {Gar{\'e}}, {Garofalo}, {Garralda}, {Gavras}, {Gerssen}, {Geyer}, {Gilmore},
  {Girona}, {Giuffrida}, {Gomes}, {Gonz{\'a}lez-Marcos},
  {Gonz{\'a}lez-N{\'u}{\~n}ez}, {Gonz{\'a}lez-Vidal}, {Granvik}, {Guerrier},
  {Guillout}, {Guiraud}, {G{\'u}rpide}, {Guti{\'e}rrez-S{\'a}nchez}, {Guy},
  {Haigron}, {Hatzidimitriou}, {Haywood}, {Heiter}, {Helmi}, {Hobbs},
  {Hofmann}, {Holl}, {Holland}, {Hunt}, {Hypki}, {Icardi}, {Irwin}, {Jevardat
  de Fombelle}, {Jofr{\'e}}, {Jonker}, {Jorissen}, {Julbe}, {Karampelas},
  {Kochoska}, {Kohley}, {Kolenberg}, {Kontizas}, {Koposov}, {Kordopatis},
  {Koubsky}, {Kowalczyk}, {Krone-Martins}, {Kudryashova}, {Kull}, {Bachchan},
  {Lacoste-Seris}, {Lanza}, {Lavigne}, {Le Poncin-Lafitte}, {Lebreton},
  {Lebzelter}, {Leccia}, {Leclerc}, {Lecoeur-Taibi}, {Lemaitre}, {Lenhardt},
  {Leroux}, {Liao}, {Licata}, {Lindstr{\o}m}, {Lister}, {Livanou}, {Lobel},
  {L{\"o}ffler}, {L{\'o}pez}, {Lopez-Lozano}, {Lorenz}, {Loureiro},
  {MacDonald}, {Magalh{\~a}es Fernandes}, {Managau}, {Mann}, {Mantelet},
  {Marchal}, {Marchant}, {Marconi}, {Marie}, {Marinoni}, {Marrese},
  {Marschalk{\'o}}, {Marshall}, {Mart{\'\i}n-Fleitas}, {Martino}, {Mary},
  {Matijevi{\v{c}}}, {Mazeh}, {McMillan}, {Messina}, {Mestre}, {Michalik},
  {Millar}, {Miranda}, {Molina}, {Molinaro}, {Molinaro}, {Moln{\'a}r},
  {Moniez}, {Montegriffo}, {Monteiro}, {Mor}, {Mora}, {Morbidelli}, {Morel},
  {Morgenthaler}, {Morley}, {Morris}, {Mulone}, {Muraveva}, {Musella},
  {Narbonne}, {Nelemans}, {Nicastro}, {Noval}, {Ord{\'e}novic},
  {Ordieres-Mer{\'e}}, {Osborne}, {Pagani}, {Pagano}, {Pailler}, {Palacin},
  {Palaversa}, {Parsons}, {Paulsen}, {Pecoraro}, {Pedrosa}, {Pentik{\"a}inen},
  {Pereira}, {Pichon}, {Piersimoni}, {Pineau}, {Plachy}, {Plum}, {Poujoulet},
  {Pr{\v{s}}a}, {Pulone}, {Ragaini}, {Rago}, {Rambaux}, {Ramos-Lerate},
  {Ranalli}, {Rauw}, {Read}, {Regibo}, {Renk}, {Reyl{\'e}}, {Ribeiro},
  {Rimoldini}, {Ripepi}, {Riva}, {Rixon}, {Roelens}, {Romero-G{\'o}mez},
  {Rowell}, {Royer}, {Rudolph}, {Ruiz-Dern}, {Sadowski}, {Sagrist{\`a}
  Sell{\'e}s}, {Sahlmann}, {Salgado}, {Salguero}, {Sarasso}, {Savietto},
  {Schnorhk}, {Schultheis}, {Sciacca}, {Segol}, {Segovia}, {Segransan},
  {Serpell}, {Shih}, {Smareglia}, {Smart}, {Smith}, {Solano}, {Solitro},
  {Sordo}, {Soria Nieto}, {Souchay}, {Spagna}, {Spoto}, {Stampa}, {Steele},
  {Steidelm{\"u}ller}, {Stephenson}, {Stoev}, {Suess}, {S{\"u}veges}, {Surdej},
  {Szabados}, {Szegedi-Elek}, {Tapiador}, {Taris}, {Tauran}, {Taylor},
  {Teixeira}, {Terrett}, {Tingley}, {Trager}, {Turon}, {Ulla}, {Utrilla},
  {Valentini}, {van Elteren}, {Van Hemelryck}, {van Leeuwen}, {Varadi},
  {Vecchiato}, {Veljanoski}, {Via}, {Vicente}, {Vogt}, {Voss}, {Votruba},
  {Voutsinas}, {Walmsley}, {Weiler}, {Weingrill}, {Werner}, {Wevers},
  {Whitehead}, {Wyrzykowski}, {Yoldas}, {{\v{Z}}erjal}, {Zucker}, {Zurbach},
  {Zwitter}, {Alecu}, {Allen}, {Allende Prieto}, {Amorim},
  {Anglada-Escud{\'e}}, {Arsenijevic}, {Azaz}, {Balm}, {Beck}, {Bernstein},
  {Bigot}, {Bijaoui}, {Blasco}, {Bonfigli}, {Bono}, {Boudreault}, {Bressan},
  {Brown}, {Brunet}, {Bunclark}, {Buonanno}, {Butkevich}, {Carret}, {Carrion},
  {Chemin}, {Ch{\'e}reau}, {Corcione}, {Darmigny}, {de Boer}, {de Teodoro}, {de
  Zeeuw}, {Delle Luche}, {Domingues}, {Dubath}, {Fodor}, {Fr{\'e}zouls},
  {Fries}, {Fustes}, {Fyfe}, {Gallardo}, {Gallegos}, {Gardiol}, {Gebran},
  {Gomboc}, {G{\'o}mez}, {Grux}, {Gueguen}, {Heyrovsky}, {Hoar}, {Iannicola},
  {Isasi Parache}, {Janotto}, {Joliet}, {Jonckheere}, {Keil}, {Kim},
  {Klagyivik}, {Klar}, {Knude}, {Kochukhov}, {Kolka}, {Kos}, {Kutka}, {Lainey},
  {LeBouquin}, {Liu}, {Loreggia}, {Makarov}, {Marseille}, {Martayan},
  {Martinez-Rubi}, {Massart}, {Meynadier}, {Mignot}, {Munari}, {Nguyen},
  {Nordlander}, {Ocvirk}, {O'Flaherty}, {Olias Sanz}, {Ortiz}, {Osorio},
  {Oszkiewicz}, {Ouzounis}, {Palmer}, {Park}, {Pasquato}, {Peltzer}, {Peralta},
  {P{\'e}turaud}, {Pieniluoma}, {Pigozzi}, {Poels}, {Prat}, {Prod'homme},
  {Raison}, {Rebordao}, {Risquez}, {Rocca-Volmerange}, {Rosen}, {Ruiz-Fuertes},
  {Russo}, {Sembay}, {Serraller Vizcaino}, {Short}, {Siebert}, {Silva},
  {Sinachopoulos}, {Slezak}, {Soffel}, {Sosnowska}, {Strai{\v{z}}ys}, {ter
  Linden}, {Terrell}, {Theil}, {Tiede}, {Troisi}, {Tsalmantza}, {Tur},
  {Vaccari}, {Vachier}, {Valles}, {Van Hamme}, {Veltz}, {Virtanen}, {Wallut},
  {Wichmann}, {Wilkinson}, {Ziaeepour} and {Zschocke}}]{gaiamission:2016}
{Gaia Collaboration}et~al., \bibinfo{year}{2016}b.
\newblock \bibinfo{title}{{The Gaia mission}}.
\newblock \bibinfo{journal}{\aap} \bibinfo{volume}{595}, \bibinfo{pages}{A1}.
\newblock \DOIprefix\doi{10.1051/0004-6361/201629272},
  \href{http://arxiv.org/abs/1609.04153}{{\tt arXiv:1609.04153}}.
%Type = Article
\bibitem[{{Gaia Collaboration} et~al.(2023b){Gaia Collaboration}, {Schultheis},
  {Zhao}, {Zwitter}, {Marshall}, {Drimmel}, {Fr{\'e}mat}, {Bailer-Jones},
  {Recio-Blanco}, {Kordopatis}, {de Laverny}, {Andrae}, {Dharmawardena},
  {Fouesneau}, {Sordo}, {Brown}, {Vallenari}, {Prusti}, {de Bruijne}, {Arenou},
  {Babusiaux}, {Biermann}, {Creevey}, {Ducourant}, {Evans}, {Eyer}, {Guerra},
  {Hutton}, {Jordi}, {Klioner}, {Lammers}, {Lindegren}, {Luri}, {Mignard},
  {Panem}, {Pourbaix}, {Randich}, {Sartoretti}, {Soubiran}, {Tanga}, {Walton},
  {Bastian}, {Jansen}, {Katz}, {Lattanzi}, {van Leeuwen}, {Bakker}, {Cacciari},
  {Casta{\~n}eda}, {De Angeli}, {Fabricius}, {Galluccio}, {Guerrier}, {Heiter},
  {Masana}, {Messineo}, {Mowlavi}, {Nicolas}, {Nienartowicz}, {Pailler},
  {Panuzzo}, {Riclet}, {Roux}, {Seabroke}, {Th{\'e}venin}, {Gracia-Abril},
  {Portell}, {Teyssier}, {Altmann}, {Audard}, {Bellas-Velidis}, {Benson},
  {Berthier}, {Blomme}, {Burgess}, {Busonero}, {Busso}, {C{\'a}novas}, {Carry},
  {Cellino}, {Cheek}, {Clementini}, {Damerdji}, {Davidson}, {de Teodoro},
  {Nu{\~n}ez Campos}, {Delchambre}, {Dell'Oro}, {Esquej},
  {Fern{\'a}ndez-Hern{\'a}ndez}, {Fraile}, {Garabato}, {Garc{\'\i}a-Lario},
  {Gosset}, {Haigron}, {Halbwachs}, {Hambly}, {Harrison}, {Hern{\'a}ndez},
  {Hestroffer}, {Hodgkin}, {Holl}, {Jan{\ss}en}, {Jevardat de Fombelle},
  {Jordan}, {Krone-Martins}, {Lanzafame}, {L{\"o}ffler}, {Marchal}, {Marrese},
  {Moitinho}, {Muinonen}, {Osborne}, {Pancino}, {Pauwels}, {Reyl{\'e}},
  {Riello}, {Rimoldini}, {Roegiers}, {Rybizki}, {Sarro}, {Siopis}, {Smith},
  {Sozzetti}, {Utrilla}, {van Leeuwen}, {Abbas}, {{\'A}brah{\'a}m}, {Abreu
  Aramburu}, {Aerts}, {Aguado}, {Ajaj}, {Aldea-Montero}, {Altavilla},
  {{\'A}lvarez}, {Alves}, {Anders}, {Anderson}, {Anglada Varela}, {Antoja},
  {Baines}, {Baker}, {Balaguer-N{\'u}{\~n}ez}, {Balbinot}, {Balog}, {Barache},
  {Barbato}, {Barros}, {Barstow}, {Bartolom{\'e}}, {Bassilana}, {Bauchet},
  {Becciani}, {Bellazzini}, {Berihuete}, {Bernet}, {Bertone}, {Bianchi},
  {Binnenfeld}, {Blanco-Cuaresma}, {Boch}, {Bombrun}, {Bossini}, {Bouquillon},
  {Bragaglia}, {Bramante}, {Breedt}, {Bressan}, {Brouillet}, {Brugaletta},
  {Bucciarelli}, {Burlacu}, {Butkevich}, {Buzzi}, {Caffau}, {Cancelliere},
  {Cantat-Gaudin}, {Carballo}, {Carlucci}, {Carnerero}, {Carrasco},
  {Casamiquela}, {Castellani}, {Castro-Ginard}, {Chaoul}, {Charlot}, {Chemin},
  {Chiaramida}, {Chiavassa}, {Chornay}, {Comoretto}, {Contursi}, {Cooper},
  {Cornez}, {Cowell}, {Crifo}, {Cropper}, {Crosta}, {Crowley}, {Dafonte},
  {Dapergolas}, {David}, {De Luise}, {De March}, {De Ridder}, {de Souza}, {de
  Torres}, {del Peloso}, {del Pozo}, {Delbo}, {Delgado}, {Delisle}, {Demouchy},
  {Diakite}, {Diener}, {Distefano}, {Dolding}, {Enke}, {Fabre}, {Fabrizio},
  {Faigler}, {Fedorets}, {Fernique}, {Figueras}, {Fournier}, {Fouron},
  {Fragkoudi}, {Gai}, {Garcia-Gutierrez}, {Garcia-Reinaldos},
  {Garc{\'\i}a-Torres}, {Garofalo}, {Gavel}, {Gavras}, {Gerlach}, {Geyer},
  {Giacobbe}, {Gilmore}, {Girona}, {Giuffrida}, {Gomel}, {Gomez},
  {Gonz{\'a}lez-N{\'u}{\~n}ez}, {Gonz{\'a}lez-Santamar{\'\i}a},
  {Gonz{\'a}lez-Vidal}, {Granvik}, {Guillout}, {Guiraud},
  {Guti{\'e}rrez-S{\'a}nchez}, {Guy}, {Hatzidimitriou}, {Hauser}, {Haywood},
  {Helmer}, {Helmi}, {Sarmiento}, {Hidalgo}, {H{\l}adczuk}, {Hobbs}, {Holland},
  {Huckle}, {Jardine}, {Jasniewicz}, {Jean-Antoine Piccolo},
  {Jim{\'e}nez-Arranz}, {Juaristi Campillo}, {Julbe}, {Karbevska}, {Kervella},
  {Khanna}, {Korn}, {K{\'o}sp{\'a}l}, {Kostrzewa-Rutkowska}, {Kruszy{\'n}ska},
  {Kun}, {Laizeau}, {Lambert}, {Lanza}, {Lasne}, {Le Campion}, {Lebreton},
  {Lebzelter}, {Leccia}, {Leclerc}, {Lecoeur-Taibi}, {Liao}, {Licata},
  {Lindstr{\o}m}, {Lister}, {Livanou}, {Lobel}, {Lorca}, {Loup}, {Madrero
  Pardo}, {Magdaleno Romeo}, {Managau}, {Mann}, {Manteiga}, {Marchant},
  {Marconi}, {Marcos}, {Marcos Santos}, {Mar{\'\i}n Pina}, {Marinoni},
  {Marocco}, {Martin Polo}, {Mart{\'\i}n-Fleitas}, {Marton}, {Mary}, {Masip},
  {Massari}, {Mastrobuono-Battisti}, {Mazeh}, {McMillan}, {Messina},
  {Michalik}, {Millar}, {Mints}, {Molina}, {Molinaro}, {Moln{\'a}r}, {Monari},
  {Mongui{\'o}}, {Montegriffo}, {Montero}, {Mor}, {Mora}, {Morbidelli},
  {Morel}, {Morris}, {Muraveva}, {Murphy}, {Musella}, {Nagy}, {Noval},
  {Oca{\~n}a}, {Ogden}, {Ordenovic}, {Osinde}, {Pagani}, {Pagano}, {Palaversa},
  {Palicio}, {Pallas-Quintela}, {Panahi}, {Payne-Wardenaar}, {Pe{\~n}alosa
  Esteller}, {Penttil{\"a}}, {Pichon}, {Piersimoni}, {Pineau}, {Plachy},
  {Plum}, {Poggio}, {Pr{\v{s}}a}, {Pulone}, {Racero}, {Ragaini}, {Rainer},
  {Raiteri}, {Ramos}, {Ramos-Lerate}, {Re Fiorentin}, {Regibo}, {Richards},
  {Rios Diaz}, {Ripepi}, {Riva}, {Rix}, {Rixon}, {Robichon}, {Robin}, {Robin},
  {Roelens}, {Rogues}, {Rohrbasser}, {Romero-G{\'o}mez}, {Rowell}, {Royer},
  {Ruz Mieres}, {Rybicki}, {Sadowski}, {S{\'a}ez N{\'u}{\~n}ez}, {Sagrist{\`a}
  Sell{\'e}s}, {Sahlmann}, {Salguero}, {Samaras}, {Sanchez Gimenez}, {Sanna},
  {Santove{\~n}a}, {Sarasso}, {Sciacca}, {Segol}, {Segovia}, {S{\'e}gransan},
  {Semeux}, {Shahaf}, {Siddiqui}, {Siebert}, {Siltala}, {Silvelo}, {Slezak},
  {Slezak}, {Smart}, {Snaith}, {Solano}, {Solitro}, {Souami}, {Souchay},
  {Spagna}, {Spina}, {Spoto}, {Steele}, {Steidelm{\"u}ller}, {Stephenson},
  {S{\"u}veges}, {Surdej}, {Szabados}, {Szegedi-Elek}, {Taris}, {Taylor},
  {Teixeira}, {Tolomei}, {Tonello}, {Torra}, {Torra}, {Torralba Elipe},
  {Trabucchi}, {Tsounis}, {Turon}, {Ulla}, {Unger}, {Vaillant}, {van Dillen},
  {van Reeven}, {Vanel}, {Vecchiato}, {Viala}, {Vicente}, {Voutsinas},
  {Weiler}, {Wevers}, {Wyrzykowski}, {Yoldas}, {Yvard}, {Zorec} and
  {Zucker}}]{schultheis:2023}
{Gaia Collaboration}et~al., \bibinfo{year}{2023}b.
\newblock \bibinfo{title}{{Gaia Data Release 3. Exploring and mapping the
  diffuse interstellar band at 862 nm}}.
\newblock \bibinfo{journal}{\aap} \bibinfo{volume}{674}, \bibinfo{pages}{A40}.
\newblock \DOIprefix\doi{10.1051/0004-6361/202243283},
  \href{http://arxiv.org/abs/2206.05536}{{\tt arXiv:2206.05536}}.
%Type = Article
\bibitem[{{Gaia Collaboration} et~al.(2021c){Gaia Collaboration}, {Smart},
  {Sarro}, {Rybizki}, {Reyl{\'e}}, {Robin}, {Hambly}, {Abbas}, {Barstow}, {de
  Bruijne}, {Bucciarelli}, {Carrasco}, {Cooper}, {Hodgkin}, {Masana},
  {Michalik}, {Sahlmann}, {Sozzetti}, {Brown}, {Vallenari}, {Prusti},
  {Babusiaux}, {Biermann}, {Creevey}, {Evans}, {Eyer}, {Hutton}, {Jansen},
  {Jordi}, {Klioner}, {Lammers}, {Lindegren}, {Luri}, {Mignard}, {Panem},
  {Pourbaix}, {Randich}, {Sartoretti}, {Soubiran}, {Walton}, {Arenou},
  {Bailer-Jones}, {Bastian}, {Cropper}, {Drimmel}, {Katz}, {Lattanzi}, {van
  Leeuwen}, {Bakker}, {Casta{\~n}eda}, {De Angeli}, {Ducourant}, {Fabricius},
  {Fouesneau}, {Fr{\'e}mat}, {Guerra}, {Guerrier}, {Guiraud}, {Jean-Antoine
  Piccolo}, {Messineo}, {Mowlavi}, {Nicolas}, {Nienartowicz}, {Pailler},
  {Panuzzo}, {Riclet}, {Roux}, {Seabroke}, {Sordo}, {Tanga}, {Th{\'e}venin},
  {Gracia-Abril}, {Portell}, {Teyssier}, {Altmann}, {Andrae}, {Bellas-Velidis},
  {Benson}, {Berthier}, {Blomme}, {Brugaletta}, {Burgess}, {Busso}, {Carry},
  {Cellino}, {Cheek}, {Clementini}, {Damerdji}, {Davidson}, {Delchambre},
  {Dell'Oro}, {Fern{\'a}ndez-Hern{\'a}ndez}, {Galluccio}, {Garc{\'\i}a-Lario},
  {Garcia-Reinaldos}, {Gonz{\'a}lez-N{\'u}{\~n}ez}, {Gosset}, {Haigron},
  {Halbwachs}, {Harrison}, {Hatzidimitriou}, {Heiter}, {Hern{\'a}ndez},
  {Hestroffer}, {Holl}, {Jan{\ss}en}, {Jevardat de Fombelle}, {Jordan},
  {Krone-Martins}, {Lanzafame}, {L{\"o}ffler}, {Lorca}, {Manteiga}, {Marchal},
  {Marrese}, {Moitinho}, {Mora}, {Muinonen}, {Osborne}, {Pancino}, {Pauwels},
  {Recio-Blanco}, {Richards}, {Riello}, {Rimoldini}, {Roegiers}, {Siopis},
  {Smith}, {Ulla}, {Utrilla}, {van Leeuwen}, {van Reeven}, {Abreu Aramburu},
  {Accart}, {Aerts}, {Aguado}, {Ajaj}, {Altavilla}, {{\'A}lvarez}, {{\'A}lvarez
  Cid-Fuentes}, {Alves}, {Anderson}, {Anglada Varela}, {Antoja}, {Audard},
  {Baines}, {Baker}, {Balaguer-N{\'u}{\~n}ez}, {Balbinot}, {Balog}, {Barache},
  {Barbato}, {Barros}, {Bartolom{\'e}}, {Bassilana}, {Bauchet},
  {Baudesson-Stella}, {Becciani}, {Bellazzini}, {Bernet}, {Bertone}, {Bianchi},
  {Blanco-Cuaresma}, {Boch}, {Bombrun}, {Bossini}, {Bouquillon}, {Bragaglia},
  {Bramante}, {Breedt}, {Bressan}, {Brouillet}, {Burlacu}, {Busonero},
  {Butkevich}, {Buzzi}, {Caffau}, {Cancelliere}, {C{\'a}novas},
  {Cantat-Gaudin}, {Carballo}, {Carlucci}, {Carnerero}, {Casamiquela},
  {Castellani}, {Castro-Ginard}, {Castro Sampol}, {Chaoul}, {Charlot},
  {Chemin}, {Chiavassa}, {Cioni}, {Comoretto}, {Cornez}, {Cowell}, {Crifo},
  {Crosta}, {Crowley}, {Dafonte}, {Dapergolas}, {David}, {David}, {de Laverny},
  {De Luise}, {De March}, {De Ridder}, {de Souza}, {de Teodoro}, {de Torres},
  {del Peloso}, {del Pozo}, {Delgado}, {Delgado}, {Delisle}, {Di Matteo},
  {Diakite}, {Diener}, {Distefano}, {Dolding}, {Eappachen}, {Edvardsson},
  {Enke}, {Esquej}, {Fabre}, {Fabrizio}, {Faigler}, {Fedorets}, {Fernique},
  {Fienga}, {Figueras}, {Fouron}, {Fragkoudi}, {Fraile}, {Franke}, {Gai},
  {Garabato}, {Garcia-Gutierrez}, {Garc{\'\i}a-Torres}, {Garofalo}, {Gavras},
  {Gerlach}, {Geyer}, {Giacobbe}, {Gilmore}, {Girona}, {Giuffrida}, {Gomel},
  {Gomez}, {Gonzalez-Santamaria}, {Gonz{\'a}lez-Vidal}, {Granvik},
  {Guti{\'e}rrez-S{\'a}nchez}, {Guy}, {Hauser}, {Haywood}, {Helmi}, {Hidalgo},
  {Hilger}, {H{\l}adczuk}, {Hobbs}, {Holland}, {Huckle}, {Jasniewicz},
  {Jonker}, {Juaristi Campillo}, {Julbe}, {Karbevska}, {Kervella}, {Khanna},
  {Kochoska}, {Kontizas}, {Kordopatis}, {Korn}, {Kostrzewa-Rutkowska},
  {Kruszy{\'n}ska}, {Lambert}, {Lanza}, {Lasne}, {Le Campion}, {Le Fustec},
  {Lebreton}, {Lebzelter}, {Leccia}, {Leclerc}, {Lecoeur-Taibi}, {Liao},
  {Licata}, {Lindstr{\o}m}, {Lister}, {Livanou}, {Lobel}, {Madrero Pardo},
  {Managau}, {Mann}, {Marchant}, {Marconi}, {Marcos Santos}, {Marinoni},
  {Marocco}, {Marshall}, {Martin Polo}, {Mart{\'\i}n-Fleitas}, {Masip},
  {Massari}, {Mastrobuono-Battisti}, {Mazeh}, {McMillan}, {Messina}, {Millar},
  {Mints}, {Molina}, {Molinaro}, {Moln{\'a}r}, {Montegriffo}, {Mor},
  {Morbidelli}, {Morel}, {Morris}, {Mulone}, {Munoz}, {Muraveva}, {Murphy},
  {Musella}, {Noval}, {Ord{\'e}novic}, {Orr{\`u}}, {Osinde}, {Pagani},
  {Pagano}, {Palaversa}, {Palicio}, {Panahi}, {Pawlak}, {Pe{\~n}alosa
  Esteller}, {Penttil{\"a}}, {Piersimoni}, {Pineau}, {Plachy}, {Plum},
  {Poggio}, {Poretti}, {Poujoulet}, {Pr{\v{s}}a}, {Pulone}, {Racero},
  {Ragaini}, {Rainer}, {Raiteri}, {Rambaux}, {Ramos}, {Ramos-Lerate}, {Re
  Fiorentin}, {Regibo}, {Ripepi}, {Riva}, {Rixon}, {Robichon}, {Robin},
  {Roelens}, {Rohrbasser}, {Romero-G{\'o}mez}, {Rowell}, {Royer}, {Rybicki},
  {Sadowski}, {Sagrist{\`a} Sell{\'e}s}, {Salgado}, {Salguero}, {Samaras},
  {Sanchez Gimenez}, {Sanna}, {Santove{\~n}a}, {Sarasso}, {Schultheis},
  {Sciacca}, {Segol}, {Segovia}, {S{\'e}gransan}, {Semeux}, {Shahaf},
  {Siddiqui}, {Siebert}, {Siltala}, {Slezak}, {Solano}, {Solitro}, {Souami},
  {Souchay}, {Spagna}, {Spoto}, {Steele}, {Steidelm{\"u}ller}, {Stephenson},
  {S{\"u}veges}, {Szabados}, {Szegedi-Elek}, {Taris}, {Tauran}, {Taylor},
  {Teixeira}, {Thuillot}, {Tonello}, {Torra}, {Torra}, {Turon}, {Unger},
  {Vaillant}, {van Dillen}, {Vanel}, {Vecchiato}, {Viala}, {Vicente},
  {Voutsinas}, {Weiler}, {Wevers}, {Wyrzykowski}, {Yoldas}, {Yvard}, {Zhao},
  {Zorec}, {Zucker}, {Zurbach} and {Zwitter}}]{smart:2021}
{Gaia Collaboration}et~al., \bibinfo{year}{2021}c.
\newblock \bibinfo{title}{{Gaia Early Data Release 3. The Gaia Catalogue of
  Nearby Stars}}.
\newblock \bibinfo{journal}{\aap} \bibinfo{volume}{649}, \bibinfo{pages}{A6}.
\newblock \DOIprefix\doi{10.1051/0004-6361/202039498},
  \href{http://arxiv.org/abs/2012.02061}{{\tt arXiv:2012.02061}}.
%Type = Article
\bibitem[{{Gaia Collaboration} et~al.(2023c){Gaia Collaboration}, {Vallenari},
  {Brown}, {Prusti}, {de Bruijne}, {Arenou}, {Babusiaux}, {Biermann},
  {Creevey}, {Ducourant}, {Evans}, {Eyer}, {Guerra}, {Hutton}, {Jordi},
  {Klioner}, {Lammers}, {Lindegren}, {Luri}, {Mignard}, {Panem}, {Pourbaix},
  {Randich}, {Sartoretti}, {Soubiran}, {Tanga}, {Walton}, {Bailer-Jones},
  {Bastian}, {Drimmel}, {Jansen}, {Katz}, {Lattanzi}, {van Leeuwen}, {Bakker},
  {Cacciari}, {Casta{\~n}eda}, {De Angeli}, {Fabricius}, {Fouesneau},
  {Fr{\'e}mat}, {Galluccio}, {Guerrier}, {Heiter}, {Masana}, {Messineo},
  {Mowlavi}, {Nicolas}, {Nienartowicz}, {Pailler}, {Panuzzo}, {Riclet}, {Roux},
  {Seabroke}, {Sordo}, {Th{\'e}venin}, {Gracia-Abril}, {Portell}, {Teyssier},
  {Altmann}, {Andrae}, {Audard}, {Bellas-Velidis}, {Benson}, {Berthier},
  {Blomme}, {Burgess}, {Busonero}, {Busso}, {C{\'a}novas}, {Carry}, {Cellino},
  {Cheek}, {Clementini}, {Damerdji}, {Davidson}, {de Teodoro}, {Nu{\~n}ez
  Campos}, {Delchambre}, {Dell'Oro}, {Esquej}, {Fern{\'a}ndez-Hern{\'a}ndez},
  {Fraile}, {Garabato}, {Garc{\'\i}a-Lario}, {Gosset}, {Haigron}, {Halbwachs},
  {Hambly}, {Harrison}, {Hern{\'a}ndez}, {Hestroffer}, {Hodgkin}, {Holl},
  {Jan{\ss}en}, {Jevardat de Fombelle}, {Jordan}, {Krone-Martins}, {Lanzafame},
  {L{\"o}ffler}, {Marchal}, {Marrese}, {Moitinho}, {Muinonen}, {Osborne},
  {Pancino}, {Pauwels}, {Recio-Blanco}, {Reyl{\'e}}, {Riello}, {Rimoldini},
  {Roegiers}, {Rybizki}, {Sarro}, {Siopis}, {Smith}, {Sozzetti}, {Utrilla},
  {van Leeuwen}, {Abbas}, {{\'A}brah{\'a}m}, {Abreu Aramburu}, {Aerts},
  {Aguado}, {Ajaj}, {Aldea-Montero}, {Altavilla}, {{\'A}lvarez}, {Alves},
  {Anders}, {Anderson}, {Anglada Varela}, {Antoja}, {Baines}, {Baker},
  {Balaguer-N{\'u}{\~n}ez}, {Balbinot}, {Balog}, {Barache}, {Barbato},
  {Barros}, {Barstow}, {Bartolom{\'e}}, {Bassilana}, {Bauchet}, {Becciani},
  {Bellazzini}, {Berihuete}, {Bernet}, {Bertone}, {Bianchi}, {Binnenfeld},
  {Blanco-Cuaresma}, {Blazere}, {Boch}, {Bombrun}, {Bossini}, {Bouquillon},
  {Bragaglia}, {Bramante}, {Breedt}, {Bressan}, {Brouillet}, {Brugaletta},
  {Bucciarelli}, {Burlacu}, {Butkevich}, {Buzzi}, {Caffau}, {Cancelliere},
  {Cantat-Gaudin}, {Carballo}, {Carlucci}, {Carnerero}, {Carrasco},
  {Casamiquela}, {Castellani}, {Castro-Ginard}, {Chaoul}, {Charlot}, {Chemin},
  {Chiaramida}, {Chiavassa}, {Chornay}, {Comoretto}, {Contursi}, {Cooper},
  {Cornez}, {Cowell}, {Crifo}, {Cropper}, {Crosta}, {Crowley}, {Dafonte},
  {Dapergolas}, {David}, {David}, {de Laverny}, {De Luise}, {De March}, {De
  Ridder}, {de Souza}, {de Torres}, {del Peloso}, {del Pozo}, {Delbo},
  {Delgado}, {Delisle}, {Demouchy}, {Dharmawardena}, {Di Matteo}, {Diakite},
  {Diener}, {Distefano}, {Dolding}, {Edvardsson}, {Enke}, {Fabre}, {Fabrizio},
  {Faigler}, {Fedorets}, {Fernique}, {Fienga}, {Figueras}, {Fournier},
  {Fouron}, {Fragkoudi}, {Gai}, {Garcia-Gutierrez}, {Garcia-Reinaldos},
  {Garc{\'\i}a-Torres}, {Garofalo}, {Gavel}, {Gavras}, {Gerlach}, {Geyer},
  {Giacobbe}, {Gilmore}, {Girona}, {Giuffrida}, {Gomel}, {Gomez},
  {Gonz{\'a}lez-N{\'u}{\~n}ez}, {Gonz{\'a}lez-Santamar{\'\i}a},
  {Gonz{\'a}lez-Vidal}, {Granvik}, {Guillout}, {Guiraud},
  {Guti{\'e}rrez-S{\'a}nchez}, {Guy}, {Hatzidimitriou}, {Hauser}, {Haywood},
  {Helmer}, {Helmi}, {Sarmiento}, {Hidalgo}, {Hilger}, {H{\l}adczuk}, {Hobbs},
  {Holland}, {Huckle}, {Jardine}, {Jasniewicz}, {Jean-Antoine Piccolo},
  {Jim{\'e}nez-Arranz}, {Jorissen}, {Juaristi Campillo}, {Julbe}, {Karbevska},
  {Kervella}, {Khanna}, {Kontizas}, {Kordopatis}, {Korn}, {K{\'o}sp{\'a}l},
  {Kostrzewa-Rutkowska}, {Kruszy{\'n}ska}, {Kun}, {Laizeau}, {Lambert},
  {Lanza}, {Lasne}, {Le Campion}, {Lebreton}, {Lebzelter}, {Leccia}, {Leclerc},
  {Lecoeur-Taibi}, {Liao}, {Licata}, {Lindstr{\o}m}, {Lister}, {Livanou},
  {Lobel}, {Lorca}, {Loup}, {Madrero Pardo}, {Magdaleno Romeo}, {Managau},
  {Mann}, {Manteiga}, {Marchant}, {Marconi}, {Marcos}, {Marcos Santos},
  {Mar{\'\i}n Pina}, {Marinoni}, {Marocco}, {Marshall}, {Martin Polo},
  {Mart{\'\i}n-Fleitas}, {Marton}, {Mary}, {Masip}, {Massari},
  {Mastrobuono-Battisti}, {Mazeh}, {McMillan}, {Messina}, {Michalik}, {Millar},
  {Mints}, {Molina}, {Molinaro}, {Moln{\'a}r}, {Monari}, {Mongui{\'o}},
  {Montegriffo}, {Montero}, {Mor}, {Mora}, {Morbidelli}, {Morel}, {Morris},
  {Muraveva}, {Murphy}, {Musella}, {Nagy}, {Noval}, {Oca{\~n}a}, {Ogden},
  {Ordenovic}, {Osinde}, {Pagani}, {Pagano}, {Palaversa}, {Palicio},
  {Pallas-Quintela}, {Panahi}, {Payne-Wardenaar}, {Pe{\~n}alosa Esteller},
  {Penttil{\"a}}, {Pichon}, {Piersimoni}, {Pineau}, {Plachy}, {Plum}, {Poggio},
  {Pr{\v{s}}a}, {Pulone}, {Racero}, {Ragaini}, {Rainer}, {Raiteri}, {Rambaux},
  {Ramos}, {Ramos-Lerate}, {Re Fiorentin}, {Regibo}, {Richards}, {Rios Diaz},
  {Ripepi}, {Riva}, {Rix}, {Rixon}, {Robichon}, {Robin}, {Robin}, {Roelens},
  {Rogues}, {Rohrbasser}, {Romero-G{\'o}mez}, {Rowell}, {Royer}, {Ruz Mieres},
  {Rybicki}, {Sadowski}, {S{\'a}ez N{\'u}{\~n}ez}, {Sagrist{\`a} Sell{\'e}s},
  {Sahlmann}, {Salguero}, {Samaras}, {Sanchez Gimenez}, {Sanna},
  {Santove{\~n}a}, {Sarasso}, {Schultheis}, {Sciacca}, {Segol}, {Segovia},
  {S{\'e}gransan}, {Semeux}, {Shahaf}, {Siddiqui}, {Siebert}, {Siltala},
  {Silvelo}, {Slezak}, {Slezak}, {Smart}, {Snaith}, {Solano}, {Solitro},
  {Souami}, {Souchay}, {Spagna}, {Spina}, {Spoto}, {Steele},
  {Steidelm{\"u}ller}, {Stephenson}, {S{\"u}veges}, {Surdej}, {Szabados},
  {Szegedi-Elek}, {Taris}, {Taylor}, {Teixeira}, {Tolomei}, {Tonello}, {Torra},
  {Torra}, {Torralba Elipe}, {Trabucchi}, {Tsounis}, {Turon}, {Ulla}, {Unger},
  {Vaillant}, {van Dillen}, {van Reeven}, {Vanel}, {Vecchiato}, {Viala},
  {Vicente}, {Voutsinas}, {Weiler}, {Wevers}, {Wyrzykowski}, {Yoldas}, {Yvard},
  {Zhao}, {Zorec}, {Zucker} and {Zwitter}}]{gaiadr3}
{Gaia Collaboration}et~al., \bibinfo{year}{2023}c.
\newblock \bibinfo{title}{{Gaia Data Release 3. Summary of the content and
  survey properties}}.
\newblock \bibinfo{journal}{\aap} \bibinfo{volume}{674}, \bibinfo{pages}{A1}.
\newblock \DOIprefix\doi{10.1051/0004-6361/202243940},
  \href{http://arxiv.org/abs/2208.00211}{{\tt arXiv:2208.00211}}.
%Type = Article
\bibitem[{{Garavito-Camargo} et~al.(2019){Garavito-Camargo}, {Besla},
  {Laporte}, {Johnston}, {G{\'o}mez} and {Watkins}}]{garavito-camargo:2019}
\bibinfo{author}{{Garavito-Camargo}, N.}, \bibinfo{author}{{Besla}, G.},
  \bibinfo{author}{{Laporte}, C.F.P.}, \bibinfo{author}{{Johnston}, K.V.},
  \bibinfo{author}{{G{\'o}mez}, F.A.}, \bibinfo{author}{{Watkins}, L.L.},
  \bibinfo{year}{2019}.
\newblock \bibinfo{title}{{Hunting for the Dark Matter Wake Induced by the
  Large Magellanic Cloud}}.
\newblock \bibinfo{journal}{\apj} \bibinfo{volume}{884}, \bibinfo{pages}{51}.
\newblock \DOIprefix\doi{10.3847/1538-4357/ab32eb},
  \href{http://arxiv.org/abs/1902.05089}{{\tt arXiv:1902.05089}}.
%Type = Article
\bibitem[{{Garavito-Camargo} et~al.(2021){Garavito-Camargo}, {Besla},
  {Laporte}, {Price-Whelan}, {Cunningham}, {Johnston}, {Weinberg} and
  {G{\'o}mez}}]{garavito-camargo:2021}
\bibinfo{author}{{Garavito-Camargo}, N.}, \bibinfo{author}{{Besla}, G.},
  \bibinfo{author}{{Laporte}, C.F.P.}, \bibinfo{author}{{Price-Whelan}, A.M.},
  \bibinfo{author}{{Cunningham}, E.C.}, \bibinfo{author}{{Johnston}, K.V.},
  \bibinfo{author}{{Weinberg}, M.}, \bibinfo{author}{{G{\'o}mez}, F.A.},
  \bibinfo{year}{2021}.
\newblock \bibinfo{title}{{Quantifying the Impact of the Large Magellanic Cloud
  on the Structure of the Milky Way's Dark Matter Halo Using Basis Function
  Expansions}}.
\newblock \bibinfo{journal}{\apj} \bibinfo{volume}{919}, \bibinfo{pages}{109}.
\newblock \DOIprefix\doi{10.3847/1538-4357/ac0b44},
  \href{http://arxiv.org/abs/2010.00816}{{\tt arXiv:2010.00816}}.
%Type = Article
\bibitem[{{Garavito-Camargo} et~al.(2023){Garavito-Camargo}, {Price-Whelan},
  {Samuel}, {Cunningham}, {Patel}, {Wetzel}, {Johnston}, {Arora}, {Sanderson},
  {Garrison} and {Horta}}]{garavito-camargo:2024}
{Garavito-Camargo}, N. et~al., \bibinfo{year}{2023}.
\newblock \bibinfo{title}{{On the co-rotation of Milky Way satellites: LMC-mass
  satellites induce apparent motions in outer halo tracers}}.
\newblock \bibinfo{journal}{arXiv e-prints} ,
  \bibinfo{pages}{arXiv:2311.11359}\DOIprefix\doi{10.48550/arXiv.2311.11359},
  \href{http://arxiv.org/abs/2311.11359}{{\tt arXiv:2311.11359}}.
%Type = Article
\bibitem[{{Garrow} et~al.(2020){Garrow}, {Webb} and {Bovy}}]{garrow:2020}
\bibinfo{author}{{Garrow}, T.}, \bibinfo{author}{{Webb}, J.J.},
  \bibinfo{author}{{Bovy}, J.}, \bibinfo{year}{2020}.
\newblock \bibinfo{title}{{The effects of dwarf galaxies on the orbital
  evolution of galactic globular clusters}}.
\newblock \bibinfo{journal}{\mnras} \bibinfo{volume}{499},
  \bibinfo{pages}{804--813}.
\newblock \DOIprefix\doi{10.1093/mnras/staa2773},
  \href{http://arxiv.org/abs/2007.13752}{{\tt arXiv:2007.13752}}.
%Type = Article
\bibitem[{{Gatto} et~al.(2020){Gatto}, {Napolitano}, {Spiniello}, {Longo} and
  {Paolillo}}]{gatto:2020}
\bibinfo{author}{{Gatto}, M.}, \bibinfo{author}{{Napolitano}, N.R.},
  \bibinfo{author}{{Spiniello}, C.}, \bibinfo{author}{{Longo}, G.},
  \bibinfo{author}{{Paolillo}, M.}, \bibinfo{year}{2020}.
\newblock \bibinfo{title}{{The COld STream finder Algorithm (COSTA). Searching
  for kinematical substructures in the phase space of discrete tracers}}.
\newblock \bibinfo{journal}{\aap} \bibinfo{volume}{644}, \bibinfo{pages}{A134}.
\newblock \DOIprefix\doi{10.1051/0004-6361/202039029},
  \href{http://arxiv.org/abs/2010.02266}{{\tt arXiv:2010.02266}}.
%Type = Article
\bibitem[{{Gialluca} et~al.(2021){Gialluca}, {Naidu} and
  {Bonaca}}]{gialluca:2021}
\bibinfo{author}{{Gialluca}, M.T.}, \bibinfo{author}{{Naidu}, R.P.},
  \bibinfo{author}{{Bonaca}, A.}, \bibinfo{year}{2021}.
\newblock \bibinfo{title}{{Velocity Dispersion of the GD-1 Stellar Stream}}.
\newblock \bibinfo{journal}{\apjl} \bibinfo{volume}{911}, \bibinfo{pages}{L32}.
\newblock \DOIprefix\doi{10.3847/2041-8213/abf491},
  \href{http://arxiv.org/abs/2011.12963}{{\tt arXiv:2011.12963}}.
%Type = Article
\bibitem[{{Gibbons} et~al.(2014){Gibbons}, {Belokurov} and
  {Evans}}]{gibbons:2014}
\bibinfo{author}{{Gibbons}, S.L.J.}, \bibinfo{author}{{Belokurov}, V.},
  \bibinfo{author}{{Evans}, N.W.}, \bibinfo{year}{2014}.
\newblock \bibinfo{title}{{`Skinny Milky Way please', says Sagittarius}}.
\newblock \bibinfo{journal}{\mnras} \bibinfo{volume}{445},
  \bibinfo{pages}{3788--3802}.
\newblock \DOIprefix\doi{10.1093/mnras/stu1986},
  \href{http://arxiv.org/abs/1406.2243}{{\tt arXiv:1406.2243}}.
%Type = Article
\bibitem[{{Gnedin} et~al.(1999){Gnedin}, {Hernquist} and
  {Ostriker}}]{gnedin:1999}
\bibinfo{author}{{Gnedin}, O.Y.}, \bibinfo{author}{{Hernquist}, L.},
  \bibinfo{author}{{Ostriker}, J.P.}, \bibinfo{year}{1999}.
\newblock \bibinfo{title}{{Tidal Shocking by Extended Mass Distributions}}.
\newblock \bibinfo{journal}{\apj} \bibinfo{volume}{514},
  \bibinfo{pages}{109--118}.
\newblock \DOIprefix\doi{10.1086/306910}.
%Type = Article
\bibitem[{{Gnedin} and {Ostriker}(1997)}]{gnedin:1997}
\bibinfo{author}{{Gnedin}, O.Y.}, \bibinfo{author}{{Ostriker}, J.P.},
  \bibinfo{year}{1997}.
\newblock \bibinfo{title}{{Destruction of the Galactic Globular Cluster
  System}}.
\newblock \bibinfo{journal}{\apj} \bibinfo{volume}{474},
  \bibinfo{pages}{223--255}.
\newblock \DOIprefix\doi{10.1086/303441},
  \href{http://arxiv.org/abs/astro-ph/9603042}{{\tt arXiv:astro-ph/9603042}}.
%Type = Book
\bibitem[{{Goldstein} et~al.(2002){Goldstein}, {Poole} and
  {Safko}}]{goldstein:2002}
\bibinfo{author}{{Goldstein}, H.}, \bibinfo{author}{{Poole}, C.},
  \bibinfo{author}{{Safko}, J.}, \bibinfo{year}{2002}.
\newblock \bibinfo{title}{{Classical mechanics}}.
%Type = Article
\bibitem[{{G{\'o}mez} et~al.(2015){G{\'o}mez}, {Besla}, {Carpintero},
  {Villalobos}, {O'Shea} and {Bell}}]{gomez:2015}
\bibinfo{author}{{G{\'o}mez}, F.A.}, \bibinfo{author}{{Besla}, G.},
  \bibinfo{author}{{Carpintero}, D.D.}, \bibinfo{author}{{Villalobos}, {\'A}.},
  \bibinfo{author}{{O'Shea}, B.W.}, \bibinfo{author}{{Bell}, E.F.},
  \bibinfo{year}{2015}.
\newblock \bibinfo{title}{{And Yet it Moves: The Dangers of Artificially Fixing
  the Milky Way Center of Mass in the Presence of a Massive Large Magellanic
  Cloud}}.
\newblock \bibinfo{journal}{\apj} \bibinfo{volume}{802}, \bibinfo{pages}{128}.
\newblock \DOIprefix\doi{10.1088/0004-637X/802/2/128},
  \href{http://arxiv.org/abs/1408.4128}{{\tt arXiv:1408.4128}}.
%Type = Article
\bibitem[{{G{\'o}mez} and {Helmi}(2010)}]{gomez:2010}
\bibinfo{author}{{G{\'o}mez}, F.A.}, \bibinfo{author}{{Helmi}, A.},
  \bibinfo{year}{2010}.
\newblock \bibinfo{title}{{On the identification of substructure in phase space
  using orbital frequencies}}.
\newblock \bibinfo{journal}{\mnras} \bibinfo{volume}{401},
  \bibinfo{pages}{2285--2298}.
\newblock \DOIprefix\doi{10.1111/j.1365-2966.2009.15841.x},
  \href{http://arxiv.org/abs/0904.1377}{{\tt arXiv:0904.1377}}.
%Type = Article
\bibitem[{{Gratton} et~al.(2019){Gratton}, {Bragaglia}, {Carretta}, {D'Orazi},
  {Lucatello} and {Sollima}}]{gratton:2019}
\bibinfo{author}{{Gratton}, R.}, \bibinfo{author}{{Bragaglia}, A.},
  \bibinfo{author}{{Carretta}, E.}, \bibinfo{author}{{D'Orazi}, V.},
  \bibinfo{author}{{Lucatello}, S.}, \bibinfo{author}{{Sollima}, A.},
  \bibinfo{year}{2019}.
\newblock \bibinfo{title}{{What is a globular cluster? An observational
  perspective}}.
\newblock \bibinfo{journal}{\aapr} \bibinfo{volume}{27}, \bibinfo{pages}{8}.
\newblock \DOIprefix\doi{10.1007/s00159-019-0119-3},
  \href{http://arxiv.org/abs/1911.02835}{{\tt arXiv:1911.02835}}.
%Type = Article
\bibitem[{{Gratton} et~al.(2012){Gratton}, {Carretta} and
  {Bragaglia}}]{gratton:2012}
\bibinfo{author}{{Gratton}, R.G.}, \bibinfo{author}{{Carretta}, E.},
  \bibinfo{author}{{Bragaglia}, A.}, \bibinfo{year}{2012}.
\newblock \bibinfo{title}{{Multiple populations in globular clusters. Lessons
  learned from the Milky Way globular clusters}}.
\newblock \bibinfo{journal}{\aapr} \bibinfo{volume}{20}, \bibinfo{pages}{50}.
\newblock \DOIprefix\doi{10.1007/s00159-012-0050-3},
  \href{http://arxiv.org/abs/1201.6526}{{\tt arXiv:1201.6526}}.
%Type = Article
\bibitem[{{Grillmair}(2006)}]{grillmair:2006-orphan}
\bibinfo{author}{{Grillmair}, C.J.}, \bibinfo{year}{2006}.
\newblock \bibinfo{title}{{Detection of a 60{\textdegree}-long Dwarf Galaxy
  Debris Stream}}.
\newblock \bibinfo{journal}{\apjl} \bibinfo{volume}{645},
  \bibinfo{pages}{L37--L40}.
\newblock \DOIprefix\doi{10.1086/505863},
  \href{http://arxiv.org/abs/astro-ph/0605396}{{\tt arXiv:astro-ph/0605396}}.
%Type = Article
\bibitem[{{Grillmair}(2009)}]{grillmair:2009}
\bibinfo{author}{{Grillmair}, C.J.}, \bibinfo{year}{2009}.
\newblock \bibinfo{title}{{Four New Stellar Debris Streams in the Galactic
  Halo}}.
\newblock \bibinfo{journal}{\apj} \bibinfo{volume}{693},
  \bibinfo{pages}{1118--1127}.
\newblock \DOIprefix\doi{10.1088/0004-637X/693/2/1118},
  \href{http://arxiv.org/abs/0811.3965}{{\tt arXiv:0811.3965}}.
%Type = Article
\bibitem[{{Grillmair}(2014)}]{grillmair:2014}
\bibinfo{author}{{Grillmair}, C.J.}, \bibinfo{year}{2014}.
\newblock \bibinfo{title}{{Two New Halo Debris Streams in the Sloan Digital Sky
  Survey}}.
\newblock \bibinfo{journal}{\apjl} \bibinfo{volume}{790}, \bibinfo{pages}{L10}.
\newblock \DOIprefix\doi{10.1088/2041-8205/790/1/L10},
  \href{http://arxiv.org/abs/1407.0397}{{\tt arXiv:1407.0397}}.
%Type = Article
\bibitem[{{Grillmair}(2017a)}]{grillmair:2017b}
\bibinfo{author}{{Grillmair}, C.J.}, \bibinfo{year}{2017}a.
\newblock \bibinfo{title}{{At a Crossroads: Stellar Streams in the South
  Galactic Cap}}.
\newblock \bibinfo{journal}{\apj} \bibinfo{volume}{847}, \bibinfo{pages}{119}.
\newblock \DOIprefix\doi{10.3847/1538-4357/aa8872},
  \href{http://arxiv.org/abs/1708.09029}{{\tt arXiv:1708.09029}}.
%Type = Article
\bibitem[{{Grillmair}(2017b)}]{grillmair:2017}
\bibinfo{author}{{Grillmair}, C.J.}, \bibinfo{year}{2017}b.
\newblock \bibinfo{title}{{Tails from the Orphanage}}.
\newblock \bibinfo{journal}{\apj} \bibinfo{volume}{834}, \bibinfo{pages}{98}.
\newblock \DOIprefix\doi{10.3847/1538-4357/834/2/98},
  \href{http://arxiv.org/abs/1612.03181}{{\tt arXiv:1612.03181}}.
%Type = Article
\bibitem[{{Grillmair}(2019)}]{grillmair:2019}
\bibinfo{author}{{Grillmair}, C.J.}, \bibinfo{year}{2019}.
\newblock \bibinfo{title}{{Detection of a 50{\textdegree} long Trailing Tidal
  Tail for the Globular Cluster M5}}.
\newblock \bibinfo{journal}{\apj} \bibinfo{volume}{884}, \bibinfo{pages}{174}.
\newblock \DOIprefix\doi{10.3847/1538-4357/ab441d},
  \href{http://arxiv.org/abs/1909.05927}{{\tt arXiv:1909.05927}}.
%Type = Article
\bibitem[{{Grillmair}(2022)}]{grillmair:2022}
\bibinfo{author}{{Grillmair}, C.J.}, \bibinfo{year}{2022}.
\newblock \bibinfo{title}{{The Extended Tidal Tails of NGC 7089 (M2)}}.
\newblock \bibinfo{journal}{\apj} \bibinfo{volume}{929}, \bibinfo{pages}{89}.
\newblock \DOIprefix\doi{10.3847/1538-4357/ac5bd7},
  \href{http://arxiv.org/abs/2203.04425}{{\tt arXiv:2203.04425}}.
%Type = Article
\bibitem[{{Grillmair} and {Carlberg}(2016)}]{grillmair:2016b}
\bibinfo{author}{{Grillmair}, C.J.}, \bibinfo{author}{{Carlberg}, R.G.},
  \bibinfo{year}{2016}.
\newblock \bibinfo{title}{{What a Tangled Web We Weave: Hermus as the Northern
  Extension of the Phoenix Stream}}.
\newblock \bibinfo{journal}{\apjl} \bibinfo{volume}{820}, \bibinfo{pages}{L27}.
\newblock \DOIprefix\doi{10.3847/2041-8205/820/2/L27},
  \href{http://arxiv.org/abs/1603.02278}{{\tt arXiv:1603.02278}}.
%Type = Article
\bibitem[{{Grillmair} et~al.(2013){Grillmair}, {Cutri}, {Masci}, {Conrow},
  {Sesar}, {Eisenhardt} and {Wright}}]{grillmair:2013}
\bibinfo{author}{{Grillmair}, C.J.}, \bibinfo{author}{{Cutri}, R.},
  \bibinfo{author}{{Masci}, F.J.}, \bibinfo{author}{{Conrow}, T.},
  \bibinfo{author}{{Sesar}, B.}, \bibinfo{author}{{Eisenhardt}, P.R.M.},
  \bibinfo{author}{{Wright}, E.L.}, \bibinfo{year}{2013}.
\newblock \bibinfo{title}{{Detection of a Nearby Halo Debris Stream in the WISE
  and 2MASS Surveys}}.
\newblock \bibinfo{journal}{\apjl} \bibinfo{volume}{769}, \bibinfo{pages}{L23}.
\newblock \DOIprefix\doi{10.1088/2041-8205/769/2/L23},
  \href{http://arxiv.org/abs/1304.1170}{{\tt arXiv:1304.1170}}.
%Type = Article
\bibitem[{{Grillmair} and {Dionatos}(2006)}]{grillmair:2006-pal5}
\bibinfo{author}{{Grillmair}, C.J.}, \bibinfo{author}{{Dionatos}, O.},
  \bibinfo{year}{2006}.
\newblock \bibinfo{title}{{A 22{\textdegree} Tidal Tail for Palomar 5}}.
\newblock \bibinfo{journal}{\apjl} \bibinfo{volume}{641},
  \bibinfo{pages}{L37--L39}.
\newblock \DOIprefix\doi{10.1086/503744},
  \href{http://arxiv.org/abs/astro-ph/0603062}{{\tt arXiv:astro-ph/0603062}}.
%Type = Article
\bibitem[{Grillmair and Dionatos(2006)}]{grillmair:2006-gd1}
\bibinfo{author}{Grillmair, C.J.}, \bibinfo{author}{Dionatos, O.},
  \bibinfo{year}{2006}.
\newblock \bibinfo{title}{Detection of a 63\textdegree{} {{Cold Stellar
  Stream}} in the {{Sloan Digital Sky Survey}}}.
\newblock \bibinfo{journal}{The Astrophysical Journal} \bibinfo{volume}{643},
  \bibinfo{pages}{L17--L20}.
\newblock \DOIprefix\doi{10.1086/505111}.
%Type = Article
\bibitem[{{Grillmair} et~al.(1995){Grillmair}, {Freeman}, {Irwin} and
  {Quinn}}]{grillmair:1995}
\bibinfo{author}{{Grillmair}, C.J.}, \bibinfo{author}{{Freeman}, K.C.},
  \bibinfo{author}{{Irwin}, M.}, \bibinfo{author}{{Quinn}, P.J.},
  \bibinfo{year}{1995}.
\newblock \bibinfo{title}{{Globular Clusters with Tidal Tails: Deep Two-Color
  Star Counts}}.
\newblock \bibinfo{journal}{\aj} \bibinfo{volume}{109}, \bibinfo{pages}{2553}.
\newblock \DOIprefix\doi{10.1086/117470},
  \href{http://arxiv.org/abs/astro-ph/9502039}{{\tt arXiv:astro-ph/9502039}}.
%Type = Article
\bibitem[{{Grillmair} et~al.(2015){Grillmair}, {Hetherington}, {Carlberg} and
  {Willman}}]{grillmair:2015}
\bibinfo{author}{{Grillmair}, C.J.}, \bibinfo{author}{{Hetherington}, L.},
  \bibinfo{author}{{Carlberg}, R.G.}, \bibinfo{author}{{Willman}, B.},
  \bibinfo{year}{2015}.
\newblock \bibinfo{title}{{An Orphan No Longer? Detection of the Southern
  Orphan Stream and a Candidate Progenitor}}.
\newblock \bibinfo{journal}{\apjl} \bibinfo{volume}{812}, \bibinfo{pages}{L26}.
\newblock \DOIprefix\doi{10.1088/2041-8205/812/2/L26},
  \href{http://arxiv.org/abs/1509.07503}{{\tt arXiv:1509.07503}}.
%Type = Article
\bibitem[{Grillmair and Johnson(2006)}]{grillmair:2006-ngc5466}
\bibinfo{author}{Grillmair, C.J.}, \bibinfo{author}{Johnson, R.},
  \bibinfo{year}{2006}.
\newblock \bibinfo{title}{The {{Detection}} of a 45\textdegree{} {{Tidal Stream
  Associated}} with the {{Globular Cluster NGC}} 5466}.
\newblock \bibinfo{journal}{The Astrophysical Journal} \bibinfo{volume}{639},
  \bibinfo{pages}{L17--L20}.
\newblock \DOIprefix\doi{10.1086/501439}.
%Type = Article
\bibitem[{{Gunn} et~al.(1998){Gunn}, {Carr}, {Rockosi}, {Sekiguchi}, {Berry},
  {Elms}, {de Haas}, {Ivezi{\'c}}, {Knapp}, {Lupton}, {Pauls}, {Simcoe},
  {Hirsch}, {Sanford}, {Wang}, {York}, {Harris}, {Annis}, {Bartozek},
  {Boroski}, {Bakken}, {Haldeman}, {Kent}, {Holm}, {Holmgren}, {Petravick},
  {Prosapio}, {Rechenmacher}, {Doi}, {Fukugita}, {Shimasaku}, {Okada}, {Hull},
  {Siegmund}, {Mannery}, {Blouke}, {Heidtman}, {Schneider}, {Lucinio} and
  {Brinkman}}]{gunn:1998}
{Gunn}, J.~E. et~al., \bibinfo{year}{1998}.
\newblock \bibinfo{title}{{The Sloan Digital Sky Survey Photometric Camera}}.
\newblock \bibinfo{journal}{\aj} \bibinfo{volume}{116},
  \bibinfo{pages}{3040--3081}.
\newblock \DOIprefix\doi{10.1086/300645},
  \href{http://arxiv.org/abs/astro-ph/9809085}{{\tt arXiv:astro-ph/9809085}}.
%Type = Article
\bibitem[{{Hansen} et~al.(2021){Hansen}, {Ji}, {Da Costa}, {Li}, {Casey},
  {Pace}, {Cullinane}, {Erkal}, {Koposov}, {Kuehn}, {Lewis}, {Mackey},
  {Simpson}, {Shipp}, {Zucker}, {Bland-Hawthorn} and {S5
  Collaboration}}]{hansen:2021}
{Hansen}, T.~T. et~al., \bibinfo{year}{2021}.
\newblock \bibinfo{title}{{S$^{5}$: The Destruction of a Bright Dwarf Galaxy as
  Revealed by the Chemistry of the Indus Stellar Stream}}.
\newblock \bibinfo{journal}{\apj} \bibinfo{volume}{915}, \bibinfo{pages}{103}.
\newblock \DOIprefix\doi{10.3847/1538-4357/abfc54},
  \href{http://arxiv.org/abs/2104.13883}{{\tt arXiv:2104.13883}}.
%Type = Article
\bibitem[{{Hansen} et~al.(2020){Hansen}, {Riley}, {Strigari}, {Marshall},
  {Ferguson}, {Zepeda} and {Sneden}}]{hansen:2020}
\bibinfo{author}{{Hansen}, T.T.}, \bibinfo{author}{{Riley}, A.H.},
  \bibinfo{author}{{Strigari}, L.E.}, \bibinfo{author}{{Marshall}, J.L.},
  \bibinfo{author}{{Ferguson}, P.S.}, \bibinfo{author}{{Zepeda}, J.},
  \bibinfo{author}{{Sneden}, C.}, \bibinfo{year}{2020}.
\newblock \bibinfo{title}{{A Chemo-dynamical Link between the Gj{\"o}ll Stream
  and NGC 3201}}.
\newblock \bibinfo{journal}{\apj} \bibinfo{volume}{901}, \bibinfo{pages}{23}.
\newblock \DOIprefix\doi{10.3847/1538-4357/ababa5},
  \href{http://arxiv.org/abs/2007.12165}{{\tt arXiv:2007.12165}}.
%Type = Article
\bibitem[{{Harris}(1996)}]{harris:1996}
\bibinfo{author}{{Harris}, W.E.}, \bibinfo{year}{1996}.
\newblock \bibinfo{title}{{A Catalog of Parameters for Globular Clusters in the
  Milky Way}}.
\newblock \bibinfo{journal}{\aj} \bibinfo{volume}{112}, \bibinfo{pages}{1487}.
\newblock \DOIprefix\doi{10.1086/118116}.
%Type = Article
\bibitem[{{Harris}(2010)}]{harris:2010}
\bibinfo{author}{{Harris}, W.E.}, \bibinfo{year}{2010}.
\newblock \bibinfo{title}{{A New Catalog of Globular Clusters in the Milky
  Way}}.
\newblock \bibinfo{journal}{arXiv e-prints} ,
  \bibinfo{pages}{arXiv:1012.3224}\DOIprefix\doi{10.48550/arXiv.1012.3224},
  \href{http://arxiv.org/abs/1012.3224}{{\tt arXiv:1012.3224}}.
%Type = Article
\bibitem[{{Hattori} et~al.(2016){Hattori}, {Erkal} and
  {Sanders}}]{hattori:2016}
\bibinfo{author}{{Hattori}, K.}, \bibinfo{author}{{Erkal}, D.},
  \bibinfo{author}{{Sanders}, J.L.}, \bibinfo{year}{2016}.
\newblock \bibinfo{title}{{Shepherding tidal debris with the Galactic bar: the
  Ophiuchus stream}}.
\newblock \bibinfo{journal}{\mnras} \bibinfo{volume}{460},
  \bibinfo{pages}{497--512}.
\newblock \DOIprefix\doi{10.1093/mnras/stw1006},
  \href{http://arxiv.org/abs/1512.04536}{{\tt arXiv:1512.04536}}.
%Type = Article
\bibitem[{{Hawkins} et~al.(2023){Hawkins}, {Price-Whelan}, {Sheffield},
  {Subrahimovic}, {Beaton}, {Belokurov}, {Erkal}, {Koposov}, {Lane}, {Laporte}
  and {Nitschelm}}]{hawkins:2023}
{Hawkins}, K. et~al., \bibinfo{year}{2023}.
\newblock \bibinfo{title}{{On the Hunt for the Origins of the Orphan-Chenab
  Stream: Detailed Element Abundances with APOGEE and Gaia}}.
\newblock \bibinfo{journal}{\apj} \bibinfo{volume}{948}, \bibinfo{pages}{123}.
\newblock \DOIprefix\doi{10.3847/1538-4357/acb698},
  \href{http://arxiv.org/abs/2205.14218}{{\tt arXiv:2205.14218}}.
%Type = Article
\bibitem[{{Hayes} et~al.(2020){Hayes}, {Majewski}, {Hasselquist}, {Anguiano},
  {Shetrone}, {Law}, {Schiavon}, {Cunha}, {Smith}, {Beaton}, {Price-Whelan},
  {Allende Prieto}, {Battaglia}, {Bizyaev}, {Brownstein}, {Cohen},
  {Frinchaboy}, {Garc{\'\i}a-Hern{\'a}ndez}, {Lacerna}, {Lane},
  {M{\'e}sz{\'a}ros}, {Bidin}, {M{\~{u}}noz}, {Nidever}, {Oravetz}, {Oravetz},
  {Pan}, {Roman-Lopes}, {Sobeck} and {Stringfellow}}]{hayes:2020}
{Hayes}, C.~R. et~al., \bibinfo{year}{2020}.
\newblock \bibinfo{title}{{Metallicity and {\ensuremath{\alpha}}-Element
  Abundance Gradients along the Sagittarius Stream as Seen by APOGEE}}.
\newblock \bibinfo{journal}{\apj} \bibinfo{volume}{889}, \bibinfo{pages}{63}.
\newblock \DOIprefix\doi{10.3847/1538-4357/ab62ad},
  \href{http://arxiv.org/abs/1912.06707}{{\tt arXiv:1912.06707}}.
%Type = Article
\bibitem[{{Helmi}(2004)}]{helmi:2004}
\bibinfo{author}{{Helmi}, A.}, \bibinfo{year}{2004}.
\newblock \bibinfo{title}{{Velocity Trends in the Debris of Sagittarius and the
  Shape of the Dark Matter Halo of Our Galaxy}}.
\newblock \bibinfo{journal}{\apjl} \bibinfo{volume}{610},
  \bibinfo{pages}{L97--L100}.
\newblock \DOIprefix\doi{10.1086/423340},
  \href{http://arxiv.org/abs/astro-ph/0406396}{{\tt arXiv:astro-ph/0406396}}.
%Type = Article
\bibitem[{{Helmi}(2020)}]{helmi:2020}
\bibinfo{author}{{Helmi}, A.}, \bibinfo{year}{2020}.
\newblock \bibinfo{title}{{Streams, Substructures, and the Early History of the
  Milky Way}}.
\newblock \bibinfo{journal}{\araa} \bibinfo{volume}{58},
  \bibinfo{pages}{205--256}.
\newblock \DOIprefix\doi{10.1146/annurev-astro-032620-021917},
  \href{http://arxiv.org/abs/2002.04340}{{\tt arXiv:2002.04340}}.
%Type = Article
\bibitem[{{Helmi} et~al.(2018){Helmi}, {Babusiaux}, {Koppelman}, {Massari},
  {Veljanoski} and {Brown}}]{helmi:2018b}
\bibinfo{author}{{Helmi}, A.}, \bibinfo{author}{{Babusiaux}, C.},
  \bibinfo{author}{{Koppelman}, H.H.}, \bibinfo{author}{{Massari}, D.},
  \bibinfo{author}{{Veljanoski}, J.}, \bibinfo{author}{{Brown}, A.G.A.},
  \bibinfo{year}{2018}.
\newblock \bibinfo{title}{{The merger that led to the formation of the Milky
  Way's inner stellar halo and thick disk}}.
\newblock \bibinfo{journal}{\nat} \bibinfo{volume}{563},
  \bibinfo{pages}{85--88}.
\newblock \DOIprefix\doi{10.1038/s41586-018-0625-x},
  \href{http://arxiv.org/abs/1806.06038}{{\tt arXiv:1806.06038}}.
%Type = Article
\bibitem[{{Helmi} and {Koppelman}(2016)}]{helmi:2016}
\bibinfo{author}{{Helmi}, A.}, \bibinfo{author}{{Koppelman}, H.H.},
  \bibinfo{year}{2016}.
\newblock \bibinfo{title}{{The Time Evolution of Gaps in Tidal Streams}}.
\newblock \bibinfo{journal}{\apjl} \bibinfo{volume}{828}, \bibinfo{pages}{L10}.
\newblock \DOIprefix\doi{10.3847/2041-8205/828/1/L10},
  \href{http://arxiv.org/abs/1606.08782}{{\tt arXiv:1606.08782}}.
%Type = Article
\bibitem[{{Helmi} and {White}(1999)}]{helmi:1999}
\bibinfo{author}{{Helmi}, A.}, \bibinfo{author}{{White}, S.D.M.},
  \bibinfo{year}{1999}.
\newblock \bibinfo{title}{{Building up the stellar halo of the Galaxy}}.
\newblock \bibinfo{journal}{\mnras} \bibinfo{volume}{307},
  \bibinfo{pages}{495--517}.
\newblock \DOIprefix\doi{10.1046/j.1365-8711.1999.02616.x},
  \href{http://arxiv.org/abs/astro-ph/9901102}{{\tt arXiv:astro-ph/9901102}}.
%Type = Article
\bibitem[{{Hendel} and {Johnston}(2015)}]{hendel:2015}
\bibinfo{author}{{Hendel}, D.}, \bibinfo{author}{{Johnston}, K.V.},
  \bibinfo{year}{2015}.
\newblock \bibinfo{title}{{Tidal debris morphology and the orbits of satellite
  galaxies}}.
\newblock \bibinfo{journal}{\mnras} \bibinfo{volume}{454},
  \bibinfo{pages}{2472--2485}.
\newblock \DOIprefix\doi{10.1093/mnras/stv2035},
  \href{http://arxiv.org/abs/1509.06369}{{\tt arXiv:1509.06369}}.
%Type = Article
\bibitem[{{Hendel} et~al.(2018){Hendel}, {Scowcroft}, {Johnston}, {Fardal},
  {van der Marel}, {Sohn}, {Price-Whelan}, {Beaton}, {Besla}, {Bono}, {Cioni},
  {Clementini}, {Cohen}, {Fabrizio}, {Freedman}, {Garofalo}, {Grillmair},
  {Kallivayalil}, {Kollmeier}, {Law}, {Madore}, {Majewski}, {Marengo},
  {Monson}, {Neeley}, {Nidever}, {Pietrzy{\'n}ski}, {Seibert}, {Sesar},
  {Smith}, {Soszy{\'n}ski} and {Udalski}}]{hendel:2018}
{Hendel}, D. et~al., \bibinfo{year}{2018}.
\newblock \bibinfo{title}{{SMHASH: anatomy of the Orphan Stream using RR Lyrae
  stars}}.
\newblock \bibinfo{journal}{\mnras} \bibinfo{volume}{479},
  \bibinfo{pages}{570--587}.
\newblock \DOIprefix\doi{10.1093/mnras/sty1455},
  \href{http://arxiv.org/abs/1711.04663}{{\tt arXiv:1711.04663}}.
%Type = Article
\bibitem[{{Hernquist} and {Quinn}(1987)}]{hernquist:1987}
\bibinfo{author}{{Hernquist}, L.}, \bibinfo{author}{{Quinn}, P.J.},
  \bibinfo{year}{1987}.
\newblock \bibinfo{title}{{Shells and Dark Matter in Elliptical Galaxies}}.
\newblock \bibinfo{journal}{\apj} \bibinfo{volume}{312}, \bibinfo{pages}{1}.
\newblock \DOIprefix\doi{10.1086/164844}.
%Type = Article
\bibitem[{{Hilmi} et~al.(2024){Hilmi}, {Erkal}, {Koposov}, {Li}, {Lilleengen},
  {Ji}, {Lewis}, {Shipp}, {Pace}, {Zucker}, {Limberg} and {Usman}}]{hilmi:2024}
{Hilmi}, T. et~al., \bibinfo{year}{2024}.
\newblock \bibinfo{title}{{Inferring dark matter subhalo properties from
  simulated subhalo-stream encounters}}.
\newblock \bibinfo{journal}{arXiv e-prints} ,
  \bibinfo{pages}{arXiv:2404.02953}\DOIprefix\doi{10.48550/arXiv.2404.02953},
  \href{http://arxiv.org/abs/2404.02953}{{\tt arXiv:2404.02953}}.
%Type = Article
\bibitem[{{Hopkins} et~al.(2018){Hopkins}, {Wetzel}, {Kere{\v{s}}},
  {Faucher-Gigu{\`e}re}, {Quataert}, {Boylan-Kolchin}, {Murray}, {Hayward},
  {Garrison-Kimmel}, {Hummels}, {Feldmann}, {Torrey}, {Ma},
  {Angl{\'e}s-Alc{\'a}zar}, {Su}, {Orr}, {Schmitz}, {Escala}, {Sanderson},
  {Grudi{\'c}}, {Hafen}, {Kim}, {Fitts}, {Bullock}, {Wheeler}, {Chan}, {Elbert}
  and {Narayanan}}]{hopkins:2018}
{Hopkins}, P.~F. et~al., \bibinfo{year}{2018}.
\newblock \bibinfo{title}{{FIRE-2 simulations: physics versus numerics in
  galaxy formation}}.
\newblock \bibinfo{journal}{\mnras} \bibinfo{volume}{480},
  \bibinfo{pages}{800--863}.
\newblock \DOIprefix\doi{10.1093/mnras/sty1690},
  \href{http://arxiv.org/abs/1702.06148}{{\tt arXiv:1702.06148}}.
%Type = Article
\bibitem[{{Hu} et~al.(2000){Hu}, {Barkana} and {Gruzinov}}]{hu:2000}
\bibinfo{author}{{Hu}, W.}, \bibinfo{author}{{Barkana}, R.},
  \bibinfo{author}{{Gruzinov}, A.}, \bibinfo{year}{2000}.
\newblock \bibinfo{title}{{Fuzzy Cold Dark Matter: The Wave Properties of
  Ultralight Particles}}.
\newblock \bibinfo{journal}{\prl} \bibinfo{volume}{85},
  \bibinfo{pages}{1158--1161}.
\newblock \DOIprefix\doi{10.1103/PhysRevLett.85.1158},
  \href{http://arxiv.org/abs/astro-ph/0003365}{{\tt arXiv:astro-ph/0003365}}.
%Type = Article
\bibitem[{{Huang} et~al.(2019){Huang}, {Chen}, {Zhang}, {Yuan}, {Xiang},
  {Wang}, {Tian} and {Liu}}]{huang:2019}
\bibinfo{author}{{Huang}, Y.}, \bibinfo{author}{{Chen}, B.Q.},
  \bibinfo{author}{{Zhang}, H.W.}, \bibinfo{author}{{Yuan}, H.B.},
  \bibinfo{author}{{Xiang}, M.S.}, \bibinfo{author}{{Wang}, C.},
  \bibinfo{author}{{Tian}, Z.J.}, \bibinfo{author}{{Liu}, X.W.},
  \bibinfo{year}{2019}.
\newblock \bibinfo{title}{{Member Stars of the GD-1 Tidal Stream from the SDSS,
  LAMOST, and Gaia Surveys}}.
\newblock \bibinfo{journal}{\apj} \bibinfo{volume}{877}, \bibinfo{pages}{13}.
\newblock \DOIprefix\doi{10.3847/1538-4357/ab158a},
  \href{http://arxiv.org/abs/1806.03748}{{\tt arXiv:1806.03748}}.
%Type = Article
\bibitem[{{Ibata} et~al.(2001a){Ibata}, {Irwin}, {Lewis} and
  {Stolte}}]{ibata:2001a}
\bibinfo{author}{{Ibata}, R.}, \bibinfo{author}{{Irwin}, M.},
  \bibinfo{author}{{Lewis}, G.F.}, \bibinfo{author}{{Stolte}, A.},
  \bibinfo{year}{2001}a.
\newblock \bibinfo{title}{{Galactic Halo Substructure in the Sloan Digital Sky
  Survey: The Ancient Tidal Stream from the Sagittarius Dwarf Galaxy}}.
\newblock \bibinfo{journal}{\apjl} \bibinfo{volume}{547},
  \bibinfo{pages}{L133--L136}.
\newblock \DOIprefix\doi{10.1086/318894},
  \href{http://arxiv.org/abs/astro-ph/0004255}{{\tt arXiv:astro-ph/0004255}}.
%Type = Article
\bibitem[{{Ibata} et~al.(2001b){Ibata}, {Lewis}, {Irwin}, {Totten} and
  {Quinn}}]{ibata:2001b}
\bibinfo{author}{{Ibata}, R.}, \bibinfo{author}{{Lewis}, G.F.},
  \bibinfo{author}{{Irwin}, M.}, \bibinfo{author}{{Totten}, E.},
  \bibinfo{author}{{Quinn}, T.}, \bibinfo{year}{2001}b.
\newblock \bibinfo{title}{{Great Circle Tidal Streams: Evidence for a Nearly
  Spherical Massive Dark Halo around the Milky Way}}.
\newblock \bibinfo{journal}{\apj} \bibinfo{volume}{551},
  \bibinfo{pages}{294--311}.
\newblock \DOIprefix\doi{10.1086/320060},
  \href{http://arxiv.org/abs/astro-ph/0004011}{{\tt arXiv:astro-ph/0004011}}.
%Type = Article
\bibitem[{{Ibata} et~al.(2021){Ibata}, {Malhan}, {Martin}, {Aubert}, {Famaey},
  {Bianchini}, {Monari}, {Siebert}, {Thomas}, {Bellazzini}, {Bonifacio},
  {Caffau} and {Renaud}}]{ibata:2021}
{Ibata}, R. et~al., \bibinfo{year}{2021}.
\newblock \bibinfo{title}{{Charting the Galactic Acceleration Field. I. A
  Search for Stellar Streams with Gaia DR2 and EDR3 with Follow-up from
  ESPaDOnS and UVES}}.
\newblock \bibinfo{journal}{\apj} \bibinfo{volume}{914}, \bibinfo{pages}{123}.
\newblock \DOIprefix\doi{10.3847/1538-4357/abfcc2},
  \href{http://arxiv.org/abs/2012.05245}{{\tt arXiv:2012.05245}}.
%Type = Article
\bibitem[{{Ibata} et~al.(2023){Ibata}, {Malhan}, {Tenachi}, {Ardern-Arentsen},
  {Bellazzini}, {Bianchini}, {Bonifacio}, {Caffau}, {Diakogiannis}, {Errani},
  {Famaey}, {Ferrone}, {Martin}, {di Matteo}, {Monari}, {Renaud},
  {Starkenburg}, {Thomas}, {Viswanathan} and {Yuan}}]{ibata:2023}
{Ibata}, R. et~al., \bibinfo{year}{2023}.
\newblock \bibinfo{title}{{Charting the Galactic acceleration field II. A
  global mass model of the Milky Way from the STREAMFINDER Atlas of Stellar
  Streams detected in Gaia DR3}}.
\newblock \bibinfo{journal}{arXiv e-prints} ,
  \bibinfo{pages}{arXiv:2311.17202}\DOIprefix\doi{10.48550/arXiv.2311.17202},
  \href{http://arxiv.org/abs/2311.17202}{{\tt arXiv:2311.17202}}.
%Type = Article
\bibitem[{{Ibata} et~al.(2020){Ibata}, {Thomas}, {Famaey}, {Malhan}, {Martin}
  and {Monari}}]{ibata:2020}
\bibinfo{author}{{Ibata}, R.}, \bibinfo{author}{{Thomas}, G.},
  \bibinfo{author}{{Famaey}, B.}, \bibinfo{author}{{Malhan}, K.},
  \bibinfo{author}{{Martin}, N.}, \bibinfo{author}{{Monari}, G.},
  \bibinfo{year}{2020}.
\newblock \bibinfo{title}{{Detection of Strong Epicyclic Density Spikes in the
  GD-1 Stellar Stream: An Absence of Evidence for the Influence of Dark Matter
  Subhalos?}}
\newblock \bibinfo{journal}{\apj} \bibinfo{volume}{891}, \bibinfo{pages}{161}.
\newblock \DOIprefix\doi{10.3847/1538-4357/ab7303},
  \href{http://arxiv.org/abs/2002.01488}{{\tt arXiv:2002.01488}}.
%Type = Article
\bibitem[{{Ibata} et~al.(2019a){Ibata}, {Bellazzini}, {Malhan}, {Martin} and
  {Bianchini}}]{ibata:2019b}
\bibinfo{author}{{Ibata}, R.A.}, \bibinfo{author}{{Bellazzini}, M.},
  \bibinfo{author}{{Malhan}, K.}, \bibinfo{author}{{Martin}, N.},
  \bibinfo{author}{{Bianchini}, P.}, \bibinfo{year}{2019}a.
\newblock \bibinfo{title}{{Identification of the long stellar stream of the
  prototypical massive globular cluster {\ensuremath{\omega}} Centauri}}.
\newblock \bibinfo{journal}{Nature Astronomy} \bibinfo{volume}{3},
  \bibinfo{pages}{667--672}.
\newblock \DOIprefix\doi{10.1038/s41550-019-0751-x},
  \href{http://arxiv.org/abs/1902.09544}{{\tt arXiv:1902.09544}}.
%Type = Article
\bibitem[{{Ibata} et~al.(1995){Ibata}, {Gilmore} and {Irwin}}]{ibata:1995}
\bibinfo{author}{{Ibata}, R.A.}, \bibinfo{author}{{Gilmore}, G.},
  \bibinfo{author}{{Irwin}, M.J.}, \bibinfo{year}{1995}.
\newblock \bibinfo{title}{{Sagittarius: the nearest dwarf galaxy}}.
\newblock \bibinfo{journal}{\mnras} \bibinfo{volume}{277},
  \bibinfo{pages}{781--800}.
\newblock \DOIprefix\doi{10.1093/mnras/277.3.781},
  \href{http://arxiv.org/abs/astro-ph/9506071}{{\tt arXiv:astro-ph/9506071}}.
%Type = Article
\bibitem[{{Ibata} et~al.(2002){Ibata}, {Lewis}, {Irwin} and
  {Quinn}}]{ibata:2002}
\bibinfo{author}{{Ibata}, R.A.}, \bibinfo{author}{{Lewis}, G.F.},
  \bibinfo{author}{{Irwin}, M.J.}, \bibinfo{author}{{Quinn}, T.},
  \bibinfo{year}{2002}.
\newblock \bibinfo{title}{{Uncovering cold dark matter halo substructure with
  tidal streams}}.
\newblock \bibinfo{journal}{\mnras} \bibinfo{volume}{332},
  \bibinfo{pages}{915--920}.
\newblock \DOIprefix\doi{10.1046/j.1365-8711.2002.05358.x},
  \href{http://arxiv.org/abs/astro-ph/0110690}{{\tt arXiv:astro-ph/0110690}}.
%Type = Article
\bibitem[{{Ibata} et~al.(2016){Ibata}, {Lewis} and {Martin}}]{ibata:2016}
\bibinfo{author}{{Ibata}, R.A.}, \bibinfo{author}{{Lewis}, G.F.},
  \bibinfo{author}{{Martin}, N.F.}, \bibinfo{year}{2016}.
\newblock \bibinfo{title}{{Feeling the Pull: a Study of Natural Galactic
  Accelerometers. I. Photometry of the Delicate Stellar Stream of the Palomar 5
  Globular Cluster}}.
\newblock \bibinfo{journal}{\apj} \bibinfo{volume}{819}, \bibinfo{pages}{1}.
\newblock \DOIprefix\doi{10.3847/0004-637X/819/1/1},
  \href{http://arxiv.org/abs/1512.03054}{{\tt arXiv:1512.03054}}.
%Type = Article
\bibitem[{{Ibata} et~al.(2017){Ibata}, {Lewis}, {Thomas}, {Martin} and
  {Chapman}}]{ibata:2017}
\bibinfo{author}{{Ibata}, R.A.}, \bibinfo{author}{{Lewis}, G.F.},
  \bibinfo{author}{{Thomas}, G.}, \bibinfo{author}{{Martin}, N.F.},
  \bibinfo{author}{{Chapman}, S.}, \bibinfo{year}{2017}.
\newblock \bibinfo{title}{{Feeling the Pull: A Study of Natural Galactic
  Accelerometers. II. Kinematics and Mass of the Delicate Stellar Stream of the
  Palomar 5 Globular Cluster}}.
\newblock \bibinfo{journal}{\apj} \bibinfo{volume}{842}, \bibinfo{pages}{120}.
\newblock \DOIprefix\doi{10.3847/1538-4357/aa7514},
  \href{http://arxiv.org/abs/1708.06360}{{\tt arXiv:1708.06360}}.
%Type = Article
\bibitem[{{Ibata} et~al.(2019b){Ibata}, {Malhan} and {Martin}}]{ibata:2019}
\bibinfo{author}{{Ibata}, R.A.}, \bibinfo{author}{{Malhan}, K.},
  \bibinfo{author}{{Martin}, N.F.}, \bibinfo{year}{2019}b.
\newblock \bibinfo{title}{{The Streams of the Gaping Abyss: A Population of
  Entangled Stellar Streams Surrounding the Inner Galaxy}}.
\newblock \bibinfo{journal}{\apj} \bibinfo{volume}{872}, \bibinfo{pages}{152}.
\newblock \DOIprefix\doi{10.3847/1538-4357/ab0080},
  \href{http://arxiv.org/abs/1901.07566}{{\tt arXiv:1901.07566}}.
%Type = Article
\bibitem[{{Ibata} et~al.(2018){Ibata}, {Malhan}, {Martin} and
  {Starkenburg}}]{ibata:2018}
\bibinfo{author}{{Ibata}, R.A.}, \bibinfo{author}{{Malhan}, K.},
  \bibinfo{author}{{Martin}, N.F.}, \bibinfo{author}{{Starkenburg}, E.},
  \bibinfo{year}{2018}.
\newblock \bibinfo{title}{{Phlegethon, a Nearby 75{\textdegree}-long Retrograde
  Stellar Stream}}.
\newblock \bibinfo{journal}{\apj} \bibinfo{volume}{865}, \bibinfo{pages}{85}.
\newblock \DOIprefix\doi{10.3847/1538-4357/aadba3},
  \href{http://arxiv.org/abs/1806.01195}{{\tt arXiv:1806.01195}}.
%Type = Article
\bibitem[{{Ishigaki} et~al.(2016){Ishigaki}, {Hwang}, {Chiba} and
  {Aoki}}]{ishigaki:2016}
\bibinfo{author}{{Ishigaki}, M.N.}, \bibinfo{author}{{Hwang}, N.},
  \bibinfo{author}{{Chiba}, M.}, \bibinfo{author}{{Aoki}, W.},
  \bibinfo{year}{2016}.
\newblock \bibinfo{title}{{Line-of-sight Velocity and Metallicity Measurements
  of the Palomar 5 Tidal Stream}}.
\newblock \bibinfo{journal}{\apj} \bibinfo{volume}{823}, \bibinfo{pages}{157}.
\newblock \DOIprefix\doi{10.3847/0004-637X/823/2/157},
  \href{http://arxiv.org/abs/1604.03188}{{\tt arXiv:1604.03188}}.
%Type = Article
\bibitem[{{Ivezic} et~al.(2008){Ivezic}, {Axelrod}, {Brandt}, {Burke},
  {Claver}, {Connolly}, {Cook}, {Gee}, {Gilmore}, {Jacoby}, {Jones}, {Kahn},
  {Kantor}, {Krabbendam}, {Lupton}, {Monet}, {Pinto}, {Saha}, {Schalk},
  {Schneider}, {Strauss}, {Stubbs}, {Sweeney}, {Szalay}, {Thaler}, {Tyson} and
  {LSST Collaboration}}]{ivezic:2008}
{Ivezic}, Z. et~al., \bibinfo{year}{2008}.
\newblock \bibinfo{title}{{Large Synoptic Survey Telescope: From Science
  Drivers To Reference Design}}.
\newblock \bibinfo{journal}{Serbian Astronomical Journal}
  \bibinfo{volume}{176}, \bibinfo{pages}{1--13}.
\newblock \DOIprefix\doi{10.2298/SAJ0876001I}.
%Type = Article
\bibitem[{Jethwa et~al.(2018)Jethwa, Torrealba, Navarrete, {Carballo-Bello},
  {de Boer}, Erkal, Koposov, Duffau, Geisler, Catelan and
  Belokurov}]{jethwa:2018}
Jethwa, P. et~al., \bibinfo{year}{2018}.
\newblock \bibinfo{title}{Discovery of a thin stellar stream in the {{SLAMS}}
  survey}.
\newblock \bibinfo{journal}{\textbackslash mnras} \bibinfo{volume}{480},
  \bibinfo{pages}{5342--5351}.
\newblock \DOIprefix\doi{10.1093/mnras/sty2226}.
%Type = Article
\bibitem[{{Ji} et~al.(2020){Ji}, {Li}, {Hansen}, {Casey}, {Koposov}, {Pace},
  {Mackey}, {Lewis}, {Simpson}, {Bland-Hawthorn}, {Cullinane}, {Da Costa},
  {Hattori}, {Martell}, {Kuehn}, {Erkal}, {Shipp}, {Wan} and
  {Zucker}}]{ji:2020}
{Ji}, A.~P. et~al., \bibinfo{year}{2020}.
\newblock \bibinfo{title}{{The Southern Stellar Stream Spectroscopic Survey
  (S$^{5}$): Chemical Abundances of Seven Stellar Streams}}.
\newblock \bibinfo{journal}{\aj} \bibinfo{volume}{160}, \bibinfo{pages}{181}.
\newblock \DOIprefix\doi{10.3847/1538-3881/abacb6},
  \href{http://arxiv.org/abs/2008.07568}{{\tt arXiv:2008.07568}}.
%Type = Article
\bibitem[{{Jin} et~al.(2023){Jin}, {Trager}, {Dalton}, {Aguerri}, {Drew},
  {Falc{\'o}n-Barroso}, {G{\"a}nsicke}, {Hill}, {Iovino}, {Pieri}, {Poggianti},
  {Smith}, {Vallenari}, {Abrams}, {Aguado}, {Antoja}, {Arag{\'o}n-Salamanca},
  {Ascasibar}, {Babusiaux}, {Balcells}, {Barrena}, {Battaglia}, {Belokurov},
  {Bensby}, {Bonifacio}, {Bragaglia}, {Carrasco}, {Carrera}, {Cornwell},
  {Dom{\'\i}nguez-Palmero}, {Duncan}, {Famaey}, {Fari{\~n}a}, {Gonzalez},
  {Guest}, {Hatch}, {Hess}, {Hoskin}, {Irwin}, {Knapen}, {Koposov}, {Kuchner},
  {Laigle}, {Lewis}, {Longhetti}, {Lucatello}, {M{\'e}ndez-Abreu}, {Mercurio},
  {Molaeinezhad}, {Mongui{\'o}}, {Morrison}, {Murphy}, {Peralta de Arriba},
  {P{\'e}rez}, {P{\'e}rez-R{\`a}fols}, {Pic{\'o}}, {Raddi}, {Romero-G{\'o}mez},
  {Royer}, {Siebert}, {Seabroke}, {Som}, {Terrett}, {Thomas}, {Wesson},
  {Worley}, {Alfaro}, {Allende Prieto}, {Alonso-Santiago}, {Amos}, {Ashley},
  {Balaguer-N{\'u} nez}, {Balbinot}, {Bellazzini}, {Benn}, {Berlanas},
  {Bernard}, {Best}, {Bettoni}, {Bianco}, {Bishop}, {Blomqvist}, {Boeche},
  {Bolzonella}, {Bonoli}, {Bosma}, {Britavskiy}, {Busarello}, {Caffau},
  {Cantat-Gaudin}, {Castro-Ginard}, {Couto}, {Carbajo-Hijarrubia}, {Carter},
  {Casamiquela}, {Conrado}, {Corcho-Caballero}, {Costantin}, {Deason}, {de
  Burgos}, {De Grandi}, {Di Matteo}, {Dom{\'\i}nguez-G{\'o}mez}, {Dorda},
  {Drake}, {Dutta}, {Erkal}, {Feltzing}, {Ferr{\'e}-Mateu}, {Feuillet},
  {Figueras}, {Fossati}, {Franciosini}, {Frasca}, {Fumagalli}, {Gallazzi},
  {Garc{\'\i}a-Benito}, {Fusillo}, {Gebran}, {Gilbert}, {Gledhill},
  {Gonz{\'a}lez Delgado}, {Greimel}, {Guarcello}, {Guerra}, {Gullieuszik},
  {Haines}, {Hardcastle}, {Harris}, {Haywood}, {Helmi}, {Hernandez}, {Herrero},
  {Hughes}, {Irsic}, {Jablonka}, {Jarvis}, {Jordi}, {Kondapally}, {Kordopatis},
  {Krogager}, {La Barbera}, {Lam}, {Larsen}, {Lemasle}, {Lewis}, {Lhom{\'e}},
  {Lind}, {Lodi}, {Longobardi}, {Lonoce}, {Magrini}, {Ma{\'\i}z Apell{\'a}niz},
  {Marchal}, {Marco}, {Martin}, {Matsuno}, {Maurogordato}, {Merluzzi},
  {Miralda-Escud{\'e}}, {Molinari}, {Monari}, {Morelli}, {Mottram}, {Naylor},
  {Negueruela}, {Onorbe}, {Pancino}, {Peirani}, {Peletier}, {Pozzetti},
  {Rainer}, {Ramos}, {Read}, {Rossi}, {R{\"o}ttgering},
  {Rubi{\~n}o-Mart{\'\i}n}, {Sabater Montes}, {San Juan}, {Sanna}, {Schallig},
  {Schiavon}, {Schultheis}, {Serra}, {Shimwell}, {Sim{\'o}n-D{\'\i}az},
  {Smith}, {Sordo}, {Sorini}, {Soubiran}, {Starkenburg}, {Steele}, {Stott},
  {Stuik}, {Tolstoy}, {Tortora}, {Tsantaki}, {Van der Swaelmen}, {van Weeren},
  {Vergani}, {Verheijen}, {Verro}, {Vink}, {Vioque}, {Walcher}, {Walton},
  {Wegg}, {Weijmans}, {Williams}, {Wilson}, {Wright}, {Xylakis-Dornbusch},
  {Youakim}, {Zibetti} and {Zurita}}]{jin:2023}
{Jin}, S. et~al., \bibinfo{year}{2023}.
\newblock \bibinfo{title}{{The wide-field, multiplexed, spectroscopic facility
  WEAVE: Survey design, overview, and simulated implementation}}.
\newblock \bibinfo{journal}{\mnras} \DOIprefix\doi{10.1093/mnras/stad557},
  \href{http://arxiv.org/abs/2212.03981}{{\tt arXiv:2212.03981}}.
%Type = Article
\bibitem[{{Johnson} et~al.(2020){Johnson}, {Conroy}, {Naidu}, {Bonaca},
  {Zaritsky}, {Ting}, {Cargile}, {Han} and {Speagle}}]{johnson:2020}
{Johnson}, B.~D. et~al., \bibinfo{year}{2020}.
\newblock \bibinfo{title}{{A Diffuse Metal-poor Component of the Sagittarius
  Stream Revealed by the H3 Survey}}.
\newblock \bibinfo{journal}{\apj} \bibinfo{volume}{900}, \bibinfo{pages}{103}.
\newblock \DOIprefix\doi{10.3847/1538-4357/abab08},
  \href{http://arxiv.org/abs/2007.14408}{{\tt arXiv:2007.14408}}.
%Type = Article
\bibitem[{{Johnston}(1998)}]{johnston:1998}
\bibinfo{author}{{Johnston}, K.V.}, \bibinfo{year}{1998}.
\newblock \bibinfo{title}{{A Prescription for Building the Milky Way's Halo
  from Disrupted Satellites}}.
\newblock \bibinfo{journal}{\apj} \bibinfo{volume}{495},
  \bibinfo{pages}{297--308}.
\newblock \DOIprefix\doi{10.1086/305273},
  \href{http://arxiv.org/abs/astro-ph/9710007}{{\tt arXiv:astro-ph/9710007}}.
%Type = Article
\bibitem[{{Johnston} et~al.(1996){Johnston}, {Hernquist} and
  {Bolte}}]{johnston:1996}
\bibinfo{author}{{Johnston}, K.V.}, \bibinfo{author}{{Hernquist}, L.},
  \bibinfo{author}{{Bolte}, M.}, \bibinfo{year}{1996}.
\newblock \bibinfo{title}{{Fossil Signatures of Ancient Accretion Events in the
  Halo}}.
\newblock \bibinfo{journal}{\apj} \bibinfo{volume}{465}, \bibinfo{pages}{278}.
\newblock \DOIprefix\doi{10.1086/177418},
  \href{http://arxiv.org/abs/astro-ph/9602060}{{\tt arXiv:astro-ph/9602060}}.
%Type = Article
\bibitem[{{Johnston} et~al.(2005){Johnston}, {Law} and
  {Majewski}}]{johnston:2005}
\bibinfo{author}{{Johnston}, K.V.}, \bibinfo{author}{{Law}, D.R.},
  \bibinfo{author}{{Majewski}, S.R.}, \bibinfo{year}{2005}.
\newblock \bibinfo{title}{{A Two Micron All Sky Survey View of the Sagittarius
  Dwarf Galaxy. III. Constraints on the Flattening of the Galactic Halo}}.
\newblock \bibinfo{journal}{\apj} \bibinfo{volume}{619},
  \bibinfo{pages}{800--806}.
\newblock \DOIprefix\doi{10.1086/426777},
  \href{http://arxiv.org/abs/astro-ph/0407565}{{\tt arXiv:astro-ph/0407565}}.
%Type = Article
\bibitem[{{Johnston} et~al.(2002){Johnston}, {Spergel} and
  {Haydn}}]{johnston:2002}
\bibinfo{author}{{Johnston}, K.V.}, \bibinfo{author}{{Spergel}, D.N.},
  \bibinfo{author}{{Haydn}, C.}, \bibinfo{year}{2002}.
\newblock \bibinfo{title}{{How Lumpy Is the Milky Way's Dark Matter Halo?}}
\newblock \bibinfo{journal}{\apj} \bibinfo{volume}{570},
  \bibinfo{pages}{656--664}.
\newblock \DOIprefix\doi{10.1086/339791},
  \href{http://arxiv.org/abs/astro-ph/0111196}{{\tt arXiv:astro-ph/0111196}}.
%Type = Article
\bibitem[{{Just} et~al.(2009){Just}, {Berczik}, {Petrov} and
  {Ernst}}]{just:2009}
\bibinfo{author}{{Just}, A.}, \bibinfo{author}{{Berczik}, P.},
  \bibinfo{author}{{Petrov}, M.I.}, \bibinfo{author}{{Ernst}, A.},
  \bibinfo{year}{2009}.
\newblock \bibinfo{title}{{Quantitative analysis of clumps in the tidal tails
  of star clusters}}.
\newblock \bibinfo{journal}{\mnras} \bibinfo{volume}{392},
  \bibinfo{pages}{969--981}.
\newblock \DOIprefix\doi{10.1111/j.1365-2966.2008.14099.x},
  \href{http://arxiv.org/abs/0808.3293}{{\tt arXiv:0808.3293}}.
%Type = Article
\bibitem[{{Kado-Fong} et~al.(2018){Kado-Fong}, {Greene}, {Hendel},
  {Price-Whelan}, {Greco}, {Goulding}, {Huang}, {Johnston}, {Komiyama}, {Lee},
  {Lust}, {Strauss} and {Tanaka}}]{kado-fong:2018}
{Kado-Fong}, E. et~al., \bibinfo{year}{2018}.
\newblock \bibinfo{title}{{Tidal Features at 0.05 < z < 0.45 in the Hyper
  Suprime-Cam Subaru Strategic Program: Properties and Formation Channels}}.
\newblock \bibinfo{journal}{\apj} \bibinfo{volume}{866}, \bibinfo{pages}{103}.
\newblock \DOIprefix\doi{10.3847/1538-4357/aae0f0},
  \href{http://arxiv.org/abs/1805.05970}{{\tt arXiv:1805.05970}}.
%Type = Article
\bibitem[{{Kallivayalil} et~al.(2006){Kallivayalil}, {van der Marel}, {Alcock},
  {Axelrod}, {Cook}, {Drake} and {Geha}}]{kallivayalil:2006}
\bibinfo{author}{{Kallivayalil}, N.}, \bibinfo{author}{{van der Marel}, R.P.},
  \bibinfo{author}{{Alcock}, C.}, \bibinfo{author}{{Axelrod}, T.},
  \bibinfo{author}{{Cook}, K.H.}, \bibinfo{author}{{Drake}, A.J.},
  \bibinfo{author}{{Geha}, M.}, \bibinfo{year}{2006}.
\newblock \bibinfo{title}{{The Proper Motion of the Large Magellanic Cloud
  Using HST}}.
\newblock \bibinfo{journal}{\apj} \bibinfo{volume}{638},
  \bibinfo{pages}{772--785}.
\newblock \DOIprefix\doi{10.1086/498972},
  \href{http://arxiv.org/abs/astro-ph/0508457}{{\tt arXiv:astro-ph/0508457}}.
%Type = Article
\bibitem[{{Kallivayalil} et~al.(2013){Kallivayalil}, {van der Marel}, {Besla},
  {Anderson} and {Alcock}}]{kallivayalil:2013}
\bibinfo{author}{{Kallivayalil}, N.}, \bibinfo{author}{{van der Marel}, R.P.},
  \bibinfo{author}{{Besla}, G.}, \bibinfo{author}{{Anderson}, J.},
  \bibinfo{author}{{Alcock}, C.}, \bibinfo{year}{2013}.
\newblock \bibinfo{title}{{Third-epoch Magellanic Cloud Proper Motions. I.
  Hubble Space Telescope/WFC3 Data and Orbit Implications}}.
\newblock \bibinfo{journal}{\apj} \bibinfo{volume}{764}, \bibinfo{pages}{161}.
\newblock \DOIprefix\doi{10.1088/0004-637X/764/2/161},
  \href{http://arxiv.org/abs/1301.0832}{{\tt arXiv:1301.0832}}.
%Type = Article
\bibitem[{{Kang} et~al.(2023){Kang}, {Lee}, {Kim} and {Beers}}]{kang:2023}
\bibinfo{author}{{Kang}, G.}, \bibinfo{author}{{Lee}, Y.S.},
  \bibinfo{author}{{Kim}, Y.K.}, \bibinfo{author}{{Beers}, T.C.},
  \bibinfo{year}{2023}.
\newblock \bibinfo{title}{{A Dynamically Distinct Stellar Population in the
  Leading Arm of the Sagittarius Stream}}.
\newblock \bibinfo{journal}{\apjl} \bibinfo{volume}{954}, \bibinfo{pages}{L43}.
\newblock \DOIprefix\doi{10.3847/2041-8213/ace32b},
  \href{http://arxiv.org/abs/2306.16748}{{\tt arXiv:2306.16748}}.
%Type = Article
\bibitem[{{Katz} et~al.(2023){Katz}, {Sartoretti}, {Guerrier}, {Panuzzo},
  {Seabroke}, {Th{\'e}venin}, {Cropper}, {Benson}, {Blomme}, {Haigron},
  {Marchal}, {Smith}, {Baker}, {Chemin}, {Damerdji}, {David}, {Dolding},
  {Fr{\'e}mat}, {Gosset}, {Jan{\ss}en}, {Jasniewicz}, {Lobel}, {Plum},
  {Samaras}, {Snaith}, {Soubiran}, {Vanel}, {Zwitter}, {Antoja}, {Arenou},
  {Babusiaux}, {Brouillet}, {Caffau}, {Di Matteo}, {Fabre}, {Fabricius},
  {Fragkoudi}, {Haywood}, {Huckle}, {Hottier}, {Lasne}, {Leclerc},
  {Mastrobuono-Battisti}, {Royer}, {Teyssier}, {Zorec}, {Crifo}, {Jean-Antoine
  Piccolo}, {Turon} and {Viala}}]{katz:2023}
{Katz}, D. et~al., \bibinfo{year}{2023}.
\newblock \bibinfo{title}{{Gaia Data Release 3. Properties and validation of
  the radial velocities}}.
\newblock \bibinfo{journal}{\aap} \bibinfo{volume}{674}, \bibinfo{pages}{A5}.
\newblock \DOIprefix\doi{10.1051/0004-6361/202244220},
  \href{http://arxiv.org/abs/2206.05902}{{\tt arXiv:2206.05902}}.
%Type = Article
\bibitem[{{Kawata} et~al.(2018){Kawata}, {Baba}, {Ciuc{\v{a}}}, {Cropper},
  {Grand}, {Hunt} and {Seabroke}}]{kawata:2018}
\bibinfo{author}{{Kawata}, D.}, \bibinfo{author}{{Baba}, J.},
  \bibinfo{author}{{Ciuc{\v{a}}}, I.}, \bibinfo{author}{{Cropper}, M.},
  \bibinfo{author}{{Grand}, R.J.J.}, \bibinfo{author}{{Hunt}, J.A.S.},
  \bibinfo{author}{{Seabroke}, G.}, \bibinfo{year}{2018}.
\newblock \bibinfo{title}{{Radial distribution of stellar motions in Gaia
  DR2}}.
\newblock \bibinfo{journal}{\mnras} \bibinfo{volume}{479},
  \bibinfo{pages}{L108--L112}.
\newblock \DOIprefix\doi{10.1093/mnrasl/sly107},
  \href{http://arxiv.org/abs/1804.10175}{{\tt arXiv:1804.10175}}.
%Type = Article
\bibitem[{{Kawata} et~al.(2021){Kawata}, {Baba}, {Hunt}, {Sch{\"o}nrich},
  {Ciuc{\u{a}}}, {Friske}, {Seabroke} and {Cropper}}]{kawata:2021}
\bibinfo{author}{{Kawata}, D.}, \bibinfo{author}{{Baba}, J.},
  \bibinfo{author}{{Hunt}, J.A.S.}, \bibinfo{author}{{Sch{\"o}nrich}, R.},
  \bibinfo{author}{{Ciuc{\u{a}}}, I.}, \bibinfo{author}{{Friske}, J.},
  \bibinfo{author}{{Seabroke}, G.}, \bibinfo{author}{{Cropper}, M.},
  \bibinfo{year}{2021}.
\newblock \bibinfo{title}{{Galactic bar resonances inferred from kinematically
  hot stars in Gaia EDR3}}.
\newblock \bibinfo{journal}{\mnras} \bibinfo{volume}{508},
  \bibinfo{pages}{728--736}.
\newblock \DOIprefix\doi{10.1093/mnras/stab2582},
  \href{http://arxiv.org/abs/2012.05890}{{\tt arXiv:2012.05890}}.
%Type = Article
\bibitem[{{Khoperskov} et~al.(2023){Khoperskov}, {Minchev}, {Libeskind},
  {Haywood}, {Di Matteo}, {Belokurov}, {Steinmetz}, {Gomez}, {Grand},
  {Hoffman}, {Knebe}, {Sorce}, {Spaare}, {Tempel} and
  {Vogelsberger}}]{khoperskov:2023}
{Khoperskov}, S. et~al., \bibinfo{year}{2023}.
\newblock \bibinfo{title}{{The stellar halo in Local Group Hestia simulations.
  I. The in situ component and the effect of mergers}}.
\newblock \bibinfo{journal}{\aap} \bibinfo{volume}{677}, \bibinfo{pages}{A89}.
\newblock \DOIprefix\doi{10.1051/0004-6361/202244232},
  \href{http://arxiv.org/abs/2206.04521}{{\tt arXiv:2206.04521}}.
%Type = Article
\bibitem[{{Klypin} et~al.(1999){Klypin}, {Kravtsov}, {Valenzuela} and
  {Prada}}]{klypin:1999}
\bibinfo{author}{{Klypin}, A.}, \bibinfo{author}{{Kravtsov}, A.V.},
  \bibinfo{author}{{Valenzuela}, O.}, \bibinfo{author}{{Prada}, F.},
  \bibinfo{year}{1999}.
\newblock \bibinfo{title}{{Where Are the Missing Galactic Satellites?}}
\newblock \bibinfo{journal}{\apj} \bibinfo{volume}{522},
  \bibinfo{pages}{82--92}.
\newblock \DOIprefix\doi{10.1086/307643},
  \href{http://arxiv.org/abs/astro-ph/9901240}{{\tt arXiv:astro-ph/9901240}}.
%Type = Article
\bibitem[{{Koposov} et~al.(2012){Koposov}, {Belokurov}, {Evans}, {Gilmore},
  {Gieles}, {Irwin}, {Lewis}, {Niederste-Ostholt}, {Pe{\~n}arrubia}, {Smith},
  {Bizyaev}, {Malanushenko}, {Malanushenko}, {Schneider} and
  {Wyse}}]{koposov:2012}
{Koposov}, S.~E. et~al., \bibinfo{year}{2012}.
\newblock \bibinfo{title}{{The Sagittarius Streams in the Southern Galactic
  Hemisphere}}.
\newblock \bibinfo{journal}{\apj} \bibinfo{volume}{750}, \bibinfo{pages}{80}.
\newblock \DOIprefix\doi{10.1088/0004-637X/750/1/80},
  \href{http://arxiv.org/abs/1111.7042}{{\tt arXiv:1111.7042}}.
%Type = Article
\bibitem[{{Koposov} et~al.(2019){Koposov}, {Belokurov}, {Li}, {Mateu}, {Erkal},
  {Grillmair}, {Hendel}, {Price-Whelan}, {Laporte}, {Hawkins}, {Sohn}, {del
  Pino}, {Evans}, {Slater}, {Kallivayalil}, {Navarro} and {Orphan Aspen
  Treasury Collaboration}}]{koposov:2019}
{Koposov}, S.~E. et~al., \bibinfo{year}{2019}.
\newblock \bibinfo{title}{{Piercing the Milky Way: an all-sky view of the
  Orphan Stream}}.
\newblock \bibinfo{journal}{\mnras} \bibinfo{volume}{485},
  \bibinfo{pages}{4726--4742}.
\newblock \DOIprefix\doi{10.1093/mnras/stz457},
  \href{http://arxiv.org/abs/1812.08172}{{\tt arXiv:1812.08172}}.
%Type = Article
\bibitem[{{Koposov} et~al.(2023){Koposov}, {Erkal}, {Li}, {Da Costa},
  {Cullinane}, {Ji}, {Kuehn}, {Lewis}, {Pace}, {Shipp}, {Zucker},
  {Bland-Hawthorn}, {Lilleengen}, {Martell} and {S5
  Collaboration}}]{koposov:2023}
{Koposov}, S.~E. et~al., \bibinfo{year}{2023}.
\newblock \bibinfo{title}{{S $^{5}$: Probing the Milky Way and Magellanic
  Clouds potentials with the 6D map of the Orphan-Chenab stream}}.
\newblock \bibinfo{journal}{\mnras} \bibinfo{volume}{521},
  \bibinfo{pages}{4936--4962}.
\newblock \DOIprefix\doi{10.1093/mnras/stad551},
  \href{http://arxiv.org/abs/2211.04495}{{\tt arXiv:2211.04495}}.
%Type = Article
\bibitem[{{Koposov} et~al.(2014){Koposov}, {Irwin}, {Belokurov},
  {Gonzalez-Solares}, {Yoldas}, {Lewis}, {Metcalfe} and
  {Shanks}}]{koposov:2014}
\bibinfo{author}{{Koposov}, S.E.}, \bibinfo{author}{{Irwin}, M.},
  \bibinfo{author}{{Belokurov}, V.}, \bibinfo{author}{{Gonzalez-Solares}, E.},
  \bibinfo{author}{{Yoldas}, A.K.}, \bibinfo{author}{{Lewis}, J.},
  \bibinfo{author}{{Metcalfe}, N.}, \bibinfo{author}{{Shanks}, T.},
  \bibinfo{year}{2014}.
\newblock \bibinfo{title}{{Discovery of a cold stellar stream in the ATLAS DR1
  data.}}
\newblock \bibinfo{journal}{\mnras} \bibinfo{volume}{442},
  \bibinfo{pages}{L85--L89}.
\newblock \DOIprefix\doi{10.1093/mnrasl/slu060},
  \href{http://arxiv.org/abs/1403.3409}{{\tt arXiv:1403.3409}}.
%Type = Article
\bibitem[{{Koposov} et~al.(2010){Koposov}, {Rix} and {Hogg}}]{koposov:2010}
\bibinfo{author}{{Koposov}, S.E.}, \bibinfo{author}{{Rix}, H.W.},
  \bibinfo{author}{{Hogg}, D.W.}, \bibinfo{year}{2010}.
\newblock \bibinfo{title}{{Constraining the Milky Way Potential with a
  Six-Dimensional Phase-Space Map of the GD-1 Stellar Stream}}.
\newblock \bibinfo{journal}{\apj} \bibinfo{volume}{712},
  \bibinfo{pages}{260--273}.
\newblock \DOIprefix\doi{10.1088/0004-637X/712/1/260},
  \href{http://arxiv.org/abs/0907.1085}{{\tt arXiv:0907.1085}}.
%Type = Article
\bibitem[{{Koppelman} et~al.(2021){Koppelman}, {Hagen} and
  {Helmi}}]{koppelman:2021a}
\bibinfo{author}{{Koppelman}, H.H.}, \bibinfo{author}{{Hagen}, J.H.J.},
  \bibinfo{author}{{Helmi}, A.}, \bibinfo{year}{2021}.
\newblock \bibinfo{title}{{Frequencies, chaos, and resonances: A study of
  orbital parameters of nearby thick-disc and halo stars}}.
\newblock \bibinfo{journal}{\aap} \bibinfo{volume}{647}, \bibinfo{pages}{A37}.
\newblock \DOIprefix\doi{10.1051/0004-6361/202039390},
  \href{http://arxiv.org/abs/2009.04849}{{\tt arXiv:2009.04849}}.
%Type = Article
\bibitem[{{Koppelman} and {Helmi}(2021)}]{koppelman:2021}
\bibinfo{author}{{Koppelman}, H.H.}, \bibinfo{author}{{Helmi}, A.},
  \bibinfo{year}{2021}.
\newblock \bibinfo{title}{{Time evolution of gaps in stellar streams in
  axisymmetric St{\"a}ckel potentials}}.
\newblock \bibinfo{journal}{\aap} \bibinfo{volume}{649}, \bibinfo{pages}{A55}.
\newblock \DOIprefix\doi{10.1051/0004-6361/202039968}.
%Type = Article
\bibitem[{{Koppelman} et~al.(2019a){Koppelman}, {Helmi}, {Massari},
  {Price-Whelan} and {Starkenburg}}]{koppelman:2019b}
\bibinfo{author}{{Koppelman}, H.H.}, \bibinfo{author}{{Helmi}, A.},
  \bibinfo{author}{{Massari}, D.}, \bibinfo{author}{{Price-Whelan}, A.M.},
  \bibinfo{author}{{Starkenburg}, T.K.}, \bibinfo{year}{2019}a.
\newblock \bibinfo{title}{{Multiple retrograde substructures in the Galactic
  halo: A shattered view of Galactic history}}.
\newblock \bibinfo{journal}{\aap} \bibinfo{volume}{631}, \bibinfo{pages}{L9}.
\newblock \DOIprefix\doi{10.1051/0004-6361/201936738},
  \href{http://arxiv.org/abs/1909.08924}{{\tt arXiv:1909.08924}}.
%Type = Article
\bibitem[{{Koppelman} et~al.(2019b){Koppelman}, {Helmi}, {Massari}, {Roelenga}
  and {Bastian}}]{koppelman:2019}
\bibinfo{author}{{Koppelman}, H.H.}, \bibinfo{author}{{Helmi}, A.},
  \bibinfo{author}{{Massari}, D.}, \bibinfo{author}{{Roelenga}, S.},
  \bibinfo{author}{{Bastian}, U.}, \bibinfo{year}{2019}b.
\newblock \bibinfo{title}{{Characterization and history of the Helmi streams
  with Gaia DR2}}.
\newblock \bibinfo{journal}{\aap} \bibinfo{volume}{625}, \bibinfo{pages}{A5}.
\newblock \DOIprefix\doi{10.1051/0004-6361/201834769},
  \href{http://arxiv.org/abs/1812.00846}{{\tt arXiv:1812.00846}}.
%Type = Article
\bibitem[{{Kremer} et~al.(2020){Kremer}, {Ye}, {Rui}, {Weatherford},
  {Chatterjee}, {Fragione}, {Rodriguez}, {Spera} and {Rasio}}]{kremer:2020}
{Kremer}, K. et~al., \bibinfo{year}{2020}.
\newblock \bibinfo{title}{{Modeling Dense Star Clusters in the Milky Way and
  Beyond with the CMC Cluster Catalog}}.
\newblock \bibinfo{journal}{\apjs} \bibinfo{volume}{247}, \bibinfo{pages}{48}.
\newblock \DOIprefix\doi{10.3847/1538-4365/ab7919},
  \href{http://arxiv.org/abs/1911.00018}{{\tt arXiv:1911.00018}}.
%Type = Article
\bibitem[{{Kroupa}(2001)}]{Kroupa:2001}
\bibinfo{author}{{Kroupa}, P.}, \bibinfo{year}{2001}.
\newblock \bibinfo{title}{{On the variation of the initial mass function}}.
\newblock \bibinfo{journal}{\mnras} \bibinfo{volume}{322},
  \bibinfo{pages}{231--246}.
\newblock \DOIprefix\doi{10.1046/j.1365-8711.2001.04022.x},
  \href{http://arxiv.org/abs/astro-ph/0009005}{{\tt arXiv:astro-ph/0009005}}.
%Type = Article
\bibitem[{{Kruijssen}(2019)}]{kruijssen:2019}
\bibinfo{author}{{Kruijssen}, J.M.D.}, \bibinfo{year}{2019}.
\newblock \bibinfo{title}{{The minimum metallicity of globular clusters and its
  physical origin - implications for the galaxy mass-metallicity relation and
  observations of proto-globular clusters at high redshift}}.
\newblock \bibinfo{journal}{\mnras} \bibinfo{volume}{486},
  \bibinfo{pages}{L20--L25}.
\newblock \DOIprefix\doi{10.1093/mnrasl/slz052},
  \href{http://arxiv.org/abs/1904.09987}{{\tt arXiv:1904.09987}}.
%Type = Article
\bibitem[{{Kruijssen} et~al.(2020){Kruijssen}, {Pfeffer}, {Chevance}, {Bonaca},
  {Trujillo-Gomez}, {Bastian}, {Reina-Campos}, {Crain} and
  {Hughes}}]{kruijssen:2020}
{Kruijssen}, J.~M.~D. et~al., \bibinfo{year}{2020}.
\newblock \bibinfo{title}{{Kraken reveals itself - the merger history of the
  Milky Way reconstructed with the E-MOSAICS simulations}}.
\newblock \bibinfo{journal}{\mnras} \bibinfo{volume}{498},
  \bibinfo{pages}{2472--2491}.
\newblock \DOIprefix\doi{10.1093/mnras/staa2452},
  \href{http://arxiv.org/abs/2003.01119}{{\tt arXiv:2003.01119}}.
%Type = Article
\bibitem[{{Kuhn} et~al.(1996){Kuhn}, {Smith} and {Hawley}}]{kuhn:1996}
\bibinfo{author}{{Kuhn}, J.R.}, \bibinfo{author}{{Smith}, H.A.},
  \bibinfo{author}{{Hawley}, S.L.}, \bibinfo{year}{1996}.
\newblock \bibinfo{title}{{Tidal Disruption and Tails from the Carina Dwarf
  Spheroidal Galaxy}}.
\newblock \bibinfo{journal}{\apjl} \bibinfo{volume}{469}, \bibinfo{pages}{L93}.
\newblock \DOIprefix\doi{10.1086/310270}.
%Type = Article
\bibitem[{{K{\"u}pper} et~al.(2015){K{\"u}pper}, {Balbinot}, {Bonaca},
  {Johnston}, {Hogg}, {Kroupa} and {Santiago}}]{kupper:2015}
\bibinfo{author}{{K{\"u}pper}, A.H.W.}, \bibinfo{author}{{Balbinot}, E.},
  \bibinfo{author}{{Bonaca}, A.}, \bibinfo{author}{{Johnston}, K.V.},
  \bibinfo{author}{{Hogg}, D.W.}, \bibinfo{author}{{Kroupa}, P.},
  \bibinfo{author}{{Santiago}, B.X.}, \bibinfo{year}{2015}.
\newblock \bibinfo{title}{{Globular Cluster Streams as Galactic High-Precision
  Scales{\textemdash}the Poster Child Palomar 5}}.
\newblock \bibinfo{journal}{\apj} \bibinfo{volume}{803}, \bibinfo{pages}{80}.
\newblock \DOIprefix\doi{10.1088/0004-637X/803/2/80},
  \href{http://arxiv.org/abs/1502.02658}{{\tt arXiv:1502.02658}}.
%Type = Article
\bibitem[{{K{\"u}pper} et~al.(2010){K{\"u}pper}, {Kroupa}, {Baumgardt} and
  {Heggie}}]{kupper:2010}
\bibinfo{author}{{K{\"u}pper}, A.H.W.}, \bibinfo{author}{{Kroupa}, P.},
  \bibinfo{author}{{Baumgardt}, H.}, \bibinfo{author}{{Heggie}, D.C.},
  \bibinfo{year}{2010}.
\newblock \bibinfo{title}{{Tidal tails of star clusters}}.
\newblock \bibinfo{journal}{\mnras} \bibinfo{volume}{401},
  \bibinfo{pages}{105--120}.
\newblock \DOIprefix\doi{10.1111/j.1365-2966.2009.15690.x},
  \href{http://arxiv.org/abs/0909.2619}{{\tt arXiv:0909.2619}}.
%Type = Article
\bibitem[{{K{\"u}pper} et~al.(2012){K{\"u}pper}, {Lane} and
  {Heggie}}]{kupper:2012}
\bibinfo{author}{{K{\"u}pper}, A.H.W.}, \bibinfo{author}{{Lane}, R.R.},
  \bibinfo{author}{{Heggie}, D.C.}, \bibinfo{year}{2012}.
\newblock \bibinfo{title}{{More on the structure of tidal tails}}.
\newblock \bibinfo{journal}{\mnras} \bibinfo{volume}{420},
  \bibinfo{pages}{2700--2714}.
\newblock \DOIprefix\doi{10.1111/j.1365-2966.2011.20242.x},
  \href{http://arxiv.org/abs/1111.5013}{{\tt arXiv:1111.5013}}.
%Type = Article
\bibitem[{{K{\"u}pper} et~al.(2008){K{\"u}pper}, {MacLeod} and
  {Heggie}}]{kupper:2008}
\bibinfo{author}{{K{\"u}pper}, A.H.W.}, \bibinfo{author}{{MacLeod}, A.},
  \bibinfo{author}{{Heggie}, D.C.}, \bibinfo{year}{2008}.
\newblock \bibinfo{title}{{On the structure of tidal tails}}.
\newblock \bibinfo{journal}{\mnras} \bibinfo{volume}{387},
  \bibinfo{pages}{1248--1252}.
\newblock \DOIprefix\doi{10.1111/j.1365-2966.2008.13323.x},
  \href{http://arxiv.org/abs/0804.2476}{{\tt arXiv:0804.2476}}.
%Type = Article
\bibitem[{{Kuzma} et~al.(2015){Kuzma}, {Da Costa}, {Keller} and
  {Maunder}}]{kuzma:2015}
\bibinfo{author}{{Kuzma}, P.B.}, \bibinfo{author}{{Da Costa}, G.S.},
  \bibinfo{author}{{Keller}, S.C.}, \bibinfo{author}{{Maunder}, E.},
  \bibinfo{year}{2015}.
\newblock \bibinfo{title}{{Palomar 5 and its tidal tails: a search for new
  members in the tidal stream}}.
\newblock \bibinfo{journal}{\mnras} \bibinfo{volume}{446},
  \bibinfo{pages}{3297--3309}.
\newblock \DOIprefix\doi{10.1093/mnras/stu2343},
  \href{http://arxiv.org/abs/1411.0776}{{\tt arXiv:1411.0776}}.
%Type = Article
\bibitem[{{Kuzma} et~al.(2022){Kuzma}, {Ferguson}, {Varri}, {Irwin}, {Bernard},
  {Tolstoy}, {Pe{\~n}arrubia} and {Zucker}}]{kuzma:2022}
\bibinfo{author}{{Kuzma}, P.B.}, \bibinfo{author}{{Ferguson}, A.M.N.},
  \bibinfo{author}{{Varri}, A.L.}, \bibinfo{author}{{Irwin}, M.J.},
  \bibinfo{author}{{Bernard}, E.J.}, \bibinfo{author}{{Tolstoy}, E.},
  \bibinfo{author}{{Pe{\~n}arrubia}, J.}, \bibinfo{author}{{Zucker}, D.B.},
  \bibinfo{year}{2022}.
\newblock \bibinfo{title}{{Forward and back: kinematics of the Palomar 5 tidal
  tails}}.
\newblock \bibinfo{journal}{\mnras} \bibinfo{volume}{512},
  \bibinfo{pages}{315--327}.
\newblock \DOIprefix\doi{10.1093/mnras/stac381},
  \href{http://arxiv.org/abs/2202.04548}{{\tt arXiv:2202.04548}}.
%Type = Article
\bibitem[{{Lane} et~al.(2020){Lane}, {Navarro}, {Fattahi}, {Oman} and
  {Bovy}}]{lane:2020}
\bibinfo{author}{{Lane}, J.M.M.}, \bibinfo{author}{{Navarro}, J.F.},
  \bibinfo{author}{{Fattahi}, A.}, \bibinfo{author}{{Oman}, K.A.},
  \bibinfo{author}{{Bovy}, J.}, \bibinfo{year}{2020}.
\newblock \bibinfo{title}{{The Ophiuchus stream progenitor: a new type of
  globular cluster and its possible Sagittarius connection}}.
\newblock \bibinfo{journal}{\mnras} \bibinfo{volume}{492},
  \bibinfo{pages}{4164--4174}.
\newblock \DOIprefix\doi{10.1093/mnras/staa095},
  \href{http://arxiv.org/abs/1905.12633}{{\tt arXiv:1905.12633}}.
%Type = Article
\bibitem[{{Lane} et~al.(2012){Lane}, {K{\"u}pper} and {Heggie}}]{lane:2012}
\bibinfo{author}{{Lane}, R.R.}, \bibinfo{author}{{K{\"u}pper}, A.H.W.},
  \bibinfo{author}{{Heggie}, D.C.}, \bibinfo{year}{2012}.
\newblock \bibinfo{title}{{The tidal tails of 47 Tucanae}}.
\newblock \bibinfo{journal}{\mnras} \bibinfo{volume}{423},
  \bibinfo{pages}{2845--2853}.
\newblock \DOIprefix\doi{10.1111/j.1365-2966.2012.21093.x},
  \href{http://arxiv.org/abs/1204.2549}{{\tt arXiv:1204.2549}}.
%Type = Article
\bibitem[{{Laporte} et~al.(2020){Laporte}, {Belokurov}, {Koposov}, {Smith} and
  {Hill}}]{laporte:2020}
\bibinfo{author}{{Laporte}, C.F.P.}, \bibinfo{author}{{Belokurov}, V.},
  \bibinfo{author}{{Koposov}, S.E.}, \bibinfo{author}{{Smith}, M.C.},
  \bibinfo{author}{{Hill}, V.}, \bibinfo{year}{2020}.
\newblock \bibinfo{title}{{Chemo-dynamical properties of the Anticenter Stream:
  a surviving disc fossil from a past satellite interaction.}}
\newblock \bibinfo{journal}{\mnras} \bibinfo{volume}{492},
  \bibinfo{pages}{L61--L65}.
\newblock \DOIprefix\doi{10.1093/mnrasl/slz167},
  \href{http://arxiv.org/abs/1907.10678}{{\tt arXiv:1907.10678}}.
%Type = Article
\bibitem[{{Laporte} et~al.(2018){Laporte}, {Johnston}, {G{\'o}mez},
  {Garavito-Camargo} and {Besla}}]{laporte:2018}
\bibinfo{author}{{Laporte}, C.F.P.}, \bibinfo{author}{{Johnston}, K.V.},
  \bibinfo{author}{{G{\'o}mez}, F.A.}, \bibinfo{author}{{Garavito-Camargo},
  N.}, \bibinfo{author}{{Besla}, G.}, \bibinfo{year}{2018}.
\newblock \bibinfo{title}{{The influence of Sagittarius and the Large
  Magellanic Cloud on the stellar disc of the Milky Way Galaxy}}.
\newblock \bibinfo{journal}{\mnras} \bibinfo{volume}{481},
  \bibinfo{pages}{286--306}.
\newblock \DOIprefix\doi{10.1093/mnras/sty1574},
  \href{http://arxiv.org/abs/1710.02538}{{\tt arXiv:1710.02538}}.
%Type = Article
\bibitem[{{Laureijs} et~al.(2011){Laureijs}, {Amiaux}, {Arduini},
  {Augu{\`e}res}, {Brinchmann}, {Cole}, {Cropper}, {Dabin}, {Duvet}, {Ealet},
  {Garilli}, {Gondoin}, {Guzzo}, {Hoar}, {Hoekstra}, {Holmes}, {Kitching},
  {Maciaszek}, {Mellier}, {Pasian}, {Percival}, {Rhodes}, {Saavedra Criado},
  {Sauvage}, {Scaramella}, {Valenziano}, {Warren}, {Bender}, {Castander},
  {Cimatti}, {Le F{\`e}vre}, {Kurki-Suonio}, {Levi}, {Lilje}, {Meylan},
  {Nichol}, {Pedersen}, {Popa}, {Rebolo Lopez}, {Rix}, {Rottgering},
  {Zeilinger}, {Grupp}, {Hudelot}, {Massey}, {Meneghetti}, {Miller}, {Paltani},
  {Paulin-Henriksson}, {Pires}, {Saxton}, {Schrabback}, {Seidel}, {Walsh},
  {Aghanim}, {Amendola}, {Bartlett}, {Baccigalupi}, {Beaulieu}, {Benabed},
  {Cuby}, {Elbaz}, {Fosalba}, {Gavazzi}, {Helmi}, {Hook}, {Irwin}, {Kneib},
  {Kunz}, {Mannucci}, {Moscardini}, {Tao}, {Teyssier}, {Weller}, {Zamorani},
  {Zapatero Osorio}, {Boulade}, {Foumond}, {Di Giorgio}, {Guttridge}, {James},
  {Kemp}, {Martignac}, {Spencer}, {Walton}, {Bl{\"u}mchen}, {Bonoli},
  {Bortoletto}, {Cerna}, {Corcione}, {Fabron}, {Jahnke}, {Ligori}, {Madrid},
  {Martin}, {Morgante}, {Pamplona}, {Prieto}, {Riva}, {Toledo}, {Trifoglio},
  {Zerbi}, {Abdalla}, {Douspis}, {Grenet}, {Borgani}, {Bouwens}, {Courbin},
  {Delouis}, {Dubath}, {Fontana}, {Frailis}, {Grazian}, {Koppenh{\"o}fer},
  {Mansutti}, {Melchior}, {Mignoli}, {Mohr}, {Neissner}, {Noddle}, {Poncet},
  {Scodeggio}, {Serrano}, {Shane}, {Starck}, {Surace}, {Taylor},
  {Verdoes-Kleijn}, {Vuerli}, {Williams}, {Zacchei}, {Altieri}, {Escudero
  Sanz}, {Kohley}, {Oosterbroek}, {Astier}, {Bacon}, {Bardelli}, {Baugh},
  {Bellagamba}, {Benoist}, {Bianchi}, {Biviano}, {Branchini}, {Carbone},
  {Cardone}, {Clements}, {Colombi}, {Conselice}, {Cresci}, {Deacon}, {Dunlop},
  {Fedeli}, {Fontanot}, {Franzetti}, {Giocoli}, {Garcia-Bellido}, {Gow},
  {Heavens}, {Hewett}, {Heymans}, {Holland}, {Huang}, {Ilbert}, {Joachimi},
  {Jennins}, {Kerins}, {Kiessling}, {Kirk}, {Kotak}, {Krause}, {Lahav}, {van
  Leeuwen}, {Lesgourgues}, {Lombardi}, {Magliocchetti}, {Maguire}, {Majerotto},
  {Maoli}, {Marulli}, {Maurogordato}, {McCracken}, {McLure}, {Melchiorri},
  {Merson}, {Moresco}, {Nonino}, {Norberg}, {Peacock}, {Pello}, {Penny},
  {Pettorino}, {Di Porto}, {Pozzetti}, {Quercellini}, {Radovich}, {Rassat},
  {Roche}, {Ronayette}, {Rossetti}, {Sartoris}, {Schneider}, {Semboloni},
  {Serjeant}, {Simpson}, {Skordis}, {Smadja}, {Smartt}, {Spano}, {Spiro},
  {Sullivan}, {Tilquin}, {Trotta}, {Verde}, {Wang}, {Williger}, {Zhao},
  {Zoubian} and {Zucca}}]{laureijs:2011}
{Laureijs}, R. et~al., \bibinfo{year}{2011}.
\newblock \bibinfo{title}{{Euclid Definition Study Report}}.
\newblock \bibinfo{journal}{arXiv e-prints} ,
  \bibinfo{pages}{arXiv:1110.3193}\DOIprefix\doi{10.48550/arXiv.1110.3193},
  \href{http://arxiv.org/abs/1110.3193}{{\tt arXiv:1110.3193}}.
%Type = Article
\bibitem[{{Law} and {Majewski}(2010)}]{law:2010}
\bibinfo{author}{{Law}, D.R.}, \bibinfo{author}{{Majewski}, S.R.},
  \bibinfo{year}{2010}.
\newblock \bibinfo{title}{{The Sagittarius Dwarf Galaxy: A Model for Evolution
  in a Triaxial Milky Way Halo}}.
\newblock \bibinfo{journal}{\apj} \bibinfo{volume}{714},
  \bibinfo{pages}{229--254}.
\newblock \DOIprefix\doi{10.1088/0004-637X/714/1/229},
  \href{http://arxiv.org/abs/1003.1132}{{\tt arXiv:1003.1132}}.
%Type = Article
\bibitem[{{Law} et~al.(2009){Law}, {Majewski} and {Johnston}}]{law:2009}
\bibinfo{author}{{Law}, D.R.}, \bibinfo{author}{{Majewski}, S.R.},
  \bibinfo{author}{{Johnston}, K.V.}, \bibinfo{year}{2009}.
\newblock \bibinfo{title}{{Evidence for a Triaxial Milky Way Dark Matter Halo
  from the Sagittarius Stellar Tidal Stream}}.
\newblock \bibinfo{journal}{\apjl} \bibinfo{volume}{703},
  \bibinfo{pages}{L67--L71}.
\newblock \DOIprefix\doi{10.1088/0004-637X/703/1/L67},
  \href{http://arxiv.org/abs/0908.3187}{{\tt arXiv:0908.3187}}.
%Type = Article
\bibitem[{{Leon} et~al.(2000){Leon}, {Meylan} and {Combes}}]{leon:2000}
\bibinfo{author}{{Leon}, S.}, \bibinfo{author}{{Meylan}, G.},
  \bibinfo{author}{{Combes}, F.}, \bibinfo{year}{2000}.
\newblock \bibinfo{title}{{Tidal tails around 20 Galactic globular clusters.
  Observational evidence for gravitational disk/bulge shocking}}.
\newblock \bibinfo{journal}{\aap} \bibinfo{volume}{359},
  \bibinfo{pages}{907--931}.
\newblock \DOIprefix\doi{10.48550/arXiv.astro-ph/0006100},
  \href{http://arxiv.org/abs/astro-ph/0006100}{{\tt arXiv:astro-ph/0006100}}.
%Type = Article
\bibitem[{{Li} et~al.(2018){Li}, {Yanny} and {Wu}}]{li-yanny:2018}
\bibinfo{author}{{Li}, G.W.}, \bibinfo{author}{{Yanny}, B.},
  \bibinfo{author}{{Wu}, Y.}, \bibinfo{year}{2018}.
\newblock \bibinfo{title}{{GD-1: The Relic of an Old Metal-poor Globular
  Cluster}}.
\newblock \bibinfo{journal}{\apj} \bibinfo{volume}{869}, \bibinfo{pages}{122}.
\newblock \DOIprefix\doi{10.3847/1538-4357/aaed29},
  \href{http://arxiv.org/abs/1811.06427}{{\tt arXiv:1811.06427}}.
%Type = Article
\bibitem[{{Li} et~al.(2022){Li}, {Ji}, {Pace}, {Erkal}, {Koposov}, {Shipp}, {Da
  Costa}, {Cullinane}, {Kuehn}, {Lewis}, {Mackey}, {Simpson}, {Zucker},
  {Ferguson}, {Martell}, {Bland-Hawthorn}, {Balbinot}, {Tavangar},
  {Drlica-Wagner}, {De Silva} and {Simon}}]{li:2022}
{Li}, T.~S. et~al., \bibinfo{year}{2022}.
\newblock \bibinfo{title}{{S $^{5}$: The Orbital and Chemical Properties of One
  Dozen Stellar Streams}}.
\newblock \bibinfo{journal}{\apj} \bibinfo{volume}{928}, \bibinfo{pages}{30}.
\newblock \DOIprefix\doi{10.3847/1538-4357/ac46d3},
  \href{http://arxiv.org/abs/2110.06950}{{\tt arXiv:2110.06950}}.
%Type = Article
\bibitem[{{Li} et~al.(2019a){Li}, {Kaplinghat}, {Bechtol}, {Bolton}, {Bovy},
  {Carleton}, {Chang}, {Drlica-Wagner}, {Erkal}, {Geha}, {Greco}, {Grillmair},
  {Kim}, {Laporte}, {Lewis}, {Makler}, {Mao}, {Marshall}, {McConnachie},
  {Necib}, {Nierenberg}, {Nord}, {Pace}, {Pawlowski}, {Peter}, {Sanderson},
  {Thomas}, {Tollerud}, {Vegetti} and {Walker}}]{li:2019b}
{Li}, T.~S. et~al., \bibinfo{year}{2019}a.
\newblock \bibinfo{title}{{Astrophysical Tests of Dark Matter with Maunakea
  Spectroscopic Explorer}}.
\newblock \bibinfo{journal}{arXiv e-prints} ,
  \bibinfo{pages}{arXiv:1903.03155}\DOIprefix\doi{10.48550/arXiv.1903.03155},
  \href{http://arxiv.org/abs/1903.03155}{{\tt arXiv:1903.03155}}.
%Type = Article
\bibitem[{{Li} et~al.(2021){Li}, {Koposov}, {Erkal}, {Ji}, {Shipp}, {Pace},
  {Hilmi}, {Kuehn}, {Lewis}, {Mackey}, {Simpson}, {Wan}, {Zucker},
  {Bland-Hawthorn}, {Cullinane}, {Da Costa}, {Drlica-Wagner}, {Hattori},
  {Martell}, {Sharma} and {S5 Collaboration}}]{li:2021}
{Li}, T.~S. et~al., \bibinfo{year}{2021}.
\newblock \bibinfo{title}{{Broken into Pieces: ATLAS and Aliqa Uma as One
  Single Stream}}.
\newblock \bibinfo{journal}{\apj} \bibinfo{volume}{911}, \bibinfo{pages}{149}.
\newblock \DOIprefix\doi{10.3847/1538-4357/abeb18},
  \href{http://arxiv.org/abs/2006.10763}{{\tt arXiv:2006.10763}}.
%Type = Article
\bibitem[{{Li} et~al.(2019b){Li}, {Koposov}, {Zucker}, {Lewis}, {Kuehn},
  {Simpson}, {Ji}, {Shipp}, {Mao}, {Geha}, {Pace}, {Mackey}, {Allam}, {Tucker},
  {Da Costa}, {Erkal}, {Simon}, {Mould}, {Martell}, {Wan}, {De Silva},
  {Bechtol}, {Balbinot}, {Belokurov}, {Bland-Hawthorn}, {Casey}, {Cullinane},
  {Drlica-Wagner}, {Sharma}, {Vivas}, {Wechsler}, {Yanny} and {S5
  Collaboration}}]{li:2019}
{Li}, T.~S. et~al., \bibinfo{year}{2019}b.
\newblock \bibinfo{title}{{The southern stellar stream spectroscopic survey
  (S$^{5}$): Overview, target selection, data reduction, validation, and early
  science}}.
\newblock \bibinfo{journal}{\mnras} \bibinfo{volume}{490},
  \bibinfo{pages}{3508--3531}.
\newblock \DOIprefix\doi{10.1093/mnras/stz2731},
  \href{http://arxiv.org/abs/1907.09481}{{\tt arXiv:1907.09481}}.
%Type = Article
\bibitem[{{Lilleengen} et~al.(2023){Lilleengen}, {Petersen}, {Erkal},
  {Pe{\~n}arrubia}, {Koposov}, {Li}, {Cullinane}, {Ji}, {Kuehn}, {Lewis},
  {Mackey}, {Pace}, {Shipp}, {Zucker}, {Bland-Hawthorn}, {Hilmi} and {S5
  Collaboration}}]{lilleengen:2023}
{Lilleengen}, S. et~al., \bibinfo{year}{2023}.
\newblock \bibinfo{title}{{The effect of the deforming dark matter haloes of
  the Milky Way and the Large Magellanic Cloud on the Orphan-Chenab stream}}.
\newblock \bibinfo{journal}{\mnras} \bibinfo{volume}{518},
  \bibinfo{pages}{774--790}.
\newblock \DOIprefix\doi{10.1093/mnras/stac3108},
  \href{http://arxiv.org/abs/2205.01688}{{\tt arXiv:2205.01688}}.
%Type = Article
\bibitem[{{Lilley} et~al.(2018){Lilley}, {Sanders}, {Evans} and
  {Erkal}}]{lilley:2018}
\bibinfo{author}{{Lilley}, E.J.}, \bibinfo{author}{{Sanders}, J.L.},
  \bibinfo{author}{{Evans}, N.W.}, \bibinfo{author}{{Erkal}, D.},
  \bibinfo{year}{2018}.
\newblock \bibinfo{title}{{Galaxy halo expansions: a new biorthogonal family of
  potential-density pairs}}.
\newblock \bibinfo{journal}{\mnras} \bibinfo{volume}{476},
  \bibinfo{pages}{2092--2109}.
\newblock \DOIprefix\doi{10.1093/mnras/sty296},
  \href{http://arxiv.org/abs/1802.03350}{{\tt arXiv:1802.03350}}.
%Type = Article
\bibitem[{{Limberg} et~al.(2023a){Limberg}, {Ji}, {Naidu}, {Chiti}, {Rossi},
  {Usman}, {Ting}, {Zaritsky}, {Bonaca}, {Borbolato}, {Speagle}, {Chandra} and
  {Conroy}}]{limberg:2024}
{Limberg}, G. et~al., \bibinfo{year}{2023}a.
\newblock \bibinfo{title}{{Extending the Chemical Reach of the H3 Survey:
  Detailed Abundances of the Dwarf-galaxy Stellar Stream Wukong/LMS-1}}.
\newblock \bibinfo{journal}{arXiv e-prints} ,
  \bibinfo{pages}{arXiv:2308.13702}\DOIprefix\doi{10.48550/arXiv.2308.13702},
  \href{http://arxiv.org/abs/2308.13702}{{\tt arXiv:2308.13702}}.
%Type = Article
\bibitem[{{Limberg} et~al.(2023b){Limberg}, {Queiroz}, {Perottoni}, {Rossi},
  {Amarante}, {Santucci}, {Chiappini}, {P{\'e}rez-Villegas} and
  {Lee}}]{limberg:2023}
{Limberg}, G. et~al., \bibinfo{year}{2023}b.
\newblock \bibinfo{title}{{Phase-space Properties and Chemistry of the
  Sagittarius Stellar Stream Down to the Extremely Metal-poor ([Fe/H]
  {\ensuremath{\lesssim}} -3) Regime}}.
\newblock \bibinfo{journal}{\apj} \bibinfo{volume}{946}, \bibinfo{pages}{66}.
\newblock \DOIprefix\doi{10.3847/1538-4357/acb694},
  \href{http://arxiv.org/abs/2212.08249}{{\tt arXiv:2212.08249}}.
%Type = Techreport
\bibitem[{Lindegren et~al.(1993)Lindegren, Perryman, Bastian
  et~al.}]{lindegren:1993}
\bibinfo{author}{Lindegren, L.}, \bibinfo{author}{Perryman, M.},
  \bibinfo{author}{Bastian, U.}, et~al., \bibinfo{year}{1993}.
\newblock \bibinfo{title}{GAIA—Proposal for a Cornerstone Mission concept
  submitted to ESA in October 1993}.
\newblock \bibinfo{type}{Technical Report}. Technical Report, Lund.
%Type = Article
\bibitem[{{Lindegren} and {Perryman}(1996)}]{lindegren:1996}
\bibinfo{author}{{Lindegren}, L.}, \bibinfo{author}{{Perryman}, M.A.C.},
  \bibinfo{year}{1996}.
\newblock \bibinfo{title}{{GAIA: Global astrometric interferometer for
  astrophysics.}}
\newblock \bibinfo{journal}{\aaps} \bibinfo{volume}{116},
  \bibinfo{pages}{579--595}.
%Type = Article
\bibitem[{{Lowing} et~al.(2011){Lowing}, {Jenkins}, {Eke} and
  {Frenk}}]{lowing:2011}
\bibinfo{author}{{Lowing}, B.}, \bibinfo{author}{{Jenkins}, A.},
  \bibinfo{author}{{Eke}, V.}, \bibinfo{author}{{Frenk}, C.},
  \bibinfo{year}{2011}.
\newblock \bibinfo{title}{{A halo expansion technique for approximating
  simulated dark matter haloes}}.
\newblock \bibinfo{journal}{\mnras} \bibinfo{volume}{416},
  \bibinfo{pages}{2697--2711}.
\newblock \DOIprefix\doi{10.1111/j.1365-2966.2011.19222.x},
  \href{http://arxiv.org/abs/1010.6197}{{\tt arXiv:1010.6197}}.
%Type = Article
\bibitem[{{Lux} et~al.(2013){Lux}, {Read}, {Lake} and {Johnston}}]{lux:2013}
\bibinfo{author}{{Lux}, H.}, \bibinfo{author}{{Read}, J.I.},
  \bibinfo{author}{{Lake}, G.}, \bibinfo{author}{{Johnston}, K.V.},
  \bibinfo{year}{2013}.
\newblock \bibinfo{title}{{Constraining the Milky Way halo shape using thin
  streams}}.
\newblock \bibinfo{journal}{\mnras} \bibinfo{volume}{436},
  \bibinfo{pages}{2386--2397}.
\newblock \DOIprefix\doi{10.1093/mnras/stt1744},
  \href{http://arxiv.org/abs/1308.2235}{{\tt arXiv:1308.2235}}.
%Type = Article
\bibitem[{{Lynden-Bell} and {Lynden-Bell}(1995)}]{lynden-bell:1995}
\bibinfo{author}{{Lynden-Bell}, D.}, \bibinfo{author}{{Lynden-Bell}, R.M.},
  \bibinfo{year}{1995}.
\newblock \bibinfo{title}{{Ghostly streams from the formation of the Galaxy's
  halo}}.
\newblock \bibinfo{journal}{\mnras} \bibinfo{volume}{275},
  \bibinfo{pages}{429--442}.
\newblock \DOIprefix\doi{10.1093/mnras/275.2.429}.
%Type = Article
\bibitem[{{Macci{\`o}} et~al.(2007){Macci{\`o}}, {Dutton}, {van den Bosch},
  {Moore}, {Potter} and {Stadel}}]{maccio:2007}
\bibinfo{author}{{Macci{\`o}}, A.V.}, \bibinfo{author}{{Dutton}, A.A.},
  \bibinfo{author}{{van den Bosch}, F.C.}, \bibinfo{author}{{Moore}, B.},
  \bibinfo{author}{{Potter}, D.}, \bibinfo{author}{{Stadel}, J.},
  \bibinfo{year}{2007}.
\newblock \bibinfo{title}{{Concentration, spin and shape of dark matter haloes:
  scatter and the dependence on mass and environment}}.
\newblock \bibinfo{journal}{\mnras} \bibinfo{volume}{378},
  \bibinfo{pages}{55--71}.
\newblock \DOIprefix\doi{10.1111/j.1365-2966.2007.11720.x},
  \href{http://arxiv.org/abs/astro-ph/0608157}{{\tt arXiv:astro-ph/0608157}}.
%Type = Article
\bibitem[{{Maiolino} and {Mannucci}(2019)}]{maiolino:2019}
\bibinfo{author}{{Maiolino}, R.}, \bibinfo{author}{{Mannucci}, F.},
  \bibinfo{year}{2019}.
\newblock \bibinfo{title}{{De re metallica: the cosmic chemical evolution of
  galaxies}}.
\newblock \bibinfo{journal}{\aapr} \bibinfo{volume}{27}, \bibinfo{pages}{3}.
\newblock \DOIprefix\doi{10.1007/s00159-018-0112-2},
  \href{http://arxiv.org/abs/1811.09642}{{\tt arXiv:1811.09642}}.
%Type = Article
\bibitem[{{Majewski} et~al.(2004){Majewski}, {Kunkel}, {Law}, {Patterson},
  {Polak}, {Rocha-Pinto}, {Crane}, {Frinchaboy}, {Hummels}, {Johnston}, {Rhee},
  {Skrutskie} and {Weinberg}}]{majewski:2004}
{Majewski}, S.~R. et~al., \bibinfo{year}{2004}.
\newblock \bibinfo{title}{{A Two Micron All Sky Survey View of the Sagittarius
  Dwarf Galaxy. II. Swope Telescope Spectroscopy of M Giant Stars in the
  Dynamically Cold Sagittarius Tidal Stream}}.
\newblock \bibinfo{journal}{\aj} \bibinfo{volume}{128},
  \bibinfo{pages}{245--259}.
\newblock \DOIprefix\doi{10.1086/421372},
  \href{http://arxiv.org/abs/astro-ph/0403701}{{\tt arXiv:astro-ph/0403701}}.
%Type = Article
\bibitem[{{Majewski} et~al.(2017){Majewski}, {Schiavon}, {Frinchaboy}, {Allende
  Prieto}, {Barkhouser}, {Bizyaev}, {Blank}, {Brunner}, {Burton}, {Carrera},
  {Chojnowski}, {Cunha}, {Epstein}, {Fitzgerald}, {Garc{\'\i}a P{\'e}rez},
  {Hearty}, {Henderson}, {Holtzman}, {Johnson}, {Lam}, {Lawler}, {Maseman},
  {M{\'e}sz{\'a}ros}, {Nelson}, {Nguyen}, {Nidever}, {Pinsonneault},
  {Shetrone}, {Smee}, {Smith}, {Stolberg}, {Skrutskie}, {Walker}, {Wilson},
  {Zasowski}, {Anders}, {Basu}, {Beland}, {Blanton}, {Bovy}, {Brownstein},
  {Carlberg}, {Chaplin}, {Chiappini}, {Eisenstein}, {Elsworth}, {Feuillet},
  {Fleming}, {Galbraith-Frew}, {Garc{\'\i}a}, {Garc{\'\i}a-Hern{\'a}ndez},
  {Gillespie}, {Girardi}, {Gunn}, {Hasselquist}, {Hayden}, {Hekker}, {Ivans},
  {Kinemuchi}, {Klaene}, {Mahadevan}, {Mathur}, {Mosser}, {Muna}, {Munn},
  {Nichol}, {O'Connell}, {Parejko}, {Robin}, {Rocha-Pinto}, {Schultheis},
  {Serenelli}, {Shane}, {Silva Aguirre}, {Sobeck}, {Thompson}, {Troup},
  {Weinberg} and {Zamora}}]{majewski:2017}
{Majewski}, S.~R. et~al., \bibinfo{year}{2017}.
\newblock \bibinfo{title}{{The Apache Point Observatory Galactic Evolution
  Experiment (APOGEE)}}.
\newblock \bibinfo{journal}{\aj} \bibinfo{volume}{154}, \bibinfo{pages}{94}.
\newblock \DOIprefix\doi{10.3847/1538-3881/aa784d},
  \href{http://arxiv.org/abs/1509.05420}{{\tt arXiv:1509.05420}}.
%Type = Article
\bibitem[{{Majewski} et~al.(2003){Majewski}, {Skrutskie}, {Weinberg} and
  {Ostheimer}}]{majewski:2003}
\bibinfo{author}{{Majewski}, S.R.}, \bibinfo{author}{{Skrutskie}, M.F.},
  \bibinfo{author}{{Weinberg}, M.D.}, \bibinfo{author}{{Ostheimer}, J.C.},
  \bibinfo{year}{2003}.
\newblock \bibinfo{title}{{A Two Micron All Sky Survey View of the Sagittarius
  Dwarf Galaxy. I. Morphology of the Sagittarius Core and Tidal Arms}}.
\newblock \bibinfo{journal}{\apj} \bibinfo{volume}{599},
  \bibinfo{pages}{1082--1115}.
\newblock \DOIprefix\doi{10.1086/379504},
  \href{http://arxiv.org/abs/astro-ph/0304198}{{\tt arXiv:astro-ph/0304198}}.
%Type = Article
\bibitem[{{Malhan} and {Ibata}(2018)}]{malhan:2018a}
\bibinfo{author}{{Malhan}, K.}, \bibinfo{author}{{Ibata}, R.A.},
  \bibinfo{year}{2018}.
\newblock \bibinfo{title}{{STREAMFINDER - I. A new algorithm for detecting
  stellar streams}}.
\newblock \bibinfo{journal}{\mnras} \bibinfo{volume}{477},
  \bibinfo{pages}{4063--4076}.
\newblock \DOIprefix\doi{10.1093/mnras/sty912},
  \href{http://arxiv.org/abs/1804.11338}{{\tt arXiv:1804.11338}}.
%Type = Article
\bibitem[{{Malhan} et~al.(2019a){Malhan}, {Ibata}, {Carlberg}, {Bellazzini},
  {Famaey} and {Martin}}]{malhan:2019}
\bibinfo{author}{{Malhan}, K.}, \bibinfo{author}{{Ibata}, R.A.},
  \bibinfo{author}{{Carlberg}, R.G.}, \bibinfo{author}{{Bellazzini}, M.},
  \bibinfo{author}{{Famaey}, B.}, \bibinfo{author}{{Martin}, N.F.},
  \bibinfo{year}{2019}a.
\newblock \bibinfo{title}{{Phase-space Correlation in Stellar Streams of the
  Milky Way Halo: The Clash of Kshir and GD-1}}.
\newblock \bibinfo{journal}{\apjl} \bibinfo{volume}{886}, \bibinfo{pages}{L7}.
\newblock \DOIprefix\doi{10.3847/2041-8213/ab530e},
  \href{http://arxiv.org/abs/1911.00009}{{\tt arXiv:1911.00009}}.
%Type = Article
\bibitem[{{Malhan} et~al.(2019b){Malhan}, {Ibata}, {Carlberg}, {Valluri} and
  {Freese}}]{malhan:2019a}
\bibinfo{author}{{Malhan}, K.}, \bibinfo{author}{{Ibata}, R.A.},
  \bibinfo{author}{{Carlberg}, R.G.}, \bibinfo{author}{{Valluri}, M.},
  \bibinfo{author}{{Freese}, K.}, \bibinfo{year}{2019}b.
\newblock \bibinfo{title}{{Butterfly in a Cocoon, Understanding the Origin and
  Morphology of Globular Cluster Streams: The Case of GD-1}}.
\newblock \bibinfo{journal}{\apj} \bibinfo{volume}{881}, \bibinfo{pages}{106}.
\newblock \DOIprefix\doi{10.3847/1538-4357/ab2e07},
  \href{http://arxiv.org/abs/1903.08141}{{\tt arXiv:1903.08141}}.
%Type = Article
\bibitem[{{Malhan} et~al.(2018a){Malhan}, {Ibata}, {Goldman}, {Martin},
  {Magnier} and {Chambers}}]{malhan:2018}
\bibinfo{author}{{Malhan}, K.}, \bibinfo{author}{{Ibata}, R.A.},
  \bibinfo{author}{{Goldman}, B.}, \bibinfo{author}{{Martin}, N.F.},
  \bibinfo{author}{{Magnier}, E.}, \bibinfo{author}{{Chambers}, K.},
  \bibinfo{year}{2018}a.
\newblock \bibinfo{title}{{STREAMFINDER II: A possible fanning structure
  parallel to the GD-1 stream in Pan-STARRS1}}.
\newblock \bibinfo{journal}{\mnras} \bibinfo{volume}{478},
  \bibinfo{pages}{3862--3870}.
\newblock \DOIprefix\doi{10.1093/mnras/sty1338},
  \href{http://arxiv.org/abs/1805.08205}{{\tt arXiv:1805.08205}}.
%Type = Article
\bibitem[{{Malhan} et~al.(2018b){Malhan}, {Ibata} and {Martin}}]{malhan:2018b}
\bibinfo{author}{{Malhan}, K.}, \bibinfo{author}{{Ibata}, R.A.},
  \bibinfo{author}{{Martin}, N.F.}, \bibinfo{year}{2018}b.
\newblock \bibinfo{title}{{Ghostly tributaries to the Milky Way: charting the
  halo's stellar streams with the Gaia DR2 catalogue}}.
\newblock \bibinfo{journal}{\mnras} \bibinfo{volume}{481},
  \bibinfo{pages}{3442--3455}.
\newblock \DOIprefix\doi{10.1093/mnras/sty2474},
  \href{http://arxiv.org/abs/1804.11339}{{\tt arXiv:1804.11339}}.
%Type = Article
\bibitem[{{Malhan} et~al.(2022a){Malhan}, {Ibata}, {Sharma}, {Famaey},
  {Bellazzini}, {Carlberg}, {D'Souza}, {Yuan}, {Martin} and
  {Thomas}}]{malhan:2022}
{Malhan}, K. et~al., \bibinfo{year}{2022}a.
\newblock \bibinfo{title}{{The Global Dynamical Atlas of the Milky Way Mergers:
  Constraints from Gaia EDR3-based Orbits of Globular Clusters, Stellar
  Streams, and Satellite Galaxies}}.
\newblock \bibinfo{journal}{\apj} \bibinfo{volume}{926}, \bibinfo{pages}{107}.
\newblock \DOIprefix\doi{10.3847/1538-4357/ac4d2a},
  \href{http://arxiv.org/abs/2202.07660}{{\tt arXiv:2202.07660}}.
%Type = Article
\bibitem[{{Malhan} et~al.(2021a){Malhan}, {Valluri} and {Freese}}]{malhan:2021}
\bibinfo{author}{{Malhan}, K.}, \bibinfo{author}{{Valluri}, M.},
  \bibinfo{author}{{Freese}, K.}, \bibinfo{year}{2021}a.
\newblock \bibinfo{title}{{Probing the nature of dark matter with accreted
  globular cluster streams}}.
\newblock \bibinfo{journal}{\mnras} \bibinfo{volume}{501},
  \bibinfo{pages}{179--200}.
\newblock \DOIprefix\doi{10.1093/mnras/staa3597},
  \href{http://arxiv.org/abs/2005.12919}{{\tt arXiv:2005.12919}}.
%Type = Article
\bibitem[{{Malhan} et~al.(2022b){Malhan}, {Valluri}, {Freese} and
  {Ibata}}]{malhan:2022b}
\bibinfo{author}{{Malhan}, K.}, \bibinfo{author}{{Valluri}, M.},
  \bibinfo{author}{{Freese}, K.}, \bibinfo{author}{{Ibata}, R.A.},
  \bibinfo{year}{2022}b.
\newblock \bibinfo{title}{{New Constraints on the Dark Matter Density Profiles
  of Dwarf Galaxies from Proper Motions of Globular Cluster Streams}}.
\newblock \bibinfo{journal}{\apjl} \bibinfo{volume}{941}, \bibinfo{pages}{L38}.
\newblock \DOIprefix\doi{10.3847/2041-8213/aca6e5},
  \href{http://arxiv.org/abs/2201.03571}{{\tt arXiv:2201.03571}}.
%Type = Article
\bibitem[{{Malhan} et~al.(2021b){Malhan}, {Yuan}, {Ibata}, {Arentsen},
  {Bellazzini} and {Martin}}]{malhan:2021b}
\bibinfo{author}{{Malhan}, K.}, \bibinfo{author}{{Yuan}, Z.},
  \bibinfo{author}{{Ibata}, R.A.}, \bibinfo{author}{{Arentsen}, A.},
  \bibinfo{author}{{Bellazzini}, M.}, \bibinfo{author}{{Martin}, N.F.},
  \bibinfo{year}{2021}b.
\newblock \bibinfo{title}{{Evidence of a Dwarf Galaxy Stream Populating the
  Inner Milky Way Halo}}.
\newblock \bibinfo{journal}{\apj} \bibinfo{volume}{920}, \bibinfo{pages}{51}.
\newblock \DOIprefix\doi{10.3847/1538-4357/ac1675},
  \href{http://arxiv.org/abs/2104.09523}{{\tt arXiv:2104.09523}}.
%Type = Article
\bibitem[{Mallat(2012)}]{mallat:2012}
\bibinfo{author}{Mallat, S.}, \bibinfo{year}{2012}.
\newblock \bibinfo{title}{Group invariant scattering}.
\newblock \bibinfo{journal}{Communications on Pure and Applied Mathematics}
  \bibinfo{volume}{65}, \bibinfo{pages}{1331--1398}.
\newblock \URLprefix
  \url{https://onlinelibrary.wiley.com/doi/abs/10.1002/cpa.21413},
  \DOIprefix\doi{https://doi.org/10.1002/cpa.21413},
  \href{http://arxiv.org/abs/https://onlinelibrary.wiley.com/doi/pdf/10.1002/cpa.21413}{{\tt
  arXiv:https://onlinelibrary.wiley.com/doi/pdf/10.1002/cpa.21413}}.
%Type = Article
\bibitem[{{Mansfield} et~al.(2023){Mansfield}, {Darragh-Ford}, {Wang}, {Nadler}
  and {Wechsler}}]{mansfield:2023}
\bibinfo{author}{{Mansfield}, P.}, \bibinfo{author}{{Darragh-Ford}, E.},
  \bibinfo{author}{{Wang}, Y.}, \bibinfo{author}{{Nadler}, E.O.},
  \bibinfo{author}{{Wechsler}, R.H.}, \bibinfo{year}{2023}.
\newblock \bibinfo{title}{{Symfind: Addressing the Fragility of Subhalo Finders
  and Revealing the Durability of Subhalos}}.
\newblock \bibinfo{journal}{arXiv e-prints} ,
  \bibinfo{pages}{arXiv:2308.10926}\DOIprefix\doi{10.48550/arXiv.2308.10926},
  \href{http://arxiv.org/abs/2308.10926}{{\tt arXiv:2308.10926}}.
%Type = Article
\bibitem[{{Martin} et~al.(2018){Martin}, {Amy}, {Newberg}, {Shelton}, {Carlin},
  {Beers}, {Denissenkov} and {Willett}}]{martin:2018}
\bibinfo{author}{{Martin}, C.}, \bibinfo{author}{{Amy}, P.M.},
  \bibinfo{author}{{Newberg}, H.J.}, \bibinfo{author}{{Shelton}, S.},
  \bibinfo{author}{{Carlin}, J.L.}, \bibinfo{author}{{Beers}, T.C.},
  \bibinfo{author}{{Denissenkov}, P.}, \bibinfo{author}{{Willett}, B.A.},
  \bibinfo{year}{2018}.
\newblock \bibinfo{title}{{An orbit fit to likely Hermus Stream stars}}.
\newblock \bibinfo{journal}{\mnras} \bibinfo{volume}{477},
  \bibinfo{pages}{2419--2430}.
\newblock \DOIprefix\doi{10.1093/mnras/sty608},
  \href{http://arxiv.org/abs/1903.02414}{{\tt arXiv:1903.02414}}.
%Type = Article
\bibitem[{{Martin} et~al.(2013){Martin}, {Carlin}, {Newberg} and
  {Grillmair}}]{martin:2013}
\bibinfo{author}{{Martin}, C.}, \bibinfo{author}{{Carlin}, J.L.},
  \bibinfo{author}{{Newberg}, H.J.}, \bibinfo{author}{{Grillmair}, C.},
  \bibinfo{year}{2013}.
\newblock \bibinfo{title}{{Kinematic Discovery of a Stellar Stream Located in
  Pisces}}.
\newblock \bibinfo{journal}{\apjl} \bibinfo{volume}{765}, \bibinfo{pages}{L39}.
\newblock \DOIprefix\doi{10.1088/2041-8205/765/2/L39},
  \href{http://arxiv.org/abs/1302.2155}{{\tt arXiv:1302.2155}}.
%Type = Article
\bibitem[{{Martin} et~al.(2014){Martin}, {Ibata}, {Rich}, {Collins}, {Fardal},
  {Irwin}, {Lewis}, {McConnachie}, {Babul}, {Bate}, {Chapman}, {Conn},
  {Crnojevi{\'c}}, {Ferguson}, {Mackey}, {Navarro}, {Pe{\~n}arrubia}, {Tanvir}
  and {Valls-Gabaud}}]{martin:2014}
{Martin}, N.~F. et~al., \bibinfo{year}{2014}.
\newblock \bibinfo{title}{{The PAndAS Field of Streams: Stellar Structures in
  the Milky Way Halo toward Andromeda and Triangulum}}.
\newblock \bibinfo{journal}{\apj} \bibinfo{volume}{787}, \bibinfo{pages}{19}.
\newblock \DOIprefix\doi{10.1088/0004-637X/787/1/19},
  \href{http://arxiv.org/abs/1403.4945}{{\tt arXiv:1403.4945}}.
%Type = Article
\bibitem[{{Martin} et~al.(2022a){Martin}, {Ibata}, {Starkenburg}, {Yuan},
  {Malhan}, {Bellazzini}, {Viswanathan}, {Aguado}, {Arentsen}, {Bonifacio},
  {Carlberg}, {Gonz{\'a}lez Hern{\'a}ndez}, {Hill}, {Jablonka}, {Kordopatis},
  {Lardo}, {McConnachie}, {Navarro}, {S{\'a}nchez-Janssen}, {Sestito},
  {Thomas}, {Venn}, {Vitali} and {Voggel}}]{martin:2022b}
{Martin}, N.~F. et~al., \bibinfo{year}{2022}a.
\newblock \bibinfo{title}{{The Pristine survey - XVI. The metallicity of 26
  stellar streams around the Milky Way detected with the STREAMFINDER in Gaia
  EDR3}}.
\newblock \bibinfo{journal}{\mnras} \bibinfo{volume}{516},
  \bibinfo{pages}{5331--5354}.
\newblock \DOIprefix\doi{10.1093/mnras/stac2426},
  \href{http://arxiv.org/abs/2201.01310}{{\tt arXiv:2201.01310}}.
%Type = Article
\bibitem[{{Martin} et~al.(2022b){Martin}, {Venn}, {Aguado}, {Starkenburg},
  {Gonz{\'a}lez Hern{\'a}ndez}, {Ibata}, {Bonifacio}, {Caffau}, {Sestito},
  {Arentsen}, {Allende Prieto}, {Carlberg}, {Fabbro}, {Fouesneau}, {Hill},
  {Jablonka}, {Kordopatis}, {Lardo}, {Malhan}, {Mashonkina}, {McConnachie},
  {Navarro}, {S{\'a}nchez-Janssen}, {Thomas}, {Yuan} and
  {Mucciarelli}}]{martin:2022a}
{Martin}, N.~F. et~al., \bibinfo{year}{2022}b.
\newblock \bibinfo{title}{{A stellar stream remnant of a globular cluster below
  the metallicity floor}}.
\newblock \bibinfo{journal}{\nat} \bibinfo{volume}{601},
  \bibinfo{pages}{45--48}.
\newblock \DOIprefix\doi{10.1038/s41586-021-04162-2},
  \href{http://arxiv.org/abs/2201.01309}{{\tt arXiv:2201.01309}}.
%Type = Article
\bibitem[{{Mart{\'\i}nez-Delgado} et~al.(2023){Mart{\'\i}nez-Delgado},
  {Cooper}, {Rom{\'a}n}, {Pillepich}, {Erkal}, {Pearson}, {Moustakas},
  {Laporte}, {Laine}, {Akhlaghi}, {Lang}, {Makarov}, {Borlaff}, {Donatiello},
  {Pearson}, {Mir{\'o}-Carretero}, {Cuillandre}, {Dom{\'\i}nguez},
  {Roca-F{\`a}brega}, {Frenk}, {Schmidt}, {G{\'o}mez-Flechoso}, {Guzman},
  {Libeskind}, {Dey}, {Weaver}, {Schlegel}, {Myers} and
  {Valdes}}]{martinez-delgado:2023}
{Mart{\'\i}nez-Delgado}, D. et~al., \bibinfo{year}{2023}.
\newblock \bibinfo{title}{{Hidden depths in the local Universe: The Stellar
  Stream Legacy Survey}}.
\newblock \bibinfo{journal}{\aap} \bibinfo{volume}{671}, \bibinfo{pages}{A141}.
\newblock \DOIprefix\doi{10.1051/0004-6361/202245011},
  \href{http://arxiv.org/abs/2104.06071}{{\tt arXiv:2104.06071}}.
%Type = Article
\bibitem[{{Mart{\'\i}nez-Delgado} et~al.(2010){Mart{\'\i}nez-Delgado},
  {Gabany}, {Crawford}, {Zibetti}, {Majewski}, {Rix}, {Fliri},
  {Carballo-Bello}, {Bardalez-Gagliuffi}, {Pe{\~n}arrubia}, {Chonis}, {Madore},
  {Trujillo}, {Schirmer} and {McDavid}}]{martinez-delgado:2010}
{Mart{\'\i}nez-Delgado}, D. et~al., \bibinfo{year}{2010}.
\newblock \bibinfo{title}{{Stellar Tidal Streams in Spiral Galaxies of the
  Local Volume: A Pilot Survey with Modest Aperture Telescopes}}.
\newblock \bibinfo{journal}{\aj} \bibinfo{volume}{140},
  \bibinfo{pages}{962--967}.
\newblock \DOIprefix\doi{10.1088/0004-6256/140/4/962},
  \href{http://arxiv.org/abs/1003.4860}{{\tt arXiv:1003.4860}}.
%Type = Article
\bibitem[{{Mastrobuono-Battisti} et~al.(2012){Mastrobuono-Battisti}, {Di
  Matteo}, {Montuori} and {Haywood}}]{mastrobuono-battisti:2012}
\bibinfo{author}{{Mastrobuono-Battisti}, A.}, \bibinfo{author}{{Di Matteo},
  P.}, \bibinfo{author}{{Montuori}, M.}, \bibinfo{author}{{Haywood}, M.},
  \bibinfo{year}{2012}.
\newblock \bibinfo{title}{{Clumpy streams in a smooth dark halo: the case of
  Palomar 5}}.
\newblock \bibinfo{journal}{\aap} \bibinfo{volume}{546}, \bibinfo{pages}{L7}.
\newblock \DOIprefix\doi{10.1051/0004-6361/201219563},
  \href{http://arxiv.org/abs/1209.0466}{{\tt arXiv:1209.0466}}.
%Type = Article
\bibitem[{{Mateu}(2023)}]{mateu:2023}
\bibinfo{author}{{Mateu}, C.}, \bibinfo{year}{2023}.
\newblock \bibinfo{title}{{galstreams: A library of Milky Way stellar stream
  footprints and tracks}}.
\newblock \bibinfo{journal}{\mnras} \bibinfo{volume}{520},
  \bibinfo{pages}{5225--5258}.
\newblock \DOIprefix\doi{10.1093/mnras/stad321},
  \href{http://arxiv.org/abs/2204.10326}{{\tt arXiv:2204.10326}}.
%Type = Article
\bibitem[{{Mateu} et~al.(2011){Mateu}, {Bruzual}, {Aguilar}, {Brown},
  {Valenzuela}, {Carigi}, {Vel{\'a}zquez} and {Hern{\'a}ndez}}]{mateu:2011}
\bibinfo{author}{{Mateu}, C.}, \bibinfo{author}{{Bruzual}, G.},
  \bibinfo{author}{{Aguilar}, L.}, \bibinfo{author}{{Brown}, A.G.A.},
  \bibinfo{author}{{Valenzuela}, O.}, \bibinfo{author}{{Carigi}, L.},
  \bibinfo{author}{{Vel{\'a}zquez}, H.}, \bibinfo{author}{{Hern{\'a}ndez}, F.},
  \bibinfo{year}{2011}.
\newblock \bibinfo{title}{{Detection of satellite remnants in the Galactic Halo
  with Gaia - II. A modified great circle cell method}}.
\newblock \bibinfo{journal}{\mnras} \bibinfo{volume}{415},
  \bibinfo{pages}{214--224}.
\newblock \DOIprefix\doi{10.1111/j.1365-2966.2011.18690.x},
  \href{http://arxiv.org/abs/1103.0283}{{\tt arXiv:1103.0283}}.
%Type = Article
\bibitem[{{Mateu} et~al.(2018){Mateu}, {Read} and {Kawata}}]{mateu:2018}
\bibinfo{author}{{Mateu}, C.}, \bibinfo{author}{{Read}, J.I.},
  \bibinfo{author}{{Kawata}, D.}, \bibinfo{year}{2018}.
\newblock \bibinfo{title}{{Fourteen candidate RR Lyrae star streams in the
  inner Galaxy}}.
\newblock \bibinfo{journal}{\mnras} \bibinfo{volume}{474},
  \bibinfo{pages}{4112--4129}.
\newblock \DOIprefix\doi{10.1093/mnras/stx2937},
  \href{http://arxiv.org/abs/1711.03967}{{\tt arXiv:1711.03967}}.
%Type = Article
\bibitem[{{McConnachie} et~al.(2009){McConnachie}, {Irwin}, {Ibata},
  {Dubinski}, {Widrow}, {Martin}, {C{\^o}t{\'e}}, {Dotter}, {Navarro},
  {Ferguson}, {Puzia}, {Lewis}, {Babul}, {Barmby}, {Bienaym{\'e}}, {Chapman},
  {Cockcroft}, {Collins}, {Fardal}, {Harris}, {Huxor}, {Mackey},
  {Pe{\~n}arrubia}, {Rich}, {Richer}, {Siebert}, {Tanvir}, {Valls-Gabaud} and
  {Venn}}]{mcconnachie:2009}
{McConnachie}, A.~W. et~al., \bibinfo{year}{2009}.
\newblock \bibinfo{title}{{The remnants of galaxy formation from a panoramic
  survey of the region around M31}}.
\newblock \bibinfo{journal}{\nat} \bibinfo{volume}{461},
  \bibinfo{pages}{66--69}.
\newblock \DOIprefix\doi{10.1038/nature08327},
  \href{http://arxiv.org/abs/0909.0398}{{\tt arXiv:0909.0398}}.
%Type = Article
\bibitem[{{McMillan}(2017)}]{mcmillan:2017}
\bibinfo{author}{{McMillan}, P.J.}, \bibinfo{year}{2017}.
\newblock \bibinfo{title}{{The mass distribution and gravitational potential of
  the Milky Way}}.
\newblock \bibinfo{journal}{\mnras} \bibinfo{volume}{465},
  \bibinfo{pages}{76--94}.
\newblock \DOIprefix\doi{10.1093/mnras/stw2759},
  \href{http://arxiv.org/abs/1608.00971}{{\tt arXiv:1608.00971}}.
%Type = Article
\bibitem[{{Meingast} and {Alves}(2019)}]{meingast:2019}
\bibinfo{author}{{Meingast}, S.}, \bibinfo{author}{{Alves}, J.},
  \bibinfo{year}{2019}.
\newblock \bibinfo{title}{{Extended stellar systems in the solar neighborhood.
  I. The tidal tails of the Hyades}}.
\newblock \bibinfo{journal}{\aap} \bibinfo{volume}{621}, \bibinfo{pages}{L3}.
\newblock \DOIprefix\doi{10.1051/0004-6361/201834622},
  \href{http://arxiv.org/abs/1811.04931}{{\tt arXiv:1811.04931}}.
%Type = Article
\bibitem[{{Mendelsohn} et~al.(2022){Mendelsohn}, {Newberg}, {Shelton},
  {Widrow}, {Thompson} and {Grillmair}}]{mendelsohn:2022}
\bibinfo{author}{{Mendelsohn}, E.J.}, \bibinfo{author}{{Newberg}, H.J.},
  \bibinfo{author}{{Shelton}, S.}, \bibinfo{author}{{Widrow}, L.M.},
  \bibinfo{author}{{Thompson}, J.M.}, \bibinfo{author}{{Grillmair}, C.J.},
  \bibinfo{year}{2022}.
\newblock \bibinfo{title}{{Estimate of the Mass and Radial Profile of the
  Orphan-Chenab Stream's Dwarf-galaxy Progenitor Using MilkyWay@home}}.
\newblock \bibinfo{journal}{\apj} \bibinfo{volume}{926}, \bibinfo{pages}{106}.
\newblock \DOIprefix\doi{10.3847/1538-4357/ac498a},
  \href{http://arxiv.org/abs/2201.03637}{{\tt arXiv:2201.03637}}.
%Type = Article
\bibitem[{{Mestre} et~al.(2020){Mestre}, {Llinares} and
  {Carpintero}}]{mestre:2020}
\bibinfo{author}{{Mestre}, M.}, \bibinfo{author}{{Llinares}, C.},
  \bibinfo{author}{{Carpintero}, D.D.}, \bibinfo{year}{2020}.
\newblock \bibinfo{title}{{Effects of chaos on the detectability of stellar
  streams}}.
\newblock \bibinfo{journal}{\mnras} \bibinfo{volume}{492},
  \bibinfo{pages}{4398--4408}.
\newblock \DOIprefix\doi{10.1093/mnras/stz3505},
  \href{http://arxiv.org/abs/1912.05592}{{\tt arXiv:1912.05592}}.
%Type = Article
\bibitem[{{Mirabal} and {Bonaca}(2021)}]{mirabal:2021}
\bibinfo{author}{{Mirabal}, N.}, \bibinfo{author}{{Bonaca}, A.},
  \bibinfo{year}{2021}.
\newblock \bibinfo{title}{{Machine-learned dark matter subhalo candidates in
  the 4FGL-DR2: search for the perturber of the GD-1 stream}}.
\newblock \bibinfo{journal}{\jcap} \bibinfo{volume}{2021},
  \bibinfo{pages}{033}.
\newblock \DOIprefix\doi{10.1088/1475-7516/2021/11/033},
  \href{http://arxiv.org/abs/2105.12131}{{\tt arXiv:2105.12131}}.
%Type = Article
\bibitem[{{Molin{\'e}} et~al.(2017){Molin{\'e}}, {S{\'a}nchez-Conde},
  {Palomares-Ruiz} and {Prada}}]{moline:2017}
\bibinfo{author}{{Molin{\'e}}, {\'A}.}, \bibinfo{author}{{S{\'a}nchez-Conde},
  M.A.}, \bibinfo{author}{{Palomares-Ruiz}, S.}, \bibinfo{author}{{Prada}, F.},
  \bibinfo{year}{2017}.
\newblock \bibinfo{title}{{Characterization of subhalo structural properties
  and implications for dark matter annihilation signals}}.
\newblock \bibinfo{journal}{\mnras} \bibinfo{volume}{466},
  \bibinfo{pages}{4974--4990}.
\newblock \DOIprefix\doi{10.1093/mnras/stx026},
  \href{http://arxiv.org/abs/1603.04057}{{\tt arXiv:1603.04057}}.
%Type = Article
\bibitem[{{Monaco} et~al.(2007){Monaco}, {Bellazzini}, {Bonifacio}, {Buzzoni},
  {Ferraro}, {Marconi}, {Sbordone} and {Zaggia}}]{monaco:2007}
\bibinfo{author}{{Monaco}, L.}, \bibinfo{author}{{Bellazzini}, M.},
  \bibinfo{author}{{Bonifacio}, P.}, \bibinfo{author}{{Buzzoni}, A.},
  \bibinfo{author}{{Ferraro}, F.R.}, \bibinfo{author}{{Marconi}, G.},
  \bibinfo{author}{{Sbordone}, L.}, \bibinfo{author}{{Zaggia}, S.},
  \bibinfo{year}{2007}.
\newblock \bibinfo{title}{{High-resolution spectroscopy of RGB stars in the
  Sagittarius streams. I. Radial velocities and chemical abundances}}.
\newblock \bibinfo{journal}{\aap} \bibinfo{volume}{464},
  \bibinfo{pages}{201--209}.
\newblock \DOIprefix\doi{10.1051/0004-6361:20066228},
  \href{http://arxiv.org/abs/astro-ph/0611070}{{\tt arXiv:astro-ph/0611070}}.
%Type = Article
\bibitem[{{Morrison} et~al.(2000){Morrison}, {Mateo}, {Olszewski}, {Harding},
  {Dohm-Palmer}, {Freeman}, {Norris} and {Morita}}]{morrison:2000}
\bibinfo{author}{{Morrison}, H.L.}, \bibinfo{author}{{Mateo}, M.},
  \bibinfo{author}{{Olszewski}, E.W.}, \bibinfo{author}{{Harding}, P.},
  \bibinfo{author}{{Dohm-Palmer}, R.C.}, \bibinfo{author}{{Freeman}, K.C.},
  \bibinfo{author}{{Norris}, J.E.}, \bibinfo{author}{{Morita}, M.},
  \bibinfo{year}{2000}.
\newblock \bibinfo{title}{{Mapping the Galactic Halo. I. The ``Spaghetti''
  Survey}}.
\newblock \bibinfo{journal}{\aj} \bibinfo{volume}{119},
  \bibinfo{pages}{2254--2273}.
\newblock \DOIprefix\doi{10.1086/301357},
  \href{http://arxiv.org/abs/astro-ph/0001492}{{\tt arXiv:astro-ph/0001492}}.
%Type = Article
\bibitem[{{Moster} et~al.(2010){Moster}, {Somerville}, {Maulbetsch}, {van den
  Bosch}, {Macci{\`o}}, {Naab} and {Oser}}]{moster:2010}
\bibinfo{author}{{Moster}, B.P.}, \bibinfo{author}{{Somerville}, R.S.},
  \bibinfo{author}{{Maulbetsch}, C.}, \bibinfo{author}{{van den Bosch}, F.C.},
  \bibinfo{author}{{Macci{\`o}}, A.V.}, \bibinfo{author}{{Naab}, T.},
  \bibinfo{author}{{Oser}, L.}, \bibinfo{year}{2010}.
\newblock \bibinfo{title}{{Constraints on the Relationship between Stellar Mass
  and Halo Mass at Low and High Redshift}}.
\newblock \bibinfo{journal}{\apj} \bibinfo{volume}{710},
  \bibinfo{pages}{903--923}.
\newblock \DOIprefix\doi{10.1088/0004-637X/710/2/903},
  \href{http://arxiv.org/abs/0903.4682}{{\tt arXiv:0903.4682}}.
%Type = Article
\bibitem[{{Munn} et~al.(2004){Munn}, {Monet}, {Levine}, {Canzian}, {Pier},
  {Harris}, {Lupton}, {Ivezi{\'c}}, {Hindsley}, {Hennessy}, {Schneider} and
  {Brinkmann}}]{munn:2004}
{Munn}, J.~A. et~al., \bibinfo{year}{2004}.
\newblock \bibinfo{title}{{An Improved Proper-Motion Catalog Combining USNO-B
  and the Sloan Digital Sky Survey}}.
\newblock \bibinfo{journal}{\aj} \bibinfo{volume}{127},
  \bibinfo{pages}{3034--3042}.
\newblock \DOIprefix\doi{10.1086/383292}.
%Type = Article
\bibitem[{{Myeong} et~al.(2017){Myeong}, {Evans}, {Belokurov}, {Koposov} and
  {Sanders}}]{myeong:2017}
\bibinfo{author}{{Myeong}, G.C.}, \bibinfo{author}{{Evans}, N.W.},
  \bibinfo{author}{{Belokurov}, V.}, \bibinfo{author}{{Koposov}, S.E.},
  \bibinfo{author}{{Sanders}, J.L.}, \bibinfo{year}{2017}.
\newblock \bibinfo{title}{{A halo substructure in Gaia Data Release 1}}.
\newblock \bibinfo{journal}{\mnras} \bibinfo{volume}{469},
  \bibinfo{pages}{L78--L82}.
\newblock \DOIprefix\doi{10.1093/mnrasl/slx051},
  \href{http://arxiv.org/abs/1704.01363}{{\tt arXiv:1704.01363}}.
%Type = Article
\bibitem[{{Myeong} et~al.(2019){Myeong}, {Vasiliev}, {Iorio}, {Evans} and
  {Belokurov}}]{myeong:2019}
\bibinfo{author}{{Myeong}, G.C.}, \bibinfo{author}{{Vasiliev}, E.},
  \bibinfo{author}{{Iorio}, G.}, \bibinfo{author}{{Evans}, N.W.},
  \bibinfo{author}{{Belokurov}, V.}, \bibinfo{year}{2019}.
\newblock \bibinfo{title}{{Evidence for two early accretion events that built
  the Milky Way stellar halo}}.
\newblock \bibinfo{journal}{\mnras} \bibinfo{volume}{488},
  \bibinfo{pages}{1235--1247}.
\newblock \DOIprefix\doi{10.1093/mnras/stz1770},
  \href{http://arxiv.org/abs/1904.03185}{{\tt arXiv:1904.03185}}.
%Type = Article
\bibitem[{{Nadler} et~al.(2023){Nadler}, {Mansfield}, {Wang}, {Du}, {Adhikari},
  {Banerjee}, {Benson}, {Darragh-Ford}, {Mao}, {Wagner-Carena}, {Wechsler} and
  {Wu}}]{nadler:2023}
{Nadler}, E.~O. et~al., \bibinfo{year}{2023}.
\newblock \bibinfo{title}{{Symphony: Cosmological Zoom-in Simulation Suites
  over Four Decades of Host Halo Mass}}.
\newblock \bibinfo{journal}{\apj} \bibinfo{volume}{945}, \bibinfo{pages}{159}.
\newblock \DOIprefix\doi{10.3847/1538-4357/acb68c},
  \href{http://arxiv.org/abs/2209.02675}{{\tt arXiv:2209.02675}}.
%Type = Article
\bibitem[{{Nadler} et~al.(2020){Nadler}, {Wechsler}, {Bechtol}, {Mao}, {Green},
  {Drlica-Wagner}, {McNanna}, {Mau}, {Pace}, {Simon}, {Kravtsov}, {Dodelson},
  {Li}, {Riley}, {Wang}, {Abbott}, {Aguena}, {Allam}, {Annis}, {Avila},
  {Bernstein}, {Bertin}, {Brooks}, {Burke}, {Rosell}, {Kind}, {Carretero},
  {Costanzi}, {da Costa}, {De Vicente}, {Desai}, {Evrard}, {Flaugher},
  {Fosalba}, {Frieman}, {Garc{\'\i}a-Bellido}, {Gaztanaga}, {Gerdes}, {Gruen},
  {Gschwend}, {Gutierrez}, {Hartley}, {Hinton}, {Honscheid}, {Krause}, {Kuehn},
  {Kuropatkin}, {Lahav}, {Maia}, {Marshall}, {Menanteau}, {Miquel}, {Palmese},
  {Paz-Chinch{\'o}n}, {Plazas}, {Romer}, {Sanchez}, {Santiago}, {Scarpine},
  {Serrano}, {Smith}, {Soares-Santos}, {Suchyta}, {Tarle}, {Thomas}, {Varga},
  {Walker} and {DES Collaboration}}]{nadler:2020}
{Nadler}, E.~O. et~al., \bibinfo{year}{2020}.
\newblock \bibinfo{title}{{Milky Way Satellite Census. II. Galaxy-Halo
  Connection Constraints Including the Impact of the Large Magellanic Cloud}}.
\newblock \bibinfo{journal}{\apj} \bibinfo{volume}{893}, \bibinfo{pages}{48}.
\newblock \DOIprefix\doi{10.3847/1538-4357/ab846a},
  \href{http://arxiv.org/abs/1912.03303}{{\tt arXiv:1912.03303}}.
%Type = Article
\bibitem[{{Naidu} et~al.(2020){Naidu}, {Conroy}, {Bonaca}, {Johnson}, {Ting},
  {Caldwell}, {Zaritsky} and {Cargile}}]{naidu:2020}
\bibinfo{author}{{Naidu}, R.P.}, \bibinfo{author}{{Conroy}, C.},
  \bibinfo{author}{{Bonaca}, A.}, \bibinfo{author}{{Johnson}, B.D.},
  \bibinfo{author}{{Ting}, Y.S.}, \bibinfo{author}{{Caldwell}, N.},
  \bibinfo{author}{{Zaritsky}, D.}, \bibinfo{author}{{Cargile}, P.A.},
  \bibinfo{year}{2020}.
\newblock \bibinfo{title}{{Evidence from the H3 Survey That the Stellar Halo Is
  Entirely Comprised of Substructure}}.
\newblock \bibinfo{journal}{\apj} \bibinfo{volume}{901}, \bibinfo{pages}{48}.
\newblock \DOIprefix\doi{10.3847/1538-4357/abaef4},
  \href{http://arxiv.org/abs/2006.08625}{{\tt arXiv:2006.08625}}.
%Type = Article
\bibitem[{{Naidu} et~al.(2021){Naidu}, {Conroy}, {Bonaca}, {Zaritsky},
  {Weinberger}, {Ting}, {Caldwell}, {Tacchella}, {Han}, {Speagle} and
  {Cargile}}]{naidu:2021}
{Naidu}, R.~P. et~al., \bibinfo{year}{2021}.
\newblock \bibinfo{title}{{Reconstructing the Last Major Merger of the Milky
  Way with the H3 Survey}}.
\newblock \bibinfo{journal}{\apj} \bibinfo{volume}{923}, \bibinfo{pages}{92}.
\newblock \DOIprefix\doi{10.3847/1538-4357/ac2d2d},
  \href{http://arxiv.org/abs/2103.03251}{{\tt arXiv:2103.03251}}.
%Type = Article
\bibitem[{{Navarrete} et~al.(2017){Navarrete}, {Belokurov} and
  {Koposov}}]{navarrete:2017}
\bibinfo{author}{{Navarrete}, C.}, \bibinfo{author}{{Belokurov}, V.},
  \bibinfo{author}{{Koposov}, S.E.}, \bibinfo{year}{2017}.
\newblock \bibinfo{title}{{The Discovery of Tidal Tails around the Globular
  Cluster NGC 7492 with Pan-STARRS1}}.
\newblock \bibinfo{journal}{\apjl} \bibinfo{volume}{841}, \bibinfo{pages}{L23}.
\newblock \DOIprefix\doi{10.3847/2041-8213/aa72e1},
  \href{http://arxiv.org/abs/1705.04324}{{\tt arXiv:1705.04324}}.
%Type = Article
\bibitem[{{Necib} et~al.(2020a){Necib}, {Ostdiek}, {Lisanti}, {Cohen},
  {Freytsis} and {Garrison-Kimmel}}]{necib:2020}
\bibinfo{author}{{Necib}, L.}, \bibinfo{author}{{Ostdiek}, B.},
  \bibinfo{author}{{Lisanti}, M.}, \bibinfo{author}{{Cohen}, T.},
  \bibinfo{author}{{Freytsis}, M.}, \bibinfo{author}{{Garrison-Kimmel}, S.},
  \bibinfo{year}{2020}a.
\newblock \bibinfo{title}{{Chasing Accreted Structures within Gaia DR2 Using
  Deep Learning}}.
\newblock \bibinfo{journal}{\apj} \bibinfo{volume}{903}, \bibinfo{pages}{25}.
\newblock \DOIprefix\doi{10.3847/1538-4357/abb814},
  \href{http://arxiv.org/abs/1907.07681}{{\tt arXiv:1907.07681}}.
%Type = Article
\bibitem[{{Necib} et~al.(2020b){Necib}, {Ostdiek}, {Lisanti}, {Cohen},
  {Freytsis}, {Garrison-Kimmel}, {Hopkins}, {Wetzel} and
  {Sanderson}}]{necib:2020b}
{Necib}, L. et~al., \bibinfo{year}{2020}b.
\newblock \bibinfo{title}{{Evidence for a vast prograde stellar stream in the
  solar vicinity}}.
\newblock \bibinfo{journal}{Nature Astronomy} \bibinfo{volume}{4},
  \bibinfo{pages}{1078--1083}.
\newblock \DOIprefix\doi{10.1038/s41550-020-1131-2},
  \href{http://arxiv.org/abs/1907.07190}{{\tt arXiv:1907.07190}}.
%Type = Book
\bibitem[{{Newberg}(2016)}]{newberg:2016}
\bibinfo{author}{{Newberg}, H.J.}, \bibinfo{year}{2016}.
\newblock \bibinfo{title}{{Tidal Streams in the Local Group and Beyond}}.
  volume \bibinfo{volume}{420} of \textit{\bibinfo{series}{Astrophysics and
  Space Science Library}}.
\newblock \DOIprefix\doi{10.1007/978-3-319-19336-6}.
%Type = Article
\bibitem[{{Newberg} et~al.(2010){Newberg}, {Willett}, {Yanny} and
  {Xu}}]{newberg:2010}
\bibinfo{author}{{Newberg}, H.J.}, \bibinfo{author}{{Willett}, B.A.},
  \bibinfo{author}{{Yanny}, B.}, \bibinfo{author}{{Xu}, Y.},
  \bibinfo{year}{2010}.
\newblock \bibinfo{title}{{The Orbit of the Orphan Stream}}.
\newblock \bibinfo{journal}{\apj} \bibinfo{volume}{711},
  \bibinfo{pages}{32--49}.
\newblock \DOIprefix\doi{10.1088/0004-637X/711/1/32},
  \href{http://arxiv.org/abs/1001.0576}{{\tt arXiv:1001.0576}}.
%Type = Article
\bibitem[{Newberg et~al.(2002)Newberg, Yanny, Rockosi, Grebel, Rix, Brinkmann,
  Csabai, Hennessy, Hindsley, Ibata, Ivezi{\'c}, Lamb, Nash, Odenkirchen, Rave,
  Schneider, Smith, Stolte and York}]{Newberg:2002}
Newberg, H.~J. et~al., \bibinfo{year}{2002}.
\newblock \bibinfo{title}{The {{Ghost}} of {{Sagittarius}} and {{Lumps}} in the
  {{Halo}} of the {{Milky Way}}}.
\newblock \bibinfo{journal}{The Astrophysical Journal} \bibinfo{volume}{569},
  \bibinfo{pages}{245--274}.
\newblock \DOIprefix\doi{10.1086/338983}.
%Type = Article
\bibitem[{{Ngan} et~al.(2015){Ngan}, {Bozek}, {Carlberg}, {Wyse}, {Szalay} and
  {Madau}}]{ngan:2015}
\bibinfo{author}{{Ngan}, W.}, \bibinfo{author}{{Bozek}, B.},
  \bibinfo{author}{{Carlberg}, R.G.}, \bibinfo{author}{{Wyse}, R.F.G.},
  \bibinfo{author}{{Szalay}, A.S.}, \bibinfo{author}{{Madau}, P.},
  \bibinfo{year}{2015}.
\newblock \bibinfo{title}{{Simulating Tidal Streams in a High-resolution Dark
  Matter Halo}}.
\newblock \bibinfo{journal}{\apj} \bibinfo{volume}{803}, \bibinfo{pages}{75}.
\newblock \DOIprefix\doi{10.1088/0004-637X/803/2/75},
  \href{http://arxiv.org/abs/1411.3760}{{\tt arXiv:1411.3760}}.
%Type = Article
\bibitem[{{Ngan} et~al.(2016){Ngan}, {Carlberg}, {Bozek}, {Wyse}, {Szalay} and
  {Madau}}]{ngan:2016}
\bibinfo{author}{{Ngan}, W.}, \bibinfo{author}{{Carlberg}, R.G.},
  \bibinfo{author}{{Bozek}, B.}, \bibinfo{author}{{Wyse}, R.F.G.},
  \bibinfo{author}{{Szalay}, A.S.}, \bibinfo{author}{{Madau}, P.},
  \bibinfo{year}{2016}.
\newblock \bibinfo{title}{{Dispersal of Tidal Debris in a Milky-Way-sized Dark
  Matter Halo}}.
\newblock \bibinfo{journal}{\apj} \bibinfo{volume}{818}, \bibinfo{pages}{194}.
\newblock \DOIprefix\doi{10.3847/0004-637X/818/2/194},
  \href{http://arxiv.org/abs/1601.04681}{{\tt arXiv:1601.04681}}.
%Type = Article
\bibitem[{{Ngan} and {Carlberg}(2014)}]{ngan:2014}
\bibinfo{author}{{Ngan}, W.H.W.}, \bibinfo{author}{{Carlberg}, R.G.},
  \bibinfo{year}{2014}.
\newblock \bibinfo{title}{{Using Gaps in N-body Tidal Streams to Probe Missing
  Satellites}}.
\newblock \bibinfo{journal}{\apj} \bibinfo{volume}{788}, \bibinfo{pages}{181}.
\newblock \DOIprefix\doi{10.1088/0004-637X/788/2/181},
  \href{http://arxiv.org/abs/1311.1710}{{\tt arXiv:1311.1710}}.
%Type = Article
\bibitem[{{Nibauer} et~al.(2023){Nibauer}, {Bonaca} and
  {Johnston}}]{nibauer:2023}
\bibinfo{author}{{Nibauer}, J.}, \bibinfo{author}{{Bonaca}, A.},
  \bibinfo{author}{{Johnston}, K.V.}, \bibinfo{year}{2023}.
\newblock \bibinfo{title}{{Constraining the Gravitational Potential from the
  Projected Morphology of Extragalactic Tidal Streams}}.
\newblock \bibinfo{journal}{\apj} \bibinfo{volume}{954}, \bibinfo{pages}{195}.
\newblock \DOIprefix\doi{10.3847/1538-4357/ace9bc},
  \href{http://arxiv.org/abs/2303.17406}{{\tt arXiv:2303.17406}}.
%Type = Article
\bibitem[{{Nikutta} et~al.(2020){Nikutta}, {Fitzpatrick}, {Scott} and
  {Weaver}}]{nikutta:2020}
\bibinfo{author}{{Nikutta}, R.}, \bibinfo{author}{{Fitzpatrick}, M.},
  \bibinfo{author}{{Scott}, A.}, \bibinfo{author}{{Weaver}, B.A.},
  \bibinfo{year}{2020}.
\newblock \bibinfo{title}{{Data Lab-A community science platform}}.
\newblock \bibinfo{journal}{Astronomy and Computing} \bibinfo{volume}{33},
  \bibinfo{pages}{100411}.
\newblock \DOIprefix\doi{10.1016/j.ascom.2020.100411}.
%Type = Article
\bibitem[{{Noether}(1918)}]{noether:1918}
\bibinfo{author}{{Noether}, E.}, \bibinfo{year}{1918}.
\newblock \bibinfo{title}{{Invariant Variation Problems}}.
\newblock \bibinfo{journal}{Nachr. D. K{\~A}{\textparagraph}nig. Gesellsch. D.
  Wiss. Zu G{\~A}{\textparagraph}ttingen} , \bibinfo{pages}{235--237}.
%Type = Article
\bibitem[{{Ochsenbein} et~al.(2000){Ochsenbein}, {Bauer} and
  {Marcout}}]{ochsenbein:2000}
\bibinfo{author}{{Ochsenbein}, F.}, \bibinfo{author}{{Bauer}, P.},
  \bibinfo{author}{{Marcout}, J.}, \bibinfo{year}{2000}.
\newblock \bibinfo{title}{{The VizieR database of astronomical catalogues}}.
\newblock \bibinfo{journal}{\aaps} \bibinfo{volume}{143},
  \bibinfo{pages}{23--32}.
\newblock \DOIprefix\doi{10.1051/aas:2000169},
  \href{http://arxiv.org/abs/astro-ph/0002122}{{\tt arXiv:astro-ph/0002122}}.
%Type = Article
\bibitem[{{Odenkirchen} et~al.(2009){Odenkirchen}, {Grebel}, {Kayser}, {Rix}
  and {Dehnen}}]{odenkirchen:2009}
\bibinfo{author}{{Odenkirchen}, M.}, \bibinfo{author}{{Grebel}, E.K.},
  \bibinfo{author}{{Kayser}, A.}, \bibinfo{author}{{Rix}, H.W.},
  \bibinfo{author}{{Dehnen}, W.}, \bibinfo{year}{2009}.
\newblock \bibinfo{title}{{Kinematics of the Tidal Debris of the Globular
  Cluster Palomar 5}}.
\newblock \bibinfo{journal}{\aj} \bibinfo{volume}{137},
  \bibinfo{pages}{3378--3387}.
\newblock \DOIprefix\doi{10.1088/0004-6256/137/2/3378}.
%Type = Article
\bibitem[{{Odenkirchen} et~al.(2001){Odenkirchen}, {Grebel}, {Rockosi},
  {Dehnen}, {Ibata}, {Rix}, {Stolte}, {Wolf}, {Anderson}, {Bahcall},
  {Brinkmann}, {Csabai}, {Hennessy}, {Hindsley}, {Ivezi{\'c}}, {Lupton},
  {Munn}, {Pier}, {Stoughton} and {York}}]{odenkirchen:2001}
{Odenkirchen}, M. et~al., \bibinfo{year}{2001}.
\newblock \bibinfo{title}{{Detection of Massive Tidal Tails around the Globular
  Cluster Palomar 5 with Sloan Digital Sky Survey Commissioning Data}}.
\newblock \bibinfo{journal}{\apjl} \bibinfo{volume}{548},
  \bibinfo{pages}{L165--L169}.
\newblock \DOIprefix\doi{10.1086/319095},
  \href{http://arxiv.org/abs/astro-ph/0012311}{{\tt arXiv:astro-ph/0012311}}.
%Type = Article
\bibitem[{{Ou} et~al.(2023){Ou}, {Necib} and {Frebel}}]{ou:2023}
\bibinfo{author}{{Ou}, X.}, \bibinfo{author}{{Necib}, L.},
  \bibinfo{author}{{Frebel}, A.}, \bibinfo{year}{2023}.
\newblock \bibinfo{title}{{Robust clustering of the local Milky Way stellar
  kinematic substructures with Gaia eDR3}}.
\newblock \bibinfo{journal}{\mnras} \bibinfo{volume}{521},
  \bibinfo{pages}{2623--2648}.
\newblock \DOIprefix\doi{10.1093/mnras/stad706},
  \href{http://arxiv.org/abs/2208.01056}{{\tt arXiv:2208.01056}}.
%Type = Article
\bibitem[{{Padmanabhan} et~al.(2008){Padmanabhan}, {Schlegel}, {Finkbeiner},
  {Barentine}, {Blanton}, {Brewington}, {Gunn}, {Harvanek}, {Hogg},
  {Ivezi{\'c}}, {Johnston}, {Kent}, {Kleinman}, {Knapp}, {Krzesinski}, {Long},
  {Neilsen}, {Nitta}, {Loomis}, {Lupton}, {Roweis}, {Snedden}, {Strauss} and
  {Tucker}}]{padmanabhan:2008}
{Padmanabhan}, N. et~al., \bibinfo{year}{2008}.
\newblock \bibinfo{title}{{An Improved Photometric Calibration of the Sloan
  Digital Sky Survey Imaging Data}}.
\newblock \bibinfo{journal}{\apj} \bibinfo{volume}{674},
  \bibinfo{pages}{1217--1233}.
\newblock \DOIprefix\doi{10.1086/524677},
  \href{http://arxiv.org/abs/astro-ph/0703454}{{\tt arXiv:astro-ph/0703454}}.
%Type = Article
\bibitem[{{Palau} and {Miralda-Escud{\'e}}(2019)}]{palau:2019}
\bibinfo{author}{{Palau}, C.G.}, \bibinfo{author}{{Miralda-Escud{\'e}}, J.},
  \bibinfo{year}{2019}.
\newblock \bibinfo{title}{{Statistical detection of a tidal stream associated
  with the globular cluster M68 using Gaia data}}.
\newblock \bibinfo{journal}{\mnras} \bibinfo{volume}{488},
  \bibinfo{pages}{1535--1557}.
\newblock \DOIprefix\doi{10.1093/mnras/stz1790},
  \href{http://arxiv.org/abs/1905.01193}{{\tt arXiv:1905.01193}}.
%Type = Article
\bibitem[{{Palau} and {Miralda-Escud{\'e}}(2023)}]{palau:2023}
\bibinfo{author}{{Palau}, C.G.}, \bibinfo{author}{{Miralda-Escud{\'e}}, J.},
  \bibinfo{year}{2023}.
\newblock \bibinfo{title}{{The oblateness of the Milky Way dark matter halo
  from the stellar streams of NGC 3201, M68, and Palomar 5}}.
\newblock \bibinfo{journal}{\mnras} \bibinfo{volume}{524},
  \bibinfo{pages}{2124--2147}.
\newblock \DOIprefix\doi{10.1093/mnras/stad1930},
  \href{http://arxiv.org/abs/2212.03587}{{\tt arXiv:2212.03587}}.
%Type = Article
\bibitem[{{Patel} et~al.(2020){Patel}, {Kallivayalil}, {Garavito-Camargo},
  {Besla}, {Weisz}, {van der Marel}, {Boylan-Kolchin}, {Pawlowski} and
  {G{\'o}mez}}]{patel:2020}
{Patel}, E. et~al., \bibinfo{year}{2020}.
\newblock \bibinfo{title}{{The Orbital Histories of Magellanic Satellites Using
  Gaia DR2 Proper Motions}}.
\newblock \bibinfo{journal}{\apj} \bibinfo{volume}{893}, \bibinfo{pages}{121}.
\newblock \DOIprefix\doi{10.3847/1538-4357/ab7b75},
  \href{http://arxiv.org/abs/2001.01746}{{\tt arXiv:2001.01746}}.
%Type = Article
\bibitem[{{Patrick} et~al.(2022){Patrick}, {Koposov} and
  {Walker}}]{patrick:2022}
\bibinfo{author}{{Patrick}, J.M.}, \bibinfo{author}{{Koposov}, S.E.},
  \bibinfo{author}{{Walker}, M.G.}, \bibinfo{year}{2022}.
\newblock \bibinfo{title}{{Uniform modelling of the stellar density of thirteen
  tidal streams within the Galactic halo}}.
\newblock \bibinfo{journal}{\mnras} \bibinfo{volume}{514},
  \bibinfo{pages}{1757--1781}.
\newblock \DOIprefix\doi{10.1093/mnras/stac1478},
  \href{http://arxiv.org/abs/2206.04241}{{\tt arXiv:2206.04241}}.
%Type = Article
\bibitem[{{Pawlowski}(2018)}]{pawlowski:2018}
\bibinfo{author}{{Pawlowski}, M.S.}, \bibinfo{year}{2018}.
\newblock \bibinfo{title}{{The planes of satellite galaxies problem, suggested
  solutions, and open questions}}.
\newblock \bibinfo{journal}{Modern Physics Letters A} \bibinfo{volume}{33},
  \bibinfo{pages}{1830004}.
\newblock \DOIprefix\doi{10.1142/S0217732318300045},
  \href{http://arxiv.org/abs/1802.02579}{{\tt arXiv:1802.02579}}.
%Type = Article
\bibitem[{{Pawlowski} and {Kroupa}(2020)}]{pawlowski:2020}
\bibinfo{author}{{Pawlowski}, M.S.}, \bibinfo{author}{{Kroupa}, P.},
  \bibinfo{year}{2020}.
\newblock \bibinfo{title}{{The Milky Way's disc of classical satellite galaxies
  in light of Gaia DR2}}.
\newblock \bibinfo{journal}{\mnras} \bibinfo{volume}{491},
  \bibinfo{pages}{3042--3059}.
\newblock \DOIprefix\doi{10.1093/mnras/stz3163},
  \href{http://arxiv.org/abs/1911.05081}{{\tt arXiv:1911.05081}}.
%Type = Article
\bibitem[{{Pawlowski} et~al.(2012){Pawlowski}, {Pflamm-Altenburg} and
  {Kroupa}}]{pawlowski:2012}
\bibinfo{author}{{Pawlowski}, M.S.}, \bibinfo{author}{{Pflamm-Altenburg}, J.},
  \bibinfo{author}{{Kroupa}, P.}, \bibinfo{year}{2012}.
\newblock \bibinfo{title}{{The VPOS: a vast polar structure of satellite
  galaxies, globular clusters and streams around the Milky Way}}.
\newblock \bibinfo{journal}{\mnras} \bibinfo{volume}{423},
  \bibinfo{pages}{1109--1126}.
\newblock \DOIprefix\doi{10.1111/j.1365-2966.2012.20937.x},
  \href{http://arxiv.org/abs/1204.5176}{{\tt arXiv:1204.5176}}.
%Type = Article
\bibitem[{{Pe{\~n}arrubia} et~al.(2010){Pe{\~n}arrubia}, {Belokurov}, {Evans},
  {Mart{\'\i}nez-Delgado}, {Gilmore}, {Irwin}, {Niederste-Ostholt} and
  {Zucker}}]{penarrubia:2010}
\bibinfo{author}{{Pe{\~n}arrubia}, J.}, \bibinfo{author}{{Belokurov}, V.},
  \bibinfo{author}{{Evans}, N.W.}, \bibinfo{author}{{Mart{\'\i}nez-Delgado},
  D.}, \bibinfo{author}{{Gilmore}, G.}, \bibinfo{author}{{Irwin}, M.},
  \bibinfo{author}{{Niederste-Ostholt}, M.}, \bibinfo{author}{{Zucker}, D.B.},
  \bibinfo{year}{2010}.
\newblock \bibinfo{title}{{Was the progenitor of the Sagittarius stream a disc
  galaxy?}}
\newblock \bibinfo{journal}{\mnras} \bibinfo{volume}{408},
  \bibinfo{pages}{L26--L30}.
\newblock \DOIprefix\doi{10.1111/j.1745-3933.2010.00921.x},
  \href{http://arxiv.org/abs/1007.1485}{{\tt arXiv:1007.1485}}.
%Type = Article
\bibitem[{{Pe{\~n}arrubia} et~al.(2016){Pe{\~n}arrubia}, {G{\'o}mez}, {Besla},
  {Erkal} and {Ma}}]{penarrubia:2016}
\bibinfo{author}{{Pe{\~n}arrubia}, J.}, \bibinfo{author}{{G{\'o}mez}, F.A.},
  \bibinfo{author}{{Besla}, G.}, \bibinfo{author}{{Erkal}, D.},
  \bibinfo{author}{{Ma}, Y.Z.}, \bibinfo{year}{2016}.
\newblock \bibinfo{title}{{A timing constraint on the (total) mass of the Large
  Magellanic Cloud}}.
\newblock \bibinfo{journal}{\mnras} \bibinfo{volume}{456},
  \bibinfo{pages}{L54--L58}.
\newblock \DOIprefix\doi{10.1093/mnrasl/slv160},
  \href{http://arxiv.org/abs/1507.03594}{{\tt arXiv:1507.03594}}.
%Type = Article
\bibitem[{{Pearson} et~al.(2015){Pearson}, {K{\"u}pper}, {Johnston} and
  {Price-Whelan}}]{pearson:2015}
\bibinfo{author}{{Pearson}, S.}, \bibinfo{author}{{K{\"u}pper}, A.H.W.},
  \bibinfo{author}{{Johnston}, K.V.}, \bibinfo{author}{{Price-Whelan}, A.M.},
  \bibinfo{year}{2015}.
\newblock \bibinfo{title}{{Tidal Stream Morphology as an Indicator of Dark
  Matter Halo Geometry: The Case of Palomar 5}}.
\newblock \bibinfo{journal}{\apj} \bibinfo{volume}{799}, \bibinfo{pages}{28}.
\newblock \DOIprefix\doi{10.1088/0004-637X/799/1/28},
  \href{http://arxiv.org/abs/1410.3477}{{\tt arXiv:1410.3477}}.
%Type = Article
\bibitem[{{Pearson} et~al.(2022){Pearson}, {Price-Whelan}, {Hogg}, {Seth},
  {Sand}, {Hunt} and {Crnojevi{\'c}}}]{pearson:2022}
\bibinfo{author}{{Pearson}, S.}, \bibinfo{author}{{Price-Whelan}, A.M.},
  \bibinfo{author}{{Hogg}, D.W.}, \bibinfo{author}{{Seth}, A.C.},
  \bibinfo{author}{{Sand}, D.J.}, \bibinfo{author}{{Hunt}, J.A.S.},
  \bibinfo{author}{{Crnojevi{\'c}}, D.}, \bibinfo{year}{2022}.
\newblock \bibinfo{title}{{Mapping Dark Matter with Extragalactic Stellar
  Streams: The Case of Centaurus A}}.
\newblock \bibinfo{journal}{\apj} \bibinfo{volume}{941}, \bibinfo{pages}{19}.
\newblock \DOIprefix\doi{10.3847/1538-4357/ac9bfb},
  \href{http://arxiv.org/abs/2205.12277}{{\tt arXiv:2205.12277}}.
%Type = Article
\bibitem[{{Pearson} et~al.(2017){Pearson}, {Price-Whelan} and
  {Johnston}}]{pearson:2017}
\bibinfo{author}{{Pearson}, S.}, \bibinfo{author}{{Price-Whelan}, A.M.},
  \bibinfo{author}{{Johnston}, K.V.}, \bibinfo{year}{2017}.
\newblock \bibinfo{title}{{Gaps and length asymmetry in the stellar stream
  Palomar 5 as effects of Galactic bar rotation}}.
\newblock \bibinfo{journal}{Nature Astronomy} \bibinfo{volume}{1},
  \bibinfo{pages}{633--639}.
\newblock \DOIprefix\doi{10.1038/s41550-017-0220-3},
  \href{http://arxiv.org/abs/1703.04627}{{\tt arXiv:1703.04627}}.
%Type = Article
\bibitem[{{Perottoni} et~al.(2019){Perottoni}, {Martin}, {Newberg},
  {Rocha-Pinto}, {Almeida-Fernandes} and {Gomes-J{\'u}nior}}]{perottoni:2019}
\bibinfo{author}{{Perottoni}, H.D.}, \bibinfo{author}{{Martin}, C.},
  \bibinfo{author}{{Newberg}, H.J.}, \bibinfo{author}{{Rocha-Pinto}, H.J.},
  \bibinfo{author}{{Almeida-Fernandes}, F.d.},
  \bibinfo{author}{{Gomes-J{\'u}nior}, A.R.}, \bibinfo{year}{2019}.
\newblock \bibinfo{title}{{A cold stellar stream in Pegasus}}.
\newblock \bibinfo{journal}{\mnras} \bibinfo{volume}{486},
  \bibinfo{pages}{843--850}.
\newblock \DOIprefix\doi{10.1093/mnras/stz869},
  \href{http://arxiv.org/abs/1903.08840}{{\tt arXiv:1903.08840}}.
%Type = Article
\bibitem[{{Perryman} et~al.(2001){Perryman}, {de Boer}, {Gilmore}, {H{\o}g},
  {Lattanzi}, {Lindegren}, {Luri}, {Mignard}, {Pace} and {de
  Zeeuw}}]{perryman:2001}
{Perryman}, M.~A.~C. et~al., \bibinfo{year}{2001}.
\newblock \bibinfo{title}{{GAIA: Composition, formation and evolution of the
  Galaxy}}.
\newblock \bibinfo{journal}{\aap} \bibinfo{volume}{369},
  \bibinfo{pages}{339--363}.
\newblock \DOIprefix\doi{10.1051/0004-6361:20010085},
  \href{http://arxiv.org/abs/astro-ph/0101235}{{\tt arXiv:astro-ph/0101235}}.
%Type = Article
\bibitem[{{Perryman} et~al.(1997){Perryman}, {Lindegren}, {Kovalevsky}, {Hoeg},
  {Bastian}, {Bernacca}, {Cr{\'e}z{\'e}}, {Donati}, {Grenon}, {Grewing}, {van
  Leeuwen}, {van der Marel}, {Mignard}, {Murray}, {Le Poole}, {Schrijver},
  {Turon}, {Arenou}, {Froeschl{\'e}} and {Petersen}}]{perryman:1997}
{Perryman}, M.~A.~C. et~al., \bibinfo{year}{1997}.
\newblock \bibinfo{title}{{The HIPPARCOS Catalogue}}.
\newblock \bibinfo{journal}{\aap} \bibinfo{volume}{323},
  \bibinfo{pages}{L49--L52}.
%Type = Article
\bibitem[{{Peter} et~al.(2013){Peter}, {Rocha}, {Bullock} and
  {Kaplinghat}}]{peter:2013}
\bibinfo{author}{{Peter}, A.H.G.}, \bibinfo{author}{{Rocha}, M.},
  \bibinfo{author}{{Bullock}, J.S.}, \bibinfo{author}{{Kaplinghat}, M.},
  \bibinfo{year}{2013}.
\newblock \bibinfo{title}{{Cosmological simulations with self-interacting dark
  matter - II. Halo shapes versus observations}}.
\newblock \bibinfo{journal}{\mnras} \bibinfo{volume}{430},
  \bibinfo{pages}{105--120}.
\newblock \DOIprefix\doi{10.1093/mnras/sts535},
  \href{http://arxiv.org/abs/1208.3026}{{\tt arXiv:1208.3026}}.
%Type = Article
\bibitem[{{Petersen} et~al.(2022){Petersen}, {Weinberg} and
  {Katz}}]{petersen:2022}
\bibinfo{author}{{Petersen}, M.S.}, \bibinfo{author}{{Weinberg}, M.D.},
  \bibinfo{author}{{Katz}, N.}, \bibinfo{year}{2022}.
\newblock \bibinfo{title}{{EXP: N-body integration using basis function
  expansions}}.
\newblock \bibinfo{journal}{\mnras} \bibinfo{volume}{510},
  \bibinfo{pages}{6201--6217}.
\newblock \DOIprefix\doi{10.1093/mnras/stab3639},
  \href{http://arxiv.org/abs/2104.14577}{{\tt arXiv:2104.14577}}.
%Type = Article
\bibitem[{{Pettee} et~al.(2024){Pettee}, {Thanvantri}, {Nachman}, {Shih},
  {Buckley} and {Collins}}]{pettee:2024}
\bibinfo{author}{{Pettee}, M.}, \bibinfo{author}{{Thanvantri}, S.},
  \bibinfo{author}{{Nachman}, B.}, \bibinfo{author}{{Shih}, D.},
  \bibinfo{author}{{Buckley}, M.R.}, \bibinfo{author}{{Collins}, J.H.},
  \bibinfo{year}{2024}.
\newblock \bibinfo{title}{{Weakly supervised anomaly detection in the Milky
  Way}}.
\newblock \bibinfo{journal}{\mnras} \bibinfo{volume}{527},
  \bibinfo{pages}{8459--8474}.
\newblock \DOIprefix\doi{10.1093/mnras/stad3663},
  \href{http://arxiv.org/abs/2305.03761}{{\tt arXiv:2305.03761}}.
%Type = Article
\bibitem[{{Pfleiderer}(1963)}]{pfleiderer:1963}
\bibinfo{author}{{Pfleiderer}, J.}, \bibinfo{year}{1963}.
\newblock \bibinfo{title}{{Gravitationseffekte bei der Begegnung zweier
  Galaxien. Mit 5 Textabbildungen}}.
\newblock \bibinfo{journal}{\zap} \bibinfo{volume}{58}, \bibinfo{pages}{12}.
%Type = Article
\bibitem[{{Pillepich} et~al.(2015){Pillepich}, {Madau} and
  {Mayer}}]{pillepich:2015}
\bibinfo{author}{{Pillepich}, A.}, \bibinfo{author}{{Madau}, P.},
  \bibinfo{author}{{Mayer}, L.}, \bibinfo{year}{2015}.
\newblock \bibinfo{title}{{Building Late-type Spiral Galaxies by In-situ and
  Ex-situ Star Formation}}.
\newblock \bibinfo{journal}{\apj} \bibinfo{volume}{799}, \bibinfo{pages}{184}.
\newblock \DOIprefix\doi{10.1088/0004-637X/799/2/184},
  \href{http://arxiv.org/abs/1407.7855}{{\tt arXiv:1407.7855}}.
%Type = Article
\bibitem[{{Pillepich} et~al.(2023){Pillepich}, {Sotillo-Ramos}, {Ramesh},
  {Nelson}, {Engler}, {Rodriguez-Gomez}, {Fournier}, {Donnari}, {Springel} and
  {Hernquist}}]{pillepich:2023}
{Pillepich}, A. et~al., \bibinfo{year}{2023}.
\newblock \bibinfo{title}{{Milky Way and Andromeda analogs from the TNG50
  simulation}}.
\newblock \bibinfo{journal}{arXiv e-prints} ,
  \bibinfo{pages}{arXiv:2303.16217}\DOIprefix\doi{10.48550/arXiv.2303.16217},
  \href{http://arxiv.org/abs/2303.16217}{{\tt arXiv:2303.16217}}.
%Type = Article
\bibitem[{{Portail} et~al.(2017){Portail}, {Gerhard}, {Wegg} and
  {Ness}}]{portail:2017}
\bibinfo{author}{{Portail}, M.}, \bibinfo{author}{{Gerhard}, O.},
  \bibinfo{author}{{Wegg}, C.}, \bibinfo{author}{{Ness}, M.},
  \bibinfo{year}{2017}.
\newblock \bibinfo{title}{{Dynamical modelling of the galactic bulge and bar:
  the Milky Way's pattern speed, stellar and dark matter mass distribution}}.
\newblock \bibinfo{journal}{\mnras} \bibinfo{volume}{465},
  \bibinfo{pages}{1621--1644}.
\newblock \DOIprefix\doi{10.1093/mnras/stw2819},
  \href{http://arxiv.org/abs/1608.07954}{{\tt arXiv:1608.07954}}.
%Type = Article
\bibitem[{{Portegies Zwart} et~al.(2010){Portegies Zwart}, {McMillan} and
  {Gieles}}]{portegies-zwart:2010}
\bibinfo{author}{{Portegies Zwart}, S.F.}, \bibinfo{author}{{McMillan},
  S.L.W.}, \bibinfo{author}{{Gieles}, M.}, \bibinfo{year}{2010}.
\newblock \bibinfo{title}{{Young Massive Star Clusters}}.
\newblock \bibinfo{journal}{\araa} \bibinfo{volume}{48},
  \bibinfo{pages}{431--493}.
\newblock \DOIprefix\doi{10.1146/annurev-astro-081309-130834},
  \href{http://arxiv.org/abs/1002.1961}{{\tt arXiv:1002.1961}}.
%Type = Article
\bibitem[{{Press} and {Schechter}(1974)}]{press:1974}
\bibinfo{author}{{Press}, W.H.}, \bibinfo{author}{{Schechter}, P.},
  \bibinfo{year}{1974}.
\newblock \bibinfo{title}{{Formation of Galaxies and Clusters of Galaxies by
  Self-Similar Gravitational Condensation}}.
\newblock \bibinfo{journal}{\apj} \bibinfo{volume}{187},
  \bibinfo{pages}{425--438}.
\newblock \DOIprefix\doi{10.1086/152650}.
%Type = Article
\bibitem[{{Price-Whelan}(2017)}]{price-whelan:2017}
\bibinfo{author}{{Price-Whelan}, A.M.}, \bibinfo{year}{2017}.
\newblock \bibinfo{title}{{Gala: A Python package for galactic dynamics}}.
\newblock \bibinfo{journal}{The Journal of Open Source Software}
  \bibinfo{volume}{2}, \bibinfo{pages}{388}.
\newblock \DOIprefix\doi{10.21105/joss.00388}.
%Type = Article
\bibitem[{{Price-Whelan} and {Bonaca}(2018)}]{price-whelan:2018}
\bibinfo{author}{{Price-Whelan}, A.M.}, \bibinfo{author}{{Bonaca}, A.},
  \bibinfo{year}{2018}.
\newblock \bibinfo{title}{{Off the Beaten Path: Gaia Reveals GD-1 Stars outside
  of the Main Stream}}.
\newblock \bibinfo{journal}{\apjl} \bibinfo{volume}{863}, \bibinfo{pages}{L20}.
\newblock \DOIprefix\doi{10.3847/2041-8213/aad7b5},
  \href{http://arxiv.org/abs/1805.00425}{{\tt arXiv:1805.00425}}.
%Type = Article
\bibitem[{{Price-Whelan} et~al.(2015){Price-Whelan}, {Johnston}, {Sheffield},
  {Laporte} and {Sesar}}]{price-whelan:2015}
\bibinfo{author}{{Price-Whelan}, A.M.}, \bibinfo{author}{{Johnston}, K.V.},
  \bibinfo{author}{{Sheffield}, A.A.}, \bibinfo{author}{{Laporte}, C.F.P.},
  \bibinfo{author}{{Sesar}, B.}, \bibinfo{year}{2015}.
\newblock \bibinfo{title}{{A reinterpretation of the Triangulum-Andromeda
  stellar clouds: a population of halo stars kicked out of the Galactic disc}}.
\newblock \bibinfo{journal}{\mnras} \bibinfo{volume}{452},
  \bibinfo{pages}{676--685}.
\newblock \DOIprefix\doi{10.1093/mnras/stv1324},
  \href{http://arxiv.org/abs/1503.08780}{{\tt arXiv:1503.08780}}.
%Type = Article
\bibitem[{{Price-Whelan} et~al.(2016a){Price-Whelan}, {Johnston}, {Valluri},
  {Pearson}, {K{\"u}pper} and {Hogg}}]{price-whelan:2016a}
\bibinfo{author}{{Price-Whelan}, A.M.}, \bibinfo{author}{{Johnston}, K.V.},
  \bibinfo{author}{{Valluri}, M.}, \bibinfo{author}{{Pearson}, S.},
  \bibinfo{author}{{K{\"u}pper}, A.H.W.}, \bibinfo{author}{{Hogg}, D.W.},
  \bibinfo{year}{2016}a.
\newblock \bibinfo{title}{{Chaotic dispersal of tidal debris}}.
\newblock \bibinfo{journal}{\mnras} \bibinfo{volume}{455},
  \bibinfo{pages}{1079--1098}.
\newblock \DOIprefix\doi{10.1093/mnras/stv2383},
  \href{http://arxiv.org/abs/1507.08662}{{\tt arXiv:1507.08662}}.
%Type = Article
\bibitem[{{Price-Whelan} et~al.(2019){Price-Whelan}, {Mateu}, {Iorio},
  {Pearson}, {Bonaca} and {Belokurov}}]{price-whelan:2019}
\bibinfo{author}{{Price-Whelan}, A.M.}, \bibinfo{author}{{Mateu}, C.},
  \bibinfo{author}{{Iorio}, G.}, \bibinfo{author}{{Pearson}, S.},
  \bibinfo{author}{{Bonaca}, A.}, \bibinfo{author}{{Belokurov}, V.},
  \bibinfo{year}{2019}.
\newblock \bibinfo{title}{{Kinematics of the Palomar 5 Stellar Stream from RR
  Lyrae Stars}}.
\newblock \bibinfo{journal}{\aj} \bibinfo{volume}{158}, \bibinfo{pages}{223}.
\newblock \DOIprefix\doi{10.3847/1538-3881/ab4cef},
  \href{http://arxiv.org/abs/1910.00595}{{\tt arXiv:1910.00595}}.
%Type = Article
\bibitem[{{Price-Whelan} et~al.(2016b){Price-Whelan}, {Sesar}, {Johnston} and
  {Rix}}]{price-whelan:2016b}
\bibinfo{author}{{Price-Whelan}, A.M.}, \bibinfo{author}{{Sesar}, B.},
  \bibinfo{author}{{Johnston}, K.V.}, \bibinfo{author}{{Rix}, H.W.},
  \bibinfo{year}{2016}b.
\newblock \bibinfo{title}{{Spending Too Much Time at the Galactic Bar: Chaotic
  Fanning of the Ophiuchus Stream}}.
\newblock \bibinfo{journal}{\apj} \bibinfo{volume}{824}, \bibinfo{pages}{104}.
\newblock \DOIprefix\doi{10.3847/0004-637X/824/2/104},
  \href{http://arxiv.org/abs/1601.06790}{{\tt arXiv:1601.06790}}.
%Type = Article
\bibitem[{{Prudil} et~al.(2021){Prudil}, {Hanke}, {Lemasle}, {Crestani},
  {Braga}, {Fabrizio}, {Koch-Hansen}, {Bono}, {Grebel}, {Matsunaga}, {Marengo},
  {da Silva}, {Dall'Ora}, {Mart{\'\i}nez-V{\'a}zquez}, {Altavilla}, {Lala},
  {Chaboyer}, {Ferraro}, {Fiorentino}, {Gilligan}, {Nonino} and
  {Th{\'e}venin}}]{prudil:2021}
{Prudil}, Z. et~al., \bibinfo{year}{2021}.
\newblock \bibinfo{title}{{Milky Way archaeology using RR Lyrae and type II
  Cepheids. I. The Orphan stream in 7D using RR Lyrae stars}}.
\newblock \bibinfo{journal}{\aap} \bibinfo{volume}{648}, \bibinfo{pages}{A78}.
\newblock \DOIprefix\doi{10.1051/0004-6361/202140422},
  \href{http://arxiv.org/abs/2102.01090}{{\tt arXiv:2102.01090}}.
%Type = Article
\bibitem[{{Qian} et~al.(2022){Qian}, {Arshad} and {Bovy}}]{qian:2022}
\bibinfo{author}{{Qian}, Y.}, \bibinfo{author}{{Arshad}, Y.},
  \bibinfo{author}{{Bovy}, J.}, \bibinfo{year}{2022}.
\newblock \bibinfo{title}{{The structure of accreted stellar streams}}.
\newblock \bibinfo{journal}{\mnras} \bibinfo{volume}{511},
  \bibinfo{pages}{2339--2348}.
\newblock \DOIprefix\doi{10.1093/mnras/stac238},
  \href{http://arxiv.org/abs/2201.11045}{{\tt arXiv:2201.11045}}.
%Type = Article
\bibitem[{{Ramos} et~al.(2018){Ramos}, {Antoja} and {Figueras}}]{ramos:2018}
\bibinfo{author}{{Ramos}, P.}, \bibinfo{author}{{Antoja}, T.},
  \bibinfo{author}{{Figueras}, F.}, \bibinfo{year}{2018}.
\newblock \bibinfo{title}{{Riding the kinematic waves in the Milky Way disk
  with Gaia}}.
\newblock \bibinfo{journal}{\aap} \bibinfo{volume}{619}, \bibinfo{pages}{A72}.
\newblock \DOIprefix\doi{10.1051/0004-6361/201833494},
  \href{http://arxiv.org/abs/1805.09790}{{\tt arXiv:1805.09790}}.
%Type = Article
\bibitem[{{Ramos} et~al.(2022){Ramos}, {Antoja}, {Yuan}, {Arentsen}, {Oria},
  {Famaey}, {Ibata}, {Mateu} and {Carballo-Bello}}]{ramos:2022}
{Ramos}, P. et~al., \bibinfo{year}{2022}.
\newblock \bibinfo{title}{{The Sagittarius stream in Gaia Early Data Release 3
  and the origin of the bifurcations}}.
\newblock \bibinfo{journal}{\aap} \bibinfo{volume}{666}, \bibinfo{pages}{A64}.
\newblock \DOIprefix\doi{10.1051/0004-6361/202142830},
  \href{http://arxiv.org/abs/2112.02105}{{\tt arXiv:2112.02105}}.
%Type = Article
\bibitem[{{Reino} et~al.(2021){Reino}, {Rossi}, {Sanderson}, {Sellentin},
  {Helmi}, {Koppelman} and {Sharma}}]{reino:2021}
\bibinfo{author}{{Reino}, S.}, \bibinfo{author}{{Rossi}, E.M.},
  \bibinfo{author}{{Sanderson}, R.E.}, \bibinfo{author}{{Sellentin}, E.},
  \bibinfo{author}{{Helmi}, A.}, \bibinfo{author}{{Koppelman}, H.H.},
  \bibinfo{author}{{Sharma}, S.}, \bibinfo{year}{2021}.
\newblock \bibinfo{title}{{Galactic potential constraints from clustering in
  action space of combined stellar stream data}}.
\newblock \bibinfo{journal}{\mnras} \bibinfo{volume}{502},
  \bibinfo{pages}{4170--4193}.
\newblock \DOIprefix\doi{10.1093/mnras/stab304},
  \href{http://arxiv.org/abs/2007.00356}{{\tt arXiv:2007.00356}}.
%Type = Article
\bibitem[{{Ren} et~al.(2020){Ren}, {Zheng}, {Valls-Gabaud}, {Duc}, {Bell},
  {Pan}, {Qin}, {Shi}, {Qiao}, {He} and {Wen}}]{ren:2020}
{Ren}, J. et~al., \bibinfo{year}{2020}.
\newblock \bibinfo{title}{{Long tidal tails in merging galaxies and their
  implications}}.
\newblock \bibinfo{journal}{\mnras} \bibinfo{volume}{499},
  \bibinfo{pages}{3399--3409}.
\newblock \DOIprefix\doi{10.1093/mnras/staa2985},
  \href{http://arxiv.org/abs/2009.11879}{{\tt arXiv:2009.11879}}.
%Type = Article
\bibitem[{{Richings} et~al.(2020){Richings}, {Frenk}, {Jenkins}, {Robertson},
  {Fattahi}, {Grand}, {Navarro}, {Pakmor}, {Gomez}, {Marinacci} and
  {Oman}}]{richings:2020}
{Richings}, J. et~al., \bibinfo{year}{2020}.
\newblock \bibinfo{title}{{Subhalo destruction in the APOSTLE and AURIGA
  simulations}}.
\newblock \bibinfo{journal}{\mnras} \bibinfo{volume}{492},
  \bibinfo{pages}{5780--5793}.
\newblock \DOIprefix\doi{10.1093/mnras/stz3448},
  \href{http://arxiv.org/abs/1811.12437}{{\tt arXiv:1811.12437}}.
%Type = Article
\bibitem[{{Riley} and {Strigari}(2020)}]{riley:2020}
\bibinfo{author}{{Riley}, A.H.}, \bibinfo{author}{{Strigari}, L.E.},
  \bibinfo{year}{2020}.
\newblock \bibinfo{title}{{The Milky Way's stellar streams and globular
  clusters do not align in a Vast Polar Structure}}.
\newblock \bibinfo{journal}{\mnras} \bibinfo{volume}{494},
  \bibinfo{pages}{983--1001}.
\newblock \DOIprefix\doi{10.1093/mnras/staa710},
  \href{http://arxiv.org/abs/2001.11564}{{\tt arXiv:2001.11564}}.
%Type = Article
\bibitem[{{Roberts} et~al.(2024){Roberts}, {Gieles}, {Erkal} and
  {Sanders}}]{roberts:2024}
\bibinfo{author}{{Roberts}, D.}, \bibinfo{author}{{Gieles}, M.},
  \bibinfo{author}{{Erkal}, D.}, \bibinfo{author}{{Sanders}, J.L.},
  \bibinfo{year}{2024}.
\newblock \bibinfo{title}{{Stellar streams from black hole-rich star
  clusters}}.
\newblock \bibinfo{journal}{arXiv e-prints} ,
  \bibinfo{pages}{arXiv:2402.06393}\DOIprefix\doi{10.48550/arXiv.2402.06393},
  \href{http://arxiv.org/abs/2402.06393}{{\tt arXiv:2402.06393}}.
%Type = Article
\bibitem[{{Rocha-Pinto} et~al.(2003){Rocha-Pinto}, {Majewski}, {Skrutskie} and
  {Crane}}]{rocha-pinto:2003}
\bibinfo{author}{{Rocha-Pinto}, H.J.}, \bibinfo{author}{{Majewski}, S.R.},
  \bibinfo{author}{{Skrutskie}, M.F.}, \bibinfo{author}{{Crane}, J.D.},
  \bibinfo{year}{2003}.
\newblock \bibinfo{title}{{Tracing the Galactic Anticenter Stellar Stream with
  2MASS M Giants}}.
\newblock \bibinfo{journal}{\apjl} \bibinfo{volume}{594},
  \bibinfo{pages}{L115--L118}.
\newblock \DOIprefix\doi{10.1086/378668}.
%Type = Article
\bibitem[{{Rocha-Pinto} et~al.(2004){Rocha-Pinto}, {Majewski}, {Skrutskie},
  {Crane} and {Patterson}}]{rocha-pinto:2004}
\bibinfo{author}{{Rocha-Pinto}, H.J.}, \bibinfo{author}{{Majewski}, S.R.},
  \bibinfo{author}{{Skrutskie}, M.F.}, \bibinfo{author}{{Crane}, J.D.},
  \bibinfo{author}{{Patterson}, R.J.}, \bibinfo{year}{2004}.
\newblock \bibinfo{title}{{Exploring Halo Substructure with Giant Stars: A
  Diffuse Star Cloud or Tidal Debris around the Milky Way in
  Triangulum-Andromeda}}.
\newblock \bibinfo{journal}{\apj} \bibinfo{volume}{615},
  \bibinfo{pages}{732--737}.
\newblock \DOIprefix\doi{10.1086/424585},
  \href{http://arxiv.org/abs/astro-ph/0405437}{{\tt arXiv:astro-ph/0405437}}.
%Type = Article
\bibitem[{Rockosi et~al.(2002)Rockosi, Odenkirchen, Grebel, Dehnen, Cudworth,
  Gunn, York, Brinkmann, Hennessy and Ivezi{\'c}}]{rockosi:2002}
Rockosi, C.~M. et~al., \bibinfo{year}{2002}.
\newblock \bibinfo{title}{A {{Matched-Filter Analysis}} of the {{Tidal Tails}}
  of the {{Globular Cluster Palomar}} 5}.
\newblock \bibinfo{journal}{The Astronomical Journal} \bibinfo{volume}{124},
  \bibinfo{pages}{349--363}.
\newblock \DOIprefix\doi{10.1086/340957}.
%Type = Article
\bibitem[{{Rodriguez} et~al.(2015){Rodriguez}, {Morscher}, {Pattabiraman},
  {Chatterjee}, {Haster} and {Rasio}}]{rodriguez:2015}
\bibinfo{author}{{Rodriguez}, C.L.}, \bibinfo{author}{{Morscher}, M.},
  \bibinfo{author}{{Pattabiraman}, B.}, \bibinfo{author}{{Chatterjee}, S.},
  \bibinfo{author}{{Haster}, C.J.}, \bibinfo{author}{{Rasio}, F.A.},
  \bibinfo{year}{2015}.
\newblock \bibinfo{title}{{Binary Black Hole Mergers from Globular Clusters:
  Implications for Advanced LIGO}}.
\newblock \bibinfo{journal}{\prl} \bibinfo{volume}{115},
  \bibinfo{pages}{051101}.
\newblock \DOIprefix\doi{10.1103/PhysRevLett.115.051101},
  \href{http://arxiv.org/abs/1505.00792}{{\tt arXiv:1505.00792}}.
%Type = Article
\bibitem[{{Roederer} and {Gnedin}(2019)}]{roederer:2019}
\bibinfo{author}{{Roederer}, I.U.}, \bibinfo{author}{{Gnedin}, O.Y.},
  \bibinfo{year}{2019}.
\newblock \bibinfo{title}{{High-resolution Optical Spectroscopy of Stars in the
  Sylgr Stellar Stream}}.
\newblock \bibinfo{journal}{\apj} \bibinfo{volume}{883}, \bibinfo{pages}{84}.
\newblock \DOIprefix\doi{10.3847/1538-4357/ab365c},
  \href{http://arxiv.org/abs/1907.03772}{{\tt arXiv:1907.03772}}.
%Type = Article
\bibitem[{{R{\"o}ser} et~al.(2019){R{\"o}ser}, {Schilbach} and
  {Goldman}}]{roser:2019}
\bibinfo{author}{{R{\"o}ser}, S.}, \bibinfo{author}{{Schilbach}, E.},
  \bibinfo{author}{{Goldman}, B.}, \bibinfo{year}{2019}.
\newblock \bibinfo{title}{{Hyades tidal tails revealed by Gaia DR2}}.
\newblock \bibinfo{journal}{\aap} \bibinfo{volume}{621}, \bibinfo{pages}{L2}.
\newblock \DOIprefix\doi{10.1051/0004-6361/201834608},
  \href{http://arxiv.org/abs/1811.03845}{{\tt arXiv:1811.03845}}.
%Type = Article
\bibitem[{{Sales} et~al.(2008){Sales}, {Helmi}, {Starkenburg}, {Morrison},
  {Engle}, {Harding}, {Mateo}, {Olszewski} and {Sivarani}}]{sales:2008}
{Sales}, L.~V. et~al., \bibinfo{year}{2008}.
\newblock \bibinfo{title}{{On the genealogy of the Orphan Stream}}.
\newblock \bibinfo{journal}{\mnras} \bibinfo{volume}{389},
  \bibinfo{pages}{1391--1398}.
\newblock \DOIprefix\doi{10.1111/j.1365-2966.2008.13659.x},
  \href{http://arxiv.org/abs/0805.0508}{{\tt arXiv:0805.0508}}.
%Type = Article
\bibitem[{{Sanders} and {Binney}(2013a)}]{sanders:2013a}
\bibinfo{author}{{Sanders}, J.L.}, \bibinfo{author}{{Binney}, J.},
  \bibinfo{year}{2013}a.
\newblock \bibinfo{title}{{Stream-orbit misalignment - I. The dangers of
  orbit-fitting}}.
\newblock \bibinfo{journal}{\mnras} \bibinfo{volume}{433},
  \bibinfo{pages}{1813--1825}.
\newblock \DOIprefix\doi{10.1093/mnras/stt806},
  \href{http://arxiv.org/abs/1305.1935}{{\tt arXiv:1305.1935}}.
%Type = Article
\bibitem[{{Sanders} and {Binney}(2013b)}]{sanders:2013b}
\bibinfo{author}{{Sanders}, J.L.}, \bibinfo{author}{{Binney}, J.},
  \bibinfo{year}{2013}b.
\newblock \bibinfo{title}{{Stream-orbit misalignment - II. A new algorithm to
  constrain the Galactic potential}}.
\newblock \bibinfo{journal}{\mnras} \bibinfo{volume}{433},
  \bibinfo{pages}{1826--1836}.
\newblock \DOIprefix\doi{10.1093/mnras/stt816},
  \href{http://arxiv.org/abs/1305.1937}{{\tt arXiv:1305.1937}}.
%Type = Article
\bibitem[{{Sanders} et~al.(2016){Sanders}, {Bovy} and {Erkal}}]{sanders:2016}
\bibinfo{author}{{Sanders}, J.L.}, \bibinfo{author}{{Bovy}, J.},
  \bibinfo{author}{{Erkal}, D.}, \bibinfo{year}{2016}.
\newblock \bibinfo{title}{{Dynamics of stream-subhalo interactions}}.
\newblock \bibinfo{journal}{\mnras} \bibinfo{volume}{457},
  \bibinfo{pages}{3817--3835}.
\newblock \DOIprefix\doi{10.1093/mnras/stw232},
  \href{http://arxiv.org/abs/1510.03426}{{\tt arXiv:1510.03426}}.
%Type = Article
\bibitem[{{Sanders} et~al.(2020){Sanders}, {Lilley}, {Vasiliev}, {Evans} and
  {Erkal}}]{sanders:2020}
\bibinfo{author}{{Sanders}, J.L.}, \bibinfo{author}{{Lilley}, E.J.},
  \bibinfo{author}{{Vasiliev}, E.}, \bibinfo{author}{{Evans}, N.W.},
  \bibinfo{author}{{Erkal}, D.}, \bibinfo{year}{2020}.
\newblock \bibinfo{title}{{Models of distorted and evolving dark matter
  haloes}}.
\newblock \bibinfo{journal}{\mnras} \bibinfo{volume}{499},
  \bibinfo{pages}{4793--4813}.
\newblock \DOIprefix\doi{10.1093/mnras/staa3079},
  \href{http://arxiv.org/abs/2009.00645}{{\tt arXiv:2009.00645}}.
%Type = Article
\bibitem[{{Sanders} et~al.(2019){Sanders}, {Smith} and {Evans}}]{sanders:2019}
\bibinfo{author}{{Sanders}, J.L.}, \bibinfo{author}{{Smith}, L.},
  \bibinfo{author}{{Evans}, N.W.}, \bibinfo{year}{2019}.
\newblock \bibinfo{title}{{The pattern speed of the Milky Way bar from
  transverse velocities}}.
\newblock \bibinfo{journal}{\mnras} \bibinfo{volume}{488},
  \bibinfo{pages}{4552--4564}.
\newblock \DOIprefix\doi{10.1093/mnras/stz1827},
  \href{http://arxiv.org/abs/1903.02009}{{\tt arXiv:1903.02009}}.
%Type = Article
\bibitem[{{Sanderson} et~al.(2019){Sanderson}, {Carlin}, {Cunningham},
  {Garavito-Camargo}, {Guhathakurta}, {Johnston}, {Laporte}, {Li}, {Sohn},
  {Anderson}, {Bellini}, {Bennett}, {Casertano}, {Fall}, {Libralato},
  {Malhotra}, {Moustakas}, {Rhodes}, {Armus}, {Choi}, {Pino}, {D'Onghia},
  {Fardal}, {Garrison-Kimmel}, {Gilbert}, {Grillmair}, {Kallivayalil}, {Kirby},
  {Li}, {Marshall}, {Price-Whelan}, {Sacchi}, {Spergel}, {Valluri} and {van der
  Marel}}]{sanderson:2019}
{Sanderson}, R. et~al., \bibinfo{year}{2019}.
\newblock \bibinfo{title}{{The Multidimensional Milky Way}}.
\newblock \bibinfo{journal}{\baas} \bibinfo{volume}{51}, \bibinfo{pages}{347}.
\newblock \DOIprefix\doi{10.48550/arXiv.1903.07641},
  \href{http://arxiv.org/abs/1903.07641}{{\tt arXiv:1903.07641}}.
%Type = Article
\bibitem[{{Sanderson} and {Helmi}(2013)}]{sanderson:2013}
\bibinfo{author}{{Sanderson}, R.E.}, \bibinfo{author}{{Helmi}, A.},
  \bibinfo{year}{2013}.
\newblock \bibinfo{title}{{An analytical phase-space model for tidal
  caustics}}.
\newblock \bibinfo{journal}{\mnras} \bibinfo{volume}{435},
  \bibinfo{pages}{378--399}.
\newblock \DOIprefix\doi{10.1093/mnras/stt1307},
  \href{http://arxiv.org/abs/1211.4522}{{\tt arXiv:1211.4522}}.
%Type = Article
\bibitem[{{Sanderson} et~al.(2015){Sanderson}, {Helmi} and
  {Hogg}}]{sanderson:2015}
\bibinfo{author}{{Sanderson}, R.E.}, \bibinfo{author}{{Helmi}, A.},
  \bibinfo{author}{{Hogg}, D.W.}, \bibinfo{year}{2015}.
\newblock \bibinfo{title}{{Action-space Clustering of Tidal Streams to Infer
  the Galactic Potential}}.
\newblock \bibinfo{journal}{\apj} \bibinfo{volume}{801}, \bibinfo{pages}{98}.
\newblock \DOIprefix\doi{10.1088/0004-637X/801/2/98},
  \href{http://arxiv.org/abs/1404.6534}{{\tt arXiv:1404.6534}}.
%Type = Article
\bibitem[{{Sandford} et~al.(2017){Sandford}, {K{\"u}pper}, {Johnston} and
  {Diemand}}]{sandford:2017}
\bibinfo{author}{{Sandford}, E.}, \bibinfo{author}{{K{\"u}pper}, A.H.W.},
  \bibinfo{author}{{Johnston}, K.V.}, \bibinfo{author}{{Diemand}, J.},
  \bibinfo{year}{2017}.
\newblock \bibinfo{title}{{Quantifying tidal stream disruption in a simulated
  Milky Way}}.
\newblock \bibinfo{journal}{\mnras} \bibinfo{volume}{470},
  \bibinfo{pages}{522--538}.
\newblock \DOIprefix\doi{10.1093/mnras/stx1268},
  \href{http://arxiv.org/abs/1705.07128}{{\tt arXiv:1705.07128}}.
%Type = Article
\bibitem[{{Santistevan} et~al.(2024){Santistevan}, {Wetzel}, {Tollerud},
  {Sanderson}, {Moreno} and {Patel}}]{santistevan:2024}
\bibinfo{author}{{Santistevan}, I.B.}, \bibinfo{author}{{Wetzel}, A.},
  \bibinfo{author}{{Tollerud}, E.}, \bibinfo{author}{{Sanderson}, R.E.},
  \bibinfo{author}{{Moreno}, J.}, \bibinfo{author}{{Patel}, E.},
  \bibinfo{year}{2024}.
\newblock \bibinfo{title}{{Modelling the orbital histories of satellites of
  Milky Way-mass galaxies: testing static host potentials against cosmological
  simulations}}.
\newblock \bibinfo{journal}{\mnras} \bibinfo{volume}{527},
  \bibinfo{pages}{8841--8864}.
\newblock \DOIprefix\doi{10.1093/mnras/stad3757},
  \href{http://arxiv.org/abs/2309.05708}{{\tt arXiv:2309.05708}}.
%Type = Article
\bibitem[{{Sawala} et~al.(2023){Sawala}, {Cautun}, {Frenk}, {Helly}, {Jasche},
  {Jenkins}, {Johansson}, {Lavaux}, {McAlpine} and {Schaller}}]{sawala:2023}
{Sawala}, T. et~al., \bibinfo{year}{2023}.
\newblock \bibinfo{title}{{The Milky Way's plane of satellites is consistent
  with {\ensuremath{\Lambda}}CDM}}.
\newblock \bibinfo{journal}{Nature Astronomy} \bibinfo{volume}{7},
  \bibinfo{pages}{481--491}.
\newblock \DOIprefix\doi{10.1038/s41550-022-01856-z},
  \href{http://arxiv.org/abs/2205.02860}{{\tt arXiv:2205.02860}}.
%Type = Article
\bibitem[{{Saydjari} et~al.(2021){Saydjari}, {Portillo}, {Slepian}, {Kahraman},
  {Burkhart} and {Finkbeiner}}]{saydjari:2021}
\bibinfo{author}{{Saydjari}, A.K.}, \bibinfo{author}{{Portillo}, S.K.N.},
  \bibinfo{author}{{Slepian}, Z.}, \bibinfo{author}{{Kahraman}, S.},
  \bibinfo{author}{{Burkhart}, B.}, \bibinfo{author}{{Finkbeiner}, D.P.},
  \bibinfo{year}{2021}.
\newblock \bibinfo{title}{{Classification of Magnetohydrodynamic Simulations
  Using Wavelet Scattering Transforms}}.
\newblock \bibinfo{journal}{\apj} \bibinfo{volume}{910}, \bibinfo{pages}{122}.
\newblock \DOIprefix\doi{10.3847/1538-4357/abe46d},
  \href{http://arxiv.org/abs/2010.11963}{{\tt arXiv:2010.11963}}.
%Type = Article
\bibitem[{{Searle} and {Zinn}(1978)}]{searle:1978}
\bibinfo{author}{{Searle}, L.}, \bibinfo{author}{{Zinn}, R.},
  \bibinfo{year}{1978}.
\newblock \bibinfo{title}{{Composition of halo clusters and the formation of
  the galactic halo.}}
\newblock \bibinfo{journal}{\apj} \bibinfo{volume}{225},
  \bibinfo{pages}{357--379}.
\newblock \DOIprefix\doi{10.1086/156499}.
%Type = Article
\bibitem[{{Sesar} et~al.(2015){Sesar}, {Bovy}, {Bernard}, {Caldwell}, {Cohen},
  {Fouesneau}, {Johnson}, {Ness}, {Ferguson}, {Martin}, {Price-Whelan}, {Rix},
  {Schlafly}, {Burgett}, {Chambers}, {Flewelling}, {Hodapp}, {Kaiser},
  {Magnier}, {Platais}, {Tonry}, {Waters} and {Wyse}}]{sesar:2015}
{Sesar}, B. et~al., \bibinfo{year}{2015}.
\newblock \bibinfo{title}{{The Nature and Orbit of the Ophiuchus Stream}}.
\newblock \bibinfo{journal}{\apj} \bibinfo{volume}{809}, \bibinfo{pages}{59}.
\newblock \DOIprefix\doi{10.1088/0004-637X/809/1/59},
  \href{http://arxiv.org/abs/1501.00581}{{\tt arXiv:1501.00581}}.
%Type = Article
\bibitem[{{Sesar} et~al.(2013){Sesar}, {Grillmair}, {Cohen}, {Bellm},
  {Bhalerao}, {Levitan}, {Laher}, {Ofek}, {Surace}, {Tang}, {Waszczak},
  {Kulkarni} and {Prince}}]{sesar:2013}
{Sesar}, B. et~al., \bibinfo{year}{2013}.
\newblock \bibinfo{title}{{Tracing the Orphan Stream to 55 kpc with RR Lyrae
  Stars}}.
\newblock \bibinfo{journal}{\apj} \bibinfo{volume}{776}, \bibinfo{pages}{26}.
\newblock \DOIprefix\doi{10.1088/0004-637X/776/1/26},
  \href{http://arxiv.org/abs/1308.0857}{{\tt arXiv:1308.0857}}.
%Type = Article
\bibitem[{{Sesar} et~al.(2016){Sesar}, {Price-Whelan}, {Cohen}, {Rix},
  {Pearson}, {Johnston}, {Bernard}, {Ferguson}, {Martin}, {Slater}, {Chambers},
  {Flewelling}, {Wainscoat} and {Waters}}]{sesar:2016}
{Sesar}, B. et~al., \bibinfo{year}{2016}.
\newblock \bibinfo{title}{{Evidence of Fanning in the Ophiuchus Stream}}.
\newblock \bibinfo{journal}{\apjl} \bibinfo{volume}{816}, \bibinfo{pages}{L4}.
\newblock \DOIprefix\doi{10.3847/2041-8205/816/1/L4},
  \href{http://arxiv.org/abs/1512.00469}{{\tt arXiv:1512.00469}}.
%Type = Article
\bibitem[{{Shao} et~al.(2021){Shao}, {Cautun}, {Deason} and
  {Frenk}}]{shao:2021}
\bibinfo{author}{{Shao}, S.}, \bibinfo{author}{{Cautun}, M.},
  \bibinfo{author}{{Deason}, A.}, \bibinfo{author}{{Frenk}, C.S.},
  \bibinfo{year}{2021}.
\newblock \bibinfo{title}{{The twisted dark matter halo of the Milky Way}}.
\newblock \bibinfo{journal}{\mnras} \bibinfo{volume}{504},
  \bibinfo{pages}{6033--6048}.
\newblock \DOIprefix\doi{10.1093/mnras/staa3883},
  \href{http://arxiv.org/abs/2005.03025}{{\tt arXiv:2005.03025}}.
%Type = Article
\bibitem[{{Sharpe} et~al.(2024){Sharpe}, {Naidu} and {Conroy}}]{sharpe:2024}
\bibinfo{author}{{Sharpe}, K.}, \bibinfo{author}{{Naidu}, R.P.},
  \bibinfo{author}{{Conroy}, C.}, \bibinfo{year}{2024}.
\newblock \bibinfo{title}{{What Is Missing from the Local Stellar Halo?}}
\newblock \bibinfo{journal}{\apj} \bibinfo{volume}{963}, \bibinfo{pages}{162}.
\newblock \DOIprefix\doi{10.3847/1538-4357/ad19ca},
  \href{http://arxiv.org/abs/2211.04562}{{\tt arXiv:2211.04562}}.
%Type = Article
\bibitem[{{Sheffield} et~al.(2021){Sheffield}, {Subrahimovic}, {Refat},
  {Beaton}, {Hasselquist}, {Hayes}, {Price-Whelan}, {Horta}, {Majewski},
  {Cunha}, {Smith}, {Fern{\'a}ndez-Trincado}, {Sobeck}, {Mu{\~n}oz},
  {Garc{\'\i}a-Hern{\'a}ndez}, {Lane}, {Nitschelm} and
  {Roman-Lopes}}]{sheffield:2021}
{Sheffield}, A.~A. et~al., \bibinfo{year}{2021}.
\newblock \bibinfo{title}{{Chemodynamically Characterizing the Jhelum Stellar
  Stream with APOGEE-2}}.
\newblock \bibinfo{journal}{\apj} \bibinfo{volume}{913}, \bibinfo{pages}{39}.
\newblock \DOIprefix\doi{10.3847/1538-4357/abee93},
  \href{http://arxiv.org/abs/2103.07488}{{\tt arXiv:2103.07488}}.
%Type = Article
\bibitem[{{Sheth} and {Tormen}(2002)}]{sheth:2002}
\bibinfo{author}{{Sheth}, R.K.}, \bibinfo{author}{{Tormen}, G.},
  \bibinfo{year}{2002}.
\newblock \bibinfo{title}{{An excursion set model of hierarchical clustering:
  ellipsoidal collapse and the moving barrier}}.
\newblock \bibinfo{journal}{\mnras} \bibinfo{volume}{329},
  \bibinfo{pages}{61--75}.
\newblock \DOIprefix\doi{10.1046/j.1365-8711.2002.04950.x},
  \href{http://arxiv.org/abs/astro-ph/0105113}{{\tt arXiv:astro-ph/0105113}}.
%Type = Article
\bibitem[{{Shih} et~al.(2023){Shih}, {Buckley} and {Necib}}]{shih:2023}
\bibinfo{author}{{Shih}, D.}, \bibinfo{author}{{Buckley}, M.R.},
  \bibinfo{author}{{Necib}, L.}, \bibinfo{year}{2023}.
\newblock \bibinfo{title}{{Via Machinae 2.0: Full-Sky, Model-Agnostic Search
  for Stellar Streams in Gaia DR2}}.
\newblock \bibinfo{journal}{arXiv e-prints} ,
  \bibinfo{pages}{arXiv:2303.01529}\DOIprefix\doi{10.48550/arXiv.2303.01529},
  \href{http://arxiv.org/abs/2303.01529}{{\tt arXiv:2303.01529}}.
%Type = Article
\bibitem[{{Shih} et~al.(2022){Shih}, {Buckley}, {Necib} and
  {Tamanas}}]{shih:2022}
\bibinfo{author}{{Shih}, D.}, \bibinfo{author}{{Buckley}, M.R.},
  \bibinfo{author}{{Necib}, L.}, \bibinfo{author}{{Tamanas}, J.},
  \bibinfo{year}{2022}.
\newblock \bibinfo{title}{{VIA MACHINAE: Searching for stellar streams using
  unsupervised machine learning}}.
\newblock \bibinfo{journal}{\mnras} \bibinfo{volume}{509},
  \bibinfo{pages}{5992--6007}.
\newblock \DOIprefix\doi{10.1093/mnras/stab3372},
  \href{http://arxiv.org/abs/2104.12789}{{\tt arXiv:2104.12789}}.
%Type = Article
\bibitem[{{Shipp} et~al.(2018){Shipp}, {Drlica-Wagner}, {Balbinot}, {Ferguson},
  {Erkal}, {Li}, {Bechtol}, {Belokurov}, {Buncher}, {Carollo}, {Carrasco Kind},
  {Kuehn}, {Marshall}, {Pace}, {Rykoff}, {Sevilla-Noarbe}, {Sheldon},
  {Strigari}, {Vivas}, {Yanny}, {Zenteno}, {Abbott}, {Abdalla}, {Allam},
  {Avila}, {Bertin}, {Brooks}, {Burke}, {Carretero}, {Castander}, {Cawthon},
  {Crocce}, {Cunha}, {D'Andrea}, {da Costa}, {Davis}, {De Vicente}, {Desai},
  {Diehl}, {Doel}, {Evrard}, {Flaugher}, {Fosalba}, {Frieman},
  {Garc{\'\i}a-Bellido}, {Gaztanaga}, {Gerdes}, {Gruen}, {Gruendl}, {Gschwend},
  {Gutierrez}, {Hartley}, {Honscheid}, {Hoyle}, {James}, {Johnson}, {Krause},
  {Kuropatkin}, {Lahav}, {Lin}, {Maia}, {March}, {Martini}, {Menanteau},
  {Miller}, {Miquel}, {Nichol}, {Plazas}, {Romer}, {Sako}, {Sanchez},
  {Santiago}, {Scarpine}, {Schindler}, {Schubnell}, {Smith}, {Smith},
  {Sobreira}, {Suchyta}, {Swanson}, {Tarle}, {Thomas}, {Tucker}, {Walker},
  {Wechsler} and {DES Collaboration}}]{shipp:2018}
{Shipp}, N. et~al., \bibinfo{year}{2018}.
\newblock \bibinfo{title}{{Stellar Streams Discovered in the Dark Energy
  Survey}}.
\newblock \bibinfo{journal}{\apj} \bibinfo{volume}{862}, \bibinfo{pages}{114}.
\newblock \DOIprefix\doi{10.3847/1538-4357/aacdab},
  \href{http://arxiv.org/abs/1801.03097}{{\tt arXiv:1801.03097}}.
%Type = Article
\bibitem[{{Shipp} et~al.(2021){Shipp}, {Erkal}, {Drlica-Wagner}, {Li}, {Pace},
  {Koposov}, {Cullinane}, {Da Costa}, {Ji}, {Kuehn}, {Lewis}, {Mackey},
  {Simpson}, {Wan}, {Zucker}, {Bland-Hawthorn}, {Ferguson}, {Lilleengen} and
  {Lilleengen}}]{shipp:2021}
{Shipp}, N. et~al., \bibinfo{year}{2021}.
\newblock \bibinfo{title}{{Measuring the Mass of the Large Magellanic Cloud
  with Stellar Streams Observed by S $^{5}$}}.
\newblock \bibinfo{journal}{\apj} \bibinfo{volume}{923}, \bibinfo{pages}{149}.
\newblock \DOIprefix\doi{10.3847/1538-4357/ac2e93},
  \href{http://arxiv.org/abs/2107.13004}{{\tt arXiv:2107.13004}}.
%Type = Article
\bibitem[{{Shipp} et~al.(2019){Shipp}, {Li}, {Pace}, {Erkal}, {Drlica-Wagner},
  {Yanny}, {Belokurov}, {Wester}, {Koposov}, {Kuehn}, {Lewis}, {Simpson},
  {Wan}, {Zucker}, {Martell}, {Wang} and {S5 Collaboration}}]{shipp:2019}
{Shipp}, N. et~al., \bibinfo{year}{2019}.
\newblock \bibinfo{title}{{Proper Motions of Stellar Streams Discovered in the
  Dark Energy Survey}}.
\newblock \bibinfo{journal}{\apj} \bibinfo{volume}{885}, \bibinfo{pages}{3}.
\newblock \DOIprefix\doi{10.3847/1538-4357/ab44bf},
  \href{http://arxiv.org/abs/1907.09488}{{\tt arXiv:1907.09488}}.
%Type = Article
\bibitem[{{Shipp} et~al.(2023){Shipp}, {Panithanpaisal}, {Necib}, {Sanderson},
  {Erkal}, {Li}, {Santistevan}, {Wetzel}, {Cullinane}, {Ji}, {Koposov},
  {Kuehn}, {Lewis}, {Pace}, {Zucker}, {Bland-Hawthorn}, {Cunningham}, {Kim},
  {Lilleengen}, {Moreno}, {Sharma}, {S Collaboration} and {FIRE
  Collaboration}}]{shipp:2023}
{Shipp}, N. et~al., \bibinfo{year}{2023}.
\newblock \bibinfo{title}{{Streams on FIRE: Populations of Detectable Stellar
  Streams in the Milky Way and FIRE}}.
\newblock \bibinfo{journal}{\apj} \bibinfo{volume}{949}, \bibinfo{pages}{44}.
\newblock \DOIprefix\doi{10.3847/1538-4357/acc582},
  \href{http://arxiv.org/abs/2208.02255}{{\tt arXiv:2208.02255}}.
%Type = Article
\bibitem[{{Shipp} et~al.(2020){Shipp}, {Price-Whelan}, {Tavangar}, {Mateu} and
  {Drlica-Wagner}}]{shipp:2020}
\bibinfo{author}{{Shipp}, N.}, \bibinfo{author}{{Price-Whelan}, A.M.},
  \bibinfo{author}{{Tavangar}, K.}, \bibinfo{author}{{Mateu}, C.},
  \bibinfo{author}{{Drlica-Wagner}, A.}, \bibinfo{year}{2020}.
\newblock \bibinfo{title}{{Discovery of Extended Tidal Tails around the
  Globular Cluster Palomar 13}}.
\newblock \bibinfo{journal}{\aj} \bibinfo{volume}{160}, \bibinfo{pages}{244}.
\newblock \DOIprefix\doi{10.3847/1538-3881/abbd3a},
  \href{http://arxiv.org/abs/2006.12501}{{\tt arXiv:2006.12501}}.
%Type = Article
\bibitem[{{Siegal-Gaskins} and {Valluri}(2008)}]{siegal-gaskins:2008}
\bibinfo{author}{{Siegal-Gaskins}, J.M.}, \bibinfo{author}{{Valluri}, M.},
  \bibinfo{year}{2008}.
\newblock \bibinfo{title}{{Signatures of {\ensuremath{\Lambda}}CDM Substructure
  in Tidal Debris}}.
\newblock \bibinfo{journal}{\apj} \bibinfo{volume}{681},
  \bibinfo{pages}{40--52}.
\newblock \DOIprefix\doi{10.1086/587450},
  \href{http://arxiv.org/abs/0710.0385}{{\tt arXiv:0710.0385}}.
%Type = Article
\bibitem[{{Simon}(2018)}]{simon:2018}
\bibinfo{author}{{Simon}, J.D.}, \bibinfo{year}{2018}.
\newblock \bibinfo{title}{{Gaia Proper Motions and Orbits of the Ultra-faint
  Milky Way Satellites}}.
\newblock \bibinfo{journal}{\apj} \bibinfo{volume}{863}, \bibinfo{pages}{89}.
\newblock \DOIprefix\doi{10.3847/1538-4357/aacdfb},
  \href{http://arxiv.org/abs/1804.10230}{{\tt arXiv:1804.10230}}.
%Type = Article
\bibitem[{{Simon} and {Geha}(2007)}]{simon:2007}
\bibinfo{author}{{Simon}, J.D.}, \bibinfo{author}{{Geha}, M.},
  \bibinfo{year}{2007}.
\newblock \bibinfo{title}{{The Kinematics of the Ultra-faint Milky Way
  Satellites: Solving the Missing Satellite Problem}}.
\newblock \bibinfo{journal}{\apj} \bibinfo{volume}{670},
  \bibinfo{pages}{313--331}.
\newblock \DOIprefix\doi{10.1086/521816},
  \href{http://arxiv.org/abs/0706.0516}{{\tt arXiv:0706.0516}}.
%Type = Article
\bibitem[{{Skrutskie} et~al.(2006){Skrutskie}, {Cutri}, {Stiening}, {Weinberg},
  {Schneider}, {Carpenter}, {Beichman}, {Capps}, {Chester}, {Elias}, {Huchra},
  {Liebert}, {Lonsdale}, {Monet}, {Price}, {Seitzer}, {Jarrett}, {Kirkpatrick},
  {Gizis}, {Howard}, {Evans}, {Fowler}, {Fullmer}, {Hurt}, {Light}, {Kopan},
  {Marsh}, {McCallon}, {Tam}, {Van Dyk} and {Wheelock}}]{skrutskie:2006}
{Skrutskie}, M.~F. et~al., \bibinfo{year}{2006}.
\newblock \bibinfo{title}{{The Two Micron All Sky Survey (2MASS)}}.
\newblock \bibinfo{journal}{\aj} \bibinfo{volume}{131},
  \bibinfo{pages}{1163--1183}.
\newblock \DOIprefix\doi{10.1086/498708}.
%Type = Article
\bibitem[{{Sohn} et~al.(2016){Sohn}, {van der Marel}, {Kallivayalil},
  {Majewski}, {Besla}, {Carlin}, {Law}, {Siegel} and {Anderson}}]{sohn:2016}
{Sohn}, S.~T. et~al., \bibinfo{year}{2016}.
\newblock \bibinfo{title}{{Hubble Space Telescope Proper Motions of Individual
  Stars in Stellar Streams: Orphan, Sagittarius, Lethe, and the New
  ``Parallel{\textquoteright} Stream''}}.
\newblock \bibinfo{journal}{\apj} \bibinfo{volume}{833}, \bibinfo{pages}{235}.
\newblock \DOIprefix\doi{10.3847/1538-4357/833/2/235},
  \href{http://arxiv.org/abs/1611.02282}{{\tt arXiv:1611.02282}}.
%Type = Article
\bibitem[{{Sollima}(2020)}]{sollima:2020}
\bibinfo{author}{{Sollima}, A.}, \bibinfo{year}{2020}.
\newblock \bibinfo{title}{{The eye of Gaia on globular clusters structure:
  tidal tails}}.
\newblock \bibinfo{journal}{\mnras} \bibinfo{volume}{495},
  \bibinfo{pages}{2222--2233}.
\newblock \DOIprefix\doi{10.1093/mnras/staa1209},
  \href{http://arxiv.org/abs/2004.13754}{{\tt arXiv:2004.13754}}.
%Type = Article
\bibitem[{{Somerville}(2002)}]{somerville:2002}
\bibinfo{author}{{Somerville}, R.S.}, \bibinfo{year}{2002}.
\newblock \bibinfo{title}{{Can Photoionization Squelching Resolve the
  Substructure Crisis?}}
\newblock \bibinfo{journal}{\apjl} \bibinfo{volume}{572},
  \bibinfo{pages}{L23--L26}.
\newblock \DOIprefix\doi{10.1086/341444},
  \href{http://arxiv.org/abs/astro-ph/0107507}{{\tt arXiv:astro-ph/0107507}}.
%Type = Article
\bibitem[{{Spergel} et~al.(2013){Spergel}, {Gehrels}, {Breckinridge},
  {Donahue}, {Dressler}, {Gaudi}, {Greene}, {Guyon}, {Hirata}, {Kalirai},
  {Kasdin}, {Moos}, {Perlmutter}, {Postman}, {Rauscher}, {Rhodes}, {Wang},
  {Weinberg}, {Centrella}, {Traub}, {Baltay}, {Colbert}, {Bennett},
  {Kiessling}, {Macintosh}, {Merten}, {Mortonson}, {Penny}, {Rozo},
  {Savransky}, {Stapelfeldt}, {Zu}, {Baker}, {Cheng}, {Content}, {Dooley},
  {Foote}, {Goullioud}, {Grady}, {Jackson}, {Kruk}, {Levine}, {Melton},
  {Peddie}, {Ruffa} and {Shaklan}}]{spergel:2013}
{Spergel}, D. et~al., \bibinfo{year}{2013}.
\newblock \bibinfo{title}{{Wide-Field InfraRed Survey Telescope-Astrophysics
  Focused Telescope Assets WFIRST-AFTA Final Report}}.
\newblock \bibinfo{journal}{arXiv e-prints} ,
  \bibinfo{pages}{arXiv:1305.5422}\DOIprefix\doi{10.48550/arXiv.1305.5422},
  \href{http://arxiv.org/abs/1305.5422}{{\tt arXiv:1305.5422}}.
%Type = Article
\bibitem[{{Spergel} and {Steinhardt}(2000)}]{spergel:2000}
\bibinfo{author}{{Spergel}, D.N.}, \bibinfo{author}{{Steinhardt}, P.J.},
  \bibinfo{year}{2000}.
\newblock \bibinfo{title}{{Observational Evidence for Self-Interacting Cold
  Dark Matter}}.
\newblock \bibinfo{journal}{\prl} \bibinfo{volume}{84},
  \bibinfo{pages}{3760--3763}.
\newblock \DOIprefix\doi{10.1103/PhysRevLett.84.3760},
  \href{http://arxiv.org/abs/astro-ph/9909386}{{\tt arXiv:astro-ph/9909386}}.
%Type = Article
\bibitem[{{Springel} et~al.(2008){Springel}, {Wang}, {Vogelsberger}, {Ludlow},
  {Jenkins}, {Helmi}, {Navarro}, {Frenk} and {White}}]{springel:2008}
{Springel}, V. et~al., \bibinfo{year}{2008}.
\newblock \bibinfo{title}{{The Aquarius Project: the subhaloes of galactic
  haloes}}.
\newblock \bibinfo{journal}{\mnras} \bibinfo{volume}{391},
  \bibinfo{pages}{1685--1711}.
\newblock \DOIprefix\doi{10.1111/j.1365-2966.2008.14066.x},
  \href{http://arxiv.org/abs/0809.0898}{{\tt arXiv:0809.0898}}.
%Type = Article
\bibitem[{{Springel} and {White}(1999)}]{springel:1999}
\bibinfo{author}{{Springel}, V.}, \bibinfo{author}{{White}, S.D.M.},
  \bibinfo{year}{1999}.
\newblock \bibinfo{title}{{Tidal tails in cold dark matter cosmologies}}.
\newblock \bibinfo{journal}{\mnras} \bibinfo{volume}{307},
  \bibinfo{pages}{162--178}.
\newblock \DOIprefix\doi{10.1046/j.1365-8711.1999.02613.x},
  \href{http://arxiv.org/abs/astro-ph/9807320}{{\tt arXiv:astro-ph/9807320}}.
%Type = Article
\bibitem[{{Starkman} et~al.(2020){Starkman}, {Bovy} and {Webb}}]{starkman:2020}
\bibinfo{author}{{Starkman}, N.}, \bibinfo{author}{{Bovy}, J.},
  \bibinfo{author}{{Webb}, J.J.}, \bibinfo{year}{2020}.
\newblock \bibinfo{title}{{An extended Pal 5 stream in Gaia DR2}}.
\newblock \bibinfo{journal}{\mnras} \bibinfo{volume}{493},
  \bibinfo{pages}{4978--4986}.
\newblock \DOIprefix\doi{10.1093/mnras/staa534},
  \href{http://arxiv.org/abs/1909.03048}{{\tt arXiv:1909.03048}}.
%Type = Article
\bibitem[{{Steinmetz} et~al.(2020){Steinmetz}, {Matijevi{\v{c}}}, {Enke},
  {Zwitter}, {Guiglion}, {McMillan}, {Kordopatis}, {Valentini}, {Chiappini},
  {Casagrande}, {Wojno}, {Anguiano}, {Bienaym{\'e}}, {Bijaoui}, {Binney},
  {Burton}, {Cass}, {de Laverny}, {Fiegert}, {Freeman}, {Fulbright}, {Gibson},
  {Gilmore}, {Grebel}, {Helmi}, {Kunder}, {Munari}, {Navarro}, {Parker},
  {Ruchti}, {Recio-Blanco}, {Reid}, {Seabroke}, {Siviero}, {Siebert}, {Stupar},
  {Watson}, {Williams}, {Wyse}, {Anders}, {Antoja}, {Birko}, {Bland-Hawthorn},
  {Bossini}, {Garc{\'\i}a}, {Carrillo}, {Chaplin}, {Elsworth}, {Famaey},
  {Gerhard}, {Jofre}, {Just}, {Mathur}, {Miglio}, {Minchev}, {Monari},
  {Mosser}, {Ritter}, {Rodrigues}, {Scholz}, {Sharma}, {Sysoliatina} and {RAVE
  Collaboration}}]{steinmetz:2020}
{Steinmetz}, M. et~al., \bibinfo{year}{2020}.
\newblock \bibinfo{title}{{The Sixth Data Release of the Radial Velocity
  Experiment (RAVE). I. Survey Description, Spectra, and Radial Velocities}}.
\newblock \bibinfo{journal}{\aj} \bibinfo{volume}{160}, \bibinfo{pages}{82}.
\newblock \DOIprefix\doi{10.3847/1538-3881/ab9ab9},
  \href{http://arxiv.org/abs/2002.04377}{{\tt arXiv:2002.04377}}.
%Type = Book
\bibitem[{{Szebehely}(1967)}]{szebehely:1967}
\bibinfo{author}{{Szebehely}, V.}, \bibinfo{year}{1967}.
\newblock \bibinfo{title}{{Theory of orbits. The restricted problem of three
  bodies}}.
%Type = Article
\bibitem[{{Takada} et~al.(2014){Takada}, {Ellis}, {Chiba}, {Greene}, {Aihara},
  {Arimoto}, {Bundy}, {Cohen}, {Dor{\'e}}, {Graves}, {Gunn}, {Heckman},
  {Hirata}, {Ho}, {Kneib}, {Le F{\`e}vre}, {Lin}, {More}, {Murayama}, {Nagao},
  {Ouchi}, {Seiffert}, {Silverman}, {Sodr{\'e}}, {Spergel}, {Strauss}, {Sugai},
  {Suto}, {Takami} and {Wyse}}]{takada:2014}
{Takada}, M. et~al., \bibinfo{year}{2014}.
\newblock \bibinfo{title}{{Extragalactic science, cosmology, and Galactic
  archaeology with the Subaru Prime Focus Spectrograph}}.
\newblock \bibinfo{journal}{\pasj} \bibinfo{volume}{66}, \bibinfo{pages}{R1}.
\newblock \DOIprefix\doi{10.1093/pasj/pst019},
  \href{http://arxiv.org/abs/1206.0737}{{\tt arXiv:1206.0737}}.
%Type = Article
\bibitem[{{Tavangar} and {Price-Whelan}({in prep.})}]{tavangar:2024}
\bibinfo{author}{{Tavangar}, K.}, \bibinfo{author}{{Price-Whelan}, A.M.},
  \bibinfo{year}{{in prep.}}
\newblock \bibinfo{title}{{Inferring the density and membership of stellar
  streams with flexible models: The GD-1 stream in Gaia Data Release 3}} .
%Type = Article
\bibitem[{{The Dark Energy Survey Collaboration}(2005)}]{des:2005}
\bibinfo{author}{{The Dark Energy Survey Collaboration}}, \bibinfo{year}{2005}.
\newblock \bibinfo{title}{{The Dark Energy Survey}}.
\newblock \bibinfo{journal}{arXiv e-prints} ,
  \bibinfo{pages}{astro--ph/0510346}\DOIprefix\doi{10.48550/arXiv.astro-ph/0510346},
  \href{http://arxiv.org/abs/astro-ph/0510346}{{\tt arXiv:astro-ph/0510346}}.
%Type = Article
\bibitem[{{The MSE Science Team} et~al.(2019){The MSE Science Team},
  {Babusiaux}, {Bergemann}, {Burgasser}, {Ellison}, {Haggard}, {Huber},
  {Kaplinghat}, {Li}, {Marshall}, {Martell}, {McConnachie}, {Percival},
  {Robotham}, {Shen}, {Thirupathi}, {Tran}, {Yeche}, {Yong}, {Adibekyan},
  {Silva Aguirre}, {Angelou}, {Asplund}, {Balogh}, {Banerjee}, {Bannister},
  {Barr{\'\i}a}, {Battaglia}, {Bayo}, {Bechtol}, {Beck}, {Beers}, {Bellinger},
  {Berg}, {Bestenlehner}, {Bilicki}, {Bitsch}, {Bland-Hawthorn}, {Bolton},
  {Boselli}, {Bovy}, {Bragaglia}, {Buzasi}, {Caffau}, {Cami}, {Carleton},
  {Casagrande}, {Cassisi}, {Catelan}, {Chang}, {Cortese}, {Damjanov}, {Davies},
  {de Grijs}, {de Rosa}, {Deason}, {di Matteo}, {Drlica-Wagner}, {Erkal},
  {Escorza}, {Ferrarese}, {Fleming}, {Font-Ribera}, {Freeman}, {G{\"a}nsicke},
  {Gabdeev}, {Gallagher}, {Gandolfi}, {Garc{\'\i}a}, {Gaulme}, {Geha},
  {Gennaro}, {Gieles}, {Gilbert}, {Gordon}, {Goswami}, {Greco}, {Grillmair},
  {Guiglion}, {H{\'e}nault-Brunet}, {Hall}, {Handler}, {Hansen}, {Hathi},
  {Hatzidimitriou}, {Haywood}, {Hern{\'a}ndez Santisteban}, {Hillenbrand},
  {Hopkins}, {Howlett}, {Hudson}, {Ibata}, {Ili{\'c}}, {Jablonka}, {Ji},
  {Jiang}, {Juneau}, {Karakas}, {Karinkuzhi}, {Kim}, {Kong}, {Konstantopoulos},
  {Krogager}, {Lagos}, {Lallement}, {Laporte}, {Lebreton}, {Lee}, {Lewis},
  {Lianou}, {Liu}, {Lodieu}, {Loveday}, {M{\'e}sz{\'a}ros}, {Makler}, {Mao},
  {Marchesini}, {Martin}, {Mateo}, {Melis}, {Merle}, {Miglio}, {Gohar
  Mohammad}, {Molaverdikhani}, {Monier}, {Morel}, {Mosser}, {Nataf}, {Necib},
  {Neilson}, {Newman}, {Nierenberg}, {Nord}, {Noterdaeme}, {O'Dea}, {Oshagh},
  {Pace}, {Palanque-Delabrouille}, {Pandey}, {Parker}, {Pawlowski}, {Peter},
  {Petitjean}, {Petric}, {Placco}, {Popovi{\'c}}, {Price-Whelan}, {Prsa},
  {Ravindranath}, {Rich}, {Ruan}, {Rybizki}, {Sakari}, {Sanderson}, {Schiavon},
  {Schimd}, {Serenelli}, {Siebert}, {Siudek}, {Smiljanic}, {Smith}, {Sobeck},
  {Starkenburg}, {Stello}, {Szab{\'o}}, {Szabo}, {Taylor}, {Thanjavur},
  {Thomas}, {Tollerud}, {Toonen}, {Tremblay}, {Tresse}, {Tsantaki},
  {Valentini}, {Van Eck}, {Variu}, {Venn}, {Villaver}, {Walker}, {Wang},
  {Wang}, {Wilson}, {Wright}, {Xu}, {Yildiz}, {Zhang}, {Zwintz}, {Anguiano},
  {Bedell}, {Chaplin}, {Collet}, {Cuillandre}, {Duc}, {Flagey}, {Hermes},
  {Hill}, {Kamath}, {Laychak}, {Ma{\l}ek}, {Marley}, {Sheinis}, {Simons},
  {Sousa}, {Szeto}, {Ting}, {Vegetti}, {Wells}, {Babas}, {Bauman}, {Bosselli},
  {C{\^o}t{\'e}}, {Colless}, {Comparat}, {Courtois}, {Crampton}, {Croom},
  {Davies}, {de Grijs}, {Denny}, {Devost}, {di Matteo}, {Driver},
  {Fernandez-Lorenzo}, {Guhathakurta}, {Han}, {Higgs}, {Hill}, {Ho}, {Hopkins},
  {Hudson}, {Ibata}, {Isani}, {Jarvis}, {Johnson}, {Jullo}, {Kaiser}, {Kneib},
  {Koda}, {Koshy}, {Mignot}, {Murowinski}, {Newman}, {Nusser}, {Pancoast},
  {Peng}, {Peroux}, {Pichon}, {Poggianti}, {Richard}, {Salmon}, {Seibert},
  {Shastri}, {Smith}, {Sutaria}, {Tao}, {Taylor}, {Tully}, {van Waerbeke},
  {Vermeulen}, {Walker}, {Willis}, {Willot} and {Withington}}]{mse:2019}
{The MSE Science Team}et~al., \bibinfo{year}{2019}.
\newblock \bibinfo{title}{{The Detailed Science Case for the Maunakea
  Spectroscopic Explorer, 2019 edition}}.
\newblock \bibinfo{journal}{arXiv e-prints} ,
  \bibinfo{pages}{arXiv:1904.04907}\DOIprefix\doi{10.48550/arXiv.1904.04907},
  \href{http://arxiv.org/abs/1904.04907}{{\tt arXiv:1904.04907}}.
%Type = Article
\bibitem[{{Thomas} and {Battaglia}(2022)}]{thomas:2022}
\bibinfo{author}{{Thomas}, G.F.}, \bibinfo{author}{{Battaglia}, G.},
  \bibinfo{year}{2022}.
\newblock \bibinfo{title}{{The Cetus-Palca stream: A disrupted small dwarf
  galaxy. A prequel to the science possible with WEAVE with precise
  spectro-photometric distances}}.
\newblock \bibinfo{journal}{\aap} \bibinfo{volume}{660}, \bibinfo{pages}{A29}.
\newblock \DOIprefix\doi{10.1051/0004-6361/202142347},
  \href{http://arxiv.org/abs/2112.03973}{{\tt arXiv:2112.03973}}.
%Type = Article
\bibitem[{{Thomas} et~al.(2018){Thomas}, {Famaey}, {Ibata}, {Renaud}, {Martin}
  and {Kroupa}}]{thomas:2018}
\bibinfo{author}{{Thomas}, G.F.}, \bibinfo{author}{{Famaey}, B.},
  \bibinfo{author}{{Ibata}, R.}, \bibinfo{author}{{Renaud}, F.},
  \bibinfo{author}{{Martin}, N.F.}, \bibinfo{author}{{Kroupa}, P.},
  \bibinfo{year}{2018}.
\newblock \bibinfo{title}{{Stellar streams as gravitational experiments. II.
  Asymmetric tails of globular cluster streams}}.
\newblock \bibinfo{journal}{\aap} \bibinfo{volume}{609}, \bibinfo{pages}{A44}.
\newblock \DOIprefix\doi{10.1051/0004-6361/201731609},
  \href{http://arxiv.org/abs/1709.01934}{{\tt arXiv:1709.01934}}.
%Type = Article
\bibitem[{{Thomas} et~al.(2023){Thomas}, {Famaey}, {Monari}, {Laporte},
  {Ibata}, {de Laverny}, {Hill} and {Boily}}]{thomas:2023}
\bibinfo{author}{{Thomas}, G.F.}, \bibinfo{author}{{Famaey}, B.},
  \bibinfo{author}{{Monari}, G.}, \bibinfo{author}{{Laporte}, C.F.P.},
  \bibinfo{author}{{Ibata}, R.}, \bibinfo{author}{{de Laverny}, P.},
  \bibinfo{author}{{Hill}, V.}, \bibinfo{author}{{Boily}, C.},
  \bibinfo{year}{2023}.
\newblock \bibinfo{title}{{Impact of the Galactic bar on tidal streams within
  the Galactic disc. The case of the tidal stream of the Hyades}}.
\newblock \bibinfo{journal}{\aap} \bibinfo{volume}{678}, \bibinfo{pages}{A180}.
\newblock \DOIprefix\doi{10.1051/0004-6361/202346650},
  \href{http://arxiv.org/abs/2309.05733}{{\tt arXiv:2309.05733}}.
%Type = Article
\bibitem[{{Thomas} et~al.(2016){Thomas}, {Ibata}, {Famaey}, {Martin} and
  {Lewis}}]{thomas:2016}
\bibinfo{author}{{Thomas}, G.F.}, \bibinfo{author}{{Ibata}, R.},
  \bibinfo{author}{{Famaey}, B.}, \bibinfo{author}{{Martin}, N.F.},
  \bibinfo{author}{{Lewis}, G.F.}, \bibinfo{year}{2016}.
\newblock \bibinfo{title}{{Exploring the reality of density substructures in
  the Palomar 5 stellar stream}}.
\newblock \bibinfo{journal}{\mnras} \bibinfo{volume}{460},
  \bibinfo{pages}{2711--2719}.
\newblock \DOIprefix\doi{10.1093/mnras/stw1189},
  \href{http://arxiv.org/abs/1605.05520}{{\tt arXiv:1605.05520}}.
%Type = Article
\bibitem[{{Toomre} and {Toomre}(1972)}]{toomre:1972}
\bibinfo{author}{{Toomre}, A.}, \bibinfo{author}{{Toomre}, J.},
  \bibinfo{year}{1972}.
\newblock \bibinfo{title}{{Galactic Bridges and Tails}}.
\newblock \bibinfo{journal}{\apj} \bibinfo{volume}{178},
  \bibinfo{pages}{623--666}.
\newblock \DOIprefix\doi{10.1086/151823}.
%Type = Article
\bibitem[{{Tremaine}(1999)}]{tremaine:1999}
\bibinfo{author}{{Tremaine}, S.}, \bibinfo{year}{1999}.
\newblock \bibinfo{title}{{The geometry of phase mixing}}.
\newblock \bibinfo{journal}{\mnras} \bibinfo{volume}{307},
  \bibinfo{pages}{877--883}.
\newblock \DOIprefix\doi{10.1046/j.1365-8711.1999.02690.x},
  \href{http://arxiv.org/abs/astro-ph/9812146}{{\tt arXiv:astro-ph/9812146}}.
%Type = Article
\bibitem[{{Usman} et~al.(2024){Usman}, {Ji}, {Li}, {Pace}, {Cullinane}, {Da
  Costa}, {Koposov}, {Lewis}, {Zucker}, {Belokurov}, {Bland-Hawthorn},
  {Ferguson}, {Hansen}, {Limberg}, {Martell}, {McKenzie} and {S5
  Collaboration}}]{usman:2024}
{Usman}, S.~A. et~al., \bibinfo{year}{2024}.
\newblock \bibinfo{title}{{Multiple Populations and a CH Star Found in the 300S
  Globular Cluster Stellar Stream}}.
\newblock \bibinfo{journal}{\mnras} \DOIprefix\doi{10.1093/mnras/stae185},
  \href{http://arxiv.org/abs/2401.02476}{{\tt arXiv:2401.02476}}.
%Type = Article
\bibitem[{{Valluri} et~al.(2010){Valluri}, {Debattista}, {Quinn} and
  {Moore}}]{valluri:2010}
\bibinfo{author}{{Valluri}, M.}, \bibinfo{author}{{Debattista}, V.P.},
  \bibinfo{author}{{Quinn}, T.}, \bibinfo{author}{{Moore}, B.},
  \bibinfo{year}{2010}.
\newblock \bibinfo{title}{{The orbital evolution induced by baryonic
  condensation in triaxial haloes}}.
\newblock \bibinfo{journal}{\mnras} \bibinfo{volume}{403},
  \bibinfo{pages}{525--544}.
\newblock \DOIprefix\doi{10.1111/j.1365-2966.2009.16192.x},
  \href{http://arxiv.org/abs/0906.4784}{{\tt arXiv:0906.4784}}.
%Type = Article
\bibitem[{{Valluri} et~al.(2012){Valluri}, {Debattista}, {Quinn},
  {Ro{\v{s}}kar} and {Wadsley}}]{valluri:2012}
\bibinfo{author}{{Valluri}, M.}, \bibinfo{author}{{Debattista}, V.P.},
  \bibinfo{author}{{Quinn}, T.R.}, \bibinfo{author}{{Ro{\v{s}}kar}, R.},
  \bibinfo{author}{{Wadsley}, J.}, \bibinfo{year}{2012}.
\newblock \bibinfo{title}{{Probing the shape and history of the Milky Way halo
  with orbital spectral analysis}}.
\newblock \bibinfo{journal}{\mnras} \bibinfo{volume}{419},
  \bibinfo{pages}{1951--1969}.
\newblock \DOIprefix\doi{10.1111/j.1365-2966.2011.19853.x},
  \href{http://arxiv.org/abs/1109.3193}{{\tt arXiv:1109.3193}}.
%Type = Book
\bibitem[{{Valtonen} and {Karttunen}(2006)}]{valtonen:2006}
\bibinfo{author}{{Valtonen}, M.}, \bibinfo{author}{{Karttunen}, H.},
  \bibinfo{year}{2006}.
\newblock \bibinfo{title}{{The Three-Body Problem}}.
%Type = Article
\bibitem[{{van Dokkum} et~al.(2019){van Dokkum}, {Gilhuly}, {Bonaca},
  {Merritt}, {Danieli}, {Lokhorst}, {Abraham}, {Conroy} and
  {Greco}}]{vandokkum:2019}
{van Dokkum}, P. et~al., \bibinfo{year}{2019}.
\newblock \bibinfo{title}{{Dragonfly Imaging of the Galaxy NGC 5907: A
  Different View of the Iconic Stellar Stream}}.
\newblock \bibinfo{journal}{\apjl} \bibinfo{volume}{883}, \bibinfo{pages}{L32}.
\newblock \DOIprefix\doi{10.3847/2041-8213/ab40c9},
  \href{http://arxiv.org/abs/1906.11260}{{\tt arXiv:1906.11260}}.
%Type = Article
\bibitem[{{Varghese} et~al.(2011){Varghese}, {Ibata} and
  {Lewis}}]{varghese:2011}
\bibinfo{author}{{Varghese}, A.}, \bibinfo{author}{{Ibata}, R.},
  \bibinfo{author}{{Lewis}, G.F.}, \bibinfo{year}{2011}.
\newblock \bibinfo{title}{{Stellar streams as probes of dark halo mass and
  morphology: a Bayesian reconstruction}}.
\newblock \bibinfo{journal}{\mnras} \bibinfo{volume}{417},
  \bibinfo{pages}{198--215}.
\newblock \DOIprefix\doi{10.1111/j.1365-2966.2011.19097.x},
  \href{http://arxiv.org/abs/1106.1765}{{\tt arXiv:1106.1765}}.
%Type = Article
\bibitem[{{Vasiliev}(2019a)}]{vasiliev:2019}
\bibinfo{author}{{Vasiliev}, E.}, \bibinfo{year}{2019}a.
\newblock \bibinfo{title}{{AGAMA: action-based galaxy modelling architecture}}.
\newblock \bibinfo{journal}{\mnras} \bibinfo{volume}{482},
  \bibinfo{pages}{1525--1544}.
\newblock \DOIprefix\doi{10.1093/mnras/sty2672},
  \href{http://arxiv.org/abs/1802.08239}{{\tt arXiv:1802.08239}}.
%Type = Article
\bibitem[{{Vasiliev}(2019b)}]{vasiliev:2019b}
\bibinfo{author}{{Vasiliev}, E.}, \bibinfo{year}{2019}b.
\newblock \bibinfo{title}{{Proper motions and dynamics of the Milky Way
  globular cluster system from Gaia DR2}}.
\newblock \bibinfo{journal}{\mnras} \bibinfo{volume}{484},
  \bibinfo{pages}{2832--2850}.
\newblock \DOIprefix\doi{10.1093/mnras/stz171},
  \href{http://arxiv.org/abs/1807.09775}{{\tt arXiv:1807.09775}}.
%Type = Article
\bibitem[{{Vasiliev}(2023)}]{vasiliev:2023}
\bibinfo{author}{{Vasiliev}, E.}, \bibinfo{year}{2023}.
\newblock \bibinfo{title}{{The Effect of the LMC on the Milky Way System}}.
\newblock \bibinfo{journal}{Galaxies} \bibinfo{volume}{11},
  \bibinfo{pages}{59}.
\newblock \DOIprefix\doi{10.3390/galaxies11020059},
  \href{http://arxiv.org/abs/2304.09136}{{\tt arXiv:2304.09136}}.
%Type = Article
\bibitem[{{Vasiliev}(2024)}]{vasiliev:2024}
\bibinfo{author}{{Vasiliev}, E.}, \bibinfo{year}{2024}.
\newblock \bibinfo{title}{{Dear Magellanic Clouds, welcome back!}}
\newblock \bibinfo{journal}{\mnras} \bibinfo{volume}{527},
  \bibinfo{pages}{437--456}.
\newblock \DOIprefix\doi{10.1093/mnras/stad2612},
  \href{http://arxiv.org/abs/2306.04837}{{\tt arXiv:2306.04837}}.
%Type = Article
\bibitem[{{Vasiliev} and {Belokurov}(2020)}]{vasiliev:2020}
\bibinfo{author}{{Vasiliev}, E.}, \bibinfo{author}{{Belokurov}, V.},
  \bibinfo{year}{2020}.
\newblock \bibinfo{title}{{The last breath of the Sagittarius dSph}}.
\newblock \bibinfo{journal}{\mnras} \bibinfo{volume}{497},
  \bibinfo{pages}{4162--4182}.
\newblock \DOIprefix\doi{10.1093/mnras/staa2114},
  \href{http://arxiv.org/abs/2006.02929}{{\tt arXiv:2006.02929}}.
%Type = Article
\bibitem[{{Vasiliev} et~al.(2021){Vasiliev}, {Belokurov} and
  {Erkal}}]{vasiliev:2021}
\bibinfo{author}{{Vasiliev}, E.}, \bibinfo{author}{{Belokurov}, V.},
  \bibinfo{author}{{Erkal}, D.}, \bibinfo{year}{2021}.
\newblock \bibinfo{title}{{Tango for three: Sagittarius, LMC, and the Milky
  Way}}.
\newblock \bibinfo{journal}{\mnras} \bibinfo{volume}{501},
  \bibinfo{pages}{2279--2304}.
\newblock \DOIprefix\doi{10.1093/mnras/staa3673},
  \href{http://arxiv.org/abs/2009.10726}{{\tt arXiv:2009.10726}}.
%Type = Article
\bibitem[{{Vesperini} and {Heggie}(1997)}]{vesperini:1997}
\bibinfo{author}{{Vesperini}, E.}, \bibinfo{author}{{Heggie}, D.C.},
  \bibinfo{year}{1997}.
\newblock \bibinfo{title}{{On the effects of dynamical evolution on the initial
  mass function of globular clusters}}.
\newblock \bibinfo{journal}{\mnras} \bibinfo{volume}{289},
  \bibinfo{pages}{898--920}.
\newblock \DOIprefix\doi{10.1093/mnras/289.4.898},
  \href{http://arxiv.org/abs/astro-ph/9705073}{{\tt arXiv:astro-ph/9705073}}.
%Type = Article
\bibitem[{{Villaescusa-Navarro} et~al.(2021){Villaescusa-Navarro},
  {Angl{\'e}s-Alc{\'a}zar}, {Genel}, {Spergel}, {Somerville}, {Dave},
  {Pillepich}, {Hernquist}, {Nelson}, {Torrey}, {Narayanan}, {Li}, {Philcox},
  {La Torre}, {Maria Delgado}, {Ho}, {Hassan}, {Burkhart}, {Wadekar},
  {Battaglia}, {Contardo} and {Bryan}}]{villaescusa-navarro:2021}
{Villaescusa-Navarro}, F. et~al., \bibinfo{year}{2021}.
\newblock \bibinfo{title}{{The CAMELS Project: Cosmology and Astrophysics with
  Machine-learning Simulations}}.
\newblock \bibinfo{journal}{\apj} \bibinfo{volume}{915}, \bibinfo{pages}{71}.
\newblock \DOIprefix\doi{10.3847/1538-4357/abf7ba},
  \href{http://arxiv.org/abs/2010.00619}{{\tt arXiv:2010.00619}}.
%Type = Article
\bibitem[{Virtanen et~al.(2020)Virtanen, Gommers, Oliphant, Haberland, Reddy,
  Cournapeau, Burovski, Peterson, Weckesser, Bright, {van der Walt}, Brett,
  Wilson, Millman, Mayorov, Nelson, Jones, Kern, Larson, Carey, Polat, Feng,
  Moore, {VanderPlas}, Laxalde, Perktold, Cimrman, Henriksen, Quintero, Harris,
  Archibald, Ribeiro, Pedregosa, {van Mulbregt} and {SciPy 1.0
  Contributors}}]{scipy}
Virtanen, P. et~al., \bibinfo{year}{2020}.
\newblock \bibinfo{title}{{{SciPy} 1.0: Fundamental Algorithms for Scientific
  Computing in Python}}.
\newblock \bibinfo{journal}{Nature Methods} \bibinfo{volume}{17},
  \bibinfo{pages}{261--272}.
\newblock \DOIprefix\doi{10.1038/s41592-019-0686-2}.
%Type = Article
\bibitem[{{Vorontsov-Velyaminov}(1959)}]{vorontsov-velyaminov:1959}
\bibinfo{author}{{Vorontsov-Velyaminov}, B.A.}, \bibinfo{year}{1959}.
\newblock \bibinfo{title}{{Atlas i Katalog Vzaimodejstvu{\^u}{\v{s}}ih
  Galakatik I.Atlas i Katalog Vzaimodejstvu{\^u}{\v{s}}ih Galakatik I.Atlas and
  catalog of interacting galaxies. 1959, Sternberg Institute, Moscow State
  University.}}
\newblock \bibinfo{journal}{Atlas and Catalog of Interacting Galaxies (1959} ,
  \bibinfo{pages}{0}.
%Type = Article
\bibitem[{{Wan} et~al.(2020){Wan}, {Lewis}, {Li}, {Simpson}, {Martell},
  {Zucker}, {Mould}, {Erkal}, {Pace}, {Mackey}, {Ji}, {Koposov}, {Kuehn},
  {Shipp}, {Balbinot}, {Bland-Hawthorn}, {Casey}, {Da Costa}, {Kafle}, {Sharma}
  and {De Silva}}]{wan:2020}
{Wan}, Z. et~al., \bibinfo{year}{2020}.
\newblock \bibinfo{title}{{The tidal remnant of an unusually metal-poor
  globular cluster}}.
\newblock \bibinfo{journal}{\nat} \bibinfo{volume}{583},
  \bibinfo{pages}{768--770}.
\newblock \DOIprefix\doi{10.1038/s41586-020-2483-6},
  \href{http://arxiv.org/abs/2007.14577}{{\tt arXiv:2007.14577}}.
%Type = Article
\bibitem[{{Wang} et~al.(2020a){Wang}, {Bose}, {Frenk}, {Gao}, {Jenkins},
  {Springel} and {White}}]{wangj:2020}
\bibinfo{author}{{Wang}, J.}, \bibinfo{author}{{Bose}, S.},
  \bibinfo{author}{{Frenk}, C.S.}, \bibinfo{author}{{Gao}, L.},
  \bibinfo{author}{{Jenkins}, A.}, \bibinfo{author}{{Springel}, V.},
  \bibinfo{author}{{White}, S.D.M.}, \bibinfo{year}{2020}a.
\newblock \bibinfo{title}{{Universal structure of dark matter haloes over a
  mass range of 20 orders of magnitude}}.
\newblock \bibinfo{journal}{\nat} \bibinfo{volume}{585},
  \bibinfo{pages}{39--42}.
\newblock \DOIprefix\doi{10.1038/s41586-020-2642-9},
  \href{http://arxiv.org/abs/1911.09720}{{\tt arXiv:1911.09720}}.
%Type = Article
\bibitem[{{Wang} et~al.(2020b){Wang}, {Iwasawa}, {Nitadori} and
  {Makino}}]{wang:2020}
\bibinfo{author}{{Wang}, L.}, \bibinfo{author}{{Iwasawa}, M.},
  \bibinfo{author}{{Nitadori}, K.}, \bibinfo{author}{{Makino}, J.},
  \bibinfo{year}{2020}b.
\newblock \bibinfo{title}{{PETAR: a high-performance N-body code for modelling
  massive collisional stellar systems}}.
\newblock \bibinfo{journal}{\mnras} \bibinfo{volume}{497},
  \bibinfo{pages}{536--555}.
\newblock \DOIprefix\doi{10.1093/mnras/staa1915},
  \href{http://arxiv.org/abs/2006.16560}{{\tt arXiv:2006.16560}}.
%Type = Article
\bibitem[{{Wang} et~al.(2023){Wang}, {Necib}, {Ji}, {Ou}, {Lisanti}, {de los
  Reyes}, {Strom} and {Truong}}]{wang:2023}
\bibinfo{author}{{Wang}, S.}, \bibinfo{author}{{Necib}, L.},
  \bibinfo{author}{{Ji}, A.P.}, \bibinfo{author}{{Ou}, X.},
  \bibinfo{author}{{Lisanti}, M.}, \bibinfo{author}{{de los Reyes}, M.A.C.},
  \bibinfo{author}{{Strom}, A.L.}, \bibinfo{author}{{Truong}, M.},
  \bibinfo{year}{2023}.
\newblock \bibinfo{title}{{High-resolution Chemical Abundances of the Nyx
  Stream}}.
\newblock \bibinfo{journal}{\apj} \bibinfo{volume}{955}, \bibinfo{pages}{129}.
\newblock \DOIprefix\doi{10.3847/1538-4357/acec4d},
  \href{http://arxiv.org/abs/2210.15013}{{\tt arXiv:2210.15013}}.
%Type = Article
\bibitem[{{Wegg} et~al.(2015){Wegg}, {Gerhard} and {Portail}}]{wegg:2015}
\bibinfo{author}{{Wegg}, C.}, \bibinfo{author}{{Gerhard}, O.},
  \bibinfo{author}{{Portail}, M.}, \bibinfo{year}{2015}.
\newblock \bibinfo{title}{{The structure of the Milky Way's bar outside the
  bulge}}.
\newblock \bibinfo{journal}{\mnras} \bibinfo{volume}{450},
  \bibinfo{pages}{4050--4069}.
\newblock \DOIprefix\doi{10.1093/mnras/stv745},
  \href{http://arxiv.org/abs/1504.01401}{{\tt arXiv:1504.01401}}.
%Type = Article
\bibitem[{{Weiss} et~al.(2018){Weiss}, {Newberg} and {Desell}}]{weiss:2018}
\bibinfo{author}{{Weiss}, J.}, \bibinfo{author}{{Newberg}, H.J.},
  \bibinfo{author}{{Desell}, T.}, \bibinfo{year}{2018}.
\newblock \bibinfo{title}{{A Tangle of Stellar Streams in the North Galactic
  Cap}}.
\newblock \bibinfo{journal}{\apjl} \bibinfo{volume}{867}, \bibinfo{pages}{L1}.
\newblock \DOIprefix\doi{10.3847/2041-8213/aae5fc},
  \href{http://arxiv.org/abs/1807.03754}{{\tt arXiv:1807.03754}}.
%Type = Article
\bibitem[{{Wetzel} et~al.(2016){Wetzel}, {Hopkins}, {Kim},
  {Faucher-Gigu{\`e}re}, {Kere{\v{s}}} and {Quataert}}]{wetzel:2016}
\bibinfo{author}{{Wetzel}, A.R.}, \bibinfo{author}{{Hopkins}, P.F.},
  \bibinfo{author}{{Kim}, J.h.}, \bibinfo{author}{{Faucher-Gigu{\`e}re}, C.A.},
  \bibinfo{author}{{Kere{\v{s}}}, D.}, \bibinfo{author}{{Quataert}, E.},
  \bibinfo{year}{2016}.
\newblock \bibinfo{title}{{Reconciling Dwarf Galaxies with
  {\ensuremath{\Lambda}}CDM Cosmology: Simulating a Realistic Population of
  Satellites around a Milky Way-mass Galaxy}}.
\newblock \bibinfo{journal}{\apjl} \bibinfo{volume}{827}, \bibinfo{pages}{L23}.
\newblock \DOIprefix\doi{10.3847/2041-8205/827/2/L23},
  \href{http://arxiv.org/abs/1602.05957}{{\tt arXiv:1602.05957}}.
%Type = Article
\bibitem[{{White} and {Rees}(1978)}]{white:1978}
\bibinfo{author}{{White}, S.D.M.}, \bibinfo{author}{{Rees}, M.J.},
  \bibinfo{year}{1978}.
\newblock \bibinfo{title}{{Core condensation in heavy halos: a two-stage theory
  for galaxy formation and clustering.}}
\newblock \bibinfo{journal}{\mnras} \bibinfo{volume}{183},
  \bibinfo{pages}{341--358}.
\newblock \DOIprefix\doi{10.1093/mnras/183.3.341}.
%Type = Article
\bibitem[{{Williams} et~al.(2011){Williams}, {Steinmetz}, {Sharma},
  {Bland-Hawthorn}, {de Jong}, {Seabroke}, {Helmi}, {Freeman}, {Binney},
  {Minchev}, {Bienaym{\'e}}, {Campbell}, {Fulbright}, {Gibson}, {Gilmore},
  {Grebel}, {Munari}, {Navarro}, {Parker}, {Reid}, {Siebert}, {Siviero},
  {Watson}, {Wyse} and {Zwitter}}]{williams:2011}
{Williams}, M.~E.~K. et~al., \bibinfo{year}{2011}.
\newblock \bibinfo{title}{{The Dawning of the Stream of Aquarius in RAVE}}.
\newblock \bibinfo{journal}{\apj} \bibinfo{volume}{728}, \bibinfo{pages}{102}.
\newblock \DOIprefix\doi{10.1088/0004-637X/728/2/102},
  \href{http://arxiv.org/abs/1012.2127}{{\tt arXiv:1012.2127}}.
%Type = Article
\bibitem[{{Willman} and {Strader}(2012)}]{willman:2012}
\bibinfo{author}{{Willman}, B.}, \bibinfo{author}{{Strader}, J.},
  \bibinfo{year}{2012}.
\newblock \bibinfo{title}{{``Galaxy,'' Defined}}.
\newblock \bibinfo{journal}{\aj} \bibinfo{volume}{144}, \bibinfo{pages}{76}.
\newblock \DOIprefix\doi{10.1088/0004-6256/144/3/76},
  \href{http://arxiv.org/abs/1203.2608}{{\tt arXiv:1203.2608}}.
%Type = Article
\bibitem[{{Woudenberg} et~al.(2023){Woudenberg}, {Koop}, {Balbinot} and
  {Helmi}}]{woudenberg:2023}
\bibinfo{author}{{Woudenberg}, H.C.}, \bibinfo{author}{{Koop}, O.},
  \bibinfo{author}{{Balbinot}, E.}, \bibinfo{author}{{Helmi}, A.},
  \bibinfo{year}{2023}.
\newblock \bibinfo{title}{{Characterization and dynamics of the peculiar stream
  Jhelum. A tentative role for the Sagittarius dwarf galaxy}}.
\newblock \bibinfo{journal}{\aap} \bibinfo{volume}{669}, \bibinfo{pages}{A102}.
\newblock \DOIprefix\doi{10.1051/0004-6361/202243266},
  \href{http://arxiv.org/abs/2202.02132}{{\tt arXiv:2202.02132}}.
%Type = Article
\bibitem[{{Wright} et~al.(2023){Wright}, {Tumlinson}, {Peeples}, {O'Shea},
  {Lochhaas}, {Corlies}, {Smith}, {Binh}, {Augustin} and
  {Simons}}]{wright:2023}
{Wright}, A.~C. et~al., \bibinfo{year}{2023}.
\newblock \bibinfo{title}{{Figuring Out Gas \& Galaxies In Enzo (FOGGIE) VII:
  The (Dis)Assembly of Stellar Halos}}.
\newblock \bibinfo{journal}{arXiv e-prints} ,
  \bibinfo{pages}{arXiv:2309.10039}\DOIprefix\doi{10.48550/arXiv.2309.10039},
  \href{http://arxiv.org/abs/2309.10039}{{\tt arXiv:2309.10039}}.
%Type = Article
\bibitem[{{Xu} et~al.(2023){Xu}, {Kang} and {Libeskind}}]{xu:2023}
\bibinfo{author}{{Xu}, Y.}, \bibinfo{author}{{Kang}, X.},
  \bibinfo{author}{{Libeskind}, N.I.}, \bibinfo{year}{2023}.
\newblock \bibinfo{title}{{A Rotating Satellite Plane around Milky Way-like
  Galaxy from the TNG50 Simulation}}.
\newblock \bibinfo{journal}{\apj} \bibinfo{volume}{954}, \bibinfo{pages}{128}.
\newblock \DOIprefix\doi{10.3847/1538-4357/ace898},
  \href{http://arxiv.org/abs/2303.00441}{{\tt arXiv:2303.00441}}.
%Type = Article
\bibitem[{{Yang} et~al.(2022a){Yang}, {Zhao}, {Ishigaki}, {Chiba}, {Yang},
  {Xue}, {Ye} and {Zhao}}]{yang:2022b}
\bibinfo{author}{{Yang}, Y.}, \bibinfo{author}{{Zhao}, J.K.},
  \bibinfo{author}{{Ishigaki}, M.N.}, \bibinfo{author}{{Chiba}, M.},
  \bibinfo{author}{{Yang}, C.Q.}, \bibinfo{author}{{Xue}, X.X.},
  \bibinfo{author}{{Ye}, X.H.}, \bibinfo{author}{{Zhao}, G.},
  \bibinfo{year}{2022}a.
\newblock \bibinfo{title}{{Existence of tidal tails for the globular cluster
  NGC 5824}}.
\newblock \bibinfo{journal}{\aap} \bibinfo{volume}{667}, \bibinfo{pages}{A37}.
\newblock \DOIprefix\doi{10.1051/0004-6361/202243976},
  \href{http://arxiv.org/abs/2208.05197}{{\tt arXiv:2208.05197}}.
%Type = Article
\bibitem[{{Yang} et~al.(2022b){Yang}, {Zhao}, {Ishigaki}, {Zhou}, {Yang},
  {Xue}, {Ye} and {Zhao}}]{yang:2022}
\bibinfo{author}{{Yang}, Y.}, \bibinfo{author}{{Zhao}, J.K.},
  \bibinfo{author}{{Ishigaki}, M.N.}, \bibinfo{author}{{Zhou}, J.Z.},
  \bibinfo{author}{{Yang}, C.Q.}, \bibinfo{author}{{Xue}, X.X.},
  \bibinfo{author}{{Ye}, X.H.}, \bibinfo{author}{{Zhao}, G.},
  \bibinfo{year}{2022}b.
\newblock \bibinfo{title}{{Revisit NGC 5466 tidal stream with Gaia, SDSS/SEGUE,
  and LAMOST}}.
\newblock \bibinfo{journal}{\mnras} \bibinfo{volume}{513},
  \bibinfo{pages}{853--863}.
\newblock \DOIprefix\doi{10.1093/mnras/stac860},
  \href{http://arxiv.org/abs/2203.13414}{{\tt arXiv:2203.13414}}.
%Type = Article
\bibitem[{Yanny et~al.(2003)Yanny, Newberg, Grebel, Kent, Odenkirchen, Rockosi,
  Schlegel, Subbarao, Brinkmann, Fukugita, Ivezic, Lamb, Schneider and
  York}]{Yanny:2003}
Yanny, B. et~al., \bibinfo{year}{2003}.
\newblock \bibinfo{title}{A {{Low-Latitude Halo Stream}} around the {{Milky
  Way}}}.
\newblock \bibinfo{journal}{The Astrophysical Journal} \bibinfo{volume}{588},
  \bibinfo{pages}{824--841}.
\newblock \DOIprefix\doi{10.1086/374220}.
%Type = Article
\bibitem[{{Yanny} et~al.(2009){Yanny}, {Rockosi}, {Newberg}, {Knapp},
  {Adelman-McCarthy}, {Alcorn}, {Allam}, {Allende Prieto}, {An}, {Anderson},
  {Anderson}, {Bailer-Jones}, {Bastian}, {Beers}, {Bell}, {Belokurov},
  {Bizyaev}, {Blythe}, {Bochanski}, {Boroski}, {Brinchmann}, {Brinkmann},
  {Brewington}, {Carey}, {Cudworth}, {Evans}, {Evans}, {Gates}, {G{\"a}nsicke},
  {Gillespie}, {Gilmore}, {Nebot Gomez-Moran}, {Grebel}, {Greenwell}, {Gunn},
  {Jordan}, {Jordan}, {Harding}, {Harris}, {Hendry}, {Holder}, {Ivans},
  {Ivezi{\v{c}}}, {Jester}, {Johnson}, {Kent}, {Kleinman}, {Kniazev},
  {Krzesinski}, {Kron}, {Kuropatkin}, {Lebedeva}, {Lee}, {French Leger},
  {L{\'e}pine}, {Levine}, {Lin}, {Long}, {Loomis}, {Lupton}, {Malanushenko},
  {Malanushenko}, {Margon}, {Martinez-Delgado}, {McGehee}, {Monet}, {Morrison},
  {Munn}, {Neilsen}, {Nitta}, {Norris}, {Oravetz}, {Owen}, {Padmanabhan},
  {Pan}, {Peterson}, {Pier}, {Platson}, {Re Fiorentin}, {Richards}, {Rix},
  {Schlegel}, {Schneider}, {Schreiber}, {Schwope}, {Sibley}, {Simmons},
  {Snedden}, {Allyn Smith}, {Stark}, {Stauffer}, {Steinmetz}, {Stoughton},
  {SubbaRao}, {Szalay}, {Szkody}, {Thakar}, {Sivarani}, {Tucker}, {Uomoto},
  {Vanden Berk}, {Vidrih}, {Wadadekar}, {Watters}, {Wilhelm}, {Wyse}, {Yarger}
  and {Zucker}}]{yanny:2009}
{Yanny}, B. et~al., \bibinfo{year}{2009}.
\newblock \bibinfo{title}{{SEGUE: A Spectroscopic Survey of 240,000 Stars with
  g = 14-20}}.
\newblock \bibinfo{journal}{\aj} \bibinfo{volume}{137},
  \bibinfo{pages}{4377--4399}.
\newblock \DOIprefix\doi{10.1088/0004-6256/137/5/4377},
  \href{http://arxiv.org/abs/0902.1781}{{\tt arXiv:0902.1781}}.
%Type = Article
\bibitem[{{Yaron} and {Gal-Yam}(2012)}]{yaron:2012}
\bibinfo{author}{{Yaron}, O.}, \bibinfo{author}{{Gal-Yam}, A.},
  \bibinfo{year}{2012}.
\newblock \bibinfo{title}{{WISeREP{\textemdash}An Interactive Supernova Data
  Repository}}.
\newblock \bibinfo{journal}{\pasp} \bibinfo{volume}{124}, \bibinfo{pages}{668}.
\newblock \DOIprefix\doi{10.1086/666656},
  \href{http://arxiv.org/abs/1204.1891}{{\tt arXiv:1204.1891}}.
%Type = Article
\bibitem[{{Yavetz} et~al.(2023){Yavetz}, {Johnston}, {Pearson}, {Price-Whelan}
  and {Hamilton}}]{yavetz:2023}
\bibinfo{author}{{Yavetz}, T.D.}, \bibinfo{author}{{Johnston}, K.V.},
  \bibinfo{author}{{Pearson}, S.}, \bibinfo{author}{{Price-Whelan}, A.M.},
  \bibinfo{author}{{Hamilton}, C.}, \bibinfo{year}{2023}.
\newblock \bibinfo{title}{{Stream Fanning and Bifurcations: Observable
  Signatures of Resonances in Stellar Stream Morphology}}.
\newblock \bibinfo{journal}{\apj} \bibinfo{volume}{954}, \bibinfo{pages}{215}.
\newblock \DOIprefix\doi{10.3847/1538-4357/ace7b9},
  \href{http://arxiv.org/abs/2212.11006}{{\tt arXiv:2212.11006}}.
%Type = Article
\bibitem[{{Yavetz} et~al.(2021){Yavetz}, {Johnston}, {Pearson}, {Price-Whelan}
  and {Weinberg}}]{yavetz:2021}
\bibinfo{author}{{Yavetz}, T.D.}, \bibinfo{author}{{Johnston}, K.V.},
  \bibinfo{author}{{Pearson}, S.}, \bibinfo{author}{{Price-Whelan}, A.M.},
  \bibinfo{author}{{Weinberg}, M.D.}, \bibinfo{year}{2021}.
\newblock \bibinfo{title}{{Separatrix divergence of stellar streams in galactic
  potentials}}.
\newblock \bibinfo{journal}{\mnras} \bibinfo{volume}{501},
  \bibinfo{pages}{1791--1802}.
\newblock \DOIprefix\doi{10.1093/mnras/staa3687},
  \href{http://arxiv.org/abs/2011.11919}{{\tt arXiv:2011.11919}}.
%Type = Article
\bibitem[{{Yoon} et~al.(2011){Yoon}, {Johnston} and {Hogg}}]{yoon:2011}
\bibinfo{author}{{Yoon}, J.H.}, \bibinfo{author}{{Johnston}, K.V.},
  \bibinfo{author}{{Hogg}, D.W.}, \bibinfo{year}{2011}.
\newblock \bibinfo{title}{{Clumpy Streams from Clumpy Halos: Detecting Missing
  Satellites with Cold Stellar Structures}}.
\newblock \bibinfo{journal}{\apj} \bibinfo{volume}{731}, \bibinfo{pages}{58}.
\newblock \DOIprefix\doi{10.1088/0004-637X/731/1/58},
  \href{http://arxiv.org/abs/1012.2884}{{\tt arXiv:1012.2884}}.
%Type = Article
\bibitem[{{York} et~al.(2000){York}, {Adelman}, {Anderson}, {Anderson},
  {Annis}, {Bahcall}, {Bakken}, {Barkhouser}, {Bastian}, {Berman}, {Boroski},
  {Bracker}, {Briegel}, {Briggs}, {Brinkmann}, {Brunner}, {Burles}, {Carey},
  {Carr}, {Castander}, {Chen}, {Colestock}, {Connolly}, {Crocker}, {Csabai},
  {Czarapata}, {Davis}, {Doi}, {Dombeck}, {Eisenstein}, {Ellman}, {Elms},
  {Evans}, {Fan}, {Federwitz}, {Fiscelli}, {Friedman}, {Frieman}, {Fukugita},
  {Gillespie}, {Gunn}, {Gurbani}, {de Haas}, {Haldeman}, {Harris}, {Hayes},
  {Heckman}, {Hennessy}, {Hindsley}, {Holm}, {Holmgren}, {Huang}, {Hull},
  {Husby}, {Ichikawa}, {Ichikawa}, {Ivezi{\'c}}, {Kent}, {Kim}, {Kinney},
  {Klaene}, {Kleinman}, {Kleinman}, {Knapp}, {Korienek}, {Kron}, {Kunszt},
  {Lamb}, {Lee}, {Leger}, {Limmongkol}, {Lindenmeyer}, {Long}, {Loomis},
  {Loveday}, {Lucinio}, {Lupton}, {MacKinnon}, {Mannery}, {Mantsch}, {Margon},
  {McGehee}, {McKay}, {Meiksin}, {Merelli}, {Monet}, {Munn}, {Narayanan},
  {Nash}, {Neilsen}, {Neswold}, {Newberg}, {Nichol}, {Nicinski}, {Nonino},
  {Okada}, {Okamura}, {Ostriker}, {Owen}, {Pauls}, {Peoples}, {Peterson},
  {Petravick}, {Pier}, {Pope}, {Pordes}, {Prosapio}, {Rechenmacher}, {Quinn},
  {Richards}, {Richmond}, {Rivetta}, {Rockosi}, {Ruthmansdorfer}, {Sandford},
  {Schlegel}, {Schneider}, {Sekiguchi}, {Sergey}, {Shimasaku}, {Siegmund},
  {Smee}, {Smith}, {Snedden}, {Stone}, {Stoughton}, {Strauss}, {Stubbs},
  {SubbaRao}, {Szalay}, {Szapudi}, {Szokoly}, {Thakar}, {Tremonti}, {Tucker},
  {Uomoto}, {Vanden Berk}, {Vogeley}, {Waddell}, {Wang}, {Watanabe},
  {Weinberg}, {Yanny}, {Yasuda} and {SDSS Collaboration}}]{york:2000}
{York}, D.~G. et~al., \bibinfo{year}{2000}.
\newblock \bibinfo{title}{{The Sloan Digital Sky Survey: Technical Summary}}.
\newblock \bibinfo{journal}{\aj} \bibinfo{volume}{120},
  \bibinfo{pages}{1579--1587}.
\newblock \DOIprefix\doi{10.1086/301513},
  \href{http://arxiv.org/abs/astro-ph/0006396}{{\tt arXiv:astro-ph/0006396}}.
%Type = Article
\bibitem[{{Yuan} et~al.(2022){Yuan}, {Malhan}, {Sestito}, {Ibata}, {Martin},
  {Chang}, {Li}, {Caffau}, {Bonifacio}, {Bellazzini}, {Huang}, {Voggel},
  {Longeard}, {Arentsen}, {Doliva-Dolinsky}, {Navarro}, {Famaey}, {Starkenburg}
  and {Aguado}}]{yuan:2022}
{Yuan}, Z. et~al., \bibinfo{year}{2022}.
\newblock \bibinfo{title}{{The Complexity of the Cetus Stream Unveiled from the
  Fusion of STREAMFINDER and StarGO}}.
\newblock \bibinfo{journal}{\apj} \bibinfo{volume}{930}, \bibinfo{pages}{103}.
\newblock \DOIprefix\doi{10.3847/1538-4357/ac616f},
  \href{http://arxiv.org/abs/2112.05775}{{\tt arXiv:2112.05775}}.
%Type = Article
\bibitem[{{Yuan} et~al.(2020){Yuan}, {Myeong}, {Beers}, {Evans}, {Lee},
  {Banerjee}, {Gudin}, {Hattori}, {Li}, {Matsuno}, {Placco}, {Smith}, {Whitten}
  and {Zhao}}]{yuan:2020}
{Yuan}, Z. et~al., \bibinfo{year}{2020}.
\newblock \bibinfo{title}{{Dynamical Relics of the Ancient Galactic Halo}}.
\newblock \bibinfo{journal}{\apj} \bibinfo{volume}{891}, \bibinfo{pages}{39}.
\newblock \DOIprefix\doi{10.3847/1538-4357/ab6ef7},
  \href{http://arxiv.org/abs/1910.07538}{{\tt arXiv:1910.07538}}.
%Type = Article
\bibitem[{{Yuan} et~al.(2019){Yuan}, {Smith}, {Xue}, {Li}, {Liu}, {Wang}, {Li}
  and {Chang}}]{yuan:2019}
\bibinfo{author}{{Yuan}, Z.}, \bibinfo{author}{{Smith}, M.C.},
  \bibinfo{author}{{Xue}, X.X.}, \bibinfo{author}{{Li}, J.},
  \bibinfo{author}{{Liu}, C.}, \bibinfo{author}{{Wang}, Y.},
  \bibinfo{author}{{Li}, L.}, \bibinfo{author}{{Chang}, J.},
  \bibinfo{year}{2019}.
\newblock \bibinfo{title}{{Revealing the Complicated Story of the Cetus Stream
  with StarGO}}.
\newblock \bibinfo{journal}{\apj} \bibinfo{volume}{881}, \bibinfo{pages}{164}.
\newblock \DOIprefix\doi{10.3847/1538-4357/ab2e09},
  \href{http://arxiv.org/abs/1902.05248}{{\tt arXiv:1902.05248}}.
%Type = Article
\bibitem[{{Zhao} et~al.(2020){Zhao}, {Ye}, {Wu}, {Yang}, {Oswalt}, {Xue},
  {Chen}, {Zhang} and {Zhao}}]{zhao:2020}
{Zhao}, J.~K. et~al., \bibinfo{year}{2020}.
\newblock \bibinfo{title}{{Two Portions of the Sagittarius Stream in the LAMOST
  Complete Spectroscopic Survey of Pointing Area at the Southern Galactic
  Cap}}.
\newblock \bibinfo{journal}{\apj} \bibinfo{volume}{904}, \bibinfo{pages}{61}.
\newblock \DOIprefix\doi{10.3847/1538-4357/abbc1f},
  \href{http://arxiv.org/abs/2009.11533}{{\tt arXiv:2009.11533}}.
%Type = Article
\bibitem[{{Zucker} et~al.(2021){Zucker}, {Simpson}, {Martell}, {Lewis},
  {Casey}, {Ting}, {Horner}, {Nordlander}, {Wyse}, {Zwitter}, {Bland-Hawthorn},
  {Buder}, {Asplund}, {De Silva}, {D'Orazi}, {Freeman}, {Hayden}, {Kos}, {Lin},
  {Lind}, {Schlesinger}, {Sharma} and {Stello}}]{zucker:2021}
{Zucker}, D.~B. et~al., \bibinfo{year}{2021}.
\newblock \bibinfo{title}{{The GALAH Survey: No Chemical Evidence of an
  Extragalactic Origin for the Nyx Stream}}.
\newblock \bibinfo{journal}{\apjl} \bibinfo{volume}{912}, \bibinfo{pages}{L30}.
\newblock \DOIprefix\doi{10.3847/2041-8213/abf7cd},
  \href{http://arxiv.org/abs/2104.08684}{{\tt arXiv:2104.08684}}.

\end{thebibliography}

\end{document}